\documentclass{iopart}
\usepackage{graphicx}
\usepackage{color}
\usepackage{amsbsy}
\expandafter\let\csname equation*\endcsname\relax
\expandafter\let\csname endequation*\endcsname\relax 
\usepackage{amsmath}

\newcommand{\mean}[1]{\left\langle #1 \right\rangle} 
\newcommand{\ps}{p^\text{s}}

\newcommand{\eps}{\varepsilon}

\def\xs{\tilde x(\tau)}

\def\la{\lambda(\tau)}
\def\l{\lambda}

\def\beq{\begin{equation}}
\def\ee{\end{equation}}
\def\bi{\begin {itemize}}
\def\ei{\end{itemize}}

\def\eps{\epsilon}

\def\na{n_{\alpha}}

\def\xs{\tilde x(\tau)}

\def\kk{(x,\tau)}

\def\la{\lambda(\tau)}
\def\l{\lambda}

\def\lt{\lambda_t}

\def\Sm{S_{\rm m}(\tau)}

\def\dSt{\dot S^{\rm tot}(\tau)}
\def\dst{\dot s^{\rm tot}(t)}
\def\dSm{\dot S^{\rm m}(\tau)}
\def\dsm{\dot s^{\rm m}(t)}
\def\m{^{\rm m}}
\def\t{^{\rm tot}}

\def\px{p(x,\tau)}

\def\wpm{w_{n_j^+n_j^-}}
\def\wmp{w_{n_j^-n_j^+}}
\def\njp{{n_j}^+}
\def\njm{{n_j}^-}

\def\tot{^{\rm tot}}

\def\beq{\begin{equation}}
\def\ee{\end{equation}}

\def\bi{\begin {itemize}}
\def\ei{\end{itemize}}

\def\eps{\epsilon}

\def\lo{\lambda_0}

\def\wmn{w_{mn}}
\def\wnm{w_{nm}}

\newcommand{\dbar}{\mathchar'26\mkern-12mu d}

\begin{document}

\title{Stochastic thermodynamics, fluctuation theorems, and molecular machines}
\author{Udo Seifert
}

\address{
{II.} Institut f\"ur Theoretische Physik, Universit\"at Stuttgart,
  70550 Stuttgart, Germany}
\begin{abstract}
Stochastic thermodynamics as reviewed here systematically provides a 
framework for extending the notions of classical thermodynamics like 
work, heat and entropy production to the level of individual trajectories 
of well-defined non-equilibrium ensembles. It applies whenever a 
non-equilibrium process is still coupled to one (or several) heat bath(s) 
of constant temperature. Paradigmatic systems are single colloidal 
particles in time-dependent laser traps, polymers in external 
flow, enzymes and molecular motors in single molecule assays, small 
biochemical networks and thermoelectric devices involving single electron 
transport. For such systems,  a first-law
like energy balance can be identified along fluctuating trajectories.
 For a basic Markovian dynamics implemented either on the continuum level 
with 
Langevin equations or on a discrete set of states as 
a master equation, thermodynamic consistency imposes a local-detailed
balance constraint on noise and rates, respectively. 
Various integral and detailed fluctuation theorems, which are derived here 
in a unifying approach from one master theorem, constrain the probability
distributions for work, heat and entropy production depending on the
nature of the system and the choice of non-equilibrium conditions. 
For non-equilibrium steady states, particularly strong results hold
 like a generalized fluctuation-dissipation theorem involving 
entropy production. Ramifications and applications of these concepts
include optimal driving between specified states in finite time, the 
role of measurement-based feedback processes and the relation between 
dissipation and irreversibility. Efficiency and, in particular, 
efficiency at maximum power, can be discussed systematically beyond
the linear response regime for two classes of 
molecular machines, isothermal ones like molecular motors, and heat 
engines like thermoelectric devices, using a common framework based on a
cycle decomposition of entropy production.
\end{abstract}


\maketitle

\setcounter{tocdepth}{3}
\tableofcontents
\vfill\eject

\def\F{{\cal F}}

\def\Fe{F^{\rm enz}}
\def\Fn{F_n^{\rm enz}}
\def\Fm{F_m^{\rm enz}}
\def\En{E_n^{\rm enz}}
\def\Em{E_m^{\rm enz}}
\def\Sm{S_m^{\rm enz}}
\def\Sn{S_n^{\rm enz}}

\def\Cn{{\cal C}_n}
\def\k{k_{\rm B}}
\def\kT{\k T}
\def\x{{\bf \xi}}
\def\peqn{p^{\rm eq}_n}
\def\peqm{p^{\rm eq}_m}
\def\peqx{p^{\rm eq}(\xi)}
\def\Ee{{\cal E}^{\rm eq}}
\def\Feq{{\cal F}^{\rm eq}}
\def\Se{{\cal S}^{\rm eq}}
\def\qmn{q_{mn}}
\def\demn{\Delta E_{mn}}
\def\Smn{\Delta S^{\rm}_{mn}}
\def\peqnt{p^{\rm eq}_{n(t)}}
\def\V{V(\x)}

\def\Fs{{F}^{\rm sol}}
\def\Es{{E}^{\rm sol}}
\def\Ss{{S}^{\rm sol}}

\def\mr{{n_\rho^-}}
\def\nr{{n_\rho^+}}
\def\ri{r^\rho_i}
\def\si{s^\rho_i}

\def\mie{\mu_i^{\rm eq}}
\def\deqr{\Delta \mu^{\rm eq}_\rho}

\def\xe{\xi^{\rm enz}}
\def\xs{\xi^{\rm sol}}
\def\xxs{\{\xs\}}
\def\vs{V^{\rm sol}(\xs)}
\def\vx{V(\xe,\xs)}
\def\vt{V^{\rm tot}(\xe,\xs)}
\def\ps{p(\xs)}
\def\pn{p(\xi|n)}
\def\sx{\sum_{{\xs}}}
\def\sn{\sum_{\xi \in {\cal C}_n}}

\def\ftn{F^{\rm }_n}
\def\etn{E^{\rm }_n}
\def\stn{S^{\rm }_n}

\def\ftm{F^{\rm }_m}

\def\fhn{\hat F_n^0}
\def\ehn{\hat E_n^0}
\def\shn{\hat S_n^0}

\def\fhm{\hat F_m^0}

\def\nt{N_T}
\def\nd{N_D}
\def\np{N_P}

\def\Ni{\{N_i\}}
\def\ci{\{c_i\}}

\def\cie{\{c_i^{\rm eq}\}}

\def\rr{_\rho}
\def\dfr{\Delta F\rr}
\def\dfer{\dfr^{\rm enz}}
\def\dfsr{\Delta{F}\rr^{\rm sol}}
\def\der{\Delta E\rr}
\def\deer{\der^{\rm enz}}
\def\desr{\Delta{E}\rr^{\rm sol}}
\def\dsr{\Delta S\rr}
\def\dser{\dsr^{\rm enz}}
\def\dssr{\Delta{ S}\rr^{\rm sol}}

\def\dmr{\Delta \mu_\rho}

\def\ss{^{\rm sys}}
\def\sys{^{\rm sys}}
\def\hk{^{\rm hk}}
\def\ex{^{\rm ex}}
\def\eq{^{\rm eq}}
\def\intr{^{\rm int}}
\def\aad{^{\rm ad}}
\def\na{^{\rm na}}
\def\med{^{\rm med}}

\def\ea{{\sl et al.}~}

\section{Introduction}
\subsection{From classical to stochastic thermodynamics}
Classical thermodynamics, at its heart, deals with general
 laws governing the
transformations of a system, in particular, those involving the exchange of heat,
work and matter with an environment. As a central result,  total entropy
production  is identified 
that in any such 
process can never decrease, leading, inter alia, to fundamental limits on the
efficiency of heat engines and refrigerators. 
The thermodynamic characterization of systems in equilibrium got its
microscopic justification from equilibrium statistical mechanics which states
that for a system in contact with a heat bath the probability to find 
it in  any specific microstate is given by the Boltzmann factor. For small
deviations from equilibrium, linear response theory allows to express
transport properties caused by small external fields through equilibrium
correlation functions. On a more phenomenological level, linear irreversible
thermodynamics provides a relation between such transport coefficients and
entropy production in terms of forces and fluxes.
Beyond this linear response regime, for a long time, no universal exact
results
were available.

During the last 20 years fresh approaches 
 have revealed general laws
applicable
to nonequilibrium system
thus pushing the range of validity of exact  thermodynamic statements
 beyond the realm of linear response deep into the 
genuine non-equilibrium region. These exact results,
which become particularly relevant for small systems with appreciable
(typically non-Gaussian) 
fluctuations,
 generically refer to
distribution functions of thermodynamic quantities like exchanged heat,
applied work or entropy production.

First, for a thermostatted shear-driven
fluid in contact with a heat bath,
  a remarkable symmetry of the probability distribution of entropy
production in the steady state  was
discovered numerically and justified heuristically
by Evans \ea \cite{evan93}. Now known as 
the (steady state) 
 fluctuation theorem (FT), it was first
proven for a large class of systems
using  concepts  from chaotic dynamics by Gallavotti and Cohen
\cite{gall95}, later for driven Langevin dynamics by Kurchan 
\cite{kurc98} and for driven diffusive dynamics by Lebowitz and Spohn
\cite{lebo99}. As a variant, a transient fluctuation theorem valid for
relaxation towards the steady state was found by Evans and Searles
\cite{evan94}.

Second, Jarzynski proved a
remarkable relation which allows to express the free
energy difference between two  equilibrium
states by a nonlinear average over the work
required to drive the system in a non-equilibrium process
from one state to the other
\cite{jarz97,jarz97a}.
By comparing probability distributions for the work spent in the original
process with the
time-reversed one, Crooks found a ``refinement'' of the 
Jarzynski relation (JR), now called the
Crooks fluctuation theorem \cite{croo99,croo00}. Both, this relation
and another refinement of the JR, the Hummer-Szabo relation \cite{humm01}
 became particularly useful for
determining free energy differences and landscapes of
biomolecules. These relations are  the most prominent ones
within a class of exact results (some of which found
even earlier \cite{boch77,boch79} and then rediscovered) 
valid for nonequilibrium systems driven by time-dependent forces.
A close analogy to the JR, which relates different equilibrium states,
 is the
Hatano-Sasa relation that applies to transitions between two different 
non-equilibrium steady
states
\cite{hata01}.

Third, for driven Brownian motion,
Sekimoto realized that two central concepts of classical
thermodynamics, namely the exchanged heat and the applied work, can be
meaningfully defined on the level of individual trajectories
\cite{seki97,seki98}. 
These quantities entering the first law 
become  fluctuating ones giving birth to what he called stochastic energetics
as described in his monograph \cite{seki10}.
Fourth, Maes emphasized that entropy production in the medium is related to
that part of the stochastic action, which determines the weight of
trajectories
that is odd under time-reversal \cite{maes03,maes03b}.

Finally, building systematically on a concept briefly noticed previously
\cite{croo99,qian02b},
a unifying perspective on these developments emerged
by   realizing that
besides the fluctuations of the entropy production in the heat bath one should
similarly assign a fluctuating, or "stochastic", entropy to the system proper
\cite{seif05a}. Once this is done, the key
 quantities known from classical thermodynamics are
defined along  individual
trajectories where they become  accessible to experimental or numerical
 measurements. This approach of taking both energy conservation, i.e., the first law,
 and
entropy production seriously on this
 mesoscopic level has been called stochastic thermodynamics \cite{seif07}, 
thus revitalizing a notion originally introduced by the Brussels school in the
mid-eighties where it was used on the ensemble level for chemical nonequilibrium
systems \cite{vdb85,mou86}.

\subsection{Main features of stochastic thermodynamics}
Stochastic thermodynamics as here understood applies to (small) systems like
colloidal particles, (bio)polymers (like DNA, RNA, and proteins), enzymes and
molecular motors. All these systems are embedded in an aqueous solution.
Three types of non-equilibrium situations can be distinguished for these
systems. First, one could prepare the system in a non-equilibrium initial
state and study the relaxation towards equilibrium. 
Second, genuine driving can be caused
by the action of time-dependent external forces, fields, flows or
unbalanced chemical reactions. Third, if the external driving is
time-independent
the system will reach a non-equilibrium steady state (NESS). For this latter
class,
particularly strong exact results exist.
In all cases, even under such nonequililibrium conditions,
the temperature of the system, which is the one of the embedding solution,
remains well-defined. This
property 
together with 
the related  necessary time-scale separation between
the observable, typically slow, degrees of freedom of the system
 and the unobservable fast
ones made up by the thermal bath (and, in the case of biopolymers, by fast
internal ones of the system)
allows  for a consistent thermodynamic description.

The collection of the relevant slow degrees of freedom makes up the state 
of the system. Since this state changes either due to the driving or due to
the ever present fluctuations, it leads to a trajectory of the system.
Such trajectories belong to an ensemble which is fully characterized by the
distribution of the initial state, by the properties of the
thermal noise acting on the system and by specifying the (possibily time-dependent) 
external driving. The thermodynamic quantities defined along the trajectory
like applied work and exchanged heat thus follow a distribution which can be
measured
experimentally or be determined in numerical simulations.

Theoretically, the time-scale separation implies that the dynamics becomes
Markovian, i.e., the future state of the system depends only on the present one
with no memory of the past. If the states are made up by continuous variables (like
position), the dynamics follows a Langevin equation for an individual system
 and a Fokker-Planck
equation for the whole ensemble. Sometimes it is more convenient to identify
discrete states with transition rates governing the dynamics which,  on
the ensemble level, leads to a master equation.

Within such a stochastic dynamics, the exact results quoted above
for the distribution functions of certain  thermodynamic quantities
follow universally for any system 
from  rather unsophisticated mathematics. It is sufficient to invoke a
``conjugate'' dynamics, typically, but not exclusively, time-reversal, to derive these theorems in
a few lines. Essentially,  they lead to universal constraints on
these distributions.
One inevitable consequence of these theorems is the occurrence of trajectories
with negative total entropy production. Such events have occasionally been
called (transient) violations of the second law. In fairness to classical
thermodynamics, however, one should emphasize that this classical theory
ignores fluctuations. If the second law is understood as refering
to the mean entropy production, it is 
indeed confirmed by these more recent exact relations.
Moreover, they show that the probability for such events become
typically exponentially small
in the relevant system size which means that one has to sample
exponentially many trajectories in order to observe these ``violations''.

Since these constraints on the distributions
are so universal, one might suspect that they are useless for 
uncovering system specific properties. Quite to the contrary,
some of them offer a surprising relation between equilibrium and
non-equilibrium properties with the JR as the most prominent 
and useful example. Moreover,
 such constraints can be used  as an obvious
 check whether the 
assumptions of the model apply to any particular system.
Finally, studying non-universal features of these distribution functions
and trying to find further common aspects in these has become
an important part of the  activities in this field.

Going beyond the thermodynamic framework, it turns out that many of the FTs hold formally
true for any kind of Markovian stochastic dynamics. The thermodynamic
interpretation 
of the involved quantities as heat and work is not mandatory
to derive such a priori surprising relationships between functionals
defined along dynamic trajectories.

\subsection{Hamiltonian, thermostatted and open quantum
dynamics}

Even though I will focus in the main part of this review on systems described
by a stochastic dynamics, it is appropriate to mention briefly alternative 
approaches as some of the FTs have originally been derived using a
deterministic framework.

Hamiltonian dynamics works, in principle, if the external driving is modelled
by a time-dependent potential arising, e.g., from a movable piston, tip of an
atomic force microscope, or
optical tweezer.
Conceptually,
 one typically  requires  thermalized initial conditions, 
then  imagines to cut off the system from the heat bath leading to the
deterministic motion and finally one has to reconnect the heat bath again.
In a second variant, the heat bath is considered to be
part of the system but one then
has to follow all degrees of freedom.  
  One disadvantage of Hamiltonian dynamics is that it cannot deal with
a genuine NESS, which is
driven by a time-indepedent external field or flow, since such a setting
inevitably
 heats up the system.

Thermostatted dynamics can deal with NESSs. Here, one keeps deterministic 
equations of motion
and introduces a friction term making sure that on average the
relevant energy (kinetic or total,
depending on the scheme) does not change \cite{evan08}.

Even though a deterministic dynamics is sometimes considered to be more
fundamental than a stochastic one, the latter has at least three
advantanges from the perspective held in this review. First, 
from a  practical point of view, in soft matter and biophysics a description 
focussing on the relevant (and measurable) degrees of freedom 
and ignoring water molecules from the outset 
has a certain economical appeal. Second, stochastic dynamics can describe 
transitions between discrete states like in (bio)chemical reactions with 
essentially the same conceptual framework used for systems with continuous
degrees
of freedom. Third, the mathematics required for deriving the exact relations
and for stating their range of validity is surpringly simple compared to
what is required for dealing with NESSs in the deterministic setting.

Open quantum systems will not be discussed explicitly in this review.
Some of the FTs can indeed be formulated for these systems, sometimes at the
cost of requiring somewhat unrealistic measurements at the
beginning and end of a process, as reviewed in \cite{espo09c,camp11}.
The results derived and discussed in the following, however, are directly applicable to
open quantum systems whenever coherences, i.e. the role of non-diagonal elements in the density
matrix, can be ignored. The dynamics of the driven or open quantum system is
then equivalent to a classical stochastic one. For the validity of the
exact relations in these cases, the quantum-mechanical
origin of the transition rates is  inconsequential.

\subsection{Scope and organization of this review}
By writing this review, apart from focussing entirely on systems governed by 
Markovian stochastic dynamics,
 I have been guided by the following principles
concerning format and content.

First, I have tried to present the field in a systematic order (and notation)
rather than
to follow the historical development which has been briefly alluded to in
the introductory section above. Such an approach leads to a more concise and
coherent presentation.
Moreover, I have tried to keep most of the more technical 
parts (some of which are original) still self-contained. Both features
should help those using this material as a basis for
courses which I
have given several times at the University of Stuttgart and at summer schools in
 Beijing, Boulder and J\"ulich.

Second, as a consequence of the more systematic presentation,
experimental, analytical and numerical case  studies of specific systems 
are mostly grouped
together and typically placed after the general theory
where  they fit best.

Third, for the exact results the notions ``theorem'', ``equality'' and ``relation''  are
used here  in no particular hierarchy. I rather try to follow the
practice established in the field so far. In particular, it is not implied that a
result called here ``theorem'' is in any sense deeper that another one
called ``relation''.

This review starts in Sect. {\ref{sec2} by introducing a paradigm for this field which is 
a  colloidal particle driven by a time-dependent force as it has been realized
in several experiments. Using this system, the main concepts of stochastic
thermodynamics, like work, heat and entropy changes along individual
trajectories, will be introduced. At the end of this
section,
 simple generalizations of driven one-dimensional motion such as
 three-dimensional
motion, coupled degrees of freedom and  motion in 
external flow are discussed. 

A general
classification and a physical 
discussion of the major fluctuation theorems
dealing with work and the various contributions to 
entropy production follows in Sect. \ref{sec3}.
In Sect. \ref{sec4}, I present a unifying perspective on basically
all known FTs for
stochastic dynamics using the concept of a conjugate
dynamics. It is shown explicitly, how these FTs follow from one
master theorem.
Sect. \ref{sec5} contains an overview of experimental, analytical
or numerical studies of Langevin-type dynamics in specific systems. 
Sect. \ref{sec6} deals with Markovian
dynamics on a discrete set of states for which FTs hold even
without assuming a thermodynamic structure.

The second part of the review deals with ramifications, consequences and
applications of these concepts. In Sect. \ref{sec-7}, the optimal driving of such processes
is discussed and the relation between time's arrow and the amount of dissipation derived.
Both concepts can then be used to discuss the role of measurements and (optimal)
feedback in these systems. Sect. \ref{sec-8} deals with generalizations of
the well-known 
fluctuation-dissipation theorem to NESSs where it is shown that stochastic entropy
plays a crucial role.

Biomolecular systems are discussed in Sect. \ref{sec-9} where special emphasis is given to
the role of time-scale separation between the fast (unobservable) degrees of freedom 
making up a well-defined
heat bath for the non-equilibrium processes of the slow variables  caused by mechanical or
chemical imbalances. From a conceptual
point of view the second essential new aspect of these  systems is that each of
the states is composed of many microstates which leads to the crucial notion
of intrinsic entropy that enters some of the exact 
relations in a non-trivial way.

Coming back to the issues that stood at the origin of thermodynamics,
the final two sections discuss the efficiency and optimization
of nano and micro engines and devices where it is useful to distinguish
isothermal engines like  molecular motors discussed in Sect. \ref{sec-10} from heat
engines like thermoelectric devices  treated in Sect. \ref{sec-11}.
A brief summary and a few perspectives are sketched in
Sect. \ref{sec-12}.

\subsection{Complementary reviews}

A selection of further reviews dealing with the topics discussed 
in the first part of this  article
 can roughly be 
grouped as follows.\footnote{Relevant reviews for the more specific topics
treated in the second part of this review will be mentioned in the
respective sections.}

The influential essay \cite{bust05} had an introductory character. More
recent non-technical accounts have been given by Jarzynski \cite{jarz11}
and van den Broeck \cite{vdb10} who both emphasize the relation of
the fluctuation theorems with irreversibility and time's arrow.
Other brief reviews by some of the main proponents include  
\cite{maes03,qian06,kurc07,impa07a} and the contributions in the 
collection \cite{klag12}.
Ritort has written a review on the role of non-equilibrium fluctuations
in small systems  with special emphasis on the applications to
 biomolecular systems \cite{rito08}. A review focussing on 
experimental work by one of the main groups working on fluctuation
theorems is \cite{cili10}.

For stochastic dynamics based on the master equation, a comprehensive
derivation of FT's has been given by Harris and Sch\"utz \cite{harr07}.
The fluctuation theorem  in the context of thermostatted dynamics 
has been systematically reviewed by Evans and Searles \cite{evan02}.
From the perspective of chaotic dynamics it is treated in 
Gallavotti's monograph
\cite{gall99a}.
The links between different approaches and rigorous mathematical
statements are surveyed in \cite{rond07,zamp07,sevi08}.

Stochastic thermodynamics focusses on a description of individual
trajectories as does an alternative approach by Attard 
introducing a ``second
entropy'' \cite{atta09}.
 On a  more coarse-grained level, phenomenological
thermodynamic theories of non-equilibrium systems have been developed
inter alia under the label of ``extended irreversible thermodynamics'' \cite{jou01},
''GENERIC'' \cite{otti04},
``mesoscopic dynamics of thermodynamic systems'' \cite{regu05}
and ``steady state thermodynamics'' \cite{oono98,sasa06}.

Nice reviews covering related recent topics in non-equilibrium
physics are \cite{frey05,chou11}.

\section{Colloidal particle as paradigm}
\label{sec2}
The main concepts of stochastic thermodynamics can be introduced using as a simple model system a
colloidal particle confined to one spatial dimension, which can arguably serve
as {\sl the} paradigm in the field.

\subsection{Stochastic dynamics}
 \label{sec:sd}
The overdamped motion $x(\tau)$ of a colloidal particle 
(or any other system with a
single continuous
degree of freedom) can be described using three 
 equivalent but complementary descriptions of stochastic dynamics, the
 Langevin equation,
the path integral, and  the Fokker-Planck equation.

The Langevin equation reads 
\begin{equation} \label{Lv}
\dot x = \mu F(x,\l) + \zeta = \mu (-\partial_xV(x,\l) + f(x,\l) ) + \zeta .
\end{equation}
The systematic force $F(x,\l)$ can arise from a conservative potential $V(x,\l)$
 and/or be
applied to the particle directly as $ f(x,\l)$. Distinguishing these two
sources
is relevant only in the case of periodic boundary conditions on a ring of
finite length.
Both sources
may be time-dependent through an external control
parameter $\la$ varied from $\l(0)\equiv \l_0$ to 
$\l(t)\equiv \l_t$ according to some prescribed 
 protocol. 

The  thermal noise has correlations
\begin{equation} \label{eq5}
\langle \zeta(\tau)\zeta(\tau')\rangle = 2 D \delta(\tau-\tau')
\end{equation}
where $D$ is the diffusion constant. In equilibrium, $D$ and the mobility $\mu$ 
are related by the Einstein relation
\begin{equation} \label{eqE}
D=T \mu
\end{equation} 
where $T$ is the temperature of the surrounding medium with Boltzmann's constant $k_B$ set to unity throughout this 
review  to make entropy dimensionless.  
In stochastic 
thermodynamics, one assumes that 
the strength of the noise is not affected by the presence of a time-dependent force. The range of validity of this  
crucial assumption  can  be 
tested experimentally or in simulations by comparing with  
theoretical results derived on the basis of this assumption.

The Langevin dynamics generates trajectories $x(\tau)$ starting at $x(0)\equiv x_0$ with a weight
\beq
p[x(\tau)|x_0]={\mathcal N}\exp [-{\cal A}([x(\tau),\lambda(\tau)])]
\label{eq:PI2}
\ee
where 
\beq
{\cal A}([x(\tau),\lambda(\tau)])\equiv  \int_0^td\tau[(\dot x -\mu F)^2/4D + \mu \partial_x F /2]
\label{eq:PI}
\ee
 is the ``action'' associated with the trajectory.
The last term arises from the Stratonovich convention for the discretization
in the Jacobian when the weight for a noise history $[\zeta(\tau)]$ is
expressed by $[x(\tau)]$. This symmetric discretization is used
implicitly throughout this review.
 Path dependent observables $\Omega[x(\tau)]$  can then
be averaged using this weight in a path integral which requires
a path-independent normalization  $\cal{N}$ such that summing the weight
(\ref{eq:PI2},\ref{eq:PI}) over all paths
is 1. Throughout the review averages using this weight and a given initial
distribution
$p_0(x)$ will be denoted by $\langle .... \rangle$ as in
\beq
\langle \Omega[x(\tau)] \rangle
\equiv \int dx_0\int d[x(\tau)] \Omega[x(\tau)] p[x(\tau)|x_0]p_0(x_0)  .
\ee

Equivalently, the Fokker-Planck equation 
for the probability $p(x,\tau)$ to find the particle at $x$ at time 
$\tau$ is
\begin{eqnarray}
\partial_\tau p(x,\tau)&=& - \partial_x j(x,\tau) 
\nonumber
\\
&=&-\partial_x \left(\mu F(x,\l)p(x,\tau) -D\partial_xp(x,\tau)\right) 
\label{eq:fp}
\end{eqnarray}
where $j(x,\tau)$ is the probability current.
This partial differential equation must be augmented by a
normalized initial
distribution $p(x,0)\equiv p_0(x)$. 
For further calculations, it is useful to define a mean local velocity
\beq
\nu(x,\tau)\equiv j(x,\tau)/p(x,\tau) .
\label{eq:nu}
\ee

More technical background concerning these three equivalent descriptions 
of Markovian  stochastics dynamics 
of a continuous degree of freedom is provided in the monographs 
\cite{gardiner,risken,chai01a,chai01b}.

\subsection{Non-equilibrium steady states (NESSs)}
\label{sec-ness}
For time-independent control parameter $\l$, any initial distribution will
finally
reach a stationary state $p^s(x,\lambda)$. For $f=0$, this stationary state is the thermal
equilibrium,
\beq
p\eq(x,\l) =\exp[-(V(x,\l)- \F(\l))/T] ,
\ee
with the free energy
\begin{equation}
\F(\l)\equiv -T\ln \int 
dx
~\exp[-V(x,\l)/T].
\label{eq:F}
\end{equation}

\begin{figure}[t]
\includegraphics[width=6cm]{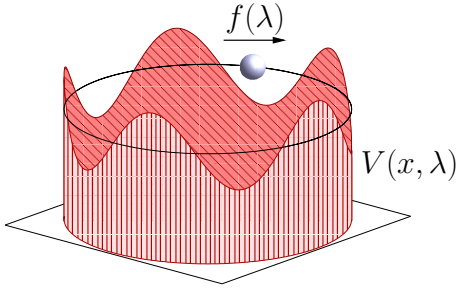}
\caption{Colloidal particle driven along a periodic potential $V(x,\l)$
by a non-conservative force $f(\l)$. In a NESS, the external parameter $\l$ is independent of time.
}
\label{fig-ring}
\end{figure}

A non-conservative force acting on a ring as shown in Fig. \ref{fig-ring} generates a
 paradigm for a genuine non-equilibrium steady state (NESS) with a
 stationary distribution
\beq
p^s(x,\l)\equiv  \exp[-\phi(x,\l)] ,
\label{eq:phi}
\ee
where $\phi(x,\l)$ is the ``non-equilibrium'' potential.
In one dimension, $p^s(x,\l)$  
can be obtained explicitly by quadratures \cite{risken}
or by an intriguing mapping to an equilibrium problem \cite{tail08}.
Characteristic for such a NESS is a steady
current
\begin{equation} \label{eq:sc}
j^s=\mu F(x) p^s(x)-D\partial_x p^s(x) \equiv v^s(x)p^s(x)
\end{equation}
with the mean local velocity 
$
v^s(x)$.
Even for time-dependent driving, one can express the total force
\beq
F(x,\lambda)=[v^s(x,\lambda)-D\partial_x\phi(x,\lambda)]/\mu
\label{eq:Fnu}
\ee
through quantities refering to the corresponding stationary state
which is sometimes helpful.

Occasionally, we will use $\langle ...\rangle\eq$ and
$\langle ...\rangle^s$ to emphasize when averages or correlation functions
are taken in genuine equilibrium and in a NESS, respectively.

\subsection{Stochastic energetics}
\subsubsection{The first law}
\label{sect:stochen}

Sekimoto suggested to endow the Langevin dynamics with a
thermodynamic interpretation by applying the notions appearing in the first law
\begin{equation}
\dbar w=dE+\dbar q
\label{eq:ebl}
\end{equation}
to an individual fluctuating trajectory \cite{seki97,seki98}.
 Throughout the article, we use the
convention that work applied to the particle (or more generally system) is
positive as is heat transferred or dissipated into the environment.

It is instructive first to identify the first law for a particle in
equilibrium, i.e. for $f=0$ and constant $\lambda$. In this case, no work is applied to
the system and hence an increase in internal energy,
defined by the position in the potential, $dE=dV=(\partial_xV) dx=-\dbar q$,
must be associated with heat taken up from the reservoir.

Applying work to the particle either requires a time-dependent potential
$V(x,\lambda(\tau))$
and (or) an external force $f(x,\lambda(\tau))$.
 The increment in work applied to the particle then reads  
\begin{equation}
\dbar w= (\partial V/ \partial \lambda)~d\lambda + f ~ dx ,
\label{eq:work}
\end{equation}
where the first term arises from changing the potential at
fixed particle position.
Consequently, the heat dissipated into the
medium must be identified with
\begin{equation}
\dbar q= \dbar w-dV= F dx  .
\label{eq:Fdx}
\end{equation}
This relation makes physical sense since in an overdamped system
the total force times the displacement corresponds to  dissipation. 
Integrated over a time interval $t$, one obtains the expressions
\begin{equation}
w[x(\tau)] = \int_0^t [(\partial V / \partial \lambda) \dot \lambda + f \dot x] \, d\tau 
\ee
and
\beq
 q[x(\tau)] = \int_0^td\tau~ \dot q= \int_0^t F \dot x \, d\tau
\label{eq:heat}
\end{equation}
and the integrated first law
\begin{equation}
 w[x(\tau)]= q[x(\tau)] + \Delta V= q[x(\tau)] + V(x_t,\lt)-V(x_0,\lo)
\label{eq:fl}
\end{equation}
on the level of an individual trajectory.

The expression for the heat requires a prescription of how to evaluate 
$F\dot x$. As above in the path integral, one has to use the
mid-point , i.e.,  Stratonovitch rule for  which  the ordinary rules of
calculus
for differentials and integrals apply.

The expression for the  heat dissipated along the trajectory
$x(\tau)$ can also be written in the form
\beq
q[x(\tau)] = T \frac{{\cal A}([x(\tau),\lambda(\tau)])}{{\cal A}([x(t-\tau),\lambda(t-\tau)])}
=T
\ln \frac{p[x(\tau),\l(\tau)]}
{p[\tilde x(\tau),\tilde\l(\tau)]}
\label{eq:q-rev}
\ee
as a ratio involving the weight (\ref{eq:PI}) for this trajectory 
given its initial point $x_0$ compared to the
weight of the time-reversed trajectory $\tilde x(\tau)\equiv x(t-\tau)$ under the
reversed protocol $\tilde\l(\tau) \equiv \l(t-\tau)$ for $\tilde x_0=x(t)\equiv x_t$. This formulation points
to the deep relation between dissipation and time reversal which repeatedly
shows up in this field.

\subsubsection{Housekeeping and ``excess'' heat}
Motivated by steady state thermodynamics, 
it will be convenient to split the dissipated heat into two
contributions \cite{oono98,hata01}
\beq 
q \equiv q\hk + q\ex .
\label{eq:q-split}
\ee
The housekeeping heat is the heat inevitably dissipated in maintaining the 
corresponding NESS. 
For a Langevin dynamics,
it reads
\beq
q\hk\equiv \int_0^td\tau \dot x (\tau) \mu^{-1}v^s(x(\tau),\l(\tau)) .
\label{eq:hk}
\ee
The ``excess'' heat
\begin{eqnarray}
q\ex &=& -(D/\mu) \int_0^td\tau \dot x (\tau)
\partial_x\phi(x,\lambda)
\nonumber\\
&=& T[-\Delta \phi + \int_0^t d\tau \dot \lambda \partial_\lambda
\phi]
\label{excess}
\end{eqnarray}
is the heat associated with changing the external control parameter
 where we have used  (\ref{eq:Fnu},\ref{eq:heat}).

\subsubsection{Heat and strong coupling}

This interpretation of the first law and, in particular, of 
heat relies on the implicit assumption that the unavoidable coupling between particle
(or, more generally, system) described by the slow variable $x$ and the degrees of freedom
making up the heat bath does neither depend crucially on $x$ nor on the control parameter $\l$.
Such an idealization may well be obeyed for a colloidal particle in a laser trap
but will certainly fail for more complex systems like biomolecules.
In the following, we first continue with this simple assumption.
In Sect. \ref{sec9-2}, we discuss the general case and point out which of the
results derived in the following will require a ramification. Roughly speaking, most of the
fluctuation theorems hold true with minor modifications
whereas infering heat correctly indeed requires one more term compared to
(\ref{eq:Fdx}).

\subsubsection{Alternative identification of work}

The definition of work (\ref{eq:work})  has  been criticised for
supposedly being in conflict with a more conventional view that work should
be given by force times displacement, see \cite{vila08}  and, for rebuttals,
\cite{peli08,peli08a,horo08}.
In principle, such a view could be 
integrated into the present scheme by splitting the potential into two
contributions,
\beq
V(x,\l)=V^0(x,\l_0)+ V\ex(x,\l),
\ee
the first being an intrinsic time-independent potential, and the second one
a time-dependent external potential used to transmit the external force. If one  defines
 work as
\beq
\dbar w\ex\equiv (-\partial_xV\ex(x,\l) + f) dx,
\label{eq:wbk}
\ee
it is trivial to check that the first law then holds in the form
\beq
\dbar w\ex = dE\ex + \dbar q
\ee
with the corresponding change in internal energy 
$dE\ex\equiv dV^0=\partial_xV^0dx$ and the  identification of heat (\ref{eq:Fdx}) 
unchanged. Clearly, within such a framework, it would be  appropriate to identify the internal energy with
changes in the intrinsic potential only. Integrated over a trajectory, this
definition of work differs from the previous one by a boundary term,
$\Delta w=\Delta w\ex + \Delta V\ex$.

It is crucial to appreciate that exchanged heat as a physical concept is,
and should be, 
independent of the convention how it is split into work and changes in
internal energy. The latter freedom is inconsequential as long as one stays
within one scheme. A clear disadvantage of this alternative scheme, however,
 is that changes
in the free energy of a system are no longer given by the quasistatic work
relating two states. In this review, we will  keep the definitions as
introduced in Sect. \ref{sect:stochen} and only 
occasionally quote results for the alternative 
expression for work introduced in this section. 

\subsection{Stochastic entropy}
Having expressed the first law along an individual trajectory, it seems
natural to ask whether entropy can be identified on this level as well.
For a simple colloidal particle,
the corresponding quantity turns out to have two contributions. First, the
heat dissipated 
into the environment should obviously be identified with an
increase in entropy of the medium
\begin{equation}
\Delta s\m[x(\tau)] \equiv q[x(\tau)]/T.
\label{eq:dsm}
\end{equation} 
Second, one identifies as a stochastic or trajectory
dependent entropy \index{stochastic entropy} of the system the quantity 
\cite{seif05a}
\begin{equation}
s(\tau) \equiv  -\ln p(x(\tau),\tau)
\label{eq:s}
\end{equation}
where the probability $p(x,\tau)$ obtained by first solving the
Fokker-Planck equation is evaluated along the stochastic 
trajectory $x(\tau)$. Thus, the stochastic entropy depends not only on the
individual trajectory but also on the ensemble. If the same trajectory $x
(\tau)$ is taken from an ensemble generated by another initial condition
$p(x,0)$, it will lead to a different value for $s(\tau)$. 

In equilibrium, i.e., for $f\equiv 0$ and constant $\l$, 
the stochastic entropy $s(\tau)$ just defined obeys the well-known 
thermodynamic
relation, $TS=E-F$, between entropy, internal energy and free energy  
in the form 
\begin{equation}
Ts(\tau)=V(x(\tau),\l)-\F(\l)  ,
\end{equation} 
now along the fluctuating trajectory at any time with the free energy 
defined in (\ref{eq:F}) above.

Using  the Fokker-Planck equation  
the rate of change of the  entropy of the system (\ref{eq:s}) follows 
 as \cite{seif05a}
\beq
\dot s(\tau)= \left. -{\partial_\tau \px\over \px}\right|_{x(\tau)} +   
 \left( {j(x,\tau)\over D \px}
-  {\mu F(x,\l)\over D} \right)_{x(\tau)} \dot x .
\ee Since
the very last term can be related to the rate of  heat  dissipation   in the 
medium (\ref{eq:heat}), using $D=T\mu$, one obtains a balance equation for 
the trajectory-dependent total entropy production as
\begin{equation}
\dst \equiv \dsm+\dot s(\tau) =  - \left. {\partial_\tau \px\over \px}\right|_{x(\tau)}+ 
\left.{j(x,\tau) \over D\px}\right|_{x(\tau)} \dot x .
\label{eq:s1}
\end{equation}
The first term on the right hand side signifies
a change in $\px$ which can be due to a time-dependent $\la$ or, even
for a
constant
 $\l_0$, due to relaxation from a non-stationary initial state
$p_0(x)\not = p^s(x,\l_0)$.

 As a variant on the trajectory level, occasionally
 $\phi(x,\lambda)=-\ln p^s(x,\l)$ has been suggested as 
a definition of system entropy. Such a
choice is physically questionable as the following example
shows. Consider diffusive relaxation of a
localized initial distribution $p_0(x)$
in a
finite region $0\leq x\leq L$.
 Since $p^s(x)=1/L$, $\phi(x)$ will not change during this
process.
On the other hand, such a diffusive relaxation should clearly lead to an
entropy
increase.
Only in cases
where one  starts in a NESS and waits for final relaxation, the change in system
entropy can also be expressed by a change in the non-equilibrium potential
according to $
\Delta s = \Delta \phi $.

\subsection{Ensemble averages}

Upon averaging, the expressions for the thermodynamic quantities along the
individual trajectory should become the ensemble quantities of non-equilibrium
thermodynamics derived previously for such Fokker-Planck systems, see, e.g.
\cite{qian02b}. 

Averages for quantities involving the position $x(\tau)$ of
the particle are most easily  performed using the probability $p(x,\tau)$.
Somewhat more delicate are averages over quantities like the heat that
involve  products of the velocity $\dot x$
 and a function $g(x)$. These can be performed in two steps.
First, one can evaluate the average 
$\langle\dot x|x,\tau\rangle$ conditioned on the position $x$ in the
spirit of the Stratonovitch
discretization as
\begin{eqnarray}
\langle\dot x|x,\tau\rangle &\equiv& \lim_{\Delta \tau\to 0}
\left(\langle x(\tau+\Delta \tau) - x(\tau)|x(\tau)=x\rangle + \right.\nonumber
\\ &~&~\left. \langle x(\tau) - x(\tau-\Delta \tau)|x(\tau)=x\rangle\right)/{(2\Delta \tau)}.
\end{eqnarray}
The averages in the bracket on the rhs can be evaluated by discretizing the path integral
(\ref{eq:PI2}, \ref{eq:PI}) for one step. The first term straightforwardly yields $\mu F(x,\tau)\Delta \tau$.
In the second one, the end-point conditioning (best implemented by Bayes' theorem) 
is crucial which leads to an additional contribution if the distribution is not uniform. The final
result is \cite{seif05a}
\beq
\langle\dot x|x,\tau\rangle = \mu F(x,\tau)-D\partial_xp(x,\tau)=\nu(x,\tau) .
\ee
Any subsequent average over position is now trivial leading to
\beq
\langle g(x)\dot x\rangle = \langle g(x)\nu(x,\tau)\rangle = \int dx g(x)j(x,\tau) .
\label{eq:gdotx}
\ee

With these relations, one obtains, e.g.,  for the averaged total entropy 
production rate from (\ref{eq:s1}) the expression 
\begin{equation}
\dSt \equiv \langle\dot s\tot(\tau)\rangle
= \int dx {j\kk^2\over D p\kk} = \langle \nu(x,\tau)^2\rangle /D  \geq 0 ,
\label{eq:ens-tot}
\end{equation}
where  equality holds in equilibrium only. In a NESS, $\nu(x,\tau)=v^s(x)$
which 
thus determines the mean dissipation rate.
Averaging the increase in entropy of the medium
 along similar lines leads to
\beq
 \dSm \equiv
 \langle  \dsm \rangle = \int dx F(x,\tau)j(x,\tau)/T .
\label{eq:ens-m}
\ee
Hence upon averaging, the increase in entropy of the system proper
 becomes 
$
\dot S(\tau)\equiv \langle \dot s(\tau)\rangle= \dSt-\dSm $.
On the ensemble level, this balance equation for the averaged quantities
can also be derived directly from the ensemble definition of the
system entropy
\begin{equation}
S(\tau) \equiv  - \int dx ~p(x,\tau) \ln p(x,\tau) 
= \langle s(\tau)\rangle 
\label{eq:ens}
\end{equation} 
by using the Fokker-Planck equation (\ref{eq:fp}).

\subsection{Simple generalizations}

\subsubsection{Underdamped motion}
\label{sect:under}
For some systems, it is necessary to keep the inertial term which
leads with mass $m$ and damping constant $\gamma$ to the Langevin equation
\beq
m\ddot x + \gamma\dot x =-\partial_x V(x,\l) + f(\l) + \xi
\label{eq:lv-under}
\ee
with the noise correlations 
$\langle \xi(\tau)\xi(\tau')\rangle =2\gamma T\delta(\tau-\tau')$.

The internal energy now must include the kinetic energy, $dE=dV + mv~dv$,
with $v\equiv \dot x$. Since the identification of work (\ref{eq:work}) remains valid, the
first law becomes
\beq
\dbar q=dW-dE=Fdx -mv~dv .
\label{eq:heatunder}\ee
Evaluating the stochastic entropy
\beq
s(\tau)\equiv -\ln p(x(\tau),v(\tau),\tau)
\ee
now requires a solution of the corresponding Fokker-Plank equation
\beq
\partial_\tau p=-\partial_x(vp)-\partial_v[[(-\gamma v +F)/m] p-
  T (\gamma/m^2)\partial_v]p 
\ee for $p=p(x,v,\tau)$ 
with an appropriate initial condition $p_0(x,v)$.

\subsubsection{Interacting degrees of freedom}
\label{sect:many}
\def\bx{{\bf x}}
\def\bz{{\boldsymbol{\zeta}}}
\def\bfeta{{\boldsymbol{\eta}}}
\def\bg{{\boldsymbol{\nabla}}}
\def\bforce{{\bf f}}
\def\bF{{\bf F}}
\def\bm{{\underline{\underline{\mu}}}}
\def\bj{{\bf j}}
\def\bu{{\bf u(r)}}
\def\buk{{\bf u(r}_k)}
\def\bgk{{\boldsymbol{\nabla}}_k}
\def\brk{\dot{\bf r}_k}
\def\bfk{{\bf f}_k}
\def\bxi{{\boldsymbol{\xi}}}
\def\bfl{{\bf f}_l}
\def\bgl{{\boldsymbol{\nabla}}_l}
\def\bK{{\bf K}}
\def\bD{{\underline{\underline{D}}}}
\def\bv{{\bf v}}
\def\br{{\bf r}}

The framework introduced for a single degree of freedom
can easily be generalized to several
degrees of freedom $\bx$ obeying the coupled Langevin equations
\beq
\dot \bx = \bm [-\bg V(\bx,\lambda) + \bforce(\bx, \lambda)] + \bz,
\label{eq:Lvbf}
 \ee  
  where $V(\bx,\lambda)$ is a potential and $\bforce(\bx,\lambda)$ a
non-conservative force.
The noise correlations 
\beq
\langle \bz(\tau):\bz(\tau')\rangle = 2 T \bm \delta (\tau- \tau') ,
\label{eq:znoise}
\ee
involve the mobility tensor $\bm $.
Simple examples of such system comprise a colloidal particle in three
dimensions, several interacting colloidal particles, or a polymer
where $\bx$ labels the positions of monomers. If hydrodynamic interactions are
relevant,
the mobility tensor will depend on the coordinates $\bx$.

The corresponding Fokker-Planck equation becomes
\beq
\partial_\tau p(\bx,\tau) = - \bg \bj = -\bg(\bm  (-\bg V + \bforce)
p-T\bm\bg p) 
\ee
with the local (probability) current $\bj$.

In particular for a NESS, there are two major formal differences compared 
to the one-dimensional case. First, the stationary current $\bj^s(\bx)$ becomes
$\bx$ dependent and, second, the stationary distribution $p^s(\bx)\equiv
\exp[-\phi(\bx)]$ 
or, equivalently,  the
non-equilibrium potential are
not known analytically except for the trivial case that the total forces are linear in $\bx$.

On a formal level, all expressions discussed above for the simple
colloidal particle can easily be generalized to this multi-dimensional case
by replacing scalar operations by the corresponding vector or matrix ones.

\subsubsection{Systems in external flow}
\label{sec-264}
So far, we have assumed that there is no overall hydrodynamic flow imposed
on the system. For colloids, however, external flow is a common situation.
Likewise, as we will see, colloids in moving traps can also be described in 
a co-moving frame as being subject to some flow. We therefor recall the
modifications required for the basic notions of stochastic thermodynamics
in the presence of external flow field $\bu$ \cite{spec08}.

The Langevin equations for $k=1, .. ,N$ coupled particles 
at positions ${\bf r}_k$ reads
\beq
\brk=\buk+\sum_l \bm_{kl}(-\bgl V+\bfl) + \bz_k
\ee
with the usual noise correlations 
\beq
\langle\bz_k(\tau)\bz_l(\tau')=2T\bm_{~kl}\delta(\tau-\tau')\rangle .
\ee
In such a system, the increments in external work and dissipated
heat are given by
\beq
\dbar w\equiv \left([\partial_\tau + \buk\bgk]V + \bfk[\brk-\buk]\right) dt
\label{eq:workflow}
\ee
and
\beq
\dbar q=\dbar w-dV=\left([\brk-\buk][-\bgk V+\bfk]\right) dt ,
\label{eq:heatflow}
\ee
respectively. Compared to the case withour flow, the two modifications involve
replacing the
partial
derivative by the convective one and  measuring the velocity
relative to the external flow velocity. These expressions  guarantee frame
invariance of stochastic thermodynamics \cite{spec08}.

For the experimentally studied case of one-dimensional colloid motion in a flow
of constant velocity $u$ discussed in Sect. \ref{sect:trap}
 below, the Langevin equation simplifies to
\beq
\dot x = u + \mu(-\partial_x V + f) + \zeta
\ee
and the ingredients of the first law  become 
\beq
\dbar w=(\partial_\tau V + u\partial_xV)dt + f(\dot x -u)dt
\ee
and
\beq
\dbar q= (\dot x -u)[-\partial_xV + f]~dt  .
\ee

\section{Fluctuation theorems (FTs)} 
\label{sec3}
Fluctuation theorems express universal properties of the probability
distribution
$p(\Omega)$ for functionals $\Omega[x(\tau)]$, like work, heat or entropy change, 
evaluated along the fluctuating trajectories taken
from ensembles with well-specified  initial distributions $p_0(x_0)$.
In this section, we  give a phenomenogical classification
into three classes according to their mathematical appearance
and point out some general mathematical consequences. 
The most prominent ones will then be discussed
in physical terms with references to their original derivation.
For proofs of these relations within stochastic dynamics from the 
present perspective, we provide in Sect. \ref{sect:ftuni}
the unifying one for all FTs  that 
also shows that there
is essentially an infinity of such relations.

\subsection{Phenomenological classification}
\subsubsection {Integral fluctuation theorems (IFTs)}

A non-dimensionalized
functional $\Omega[x(\tau)]$ with probability distribution function $p(\Omega)$
obeys an IFT if
\beq
\langle \exp(-\Omega)\rangle \equiv \int d\Omega~ p(\Omega) \exp(-\Omega)  = 1  .
\label{eq:IFT}
\ee
The convexity of the exponential functions then 
implies the inequality
\beq
\langle \Omega\rangle \geq 0
\ee which often represents a  well-known thermodynamic law.
With the exception of the degenerate case, $p(\Omega) = \delta (\Omega)$, the IFT implies that
there are trajectories for which $\Omega$ is negative. Such events have sometimes
then been characterized as ``violating'' the corresponding
thermodynamic law. Such a formulation is controversial since classical
thermodynamics, which ignores fluctuations from the very beginning,
is silent on issues beyond its range of applicability.
 The probability of such events quickly diminishes
for negative $\Omega$. Using (\ref{eq:IFT}), it is easy to derive for $\omega
>0$
\cite{jarz08}
\beq
{\rm
  prob}[\Omega<-\omega]\leq\int_{-\infty}^{-\omega}d\Omega~p(\Omega)~e^{-\omega-\Omega}\leq
e^{-\omega} .\ee
This estimate shows that relevant ``violations'' occur for $\Omega$ of order
1.
Restoring the dimensions in a system with $N$ relevant degrees of
freedom,  $\Omega$ will typically be of order  $N k_BT$
which implies that in a large system such events are exponentially small,
i.e., occur
exponentially rarely. This observation essentially reconciles the effective
validity of thermodynamics on the macro-scale with the still correct
mathematical
statements that even for large systems, in principle, such events must occur.

An IFT represents one constraint on the probability distribution $p(\Omega)$. If it
is somehow known  that $p(\Omega)$ is a Gaussian, the IFT implies the
relation
\beq
\langle (\Omega-\langle \Omega \rangle)^2 \rangle = 2 \langle \Omega\rangle
\ee
between variance and mean of $\Omega$.

\subsubsection{Detailed fluctuation theorems (DFTs)}
A detailed fluctuation theorem corresponds to   the stronger relation
\beq
p(-\Omega)/p(\Omega) = \exp(-\Omega)
\label{eq:dft}
\ee
for the pdf $p(\Omega)$. Such a symmetry constrains ``one half'' of the pdf
which means, e.g.,  that the even moments of $\Omega$ can be expressed by the
odd ones and vice versa. A DFT implies the corresponding IFT trivially.
Further statistical properties of $p(\Omega)$ following from the validity
of the DFT (and some from the IFT) are derived in \cite{merh10}.

Depending on the physical situation, a variable obeying the DFT has often been
called to obey either a transient FT (TFT) or a steady state FT (SSFT).
These notions will be explained below for the specific cases. 

\subsubsection{(Generalized) Crooks fluctuation theorems (CFTs)}
\def\dg{^\dagger}
These  relations compare the pdf $p(\Omega)$ of the original process one is 
interested in with the pdf $p\dg(\Omega)$ of the {\it same} physical
quantity for a ``conjugate''
(mostly the time-reversed) 
process. The general statement then is that
\beq
p\dg(-\Omega)=p(\Omega)e^{-\Omega}
\label{eq:cft}
\ee which  implies the IFT (but not the DFT) for $\Omega$
since $p\dg$ is normalized.

\subsection{Non-equilibrium work theorems}
\label{sect:ftwork}

These relations deal with the probability distribution $p(w)$ 
for work spent in driving
the system 
from a (mostly equilibrium) initial state to another (not necessarily
equilibrium) state. They require only a notion of work defined along the
trajectory but not yet the concept of stochastic entropy.

\subsubsection{Jarzynski relation (JR)}
In 1997, Jarzynski showed that the work spent in 
 driving the system from an initial
equilibrium state at $\lambda_0$ via a time-dependent potential 
$V(x,\lambda(\tau))$ for a time $t$ obeys  \cite{jarz97}
\beq
\langle \exp(-w/T) \rangle = \exp-(\Delta \F/T)
\label{eq:JR}
\ee where 
$\Delta \F \equiv \F(\lambda_t)-\F(\lambda_0)$ is the free energy 
difference between the 
equilibrium state corresponding to
the final value  $\lambda_t$ of the control parameter  and the initial state. In the
classification
scheme proposed here, it  can technically be viewn as the IFT for the  (scaled)
dissipated work
\beq
w_d\equiv(w - \Delta \F)/T .
\ee

The paramount relevance of this relation -- and its originally so
surprising feature -- is that it allows to determine the free energy difference,
which is a genuine equilibrium property, from non-equilibrium measurements
(or simulations). It represents a strengthening of the familiar second law
 $\langle w\rangle \geq \Delta F$ which follows as the corresponding
inequality. It has orginally been derived using a Hamiltonian
dynamics (which requires decoupling from and coupling to a heat bath at the
beginning and end of the process, respectively) but soon been shown to
hold for stochastic dynamics as well \cite{jarz97a,croo99,croo00}.
Its validity requires that one starts in 
the equilibrium distribution but not that the system has relaxed
at time $t$ into the new equilibrium. In fact, the actual distribution at the
end will be $p(x,t)$ but any further relaxation at constant $\l$ would not
contribute to the work anyways.

Within stochastic dynamics, the validity of the JR (as of any other FT
with a thermodynamic interpretation) essentially rests on
assuming that the noise in the Langevin equation (\ref{Lv})
is not affected by the driving. A related issue arises in the Hamiltonian
derivation of the JR which requires some care in identifying the proper role
of the
heat bath during the process \cite{cohe04,jarz04}.

The JR has been studied for many systems analytically, numerically,
and experimentally. Specific case studies for stochastic dynamics will be
classified
and quoted below in Sect. \ref{sec5}.  
As an important application, based on a generalization introduced by Hummer
and Szabo \cite{humm01}, the Jarzynski relation can be used to 
reconstruct the free energy landscape of a biomolecule as discussed in
Sect. \ref{sec9-3}
below. 

\subsubsection{Bochkov-Kuzolev relation (BKR)}
The Jarzynski relation should be distinguished from
 an earlier relation derived by Bochkov and Kuzolev \cite{boch77,boch79}. For a system initially in equilibrium in a 
time-independent potential $V_0(x)$ and for $0\leq \tau\leq t$ subject to an additional space and
time-dependent force (possibly arising from an additional  potential), the
work (\ref{eq:wbk}) 
integrated over a trajectory obeys
 the Bochkov-Kuzolev relation (BKR) 
\begin{equation} \label{eq:bk}
\langle \exp[- w\ex/T]\rangle = 1 .
\end{equation}
Contrary to some claims, the BKR is different 
from  the Jarzynski relation 
since they apply a priori to somewhat different situations \cite{seif07,jarz07,horo07}. 
The JR as discussed above
applies to processes in a time-dependent potential, whereas the BKR
applies to a process in a constant potential
with some additional force.
If, however,  in the latter case, 
this explicit force
arises from a potential as well, both the BKR and the JR (\ref{eq:JR}) hold
for the respective forms of work.

\subsubsection{Crooks fluctuation theorem (CFT)}

In the Crooks relation, the pdf for work $p(w)$ spent in the original
(the ``forward'') process is related to the pdf for work $\tilde p(w)$
applied
in the reversed process where the control parameter is driven according to
$\tilde \lambda(\tau) = \lambda (t-\tau)$ and one starts in the
equilibrium distribution corresponding to $\tilde \lambda_0=\lambda_t$. 
These two pdfs obey \cite{croo99,croo00}
\beq
\tilde p(-w)/p(w) = \exp[-(w-\Delta \F)/T].
\label{eq:cft2}
\ee
Hence, $\Delta \F$ can be obtained by locating the crossing of the two pdfs
which for biomolecular applications 
turned out to be a more reliable method than using the JR. Clearly, the
Crooks relation implies the JR since $\tilde p(w)$ is normalized.
Technically, the Crooks relation is of the type (\ref{eq:cft})
 for $R=w_d$ with the conjugate
process being the reversed one.

\subsubsection{Further general results on $p(w)$}

Beyond the JR and the CFT, further exact results on $p(w)$ are scarce.
For systems with linear 
equation of motion, the pdf for work (but not for heat) is a Gaussian for 
arbitrary time-dependent driving \cite{mazo99,spec05}. For slow driving, 
i.e., for $t_{\rm
  rel}/t\ll 1$ 
where $t_{\rm rel} $ is the typical relaxation time of the system at fixed
$\lambda$
and $t$ the duration of the process,
an expansion based on this time-scale separation  yields
 a Gaussian for  any potential \cite{spec04}. Such a result has previously
been expected \cite{herm91,wood91} or justified by
 invoking arguments based on the central limit theorem \cite{hend01}. Two 
observations show, however, that such an expansion is somewhat delicate.
First, even in simple examples there occur terms that are non-analytic in
$t_{\rm
  rel}/t$ \cite{spec04}. Second, for the special case of a ``breathing
parabola'', $V(x,\l)=\l(\tau)x^2/2$, any protocol with $\dot\l>0$ leads
to $p(w)\equiv 0$ for $w<0$ which is obviously violated by a Gaussian. 
How the latter effectively emerges in the limit of slow driving is
investigated in \cite{spec11}.

From another perspective, Engel \cite{enge09,nick11} investigated the 
asymptotic behavior of $p(w)$ for small $T$ using a saddle point analysis. 
The value of this approach is that it can provide exact results for the tail 
of the distribution. Specific examples show an exponential decay.
Saha \ea \cite{saha11} suggest that the work distribution for quite different systems
can be mapped to a class of universal distributions.

\subsection{FTs for entropy production}
\label{sect:ftent}

\subsubsection{IFT}
The total entropy production along a trajectory as given by
\begin{equation}
\Delta s\tot \equiv \Delta s\m + \Delta s ,
\end{equation} 
with
\begin{equation} \Delta s \equiv -\ln p(x_t,\lt)+\ln p(x_0,\lo)  
\label{eq:with}
\end{equation}
and $\Delta s\m$ defined in (\ref{eq:dsm}),
obeys the IFT \cite{seif05a} 
\begin{equation}
\langle \exp(-\Delta s\tot) \rangle =1  
\label{eq:R3}
\end{equation}
for arbitrary
initial distribution $p(x,0)$, arbitrary time-dependent driving $\l(\tau)$ and
an arbitrary length $t$ of the process. 

Formally, this IFT can be considered as a 
refinement of the second law, $\langle \Delta s\tot\rangle \geq 0$,
 which is the corresponding inequality.
Physically, however, it must be stressed that by using the Langevin equation
a fundamental irreversibility has been implemented from the very beginning.
Thus,  this IFT should definitely  not be considered to constitute
 a fundamental proof of the
second law.

\subsubsection{Steady-state fluctuation theorem (SSFT)}

In a NESS with fixed $\l$, the total entropy production obeys the stronger
SSFT 
\beq
p(-\Delta s\tot)/p(\Delta s\tot )= \exp(-\Delta s\tot)
\label{eq:dft2}
\ee
again for arbitrary length $t$. This relation corresponds to
 the genuine ``fluctuation theorem''. It
has first been found in simulations of
two-dimensional sheared fluids \cite{evan93} 
and then been proven by Gallavotti and
Cohen \cite{gall95} using assumptions about chaotic dynamics.
For stochastic diffusive dynamics as considered specifically in this
review,  it has been proven
 by Kurchan \cite{kurc98} and Lebowitz and Spohn \cite{lebo99}.
 Strictly speaking, in these early works the relation
holds only asymptotically in the long-time limit since entropy production
 had been associated with what is here called entropy production in the
medium. 
If one includes the entropy change of the system (\ref{eq:s}), 
the
SSFT holds even for finite times in the steady state \cite{seif05a}.

\subsubsection{Hatano-Sasa relation}

The Hatano-Sasa relation applies to systems with steady states 
$p^s(x,\lambda)=\exp[-\phi(x,\lambda)]$. With the splitting of the dissipated
heat
into a housekeeping and excess one (\ref{eq:q-split}), the IFT \cite{hata01}
\beq
\langle \exp[-(\Delta \phi + q\ex/T)]\rangle = 1
\label{eq:hs}
\ee
holds for any length of trajectory with $\Delta \phi \equiv
\phi(x_t,\lambda_t)-\phi(x_0,\lambda_0)$. The corresponding inequality
\beq
\langle \Delta \phi \rangle \geq - \langle q\ex\rangle
/T
\label{eq:inhs}
\ee
allows an interesting thermodynamic interpretation. The left-hand side can 
be seen as ensemble entropy change of the system in a  transition from one
steady state to another.  Within the framework discussed in this review,
this interpretation is literally true provided one waits for final relaxation
at constant $\lambda_t$ since then $\Delta s = \Delta \phi$. 
A recent generalization of the HS relation leads to a variational scheme for
approximating the stationary state \cite{pere11}.

With the interpretation of the left hand side as entropy change in the system, 
the inequality (\ref{eq:inhs}) provides for transitions between NESSs
what the famous Clausius inequality does for transitions between 
equilibrium states. The entropy change in the system is at least as big as
the excess heat flowing into the system.
 For transitions between NESSs, the
 inequality (\ref{eq:inhs}) is sharper than
the Clausius one (which still applies in this case and becomes just $\langle
\Delta s\rangle \geq -q/T$) since $q$ scales with the
transition
time whereas $q\ex$ can remain bounded and can actually approach equality in
(\ref{eq:inhs})
for quasistatic transitions.

Experimentally, the Hatano-Sasa relation has been verified for a colloidal particle
pulled through a viscous liquid at different velocities which
corresponds to different steady states \cite{trep04}.

\subsubsection{IFT for housekeeping heat}

Finally, it should be noted that the second contribution to heat, the
housekeeping heat also obeys an IFT \cite{spec05a}
\beq
\langle \exp[- q\hk/T]\rangle = 1
\ee
for arbitrary initial state, driving and length of trajecories.

\section{Unification of FTs}
\label{sect:ftuni}
\label{sec4}

Originally, the FTs have been found and derived on a case by case
approach. However, it has soon become clear  that within stochastic
dynamics a unifying strategy  is to investigate the behaviour of the
system under time-reversal. Subsequently, it turned out that
 comparing the 
dynamics to its ``dual'' one \cite{hata01,cher06}, eventually also in connection with 
time-reversal, allows a further unification. In this section, 
we outline this general approach and show how the prominent FTs discussed
above (and a few further ones mentioned below) fit into, or derive from,
this framework. Even though this section is inevitably somewhat technical and
dense, it is self-contained. 
It could be skipped by readers not interested in the proofs or
systematics of the FTs.
For related mathematically rigorous  
approaches to derive FTs for diffusive dynamics, see
\cite{chet08a,ge08a,liu09a,sugh11}.

\subsection{Conjugate dynamics}

\def\px{p[x(\tau)|x_0]}
\def\dg{^\dagger}
\def\eaa{\epsilon_\alpha}

\def\dshk{\Delta  s\hk}
\def\dsex{\Delta  s\ex}

FTs for the original process with trajectories $x(\tau)$, $0\leq\tau\leq t$,
 an
initial distribution $p_0(x_0)$ and a conditional weight $\px$
are most generally derived by formally 
invoking a ``conjugate'' dynamics
for trajectories $x\dg(\tau)$. These are supposed to
 obey a Langevin equation
\beq
\ \label{eq:Lv2}
\dot x\dg = \mu\dg F\dg(x\dg,\l\dg) + \zeta\dg .
\ee with $\langle \zeta\dg(\tau)\zeta\dg(\tau')\rangle = 2 \mu\dg T\dg
\delta(\tau-\tau')$. The trajectories 
with weight  $p\dg[x\dg(\tau)|x\dg_0]$
run over a time $t$ and start with 
an initial distribution
$p\dg(x\dg_0)$. Averages of the conjugate
dynamics will be denoted by $\langle ...\rangle\dg$.

This conjugate dynamics is related to the original process by a one-to-one
mapping
\beq
\{x(\tau),\l(\tau),F,\mu,T\} \to  \{x\dg(\tau),\l\dg(\tau),F\dg,\mu\dg,T\dg\} 
\ee which allows to express all quantities occuring in the conjugate
dynamics in terms of the orginal ones.

The crucial quantity leading to the FTs 
is a master functional given by the log-ratio of
the unconditioned path weights
\begin{eqnarray}
R[x(\tau)] &\equiv& \ln\frac{p[x(\tau)]}{p\dg[x\dg(\tau)]}
\nonumber\\
&=&  \ln\frac{p_0(x_0)}{p_0\dg(x\dg_0)} +\ln\frac{\px}{
  p\dg[x\dg(\tau)|x\dg_0]}  \equiv R_0 + R_1  
\label{eq:R}
\end{eqnarray}
 that consists  of a ``boundary'' term $R_0$
coming from the two initial distributions
 and a ``bulk'' term $R_1$.

Three choices for the conjugate dynamics and the associated mapping
have been considered so far. In all cases, neither the temperature
nor the functional form of the mobilities have been
changed for the conjugate dynamics, i.e.,  $T\dg=T$ and $\mu\dg=\mu$.

{\sl (i) Reversed dynamics:} This choice corresponds to
  ``time-reversal''. The mapping reads
\beq
x\dg(\tau)\equiv x(t-\tau) {~~~\rm and~~~} \l\dg(\tau)\equiv \l(t-\tau)
\label{eq:tr}
\ee with no changes at the functional dependence of the force from its
arguments, i.e., $F\dg(x\dg,\l\dg)=F(x\dg,\l\dg)$.

 The
weight of the
conjugate trajectories is
easily calculated using the mapping ({\ref{eq:tr})
in the weight (\ref{eq:PI2},\ref{eq:PI})
leading to
\beq
R_1={\cal A}([x\dg(\tau),\lambda\dg(\tau)])-{\cal A}([x(\tau),\lambda(\tau)])]= \Delta
s\m=q/T, 
\label{eq:R1-time}
\ee which is  the part of the action ${\cal A}([x(\tau),\lambda(\tau)])$ that is 
odd under time-reversal.

This relation allows a deep physical interpretation. For given initial
point $x_0$ and final point $x_t$, the log ratio between the probability to
observe a certain forward trajectory and the  probability to
observe the time-reversed  trajectory is given by the heat dissipated
along the forward trajectory.

{\sl (ii) Dual dynamics:} This choice alters the equations of motion for the
$x\dg(\tau)$ trajectories such that (i) the stationary distribution remains
the same for both processes and that (ii) the stationary current for the dual
dynamics
is minus the original one. Specifically, this mapping reads \cite{cher06}
\beq
F\dg(x\dg,\l\dg)=F(x\dg,\l\dg)-2 v^s(x\dg,\l\dg)/\mu
\label{eq:duald}
\ee
which enters the conjugate Langevin equation (\ref{eq:Lv2})
and no modification for $x$ and
$\l$, i.e., $x\dg(\tau)\equiv x(\tau)$ and $\l\dg(\tau)\equiv \l(\tau)$.

Calculating the action for the dual dynamics (\ref{eq:Lv2}), the functional $R_1$ becomes
\beq
R_1=q\hk/T\equiv \dshk.
\ee

{\sl (iii) Dual-reversed dynamics:}
For this choice, the dual dynamics  is driven with the time-reversed
protocol, i.e., the mapping of the force  (\ref{eq:duald}) is combined with
the
time-reversal (\ref{eq:tr}). In this case, the functional $R_1$
becomes \cite{cher06}
\beq
R_1=   q\ex/T\equiv \dsex .
\ee

In summary, depending on the form of the conjugate dynamics, different
parts of the dissipated heat form the functional $R_1$. For later reference,
we have introduced in the last two equations for 
the scaled contributions to the dissipated heat  the corresponding entropies. 

\subsection{The master FT}
\label{sect:master-ft}
\subsubsection{Functionals with definite parity}

The FT's apply to  functionals $S_\alpha[x(\tau)]$ of the original dynamics
that map with a definite parity $\eaa=\pm1$ to the conjugate dynamics
according to
\beq S\dg_\alpha\left([x\dg(\tau)],\l\dg,F\dg\right)=\eaa 
S_\alpha\left([x(\tau)],\l,F\right)
\label{eq:SS}
 \ee such that $ S\dg_\alpha[x\dg(\tau)]$ represents the {\it same}
physical
quantity for the conjugate dynamics as $S_\alpha[x(\tau)]
$ does for the original one.

Examples for such functionals are
work and heat that both are odd ($\eaa=-1$) for the reversed dynamics. For dual or
dual reversed dynamics, however, these two functionals have no definite
parity since both cases involve a different dynamics.
Explicitly, the heat behaves under time-reversal
as $q\dg\equiv\int_0^t d\tau
\dot x\dg F\dg=- \int_0^t d\tau \dot x F=-q$. For dual dynamics, the heat
transforms as $q\dg\equiv \int_0^t d\tau
\dot x\dg F\dg= \int_0^t d\tau \dot x (F - 2v^s/\mu) =q-2q\hk$ which has,
in general,  no
definite parity.
On the other hand,
the housekeeping heat is odd for the dual dynamics and even for both
the reversed and the dual-reversed dynamics.

The stochastic  entropy $\Delta s$, in general, has no definite parity 
under time-reversal since $s(\tau)$
is defined through the solution  $p(x,\tau)$
of the Fokker-Planck equation which is not odd under time-reversal.
In particular,  $p(x,t-\tau)$ does not solve the Fokker-Planck equation
for the time-reversed process even if one starts the reversed
process with the final distribution $p(x,t)$ of the original process.
The change in the non-equilibrium potential $\Delta \phi$, however, is odd
under time-reversal. This difference between $\Delta s$ and $\Delta \phi$
implies that $\Delta \phi$ occurs more frequently in FTs.

\subsubsection{Proof}

\label{sec:proof}

With these preparations, one can easily derive the master FT
\begin{eqnarray}
 \langle g(\{&\eaa&S\dg_\alpha[x\dg(\tau)]\})\rangle\dg
\nonumber\\&=&
\int dx_0\dg\int d[x\dg(\tau)]p\dg_0(x\dg_0)p[x\dg(\tau)|x\dg_0] 
g(\{\eaa S\dg_\alpha\})\nonumber\\
&=&\int dx_0\dg\int d[x\dg(\tau)]p_0(x_0)p[x(\tau)|x_0]\exp[-R] 
g(\{S_\alpha\})\nonumber\\
&=&\int dx_0\int d[x(\tau)]p_0(x_0)p[x(\tau)|x_0]\exp[-R] 
g(\{S_\alpha\}\nonumber)\\
&=&
\langle
g(\{S_\alpha[x(\tau)]\})\exp(-R[x(\tau])\rangle 
\label{eq:gft}
\end{eqnarray}
 for any function $g$ depending on an arbitrary number of such
functionals $S_\alpha$. For the second equality, we use the definitions 
(\ref{eq:R}) and the parity relation (\ref{eq:SS});
for the third we recognize that summing over all daggered
trajectories is equivalent to summing over all original ones  both
for $x\dg(\tau)=x(\tau)$ and $x\dg(\tau)=x(t-\tau)$.
With the choice  $g\equiv 1$, this FT leads to the most general
IFT $\langle e^{-R}\rangle =1$ from which all known IFT-like
relations follow as shown in Sect. \ref{sect:ift}.

By choosing for $g$ the characteristic function, one obtains
a generalized FT for joint probabilities in the form
\beq
\frac {p\dg(\{S\dg_\alpha = \eaa s_\alpha\})}{p(\{S_\alpha = s_\alpha\})}=
\langle\ \exp(-R)|\{S_\alpha\}=\{s_\alpha\}\rangle
\label{eq:gft2}
\ee
that relates the pdf for the conjugate process to the pdf of the original
one and a conditional average.
All known DFTs for stochastic dynamics follow as 
special cases
of this general theorem as shown in Sect. \ref{sect:dft}. The key
point is to (i) select the appropriate conjugate process
for which the quantity
of interest $\Omega$ has a unique parity,  which
is most often just the reversed dynamics, (ii) identify for the generally 
free initial distribution $p_0\dg(x)$  an appropriate function, 
and (iii) express the functional $R$ using physical quantities,
preferentially the quantity of interest $\Omega$.

\subsection{General IFTs}
\label{sect:ift}
The simplest choice for the function $g$ in (\ref{eq:gft})
is the identity, $g=1$, leading to
the IFT $\langle e^{-R}\rangle =1 $. Explicitly, one obtains for
 the three types of conjugate
dynamics:

(i) By choosing the reversed dynamics (\ref{eq:tr}) and with 
(\ref{eq:R1-time}), the class of IFTs
\beq
\left\langle \frac {p_1(x_t)}{p_0(x_0)}  \exp[{-\Delta s\m}] \right\rangle =
1
\label{eq:gift}
\ee
follows for any initial condition $p_0(x_0)$,
any length of trajectories $t$,  and
any normalized function $p_1(x_t)=p\dg_0(x_t)$ \cite{seif05a}. 
By specializing the latter
to the solution of the Fokker-Planck equation for $\tau=t$
one obtains the IFT for
total entropy production (\ref{eq:R3}).

For a system in a time-dependent potential
$V(x,\lambda)$ and by starting in an initial distribution given by the
corresponding Boltzmann factor, $p_0(x)=\exp[-(V(x,\lambda_0)-
  \F(\lambda_0))/T], 
$ one obtains the JR (\ref{eq:JR}) for the choice
$p_1(x_t) = \exp[-(V(x,\lambda_t)-{\F}(\lambda_t))/T] 
$ corresponding to the Boltzmann distribution for the
final value of the control parameter.

A variety of ``end-point'' relations can be generated
from (\ref{eq:gift}) as follows.
By choosing $p_1(x)=p(x,t)g(x)/\langle g(x_t) \rangle $, one obtains
\beq
\langle g(x_t)\exp[-\Delta s\t]\rangle = \langle g(x_t) \rangle 
\label{eq:e1}
\ee for any function $g(x)$ \cite{schm06a}.
Likewise, for $f\equiv 0$ and $V(x,\la)$, by choosing
$p_1(x)=g(x)\exp[-(V(x,\lt)-\F(\lt))/T]/\langle g(x)\rangle\eq_{\lt}$,
one obtains
\beq
\langle g(x_t)\exp[-(w-\Delta \F)/T]\rangle = 
\langle g(x)\rangle\eq_{\lt}
\label{eq:e2}
\ee
which has been first derived by Crooks \cite{croo00}. Here, the average
on the right hand side is the equilibrium average at the final
value of the
control parameter.
In the same fashion, one can derive
\beq
\langle  g(x_t)\exp[-w\ex /T]\rangle =  
\langle g(x)\rangle\eq_{\lo}
\label{eq:e3}
\ee
by choosing $ p_1(x) = g(x)\exp[-(V(x,\lo)-\F(\lo)))/T]/\langle g(x)\rangle\eq_{ \lo}
$
for a time-independent potential and arbitrary force $f(x,\tau)$ which
is the end-point relation corresponding  to the BKR (\ref{eq:bk}).
The latter follows trivially by choosing $g(x_t)=1$.

For processes with feedback control as discussed in Sect. 
\ref{sec7-3} below, it will be convenient to exploit the
end-point conditioned average
\beq
\left\langle \frac {1}{p_0(x_0)}  \exp[{-\Delta s\m}]\left.\right|x_t=x \right\rangle
p(x,t) =
1
\label{eq:gift-feed}
\ee valid for any $x$ which follows from (\ref{eq:gift}) by choosing 
$p_1(x_t)=\delta(x_t-x)$. Equivalently, by choosing $p_1(x,t)=p(x,t)$,
\beq
\int dx_0\langle \exp[{-\Delta s\m}] p(x_t,t)|x_0\rangle
=
1
\label{eq:gift-feed2}
\ee holds for summing over the initial point conditioned average.

(ii) By using the dual dynamics with $p\dg_0(x_0) = p_0(x_0)$, 
the IFT for the housekeeping heat \cite{spec05a}
\beq
\langle \exp[{-q\hk/T}]\rangle = 1
\ee
 valid for
any initial distribution follows.

(iii) For the dual-reversed dynamics, one gets the class of IFTs
from 
\beq
\left\langle \frac {p_1(x_t)}{p_0(x_0)}\exp[-q\ex/T] \right\rangle =  1
\label{eq:gift2}
\ee
valid for any initial distribution $p_0(x_0)$
and any normalized function $p_1(x_t)$. By choosing $p_0(x_0)=
\exp[-\phi(x_0,\lambda_0)]$
 and
$p_1(x_t)=\exp[-\phi(x_t,\lambda_t)]$, one obtains the Hatano-Sasa relation
(\ref{eq:hs}).
Similarly, another class
of  end-point relations
could be generated starting from (\ref{eq:gift2}).

Finally, since the IFTs, $\langle \exp[-\Omega]\rangle =1$, do not explicitly involve 
the conjugate process, one
might wonder whether they can be derived in an alternative way. Indeed,
some of them can be obtained by deriving an 
appropriate Fokker-Planck-type equation  for the joint
pdf $p(\Omega,x,\tau)$ and then 
showing $\partial_\tau \langle e^{-\Omega}\rangle =\partial_\tau\int d\Omega\int dx~e^{-\Omega}
p(\Omega,x,\tau)=0 $ 
directly, see for the Jarzynski relation \cite{jarz97a}, for the 
house-keeping heat \cite{spec05a},  and for another large class
of IFTs \cite {cher06}.

\subsection{ FTs derived from time-reversal}
\label{sect:dft}
In this section, the FTs following from using time-reversal
as conjugate dynamics are derived systematically from (\ref{eq:gft2})
by specializing to the
various scenarii concerning initial conditions and type
of driving. More or less reversing the chronological development,
we start with the more general cases and end with the more specific ones,
for which the strongest constraints on these pdfs follow.

\subsubsection{CFTs involving reversed dynamics}
By starting  both, original and reversed
dynamics, in the respective stationary state, the functional $R$ becomes
\beq
R=\Delta \phi + \Delta s\m  .
\ee
Hence, 
 one gets from (\ref{eq:gft2}) the FT
\beq
\frac{p\dg(\{S\dg_\alpha = \eaa s_\alpha\})}{ p(\{S_\alpha = s_\alpha\})}=
\langle\ e^{-(\Delta \phi + \Delta s\m)}|\{S_\alpha\}=\{s_\alpha\}\rangle .
\label{eq:dft3}
\ee For the special case that $\sum_\alpha S_\alpha = \Delta \phi + \Delta
s\m$, this relation has first been derived 
 by Garcia-Garcia \ea \cite{garc10}. Note that in general the change in 
stochastic entropy
 $\Delta s$ is not an admissible
choice for $S_\alpha$ since it lacks definite parity under time-reversal.

By choosing for $S_\alpha$ the work $w$, one gets
\beq
p\dg(-w)=p(w)\langle e^{-(\Delta \phi + \Delta s\m)}|w\rangle.
\label{eq:dft4}
\ee From this relation, the Crooks FT (\ref{eq:cft}) follows for a 
time-dependent $V(x,\lambda(\tau))$
and $f=0$, 
if one samples both processes from the respective initial equilibria,
since then  $\Delta \phi= \Delta (V-\F)/T$ and hence $R=(w-\Delta \F)/T$.

Likewise, by choosing $S_\alpha= w~\chi_A(x_0)\chi_b(x_t)$, where $\chi_{A,B}\equiv 1(0)$
if $x\in(\notin)A,B$ 
are the  characteristic functions of two regions $A$ and $B$, one obtains the variants 
derived and discussed in \cite{mara08,juni09} which allow to extend the
CFT to ``partially equilibrated'' initial and final states. These variants 
have become useful in
recovering free energy branches in single molecule experiments.

As another variant, by choosing for $S_\alpha$ the work $w\ex$ and by starting the
reverse process in the initial equilibrium, one gets 
with $\Delta \phi=\Delta V^0$
and $R=(\Delta V^0 + q)/T=w\ex$ the Crooks relation for 
$w\ex$ \cite{horo07}
\beq
p\dg(-w\ex)/p(w\ex)=\exp[-w\ex/T].
\ee

\subsubsection{DFTs for symmetric and periodic driving}

For symmetric driving, $\lambda(\tau) = \lambda(t-\tau)$, and for
$p\dg_0(x\dg_0)=p_0(x_0)$, the reversed dynamics becomes the
original one. Hence, the FTs (\ref{eq:dft3},\ref{eq:dft4})
 derived in the previous subsection
remain  valid in this case  if one replaces $p\dg$ on the left hand side
with $p$.
In this case, as in those in the following subsections, the FTs no longer
involve the conjugate dynamics explicitly which thus has become  a mere
mathematical tool to derive these relations most efficiently.
In particular, for starting in the initial equilibrium, $R=w/T$
and one gets for the pdf of work \cite{schu05,baie06}
\beq
p(-w)/p(w)=\exp[-w/T].
\label{eq:cftw}
 \ee

Likewise, for a periodically driven system with an integer number of symmetric periods
of length $t_p$, i.e., $\lambda(t_p-\tau)=\lambda(\tau)$, the reversed dynamics
is the original one. If the distribution has settled into a periodic 
stationary state, one has the DFT-like relation for total entropy production
\cite{tiet06,shar09}. Note that it
is crucial to choose not only a periodic but also a symmetric protocol since
otherwise the reversed dynamics is not the original one.

\subsubsection{SSFTs for NESSs}

For a NESS, i.e., for time-independent driving and starting in 
the stationary state, 
the reversed dynamics becomes the original one
and thus  $R=\Delta s\t$. Then (\ref{eq:gft2}) 
implies the
generalized SSFT for joint probabilities in the form 
\beq
\frac{p(\{S_\alpha = \eaa s_\alpha\})}{p(\{S_\alpha = s_\alpha\})}=
\langle\ e^{- \Delta s_{\rm tot}   }|\{S_\alpha\}=\{s_\alpha\}\rangle .
\label{eq:gft3}
\ee For this case, system entropy is indeed odd, and hence
one also has, in particular, by choosing $S_\alpha=\Delta s\t$
 the genuine SSFT (\ref{eq:dft2}) for total entropy production
and  arbitrary length $t$. As variants, illustrating the potency
of the general theorem,  one easily gets from (\ref{eq:gft3})
\beq
p(-\Delta s)= p(\Delta s)e^{-\Delta s}\langle e^{-\Delta s\m}|\Delta s \rangle
\ee
and
\beq
p(-\Delta s\m)= p(\Delta s\m)e^{-\Delta s\m}\langle e^{-\Delta s}|\Delta s\m \rangle
\ee
involving  conditional averages. Such relations seem not to have been explored in
specific systems yet.

\subsubsection{Expression for the NESS distribution}

By using an initial and end-point conditioned variant of (\ref{eq:gft2}),
Komatsu \ea manage to express the stationary distribution $p^s(x)$ in a
NESS by non-linear averages over the difference in ``excess'' heat required
either to reach $x$ from the steady state or to reach the NESS starting in $x$
\cite{koma08,koma08a,koma09} which leads to Clausius-type relations for
NESSs \cite{koma11}.  For a related expression for $p^s(x)$ in terms of an expansion around
a corresponding equilibrium state, see \cite{cola11} which contains a valuable
introduction into the 
history of such approaches.

\subsubsection{Transient fluctuation theorems TFTs}
\label{sect:tft}
This  relation applies  to {\sl time-independent} driving and arbitrary
initial condition $p_0(x_0)$. If the reversed dynamics is sampled using the
same initial condition, $p\dg_0(x\dg_0)=p_0(x_0)$, then the functional
$R$ becomes
\beq
R=-\ln[p_0(x_t)/p_0(x_0)] + q/T \equiv \Omega_t
\ee which has been called dissipation functional by Evans and Searles \cite{evan02}. 
Physically, $\Omega_t$ corresponds to the log-ratio between
 the probability to observe
the original trajectory and the one for observing the time-reversed one.
Since under these
conditions $\Omega_t$ is odd and
the reversed dynamics is equivalent to the original one, one has 
from (\ref{eq:gft2})
the TFT
\beq
p(- \Omega_t)/p(\Omega_t)=\exp[- \Omega_t] 
\label{eq:TFT}
\ee
valid for any length $t$ and initial condition.

If the  system is originally
equilibrated
in a potential $V_0(x)$ and then suddenly subject to 
either another 
time-independent potential $V_1(x)$ or a force $f(x)$ 
the dissipation functional becomes
\beq
\Omega_t=V_0(x_t)-V_1(x_t)+ V_1(x_0)-V_0(x_0) ,
\label{eq:TFTR}
\ee
or $\Omega_t=w$, respectively. In the latter case, the TFT holds
for work.

\subsection{FTs for variants}

\subsubsection{FTs for underdamped motion}

For underdamped motion as introduced in Sect. \ref{sect:under},
the functional $R_1$ defined in (\ref{eq:R}) under time-reversal
is still given by the dissipated heat, i.e., by (\ref{eq:heatunder}) 
integrated over the
trajectory \cite{impa06}.
 This fact follows by directly
evaluating the  action for the path integral
corresponding to the underdamped Langevin equation
 (\ref {eq:lv-under}). Hence, all FTs based on time-reversal
hold true also for underdamped dynamics with the obvious
modification that initial (and daggered) distributions
now depend on $x$ and $v$.

\subsubsection{FTs in the presence of external flow}

In the presence of flow, one has to specify how the flow changes in
the conjugate dynamics. For genuine time-reversal, the physically appropriate
choice is $\bu\dg=-\bu$ which leads with the definitions of work and
heat (\ref{eq:workflow},\ref{eq:heatflow}) to an odd parity for these
two functionals. Consequently, the FT's then hold as in the case without
flow. Formally, however, one could also keep the flow unchanged for
the conjugate dynamics, 
$\bu\dg=\bu$,
which would lead to another class of FTs. For a specific example
illustrating this freedom, see the 
discussion in \cite{spec08} for a dumbell in shear flow first investigated
in \cite{turi07}.

\subsubsection{FT with magnetic field}
In the presence of a (possibly time-dependent) magnetic field, the FTs
hold true essentially unchanged as proven in great generality for the 
JR and the CFT for interacting particles on a curved surface 
\cite{prad10a}. This work generalized earlier case studies on the validity
of the JR for specific sitations involving a magnetic field as mentioned
in Sect. \ref{sec5-2} below. A second motivation for this work was to
refute  earlier claims  based on simulations that the Bohr-van-Leeuven 
theorem stating the absence of classical diamagnetism could fail for 
a closed topology \cite{kuma09}.

\subsubsection{Further ``detailed theorems''}

Esposito and van der Broeck have derived what they call detailed fluctuation
theorems for $\Delta s\tot,\Delta s\hk,$ (called ``adiabiatic'' entropy change $\Delta s\aad$)
 and for the ``non-adiabiatic'' entropy change $\Delta s\na\equiv \Delta s\ex + \Delta s$ under even more
general
conditions \cite{espo10b,espo10d,vdb10a}. Their relations are beyond the realm of the
present systematics since they compare the pdf for {\sl different} physical
quantities for the original and the conjugate process whereas we 
always compare the pdfs of the {\sl same} physical quantities.\footnote{The IFT 
$\langle \exp[-\Delta s\na]\rangle = 1$, however, follows directly
from (\ref{eq:gift2}) by choosing $p_1(x_t)=p(x,t)$.}
 A unification of FTs within this broader sense
involving joint distributions
of these decompositions of entropy production is achieved in \cite{garc12} and a generalization of the
Hatano-Sasa relation in \cite{verl12}.

\subsection{FTs for athermal systems}

\subsubsection{General Langevin systems}

The derivation of the master FT in Sect. \ref{sect:master-ft} shows that for
obtaining these mathematical relations the main requirement is the existence
of a conjugate dynamics 
such as time reversal. Therefore, imposing a relation between the strength of
the noise and the 
mobility as done in Sect. \ref{sec:sd} for colloidal particles is not really
necessary. 
Neither  is it necessary to
interpret the Langevin equation using concepts of work and heat.
 We therefore sketch in this
section 
the general FT for a system of Langevin equations 
\beq
\dot \bx =\bK(\bx,\l) + \bz
\ee
with arbitrary ``force'' $\bK$ and  noise correlations
\beq
\langle \bz(\tau):\bz(\tau')\rangle = 2  \bD \delta (\tau- \tau') .
\ee
The corresponding Fokker-Planck equation becomes
\beq
\partial_\tau p(\bx,\tau) = - \bg \bj = -\bg(\bK p-\bD\bg p) 
\ee
and  the local (probability) current $\bj(\bx,\l)$.
For constant $\l$, one has the local mean velocity
\beq
\bv^s(\bx,\l)\equiv \bj^s(\bx,\l)/p^s(\bx,\l) = \bK(\bx,\l) - \bD \bg \ln p^s  .
\ee 

For time reversal as conjugate dynamics, by evaluating the corresponding
weight 
one obtains for the master functional
(\ref{eq:R1-time}) 
\beq
R_1=\int_0^t d\tau \dot \bx \bD^{-1}\bK \equiv \Delta s\m
\ee
where the identification with  $\Delta s\m$
is now purely formal. If one adds the stochastic entropy change along a trajectory
\beq
\Delta s \equiv  -\ln [p(\bx_t,\l_t)/p(\bx_0,\l_0)]
\ee
one obtains the total entropy production $\Delta s\t$.
Likewise, in analogy to the colloidal case, $\Delta s\m$
 can be split into
\beq
\dshk\equiv \int_0^t d\tau \dot x \bD^{-1} \bv^s
\ee
and
\beq
\dsex \equiv  \Delta s\m-\dshk = \int_0^td\tau \dot \bx \bg \ln p^s = -\Delta \phi +
\int_0^td\l
\dot\l\partial_\l \phi.
\ee
With these identifications all FTs involving the various forms of
entropy production
derived and discussed in Sects. \ref{sect:ftent} and
\ref{sect:ftuni}  hold true for such Langevin systems as well.

The identification of a generalized work makes immediate sense only if $\bK
=-(\bD/T)\bg V(\bx,\l)$ with some potential 
$V(\bx,\l)$ and effective temperature $T$
 in which case one is back at the
thermal model with interacting 
degrees of freedom introduced in Sect. {\ref{sect:many}. If $\bK$ cannot be derived in this way from a gradient
field
 there seems to be no gain by 
trying to impose a genuine thermodynamic interpretation
without further physical input.

\subsubsection{Stochastic fields}

The generalization of the results in the previous section for coupled Langevin
equations 
to stochastic field equations is trivially possible \cite{seif07}. Consider
a scalar field
$\Psi(\br,\tau)$ that obeys
\beq
\partial_\tau\Psi(\br,\tau)=K[\Psi(\br,\tau),\l(\tau)] + \zeta(\br,\tau)
\ee
with some functional $K[\Psi(\br,\tau),\l(\tau)]$ and  
\beq
\langle\zeta(\br,\tau)\zeta(\br',\tau')\rangle = 2 D(\br-\br')\delta(\tau-\tau').
\ee
with arbitrary spatial correlation $D((\br-\br')$.
The expressions for the entropy terms can easily be inferred; e.g., the analogy
of the entropy change in the medium becomes
\beq
\Delta s\m\equiv \int_0^t d\tau\int d\br \int d\br'
\partial_\tau\Psi(\br,\tau) D^{-1}(\br-\br')K[\Psi(\br',\tau),\l(\tau)].
\ee
 By now, it should  be obvious how to derive the corresponding FTs 
and how to generalize all these also to  the case when
$\Psi(\br,\tau)$  is a multi-component field. Likewise, it would be a trivial task
to specialize all this to driven or relaxing  ``thermal'' field theories for which
$\bK$ includes the derivative of  some Landau-Ginzburg type free energy and 
where the noise
obeys an FDT \cite{kard07}.

An interesting  applications concerns 
enstrophy dissipation  in two-dimensional turbulence \cite{baie05}. 
Field-theoretic techniques are used in \cite{mall11} to derive
generalized JRs and to explore the role of super-symmetry in this context.
Quite generally, it will be interesting to investigate stochastic versions of
the field equations of active matter \cite{rama10} from
this perspective.

\subsubsection{FTs in evolutionary dynamics}

The framework of FTs has recently been applied to the stochastic evolution
of molecular biological systems where it leads to an IFT for fitness flux
\cite{must10}.

\section{Experimental, analytical  and numerical work
for specific systems with continuous degrees of freedom}
\label{sec5}
\subsection{Principal aspects}

The various relations derived and discussed above have the status of
mathematically exact statements. As  such they require neither a 
``test'' nor a ``verification''. The justification for,  and 
the value of, 
the large body of experimental and numerical work that appeared
in this field during the last decade rather arises from the following
considerations.

First, experimental and numerical measurements of the distributions
$p(\Omega)$ entering the theorems provide non-trivial information about
the specific system under consideration. Integral and detailed
theorems give only one constraint on, and constrain only one half of,
the distribution, respectively. Beyond  the constraints imposed
by the exact relations, the distributions are non-universal
in particular for short times.

Second, the theorems involve non-linear averages. The necessarily
limited number of data entering experimental or numerical estimates
can cause deviations from the predicted exact behavior. It is
important to get experience how large such statistical errors are.
Systematic theoretical investigations concerning the error
due to finite sampling  are mentioned in Sect. \ref{sec9-3} below.

Third, the thermodynamic interpretation of the mathematical relations
in terms of work, heat and entropy rests on the crucial assumption that
the noise in the Langevin equation is not affected by the driving.
While this condition
 can trivially be guaranteed in simulations, it could
be violated in experiments. A statistically significant deviation
of experimental results from a theoretical prediction could be
rooted in the violation of this assumption.

These remarks apply to systems where one expects at least
in principle that a stochastic description of the relevant
degrees of freedom well-separated in time-scale from an
equilibrated heat bath is applicable. There
are, however, systems that a priori do not belong to this
class like sheared molecular fluids, shaken
granular matter and alike.
The proof of fluctuation theorems given above will not apply
to such systems. Still, fluctuation theorems have been proved
for other types of dynamics and experimentally investigated
in such systems.

In the following we first focus on a review of experimental and
numerical work of the first category and then briefly mention
systems for which it is less clear whether they comply with
the assumptions of a stochastic dynamics.
When refering to the experiments and the numerical work,
we will use the notions and
notations established in this review, which may occasionally differ 
from those given by the original authors.

\subsection{Overdamped motion: Colloidal particles and other systems}
\label{sec5-2}
\subsubsection{Equilibrium pdf for heat}

Even in equilibrium, explicit calculations of the pdf for heat
are typically
non-trivial.
In the long time $t\to \infty$, low temperature
$T\to 0$
limit, it has been calculated for an arbitrary potential with multiple minima \cite{foge09}.
For a harmonic potential and any $t$ and $T$, it is given by an
expression involving a Bessel function \cite{chat10}. It has also been derived analytically
in the presence of a magnetic field \cite{chat11}.

\subsubsection{Moving harmonic traps and electric circuits}
\label{sect:trap}

Wang \ea \cite{wang02} 
measured the distribution of what amounts to work (called
$\Sigma_t$ in their eq. 2) for a colloidal particle
initially in equilibrium in a harmonic trap which was then displaced
with constant velocity. The authors found that the pdf obeys a
relation corresponding
to the TFT which is strictly speaking the correct interpretation only
within the co-moving frame. Interpreted in the lab frame, 
 the driving is time-dependent. However, since for
linear forces the work distribution is a Gaussian which moreover has
to obey the JR with $\Delta \F =0$, it is clear that such a Gaussian 
also obeys the  TFT formally.

In a sequel, Wang \ea \cite{wang05} 
considered the same set-up for a quasi-steady state
situation at constant velocity. The authors showed in particular that
a quantity ($\Omega_t({\bf r})$ 
as defined in their eq. 19)
 which is 
equal to $ \Delta s\t$ obeys the DFT also for short times
as it should since this set-up  seen in the co-moving frame corresponds
to a genuine NESS.

For traps moving with constant velocity, explicit expressions for
the Gaussian work distribution, i.e. for its mean and variance,
 have been calculated in \cite{mazo99,zon03}. In all cases,
the DFT type relation is fulfilled. In contrast, the pdf for the
dissipated heat is non  Gaussian (essentially because it involves
the distribution of $(x_t-\l_t)^2$ from the internal energy) with
exponential tails. An explicit expression is not available, but its
characteristic function and in consequence its large deviation form
can be determined analytically \cite{zon03a,zon04a}. 
The pdf for work in a moving trap (and the pdf for heat in a stationary trap) 
were also measured and compared to theoretical results by
Imparato \ea \cite{impa07c}.
For a harmonically bound particle subject to a time-dependent force,
Saha \ea calculated pdfs for total entropy production, in particular,
for non-equilibrated initial conditions \cite{saha09}.

For a charged particle in a harmonic trap, work fluctuations and the
JR have been studied theoretically for a time-independent magnetic field and a 
moving trap or time-dependent electric field
 in \cite{jaya07,jime09,jime10,jime10a,jime11}, and for a time-dependent 
magnetic field in \cite{saha08}, respectively.

Trepagnier \ea \cite{trep04} studied experimentally
the transition from one 
NESS to another by changing the speed of the moving trap. If interpreted
in the co-moving frame, their experiment constituted the first 
experimental verification of the Hatano-Sasa relation (\ref{eq:hs}).

Simple electric circuits can formally be mapped to the dragged
colloidal particle. Corresponding FTs and pdfs have been investigated 
by Ciliberto and co-workers in \cite{zon04,garn05,joub08}, by Falcon and Falcon
in \cite{falc09} and by Bonaldi \ea \cite{bona09} for actively cooled resonators 
used in a gravitational wave detector. A similar mapping was used by Berg to study the JR 
applied to gene expression dynamics \cite{berg08}.

\subsubsection{Harmonic traps with changing stiffness}

Carberry \ea \cite{carb04} investigated the motion of a colloidal particle
in a harmonic trap whose stiffness is suddenly changed from one value
to another. They thus verified the TFT (\ref{eq:TFT}) for $\Omega_t$ given by 
(\ref{eq:TFTR}). For strongly localized initial conditions, this TFT has
been verified experimentally in \cite{khan11}.
Gomez-Solano \ea inferred the fluctuations of the heat exchanged between a 
colloidal particle and an ageing gel which bears some similarity to a time-dependent
stiffness \cite{gome11a}.

\subsubsection{Non-linear potentials}

Blickle \ea \cite{blic06} measured the work distribution for a colloidal
particle pushed periodically by a laser towards a repulsive substrate.
This experimental set-up was the first one
for colloidal particles that used
effectively non-harmonic potentials. The pdf    for work is distinctly 
non-Gaussian but
still in good agreement with theoretical predictions based on solving the
Fokker-Planck equation. This agreement justifies a posteriori the crucial
assumption that the noise correlations are not affected by the time-dependent
driving. Moreover, a DFT for $p(w)$ ({\ref{eq:cftw})
was checked for this periodic driving with a
symmetric protocol.
Sun determines  $p(w)$ for a potential that switches between a single well
and a double well \cite{sun03}.

\subsubsection{Stochastic resonance}

For a colloidal particle in a double well potential that is 
additionally subject to a modulated linear potential to generate 
conditions of stochastic resonance \cite{gamm98}, distributions
for work, heat and entropy were measured and calculated in
\cite{jop08,impa08}. Other numerical work using the concepts
of stochastic thermodynamics to investigate stochastic resonance
includes \cite{iwai01,dan05,lahi09}.

\subsubsection{NESS in a periodic potential}

In an experiment for this paradigmatic geometry shown in Fig. \ref{fig-ring}
which correspond to the
Langevin equation (\ref{Lv}), the  pdf for total entropy production
has been measured
and compared to theoretical predictions  \cite{spec07}, see Fig \ref{fig-ring-ent}.
Characteristically,
for short  times, this pdf exhibits several peaks
corresponding to the number of barriers the particle has surmounted.
Examples for the pdf for entropy production have also been calculated
in \cite{mari06}.
 For long
times, the asymptotic behaviour of this pdf in the form of a large
deviation function has been calculated numerically in \cite{mehl08}
where an interesting kink-like singularity was found. The same
quantity has also been derived using a variational principle
\cite{nemo11}.

\begin{figure}[t]
\includegraphics[width=6cm]{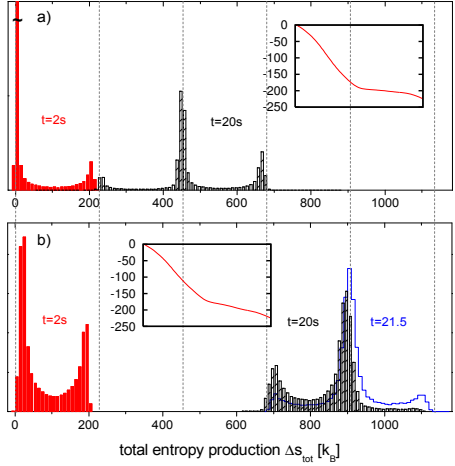}
\caption{Distribution of entropy production $p(\Delta s\t)$ for a colloidal particle driven 
along a periodic potential for two different values of external force $f$\cite{spec07}. 
The insets show the total potential
$v(x) -fx$. The different histograms refer to different trajectory lengths.
}
\label{fig-ring-ent}
\end{figure}

\subsection{Underdamped motion}

\subsubsection{Torsion pendulum as experimental realization}
A driven torsion pendulum differs fundamentally
 from  colloidal particles since here
the inertia term becomes accessible.
In a series of experiments reviewed in \cite{cili10}, 
 Ciliberto and co-workers have investigated
the pdf for various quantities with the aim of checking which ones obey
FT like relationships.

In \cite{doua05,doua05a}, the JR and the Crooks relation were used to
determine the ``free energy difference'' for the torsion 
pendulum for linear and periodic forcing. In \cite{doua06}, pdfs for
the external work were determined for three different types of protocols
for the time-dependent force in (\ref{eq:lv-under}). (i) For a linearly ramped force, the
pdf for starting in equilibrium was found to obey a TFT relation
even for short times. Since a linear ramp corresponds to a time-dependent 
driving, one would, in fact, not expect a TFT. It is found here only
because the work distribution for this linear system is Gaussian and
should obey the JR which constrains mean and variance such that the TFT
is valid. Alternatively, an explicit calculation of the pdf shows the
same result. (ii) Starting in
a quasi-steady state of the linear ramp, the pdf for $w\ex$ no longer
obeys the TFT for short times as expected. (iii) Likewise, for periodic 
driving, a DFT type for the work distribution is found only in the long 
time limit as expected. For the latter two cases, the finite time
corrections have been calculated. 

The same group investigated the pdf for the heat for similar protocols
\cite{joub07a}.
In agreement with both the general theoretical expectations and their
explicit calculations the pdf for heat does neither obey a TFT for
short times nor even the DFT asymptotically for long times since
the internal energy is not bounded for an harmonic oscillator.

The pdf for changes in the stochastic entropy and the total one were
reported for periodic driving of this system in \cite{joub08}. Once
the system has settled in the periodic steady state, for an integer
multiple of the period the functional $R$ becomes the total entropy
change which thus fulfills a DFT as found experimentally. 

\subsubsection{General theoretical results}

Using path integral techniques, Farago determined the statistics of the
power injected by the thermal forces into an underdamped particle and found it
to be
independent of an underlying confining potential \cite{fara02}.

In a series of papers for a moving trap, 
Taniguchi and Cohen investigated pdfs for work and heat as well as various FTs
using the path integral representation \cite{tani07a,tani08,tani07,cohe08}.
They also point out the ambiguity (or freedom) to define time-reversal
in this particular system.

For both a moving and a breathing trap, Minh \ea calculated work weighted
propagators for underdamped motion \cite{minh09a}.
FTs for underdamped Brownian motion 
were studied by Lev and Kiselev by transforming 
from the momentum to the energy variable \cite{lev10}, by Fingerle for the
relativistic version \cite{fing07}
and by Iso \ea for motion near black holes \cite{iso11}.
Sabhapandit determined the work fluctuations of a randomly driven 
harmonic oscillator \cite{sabh11} which was studied experimentally in \cite{gome10}.
FTs for underdamped motion in the presence of two heat baths were 
investigated in \cite{visc06,tome10,foge11}.

\subsection{Other systems}

Ever since the DFT for entropy production has been formulated, there 
have been attempts to show whether it is fulfilled in a specific
system both numerically and experimentally. Beyond the ``clean'' cases
discussed above for which the dynamics of the relevant degrees of 
freedom is clearly compatible with a stochastic Markovian dynamics,
there are a number of studies for other system where a theoretical
understanding is more challenging. Such
studies include an early theoretical work with three simple dissipative
deterministic models \cite{auma01},  
experimental
\cite{feit04,kuma11,naer11,wils11} and numerical 
\cite{visc05,pugl05a,pugl06,pugl06a,sarr10,chon10,droc11,grad11}}
 work for granular matter or dense colloids, experimental work for turbulence 
\cite{cili98,cili04,shan05}, 
numerical work for a shell model in turbulence \cite{gilb04}
and one on the role of hydrodynamic interactions \cite{belu11},
and experimental work for liquid crystals \cite{gold01,joub09},
a vibrating plate \cite{cado08}, and self-propelled particles \cite{suzu11}. Characteristically for 
the experimental systems just mentioned, one cannot necessarily
expect that these can be described by a stochastic Markovian dynamics 
for the relevant variables. 
Similarly, in the numerical works either the equations of motion 
 are not of the Langevin type (or deterministic thermostatted 
ones) or, if they are, not all variables are monitored
which effectively amounts to some coarse-graining.  Since for these cases, the FT's have not been proven, 
such tests give valuable hints on possible extensions beyond their 
established realm of validity.

Two general aspects for putting results of such case studies in
perspective are the following ones.
First, the putative validity of a DFT for the quantity $R$ is 
typically cast in the
form of checking for a constant slope of the quantity $\lim_{t\to \infty}
[\ln [p(\rho_t)/p(-\rho_t)]/t$ where $\rho_t$ is the time averaged rate 
corresponding to the quantity $R=t \rho_t$  for which we have formulated
the FT.
Since such a plot  is necessarily asymmetric in $\rho_t$ \cite{pugl05a}, for a non-trivial
statement the contribution  of higher order terms 
like $\rho_t^3$ 
must be shown to be negligible which requires a large enough range of
studied $\rho_t$-values. Moreover, a large enough $t$ is necessary.  
Second, in bulk systems often only ``local'' quantities 
can be investigated which would require local forms of the FT's. From the
perspective of a stochastic dynamics, this amounts to integrating out
other slow degrees of freedom or some type of coarse-graining
under which one can not expect the FT's to hold necessarily. We will
come back to this issue at the very end of this review.

\section{Dynamics on a discrete set of states}
\label{sec6}
\def\dst{\Delta s^{\rm tot}}
\def\dsm{\Delta s^{\rm m}}
\def\dshk{\Delta  s^{\rm hk}}
\def\dsex{\Delta  s^{\rm ex}}
\def\wmnl{w_{mn}(\lambda)}
\def\wnml{w_{nm}(\lambda)}

\subsection{Master equation dynamics}

The derivation of the FTs in Sect. \ref{sec4}
is based on the behavior of the weight
for a stochastic trajectory under time reversal or the  other
operations generating the conjugate dynamics. Therefore, they hold
for any kind of stochastic dynamics, in particular, for a master
equation type of dynamics   \cite{vankampen,schn76}  
on a discrete set of states $\{n\}$, see Fig. \ref{fig-cycle}.

\begin{figure}[t]
\includegraphics[width=6cm]{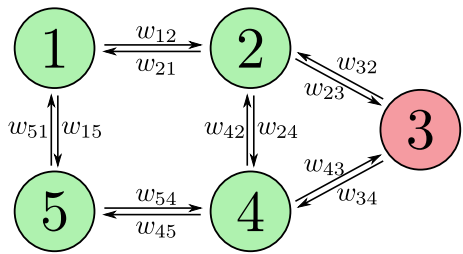}
\caption{Network with five states comprising three cycles (1245), (234), and (12345)
and the corresponding transition rates.
 Without the state 3, this network would be a unicyclic one.}
\label{fig-cycle}\end{figure}

Examples of such systems include random walks and, more
generally,  diffusive processes on a lattice, birth-death processes,
growth processes on a lattice, conformational changes between 
discrete states of a biomolecule or chemical reaction networks. 
The latter two classes of systems differ from the
previous ones since they typically occur in a well-defined
thermal environment. This feature imposes additional constraints on
the dynamics as discussed in Sect. \ref{sec-9} below.
In this section, we focus on the stochastic dynamics with
arbitrary transition rates.

\subsubsection{Transition rates and probability currents}

Transitions from a state $m$ to a state $n$ occur with a rate
 $\wmnl$ which may be
time-dependent according to the external protocol $\lambda (\tau)$.
For ease of notation, we will write $\wmn(\tau)$ or $\wmn(\l)$
 for the 
more explicit
$\wmn(\lambda(\tau))$. In principle, one could distinguish
different transitions or ``channels'' connecting the same two
states from which we refrain here for notational simplicity.

 The probability
$p_n(\tau)$ to find the system at time $\tau$ in state $n$ evolves
according to the master equation
\begin{equation}
  \partial_\tau p_n(\tau) = \sum_{m \not = n}
  [w_{mn}(\tau) p_m(\tau) - w_{nm}(\tau) p_n(\tau)] 
\label{eq:me}
\end{equation}
given an initial distribution $p_n(0)$. To each link $(mn)$ one can
associate a (directed) probability current
\beq
j_{mn}(\tau)=   p_m(\tau)w_{mn}(\tau)- p_n(\tau) w_{nm}(\tau)=-j_{nm}(\tau).
\ee

\subsubsection{Two classes of steady states}

For time-independent $\lambda$,
the system eventually reaches a steady state
provided the network is ``ergodic'', i.e., any two states
are connected through a series of links as we will always assume
in the following.
The time-independent stationary 
probabilities can be written as 
\beq p_n^s(\lambda)\equiv \exp[-\phi_n(\lambda)]
\ee
which defines the analogue of the non-equilibrium potential.
For small networks, there is an elegant graphical method to determine
$p^s_n$ from given rates \cite{schn76,hill}.

 Steady states fall into two classes depending on whether or not
 the
detailed balance condition (DBC), 
\beq
p_n^s(\lambda)\wnml=p_m^s(\lambda)\wmnl
\label{eq:db} ,
\ee is fulfilled. The first case corresponds to genuine
equilibrium, the second one to a NESS. In the latter case, there are
non-vanishing steady state probability currents
\beq
j_{mn}^s \equiv  p_m^sw_{mn}- p_n^s w_{nm} =-j_{nm}^s .
\ee

Since knowing $p^s_n$ is not sufficient to distinguish a genuine
 equilibrium from a NESS, Zia and Schmittmann \cite{zia06,zia07}
suggested to characterize a NESS by its stationary distribution
$p^s_n$ {\it and} its stationary currents $j^s_{mn}$. Then the same
NESS could be generated by a whole equivalence class of possible
rates $\wmn$ since  any two sets of rates with
\beq
p^s_m(\wmn -\wmn')=p^s_n(\wnm-\wnm')
\ee
would lead to the same NESS, i.e., 
the same stationary distribution and currents.
In the following, we will adapt the view that a system
is characterized by a definite set of rates
$\wmn(\l)$ and a protocol $\l(\tau)$ from
which all
other quantities can, in principle, 
be derived.\footnote{For completeness, 
we mention a complementary approach
where rates are derived by imposing mean currents as
 constraints \cite{evan04,baul08,baul10,mont11}. For a relation
to the minimum entropy production principle, see \cite{poll11}
and references therein.} 

A
``distance''  from equilibrium quantifying the amount of
violation of the DBC has been introduced in \cite{plat11}. 
A Lyapunov function for relaxation towards a NESS
is discussed in \cite{maes11}. Master equation dynamics as a gauge
theory is formulated in \cite{poll12}.

\subsubsection{Path weight and dynamical action}

\begin{figure}[t]
\includegraphics[width=6cm]{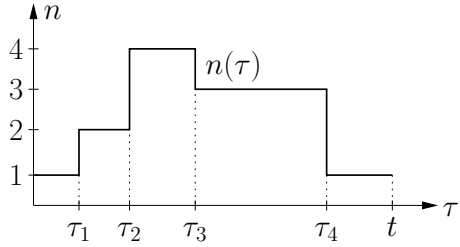}
\caption{Trajectory $n(\tau)$ of total length $t$ jumping
at discrete times $\{\tau_j\}$ between states.
}
\label{fig-traj}\end{figure}

We first characterize the fluctuating trajectories.
A trajectory $n(\tau)$ with $0\leq\tau\leq t$
starts at $n_0$
and jumps at times $\tau_j$ from $n_j^-$ to
$n_j^+$ ending up at state $n_t$ after $J$ jumps, see Fig. \ref{fig-traj}.
Defining for each state the instantaneous total exit rate
\beq
r_n(\tau) \equiv \sum_{m\not=n}w_{nm}(\tau)
\ee
the conditional weight
for a trajectory exhibiting no jump at all is given by
\beq  
p[n(\tau)=n_0|n_0]=\exp[-\int_0^t d\tau ~r_{n_0}(\tau)]  .
\label{eq:pnt0}
\ee The weight for a trajectory with $J\geq1$ jumps 
at times $\{\tau_j\}$
is given
by
\begin{eqnarray}
p[n(\tau)|n_0]= \exp[-\int_0^{\tau_1} d\tau ~r_{n_0}(\tau)] \times~~~~~~~~~~~
\nonumber\\
\prod_{j=1}^J w_{n_j^-n_j^+}(\tau_j)\exp\left[-\int_{\tau_j}^{\tau_{j+1}}
  d\tau ~r_{n_j^+}(\tau)
\right]
\label{eq:pnt}
\end{eqnarray}
 with $\tau_{J+1}\equiv t$.
Averages with these weights will be denoted by $\langle ...\rangle$ in the
following.

In analogy with the continuous case (\ref{eq:PI}), these expressions define an ``action''
\begin{eqnarray}
{\cal A} [n(\tau)]&\equiv& -\ln p[n(\tau)|n_0]
= \int_0^{\tau_1} d\tau ~r_{n_0}(\tau)
\nonumber \\
&-& \sum_{j=1}^J \left[\ln
w_{n_j^-n_j^+}(\tau_j) -\int_{\tau_j}^{\tau_{j+1}}
  d\tau ~r_{n_j^+}(\tau)\right]  .
\label{eq:dis-action}
\end{eqnarray}
\subsection{Entropy production}
\label{sec6-2}
\subsubsection{Stochastic entropy}
The concept of stochastic entropy can be transferred  immediately from the Langevin case
to the  discrete one as
\cite{seif05a}
\beq
s(\tau)\equiv -\ln p_{n(\tau)}(\tau) .
\ee
It is obtained by first solving the master equation 
(\ref{eq:me}) for $p_n(\tau)$
with a given initial distribution  and then plugging into it the
specific trajectory $n(\tau)$ taken from this ensemble.

The equation of motion for stochastic entropy becomes
\beq
\dot s(\tau) = -\left. \frac{\partial_\tau p_n(\tau)}{p_n(\tau)}\right|_{n(\tau)}
 - \sum_j\delta(\tau-\tau_j) \ln{p_{\njp}(\tau_j)\over p_{\njm}(\tau_j)} .
\label{eq:dsm-tau}
\ee
The first term shows that even if the system remains in the same state,
stochastic entropy will change whenever the ensemble is time-dependent either due
to a non-equilibrated initial state or due to time-dependent rates. The second term
shows the contributions from each transition.
 The change
in stochastic entropy during time $t$ is  given by
\beq
\Delta s=\int_0^td\tau\dot s(\tau)=-\ln p_{n_t}(t) + \ln p_{n_0}(0) .
\label{eq:dssys}\ee

\subsubsection{Time reversal}
Time-reversal as a choice for the  conjugate dynamics works analogously
to  the Langevin
case
with $n\dg(\tau)=n(t-\tau)$ and $\l\dg(\tau)=\l(t-\tau)$.
Using the weights (\ref{eq:pnt0},\ref{eq:pnt}), it is  trivial to check
that
\beq
R_1\equiv \ln \frac{p[n(\tau)|n_0]}{p\dg[n\dg(\tau)|n\dg_0]}=   \sum_j
\ln{\wmp(\tau_j)\over\wpm(\tau_j)} \equiv \Delta s\m  .
\label{eq:sm}
\ee

There are  two justifications for identifying $R_1$ with the
entropy change of the surrounding medium $\dsm$. The first is the
analogy to the Langevin case, where this term turned out to be the
dissipated heat (divided by temperature). In the absence of a first
law for the master equation dynamics, which would require further
physical input not available at this general stage, this identification
is by analogy only. Second, it turns out that the sum of the system
entropy change as defined in (\ref{eq:dssys}) and the so identified medium entropy,
\beq
\dst\equiv \Delta s + \dsm ,
\label{eq:stot}
\ee
will have all the properties required from a total entropy production.
From the expression (\ref{eq:sm}) an 
instantaneous entropy production rate in the medium can be identified as
\beq
\dot s\m(\tau) = \sum_j\delta(\tau-\tau_j) \ln{\wmp(\tau_j)\over\wpm(\tau_j)}
\ee
that makes the contributions from the individual jumps obvious.
Combining this relation with (\ref{eq:dsm-tau}), one obtains 
\beq
\dot s\t(\tau) = -\left. \frac{\partial_\tau p_n(\tau)}{p_n(\tau)}\right|_{n(\tau)}
 + \sum_j\delta(\tau-\tau_j) \ln{p_{\njm}(\tau_j)\wmp(\tau_j)    \over
   p_{\njp}(\tau_j)\wpm(\tau_j)
} .
\label{eq:dstot-tau}
\ee

\subsubsection{Ensemble level}

By averaging these trajectory-dependent contributions for entropy production, 
one obtains expressions
derived much earlier within the ensemble approach 
\cite{mou86,schn76,hill,luo84}. On a technical level,
using
\beq
\left\langle \sum_j\delta(\tau-\tau_j) d_{\njm\njp}(\tau)\right\rangle = \sum_{mn}p_m(\tau)\wmn(\tau) d_{mn}(\tau)
\ee
valid for any set of quantities
$\{d_{mn}\}$, one obtains

\beq
\dot S\tot(\tau)\equiv \langle \dot s\tot\rangle=\sum_{mn}p_m(\tau)\wmn(\tau)\ln\frac{p_m(\tau)\wmn(\tau)}{p_n(\tau)\wnm(\tau)} ,
\ee
and
\beq
\dot S\m(\tau)\equiv \langle \dot s\m\rangle=\sum_{mn}p_m(\tau)\wmn(\tau)\ln\frac{\wmn(\tau)}{\wnm(\tau)}
\ee
which should be compared with (\ref{eq:ens-tot}) and (\ref{eq:ens-m}),
respectively.

\subsubsection{Splitting entropy production}

Following the Langevin case, we can split the  entropy
production in the medium (\ref{eq:sm})
into
two contributions
\beq
\dsm = \dshk + \dsex 
\ee
with
\beq
\dshk\equiv \sum_j\ln{
p_{{n_j^-}}^s(\l_j)\wmp(\l_j)\over p^s_{{n_j^+}}(\l_j)\wpm(\l_j)}
\ee
where $\l_j\equiv \l(\tau_j)$ and
\beq
\dsex\equiv - \sum_j\ln{
p_{{n_j^-}}^s(\l_j)\over p^s_{{n_j^+}}(\l_j)}
\ee
characterizing the entropy change associated with maintaining the
corresponding steady state and the one associated with time-dependent
driving, respectively. It is simple  to rewrite $\dsex$ as the discretized 
version of $-\Delta\phi + \partial_\l\phi$ as in (\ref{excess}). This excess
entropy production has a nice geometrical interpretation along a
path in the parameter
space analogous to the Berry phase in quantum mechanics \cite{saga11b}.

A somewhat different splitting of total entropy production on the trajectory level
was introduced in \cite{espo10b,espo10d,vdb10a} writing
\beq
\Delta s\tot=\Delta s\aad + \Delta s\na
\ee
with the adiabatic entropy change $\Delta s\aad\equiv \Delta s\hk$  and the non-adiabatic one $\Delta s\na\equiv \Delta s\ex+\Delta s$.

\subsubsection{Dual dynamics}

The dual dynamics is defined by rates
\beq
\wmn\dg(\l)\equiv \wnm(\l)p^s_n(\l)/p^s_m(\l)  .
\ee These rates lead to the same stationary state as the original dynamics,
${p\dg}^s_m=p^s_m$. 
However, the stationary  currents are reversed according to 
\beq 
{j\dg}^s_{mn}(\l)=-j_{nm}^s(\l)   .
\ee
In complete analogy to the Langevin case, by comparing the weights for the original with the
dual and the dual-reversed dynamics, one obtains 
\beq
R_1 = \dshk  {~~~~~~\rm and~~~~~~~} R_1=\dsex  ,
\ee respectively.

\subsection {FTs for entropy production and ``work''}
 
\subsubsection{General validity}
With these identifications, all FTs from Sects. \ref{sect:ftent} and
\ref{sect:ftuni} involving the various forms of
entropy changes apply under exactly the same conditions as stated
there provided the occasional $x$ (and $x_0,x_t$) is 
trivially replaced by $n$ (and $n_0,n_t$).

The only variable not defined yet is the analogue of work. For networks
that at constant $\lambda$ fulfill the DBC (\ref{eq:db}), one
can identify a (dimensionless) internal energy as
$
\phi_n(\lambda)\equiv -\ln p_n^s(\l)
$
that plays the role of the potential $V(x,\l)/T$  in the Langevin case
with a  free energy identical to zero.
At this stage, there is indeed
no point in identifying a non-trivial  $\lambda$-dependent free energy.
Consequently, 
work along a trajectory corresponds to dissipated
work and can be identified in  analogy to (\ref{eq:work}) with
\beq
w\equiv \int_0^t ~d\tau~ \partial_\lambda\phi_n(\lambda)_{|n(\tau)}
 \dot \lambda  .
\label{eq:ein}
\ee
With this identification of work, the FTs from Sects. \ref{sect:ftwork} and
\ref{sect:ftuni}  
involving work hold for this master
equation dynamics as well (setting there $T=1$). It should be stressed, however, that 
without a more physical microscopic understanding of the network, 
this concept of work (and heat, if one wanted to promote $\dsm$
to this status) is a purely formal one without real physical meaning.

For networks that do not fulfill the DBC (\ref{eq:db}),
there is no unique way of assigning internal energy to a state  without
further physical input,
and, hence, no sensible 
way of identifying work even formally. Naively keeping (\ref{eq:ein}),
as sometimes suggested {\cite{ge10},
fails as the counter-example of a discretized
version of the driven
overdamped motion on a flat ring  easily shows,
since $\phi_n={\rm const}$ implies $w=0$. 

The statistics of rare events contributing to these FTs can also be studied
through a ``mapping''  of the master equation to a Schr\"odinger equation and
then analyzing the corresponding path integral \cite{altl10,altl10a}.
Finally, a somewhat formal general IFT was derived in \cite{liu09}.

\subsubsection{Experimental case studies}

Experimental work measuring the distributions of these quantities with
the perspective of ``testing'' the FTs is yet scarce. The arguably 
simplest non-trivial network is a two-state system with time-dependent 
rates. Any such two-state system necessarily obeys the DB condition.
Such a two-state system was realized experimentally by driving an
optical defect center in diamond with two lasers. The distributions
for work (\ref{eq:ein})
and for the entropy change of system, medium and total, were
measured and compared to the theoretical predictions \cite{schu05,tiet06}. 
Characteristically, for comparably short times
these distributions show quite intricate,
distinctively non-Gaussian, features, see Fig. \ref{fig-ent-diamond}.

\begin{figure}[t]
\includegraphics[width=6cm]{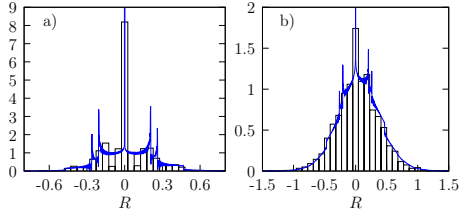}
\caption{Probability distribution for the ``work''
(\ref{eq:ein}), denoted here $R$,
in a driven two-level system for two different lengths of trajectories.
The histogram are experimental data, the full curve a theoretical
calculation, for
more details, see \cite{schu05}.
}
\label{fig-ent-diamond}
\end{figure}

\subsubsection{Analytical and numerical case studies}

Entropy production and the FT for the simplest discrete system which
is in essence a random walk biased in one direction by applying an
external field was studied in the context of a ratchet model in
\cite{monn04}, for a rotary motor in \cite{seif05,saka06} and for transport through
a membrane channel in \cite{bere08}.  Simple three and
four state systems were investigated in \cite{sun10,kuma11a}.
The statistics of dissipated heat for a driven two-level system
modelling single electron transport has been calculated in \cite{aver11}.

Entropy production on a lattice model both for a simple reaction-diffusion 
scheme and for transport was investigated in \cite{doro11}
with an attempt to clarify the occurence of a kink in the rate function
at zero entropy production. In another variant of the reaction-diffusion
scheme, the violation of an FT caused by breaking microscopic reversibility
in the sense that some backward transitions are forbidden were studied in
\cite{doro09,bena11}.  Entropy production for a model of cyclic
population dynamics was investigated in \cite{andr10}
and for effusion of a relativistic ideal gas in
\cite{cleu06}.

The analogue of work distributions for a spin system in time-dependent
magnetic fields  was investigated in 
\cite{mara05,eina09} and for time-dependent coupling constants in \cite{ohze10}. 
The surface tension in the three-dimensional Ising model is determined through 
simulations using the analogy of the JR in \cite{chat07}. 

The interplay between a  non-equilibrium phase transition and singularities in the
entropy production has been investigated for a majority vote model \cite{croc05},
for driven lattice gases \cite{deol11,tome12}, for wetting \cite{bara12} and
for kinetically constrained lattice models \cite{spec11a}.

\subsection {FT for currents}
\label{sec6-4}

\def\jj{{\scriptstyle {\cal J }}}

For a network in a NESS, currents obey an FT as first derived by Gaspard 
and Andrieux \cite{gasp04,andr04,andr06,andr07c}
exploiting the decomposition of such a network in cycles
as introduced by Schnakenberg \cite{schn76} and using concepts from large deviation theory
\cite{touc09}. For a concise derivation of the current
FDT using the formalism developed in
Sect. \ref{sec4}, we write the total entropy production along a trajectory
in the form
\beq
\dst=\sum_a n_a \Delta S_a + \Delta s_r.
\ee
Here,  $n_a$ is the number of times a cycle $a$ has been completed in
clockwise ($n_a>0$) or anti-clockwise ($n_a<0$) direction during
this trajectory leading to a fluctuating current
$\jj_a\equiv n_a/t$ for each cycle, see Fig. \ref{fig-cycle}
for an example of cycles in a network. The entropy production associated with each cycle
\beq
\Delta S_a=\sum_{(mn)\in a} \ln\frac{\wmn}{\wnm}
\label{eq:cycl-aff}
\ee
is also called the affinity of this cycle. The remainder $\Delta s_r$  collects 
the contributions arising from those parts of the trajectory that do not 
contribute to a full cycle. Clearly, the current $\jj_a$ is
uneven under time-reversal and hence qualifies as a possible variable
$S_\alpha$ with $\epsilon_\alpha=-1$ for the general SSFT (\ref{eq:gft3}) which thus becomes
\beq
\frac{p(\{-\jj_a\})}{p(\{\jj_a\})}
=\exp[- t \sum_a \Delta S_a \jj_a] ~\langle e^{- \Delta s_r}|\{\jj_a\}\rangle .
\ee
For large $t$,  $\Delta s_r$ and hence the second factor
remains of order 1. It can thus be ignored when taking the
logarithm in the long-time limit leading to the current FT
\beq
\lim_{t\to \infty}\frac{1}{t}\ln\frac {p(\{\jj_a\})}{p(\{-\jj_a\})}     =  \sum_a\Delta S_a \jj_a .
\ee
For a network coupled to different reservoirs at different temperatures or
chemical potentials, the cycle affinites $\Delta S_a$ arise from externally imposed affinites 
${\cal F}_k$ as discussed in more detail in Sect. \ref{sec10-2} below. These affinites
give rise to mesoscopic currents
\beq
\jj_k=\sum_a \jj_a d^k_a
\ee
where $d^k_a$ are a generalized distance counting how much each cycle 
contributes to the respective current. Expressed in these currents,
the FT reads
\beq
\lim_{t\to \infty}\frac{1}{t}\ln\frac {p(\{\jj_k\})}{p(\{-\jj_k\})}     =  \sum_k{\cal F}_k \jj_k.
\ee
In this derivation, it is crucial that all currents that contribute to the entropy
production (either on the cycle or on the mesoscopic level) are included.

Generalizations of such an FT have been derived for just one current in \cite{andr07a},
for networks with multiple transitions between states in
\cite{fagg11}, to a current not related to entropy production 
for a growth model on a lattice in \cite{bara10} and for 
 networks with a particular topology in \cite{bara11}.
Geometrical and topologial aspects were studied in \cite{sini07,cher09,ohku08,ohku10}. Periodically driven systems
were investigated in \cite{sing10} and a relation to supersymmetry is made in
\cite{sini11a}.
In another extension, Hurtado \ea have derived an ``isometric
fluctuation relation'' that compares  pdfs for currents with
different orientations \cite{hurt11}.

The FT for entropy production has been discussed for various chemical reaction networks in the papers by Gaspard and
Andrieux quoted above, and, more recently, be applied to transport in mesoscopic devices which,
despite their quantum character, can often still be described by a master equation whenever  coherences can be,
or are, ignored. For a few recent examples, see, e.g., \cite{altl10,altl10a,sait08,sanc10,monn10,utsu10,golu11,krau11,nico11,nico11a,cuet11,gane11}.

There is a large literature arising from the recent progress of understanding current fluctuations in NESS 
in general, not necessarily related to the FT, for which the review by Derrida could serve as a point of departure
\cite{derr07}.

\def\vs{~~\\~~\\~\\}

\def\a{_{a}}
\def\jpa{j^+\a}
\def\jma{j^-\a}

\def\jaeq{j^{+\rm{eq}}_a}
\def\jbeq{j^{+\rm{eq}}_b}
\def\jeq{j^{+\rm{eq}}}

\def\Foi{\F_{o,i}}
\def\Joi{J_{o,i}}
\def\Poi{P_{o,i}}
\def\eoi{\epsilon_{o,i}}
\def\woi{w_{o,i}}
\def\xoi{x_{o,i}}

\def\win{{W^{in}}}
\def\wout{{W^{out}}}
\def\sa{\Delta S\a}
\def\tp{\tau^+}
\def\tm{\tau^-}
\def\Ai{A_{\rm in}}
\def\Ao{A_{\rm out}}

\def\dfsa{\Delta{F}^{\rm sol}\a}
\def\desa{\Delta{E}^{\rm sol}\a}
\def\dssa{\Delta{S}^{\rm sol}\a}

\def\wma{w_a^{\rm mech}}
\def\wca{w_a^{\rm chem}}
\def\woia{w_{\{o,i\}a}}
\def\woa{w_{o,a}}
\def\wia{w_{i,a}}

\def\wo{w_o}
\def\bwo{\widehat w_{\rm out}}
\def\wi{w_i}
\def\xo{x_o}
\def\xii{x_i}
\def\etai{\eta^*_{\rm in}}
\def\etao{\eta^*_{\rm out}}
\def\etass{\eta^{**}}

\def\Po{P_o}
\def\Pii{P_i}
\def\hi{h^{\rm in}}
\def\di{d^{\rm in}_{mn}}
\def\he{h^{\rm out}}
\def\de{d^{\rm out}_{mn}}

\def\ds{\Delta S}
\def\sol{^{\rm sol(1)}}
\def\soll{^{\rm sol(2)}}

\section{Optimization, irreversibility, information  and feedback}
\label{sec-7}

\subsection{Optimal protocols}

\subsubsection{General aspects 
}
The integral fluctuation theorems like the Jarzynski relation
hold for any external protocol $\l(\tau)$ and any time interval $t$.
An optimal protocol $\l^*(\tau)$ is the  one that extremizes
the mean of a functional of the trajectory like work or heat
for given initial value $\l_i\equiv \l(0)$ and final value 
$\l_f\equiv \l(t)$ of a control parameter and a fixed total time $t$
allocated to this process. 

Mean work as
objective functions is arguably the most relevant case.
For $t \to \infty$, the minimal mean work required for 
a transition is given by the free energy difference 
$\Delta \F\equiv \F(\l_f)-\F(\l_i)$. For any finite time $t$, the mean work
should be larger and the question for the optimal protocol becomes
non-trivial. Understanding this problem 
will allow to extract the maximum amount of work
 from a given free energy difference in finite time.

Formulated as a variational problem, the optimal protocol obeys
a quite complicated Euler-Lagrange equation which is non-local in time
since changing the control parameter at time $t_1$ affects the work increment
at all later times $t_2$.  Crucial insight into general features of the
solution, however, has been obtained by investigating case studies involving
harmonic potentials \cite{schm07}. As a general feature, jumps of the optimal protocol were
found
that are absent in a linear response treatment \cite{koni05}.

A second motivation for minimizing the work could be an attempt to
improve convergence of the Jarzynski estimate to obtain free energy
differences since one might expect that a small mean work may also to lead
to a smaller variance. Due to the non-linearity of $\exp[-w/T]$, however,
one should rather find the optimal protocol for minimizing $\langle \exp[-2w/T]
\rangle$ which turns out to have jumps as well \cite{geig10}.

\subsubsection{Overdamped dynamics}

The generic jumps in the optimal protocol were first found in case
studies involving harmonic potentials \cite{schm07}.
The simplest case is a process where the center of a harmonic potential
$V(x,\l)=(x-\l)^2/2$
is shifted from $\l_i=0$ to $\l_f$ in a finite time $t$. Such a
shift does not involve any free energy difference. Hence, the mean work
required for this task will approach 0 for $t\to \infty$. For a finite time,
the optimal protocol can be calculated analytically by expressing the mean
work
as a functional of the mean position of the particle which renders the problem
local in time. The optimal protocol (in dimensionless units for time) 
\beq
\l^∗(\tau) = \l_f(\tau + 1)/(t + 2)
\ee
involves two jumps 
\beq
\Delta \l \equiv  \l(0+)- \l_0 = \l_f - \l(t-) = \l_f/(t + 2)
\ee
at the beginning and the end of the process. 
The physical
reason for, e.g.,  the first jump is the fact that with this jump the dissipation rate
is constant throughout the process. If the trap moved with constant speed
without
initial jump, the friction would slowly build up at the beginning of the 
process which ultimately would imply stronger dissipation. 
The size of the in this case symmetric jumps at beginning and end 
 vanishes as $t\to 0$.

Similar jumps have also been found in a second case study where the stiffness of a harmonic
potential $V(\l,x)=\l x^2/2$ is changed in finite time from an initial value 
$\l_i$ to $\l_f$ \cite{schm07}. For overdamped motion of a dipol in a magnetic
field
that switches the orientation, the optimal protocol can even show a degeneracy
\cite{then08}.
Further examples for optimal protocols involving
 non-linear potentials were studied numerically in
\cite{geig10}.

An intriguing mapping of this optimization problem  to deterministic
optimal transport like mass transport by a Burgers velocity field has been discussed in
\cite{aure11,aure12}. For total entropy production as objective function, turning an earlier
scaling argument \cite{schm08} into a mathematical proof, a general bound can thus be derived, 
$\Delta S\tot \geq C/t $ valid for any $t$, where $C$ depends on the given initial and final 
distribution \cite{aure12a}. The key point is that this optimization takes place in the
space of all probability distributions rather than in a restricted space of driving potentials
with a few variational parameters. In the latter case, the $\sim 1/t$ behavior will hold
only for long times. 
 
\subsubsection{Underdamped dynamics}

For under-damped dynamics, the optimal protocol involves even
stronger singularities at beginning and end of the process given by
additional $\delta$-peaks in the protocol \cite{gome08}. 
Physically, these terms guarantee
that the particle  at the beginning acquires, and at the end looses,  a finite 
mean momentum instantaneously which again minimizes total dissipation.

\subsubsection{Discrete dynamics}

Optimal protocols can also be investigated for master equation dynamics on a 
discrete set of states. A simple case has been studied as a model for a
quantum dot with a single  energy level 
$E$ connected to a reservoir with chemical potential $\mu\equiv E
-\eps(\tau)$. The optimal protocol for an externally controllable 
gap $\eps(\tau)$
and given $\eps(0)$ and $\eps(t)$ minimizing the mean work 
$W\equiv \int_0^td\tau p(\tau)\dot \eps(\tau)$, where $p(\tau)$ is the probability
that the energy level is occupied, shows jumps in
the optimal $\eps^*(\tau)$ at beginning
and end which are nicely explained in physical terms in  \cite{espo10a}.
A more general approach to the optimal protocol connecting 
arbitrary given initial and final
distributions is given in \cite{mura12}. 

\subsection{Quantifying time's arrow}

The concepts developed for deriving the FTs can
also lead to a more
quantitative understanding of  irreversibility. In Sect. \ref{sec4}, the 
time-reversed process was 
introduced as a mere tool for deriving the FTs. By considering
such a time-reversed process seriously and comparing it to the original
time-forward process, one can indeed derive an
interesting relation between dissipation
and irreversibility. An essential tool in this analysis is the
relative entropy or Kullback-Leibler distance 
\beq
D[p||q]\equiv \int dy~p(y) \ln [p(y)/q(y)]\geq 0
\ee
between two
distributions $p(y)$ and $q(y)$. Essentially, this quantity
measures how distinct the two distributions are \cite{cove06}.

We present here the stochastic version of the basic idea first introduced
using Hamiltonian dynamics \cite{kawa07,parr09}. For a process with a time-dependent
potential $V(\bx,\l)$ and its time-reversal,
which start and end in the respective equilibrium states,
the quantity $R$ in the generalized FT (\ref{eq:gft2}) becomes $R=w-\Delta\F$. 
This FT thus implies 
\beq
\langle \exp[-(w-\Delta F)]|s_\alpha\rangle = 
p\dg(\epsilon_\alpha s_\alpha) /p(s_\alpha)
\label{eq:d1}
\ee where the average is conditioned on the value $s_\alpha$
of an arbitrary  functional 
$S_\alpha[\bx(\tau)]$ along the trajectory with a definite parity $\epsilon_\alpha$
under time-reversal.
By choosing $S_\alpha= w-\Delta \F$, averaging the logarithm of (\ref{eq:d1}) 
yields
\beq
\langle w\rangle-\Delta \F = D[p(w-\Delta \F)||p\dg(-(w-\Delta \F)] .
\label{eq:diss}
\ee
This relation, which can be seen as a consequence  of the CFT (\ref{eq:cft2}), shows that
the dissipated work determines how different the distributions for
this quantity along the forward and the backward paths are. Likewise,
a large difference of these two distribution
implies a substantial dissipated work.

By choosing  $S_\alpha=\bx(t_1) = \bx\dg(t-t_1)$,
i.e., the state of the system at any intermediate time
$t_1$, one gets from (\ref{eq:d1})
 the lower bound
\beq
\langle w\rangle-\Delta \F \geq D[p(\bx(t_1))||p\dg(\bx\dg(t_1))] .
\label{eq:wfd}
\ee on the dissipated work. In contrast to this stochastic case,
for a Hamiltonian dynamics, one obtains an equality in
(\ref{eq:wfd}) due to Liouville's theorem \cite{kawa07}. 
For both
types of dynamics, further coarse graining, i.e.,
looking at the distributions for a  variable $y=y(\bx(t_1))$, 
leads to a lower bound on the dissipated work 
since relative entropy decreases under coarse graining \cite{cove06}
as nicely illustrated in the present context in \cite{gome08a}.

Similarly, by choosing $t_1=t$, one immediately
obtains the inequality
\beq
\langle w\rangle -\Delta \F \geq D[p(\bx(t))||p\eq(\bx(t))] 
\label{eq:d4}
\ee which bounds dissipation by the distinguishability of
the instantaneous distribution with the corresponding
equilibrium distribution as derived and discussed in \cite{vaik09}.

It is trivial to derive similar relations for processes involving
genuine steady states that at constant control parameter reach a NESS 
rather than equilibrium as pointed out in \cite{blyt08}. Essentially, in 
(\ref{eq:d1}--\ref{eq:d4}), 
one has to replace $\langle w\rangle -\Delta \F$ by $\langle
\Delta s\m + \Delta \phi\rangle$ where $\phi$ is the non-equilibrium
potential (\ref{eq:phi}).

Related inequalities have been discussed for transitions between
specified initial and final states \cite{espo11}.
Relations between other information theoretic measures and
non-linear averages of work and entropy along non-equilibrium trajectories
have been derived in \cite{feng08,feng09,croo11}. An intriguing relation
between generating information and dissipation has been made for DNA replication
in \cite{andr08} with a corresponding pedagogical introduction in \cite{jarz08a}.

\subsection{Measurement and feedback}
\label{sec7-3}

\def\ii{{\scriptstyle {\cal I }}}
\def\I{{\cal I}}
\def\out{^{\rm out}}
\def\inn{^{\rm in}}

\subsubsection{Feedback and the second law}

According to the Kelvin-Planck formulation of the second law, 
one cannot extract work from an equilibrated system at
constant temperature without leaving any traces of this process somewhere
else. The situation becomes apparently
different if information about the 
state of the system during this process becomes available through a measurement
as the classical example of Maxwell's demon and the Szilard engine
reviewed in \cite{leff03,maru09} 
demonstrate.\footnote{For an instructive criticism of one of the assumptions of the 
Szilard engine, see \cite{hond07}.}
Based on the result of a measurement, one can choose a particular protocol 
for a control parameter which will indeed allow either to extract work in a cyclic process
or, in a non-cyclic process, to extract more work than the corresponding free energy difference of initial
and final equilibrium state. These  statements
are still compatible with the second law since erasing the information acquired 
will also cost free energy according to Landauer's principle. Taking this
additional effect into account, the ordinary second law is restored. 
Typically, for discussing these processes within stochastic thermodynamics
the cost of 
measurement and  erasure process is first ignored in the problem of how to convert the 
acquired information
into work (most efficiently)  as it also will be ignored in the following
discussion of the main concept where we use an approach based on FTs. Related
earlier work will be briefly mentioned in Sect. \ref{sec-early}.

\subsubsection{Measurement and information}

For a quantitative description, we assume a system evolving according to a 
master equation as introduced in Sect. \ref{sec6}. If the system at time $t_1$ is in 
state $n_1=n(t_1)$, a measurement yields a 
result $y_1$ with the probability $p(y_1|n_1)=p(n_1|y_1)p(y_1)/p(n_1,t_1)$. 
Here, $p(n,t)$ is the ordinary solution of the master equation for the
given initial condition and $p(y_1)$ the probability for obtaining the result $y_1$ 
irrespective of $n_1$.
The (trajectory-dependent) information acquired in
this measurement is \cite{saga10}
\beq 
\ii(n_1,y_1)\equiv \ln[p(n_1|y_1)/p(n_1,t_1)]=\ln[p(y_1|n_1)/p(y_1)] .
\label{eq:info}
\ee
Upon averaging with $p(y_1|n_1)$, this trajectory dependent
information becomes the relative entropy $D[p(n_1|y_1)||p(n_1)]$
which still depends on the result $y_1$ of the measurement.
Further averaging over $y_1$ leads to the mutual information
\begin{eqnarray}
\I&\equiv& \int dy_1p(y_1) D[p(n_1|y_1)||p(n_1)]\\
&=& \int dy_1\int dn_1 p(n_1,y_1)\ii(n_1,y_1).
\end{eqnarray}

\subsubsection{Sagawa-Ueda equality (SUE) and a generalization}

After a measurement, the control parameter
$\l(\tau,y_1)$ for the subsequent
evolution  $t_1\leq\tau\leq t$ depends uniquely on the outcome $y_1$ 
leading to the probability distribution $p_1(n,\tau|y_1)$.
For a system with a time-dependent potential $V(n,\l)$, i.e., a system
that at any fixed $\l$ reaches a genuine equilibrium state, Sagawa and Ueda
have generalized the JR (\ref{eq:JR})
to this feedback-driven process in the form \cite{saga10,saga12}
\beq
\langle \exp [-(w-\Delta \F+ \ii)] \rangle =1
\label{eq:su}
\ee which implies for the maximal mean extractable work $W\out\equiv -\langle w\rangle$ the
bound
\beq
W\out\leq -\Delta \F +  \I 
\label{eq:su-bound}
\ee with $\I\equiv \langle\ii\rangle$. Thus, acquiring information through a measurement allows to extract
more work than what one would get from a process without feedback.

The original formulation of the SUE requires the notion of
a free energy difference for initial and final state. 
For transitions
involving genuine non-equilibrium states, i.e., those that at constant
control parameter reach a NESS, the analogous relation
\beq
\langle \exp [-(\Delta s\tot+ \ii)] \rangle =1
\label{eq:su2}
\ee
with the inequality
\beq
\langle \Delta s\tot\rangle \geq - \I
\label{ineq:su2}
\ee
hold true as 
well.\footnote{ Note that even for detailed balanced systems, the equalities
(\ref{eq:su}) and (\ref{eq:su2}) are different since, in general
$w-\Delta\F\not=\Delta s\tot$.}

A concise proof of (\ref{eq:su2}), which will be valid with a minor modification
for (\ref{eq:su}) as 
well\footnote{For proving the SUE (\ref{eq:su}), one only needs to
replace  $p_1(n_t,t|y_1)$ by $p\eq(n_t,t|y_1)$ in (\ref{eq:ave}).},
 not requiring explicit time-reversal
 follows from exploiting the IFTs (\ref{eq:gift-feed},\ref{eq:gift-feed2}) \cite{abre11a}. 
By using $\Delta s\tot=\Delta s + \Delta s\m$, splitting the last
term
into the two contributions associated with the two time intervals
$i=(0\leq\tau<t_1)$ and $ii=(t_1\leq \tau\leq t)$, making the total 
entropy change of
the system explicit with
$\Delta s= -\ln p(n_t,t|y)+\ln p_0(n_0,0)$, and the
specific expression for the information (\ref{eq:info}),
the lhs of (\ref{eq:su2})
can be written as 
\begin{eqnarray}
\left\langle \frac{1}{p_0(n_0)}e^{-\Delta s\m_i}\frac{p(n_1,t_1)}{p_1(n_1,t_1|y_1)}
e^{-\Delta s\m_{ii}}p_1(n_t,t|y_1) \right\rangle \nonumber \\
=\sum_{m_1}\underline{\left\langle\frac{1}{p_0(n_0)}e^{-\Delta s\m_i}|n_1=m_1 \right\rangle_i ~p(n_1,t_1)}\times\nonumber\\ 
\times \left\langle e^{-\Delta s\ex_{ii}}  p_1(n_t,t|y_1)   |n_1\right\rangle_{ii}  ~=1 .\label{eq:ave}
\end{eqnarray}
Introducing conditioned averages on the two intervals $i$ and $ii$  eliminates the explicit factor $1/p_1(n_1,t_1|y_1)$. 
The underlined term is $1$ for any $m_1$ due to (\ref{eq:gift-feed}). 
Likewise, the subsequent summation over $m_1$
is $1$ due to  (\ref{eq:gift-feed2}). The SUE thus holds force
for trajectory-averages still
conditioned on the results  $y_1$. Of course, further
averaging over all possible outcomes $y_1$ is allowed. 
This proof (as the original one) is easily extended to multiple 
measurements. Thus, the Sagawa-Ueda equality and its variant (\ref{eq:su2}) with
the corresponding inequalities 
hold for any number 
of measurements \cite{saga12,abre11a,horo10,ponm10,lahi12}. 

For processes involving genuine
non-equilibrium states,  the generalization of the Hatano-Sasa relation
(\ref{eq:hs})  to
processes with feedback in the form \cite{abre11a}
\beq
\langle \exp [-(\Delta s\tot - \Delta s\hk+ \ii)] \rangle =1
\label{eq:su3}
\ee with the inequality 
\beq
\langle \Delta s\tot\rangle \geq  \langle \Delta s\hk\rangle -  \I
\label{ineq:su3}
\ee
follows as easily starting with  conditioned variants of (\ref{eq:gift2}). The bound (\ref{ineq:su3})
 is
much stronger than (\ref{ineq:su2}) since $\langle \Delta s\hk\rangle$ will
typically scale with the total time $t$. For systems that at constant $\l$ exhibit detailed balance, $\Delta s\hk=0$, 
in which case (\ref{eq:su3}, \ref{ineq:su3}) become (\ref{eq:su2}, \ref{ineq:su2}).

\subsubsection{Efficiency of Brownian information machines}
For a cyclically operating  information machine, where measurements are repeated at
regular intervals separated by $t_m$ \cite{baue12}, the inequality (\ref{ineq:su2}) implies
that one can extract at most a mean power $\dot W\out$
bounded by
\beq
\dot W\out\leq   \dot \I ,
\ee where $\dot \I $ is the rate with which information is acquired.
Likewise, for processes involving transitions between genuine non-equilibrium states,
the inequality (\ref{ineq:su3}) implies
\beq
\dot W\out\leq \dot W\inn -\langle \dot q\hk \rangle  +  \dot \I .
\ee
If the rate of acquiring information is large enough,
i.e., if $   \dot \I > \langle \dot q\hk \rangle $, the extracted power
can  exceed the power $\dot W\inn$ required to sustain these
non-equilibrium steady states as demonstrated explicitly with a simple
example in \cite{abre11a}. Characteristically, the power extracted from such a machine
becomes larger, the smaller the intervals $t_m$ between the measurements are.

 A quite natural definition for the efficiency \cite{baue12,cao09}
of such a Brownian information machines obeying  $0\leq \eta\leq 1$
is in the first case
\beq
\eta\equiv  \dot W\out/  \dot \I
\ee
 and, analogously, 
\beq
\eta\equiv  \dot W\out/ [\dot W\inn -\langle \dot q\hk \rangle+ \dot \I]
\ee
for the second case.

\subsubsection{Further theoretical work and case studies}
\label{sec-early}
Several theoretical studies have investigated various aspects of such feedback
driven processes for stochastic dynamics. Kim and Qian have considered an underdamped particle
controlled by a velocity-dependent force \cite{kim04,kim07}. This problem has been analyzed
from the perspective of total entropy production in \cite{muna12}.
Suzuki and Fujitani investigate Brownian motion both 
under a time-dependent force \cite{suzu09} and
for linear systems more generally \cite{fuji10}.
Similarly, Sagawa and Ueda illustrated their
concept using a particle that is transported in a movable 
harmonic trap and can still extract work  from the surrounding heat bath
\cite{saga10}.
Feedback driven transport for ratchet-type systems
has been optimized in \cite{cao04,dini05,feit09}. Maximum power 
for such a model has been studied in \cite{feit07}. Information-theoretic
and thermodynamic concepts have been combined in \cite{touc00,cao09,cao09a}.
The thermodynamic cost of  a measurement has been modelled in \cite{gran11}.

The optimal protocol for extracting the maximal work from cyclic processes
 for particles in harmonic traps
with adjustable center and stiffness based on imperfect positional measurements
has been calculated in \cite{abre11} where it was shown that the bound (\ref{eq:su-bound})
cannot be saturated if only the center of the trap is under control. Only
by additionally adjusting the stiffness, all the information can be recovered 
provided
 an infinite time is allocated to the process. For such a machine, the efficiency
at finite cycle time has been calculated in \cite{baue12}. The issue of
saturating this  bound has been investigated in more depth introducing the
 notion of ``reversible'' feedback by Horowitz and Parrondo \cite{horo11,horo11a}.
A model for the cost of erasing information using a Brownian particle in
a double well potential was discussed in \cite{dill09}.

\subsubsection{Experimental illustrations}
Experimentally, the Sagawa-Ueda equality  has been demonstrated
by using an ingenious set-up involving electric fields that upon
measuring the position of a colloidal particle on a ``stair'' prevent that
the particle slides down a step that it has just climbed by thermal excitation
\cite{toya10a}. In another experiment, 
Landauer's principle has been illustrated using a colloidal particle
trapped in a modulated double-well potential.
The mean dissipated heat indeed saturates at the Landauer bound in the limit of long erasure cycles
\cite{beru12}.

\subsubsection{Hamiltonian dynamics for microcanonical initial conditions}
\label{sec-Ham}
Deviating from the restriction to stochastic dynamics as applied generally 
in this review, I mention a few recent studies that use  Hamiltonian dynamics and
feedback since they provide an additional perspective on what has just been described.
The Kelvin-Planck statement of the second law does not hold for micro-canonical initial conditions
which indeed allow to extract work, i.e., to decrease the mean energy from a Hamiltonian system
by manipulating an external control parameter \cite{alla02}. Specific examples have been given
for a harmonic oscillator in \cite{sato02a}, for a particle between movable walls \cite{mara10}
and for motion in a double well potential \cite{vaik11a}. While for such  microcanonical initial
conditions  no measurement is necessary, these results could be applied to an
initially canonical ensemble if the energy of the system is measured with subsequent adaption
of the protocol of  the control parameter in a feedback process. As shown explicitly in \cite{vaik11a},
the full analysis including the cost of erasing information exorcizes this 
``demon'' and restores the
ordinary second law.

\section{Fluctuation-dissipation theorem (FDT) in a NESS}
\label{sec-8}
\subsection{Overview}
\subsubsection{FDT in equilibrium}
Equilibrium systems react to small perturbations in a quite predictable way
formally expressed by the FDT, see, e.g.,  \cite{kubo}. The response of an observable $A$ at time
$\tau_2$ to  a perturbation $h$ applied at time $\tau_1$ can be written in the
form of an equilibrium correlation function as
\beq
T\delta\langle A(\tau_2)\rangle/\delta h(\tau_1)_{|h=0}\equiv
T R_A\eq(\tau_2-\tau_1)=\partial_{\tau_1}\langle A(\tau_2)B(\tau_1)\rangle\eq ,
\label{eq:fdt-eq}
\ee 
where 
the conjugate variable 
\beq
B=-\partial_hE
\label{eq:fdt-B}
\ee
follows from the energy $E(h)$ of the system. Here it is assumed  that
for any small fixed $h$ the energy of the system is still well-defined. This FDT
is the formalization 
and generalization of Onsager's regression hypothesis that  states that the decay
of an excitation 
is independent of whether it has been generated externally by a force (or field)
$h$ or by a thermal fluctuation. This theorem is of great practical significance 
since it allows to predict the response to a perturbation without ever applying
one just by sampling the corresponding equilibrium fluctuations either in
experiments or in simulations. Characteristically,  the same $B$
holds for any 
$A$  and any time difference $\tau
_2-\tau_1$. 

\subsubsection{FDT in a NESS}
Whether a similarly universal relation exists for NESSs has been addressed using
various approaches since the seventies. For an underlying stochastic dynamics,
Agarwal has expressed the response function by a correlation function
involving the typically unknown stationary distribution
\cite{agar72}. Bochkov and Kuzovlev \cite{boch79,boch81,boch81a}
and H\"anggi and Thomas \cite{hang82} have derived a variety of formal
expressions for stochastic processes. A  comprehensive review of the general relation between fluctuations and
response including,  in particular, 
deterministic chaotic systems is given in \cite{marc08}.

More recently, taking up a theme introduced earlier \cite{cugl97},
 Harada and Sasa derived a relation where the ``violation'' of the
 equilibrium FDT in a NESS was related to the rate of energy dissipation
for a Langevin system \cite{hara05,hara06} later generalized to a description
in terms of a density field \cite{hara09}. For the special case of a driven colloidal particle it
was shown 
in \cite{spec06} that the FDT in the NESS could be obtained from the
equilibrium FDT (\ref{eq:fdt-eq}) by 
subtracting from the right-hand side a second correlation function involving the
local mean velocity. 
This result suggested that in the locally co-moving frame the Onsager
hypothesis could be restored which was later extended to sheared systems
\cite{spec09} and proven for general
diffusive dynamics in \cite{chet08,chet09}. Thus, for these systems, the decay
of an 
excitation around a local NESS is still the same whether generated externally
or by the thermal fluctuations still present in the NESS.

A concise formal derivation and discussion of the general FDT in a NESS has been given by
Baiesi \ea \cite{baie09,baie09a,baie09b}. The response of particular observables was treated
at the same time  by Prost \ea \cite{pros09}. In \cite{seif09}, it was then
shown that the latter result holds indeed for any observable and that the FDT 
for a NESS becomes particularly transparent when using the concept of
stochastic entropy 
with its  splitting into a total and a medium one. In this latter work, the apparent
multitude of  FDTs in a NESS was rationalized in terms of an
equivalence relation holding for observables in NESS correlation functions.

An elegant synthesis using mathematically somewhat more demanding concepts has
just been given in \cite{chet11}. An extension of these concepts
to obtain an FDT around
non-stationary non-equilibrium states is derived in \cite{verl11}.
A connection with gauge fields is made in the geometrical approach  of
\cite{feng11}.

\subsubsection{Effective temperature}

\def\eff{^{\rm eff}}
\def\ac{^{\rm ac}}

The derivation of recent exact versions of the FDT for a NESS which
as a result typically express the response function by a sum of two correlation 
functions should be distinguished from the phenomenological concept of an
effective temperature that has been reviewed in \cite{cugl11}. Originally introduced in the
context of ageing systems, it can be formulated also for a NESS.  Simply stated, 
guided by the equilibrium form (\ref{eq:fdt-eq}) an effective temperature is defined as
\beq
T\eff(A,\tau_2-\tau_1)
\equiv \partial_{\tau_1}\langle A(\tau_2)B(\tau_1)\rangle^s/R_A(\tau_2-\tau_1)
\ee 
where $R_A$ is now taken in the NESS. 
In general, $T\eff$ will depend on both the observable $A$ and the time-difference 
$\tau_2-\tau_1$ which upon Fourier transformation corresponds to frequency $\omega$.
Obviously, this concept can become meaningful only if these dependencies are not too pronounced.

From a theoretical point of view, a strictly observable and frequency-independent
$T\eff$ follows for a Langevin system like (\ref{eq:Lvbf}) with $\bforce=0$ if
the non-equilibrium conditions are 
 caused by an additional ``active'' noise $\bfeta$ with correlations 
$\langle \bfeta(\tau_2)\bfeta(\tau_1)\rangle = 2 (T\eff-T) \bm\delta(\tau_2-\tau_1)$.
For a linear system and active noise correlated on a scale $\tau\ac$, $T\eff$
will depend on frequency. For $\omega\tau\ac\gg1$,
one gets the ordinary temperature, whereas for $\omega\tau\ac\ll 1$ the 
enhanced fluctuations lead to a larger $T\eff$.

On the other hand, if the non-equilibrium is generated by a non-conservative force of field,
such a simple reasoning is no longer possibe. Still, in interacting systems
one often finds numerically good agreement with the concept of an effective temperature
as briefly mentioned  in Sect. \ref{sec-shearsusp} below for sheared suspensions. A fundamental
 understanding when and why this is the case in general seems still to be missing.

For insight into the frequency and observable dependence for specific models and systems, 
see, e.g,  \cite{bert02,oher04} for a binary Lennard-Jones mixture in a simple shear flow, 
\cite{fiel02} for a glassy model system, 
\cite{sreb04,shok06,mart09,mart10} for
simple interacting model systems, \cite{cugl94,cugl97a,zamp05} for simple Langevin systems,
and  \cite{cala04}
for field-theoretical models.

 Examples of investigating  biophysical systems using an effective temperature
include
\cite{mart01} for hair bundle oscillations, \cite{mizu07,levi09} for the cytoskeleton,
 \cite{lego02,kiku09} for filament oscillations in an active medium, 
\cite{loi08} for self-propelled particles, \cite{mann01} for
vesicle and \cite{betz09,beni11} for
red-blood cell fluctuations.  If the response of such a system acts effectively
``against'' the perturbation, the effective temperature becomes negativ
as occasionally found in these studies. Such an observation shows 
that this concept should not be taken too literally.

An intriguing, apparently simple system in this context is ``hot Brownian motion''
where a diffusing particle  heated by a laser acts as a local heat source
\cite{radu09,ring10,ring11,chak11} for which in simulations various ``effective temperatures''
were determined \cite{joly11}. Local heating in the context of the FT has been discussed theoretically
in \cite{prad08}.

\subsection{Derivation and discussion}

In this section, we sketch the derivation of the various forms of the FDT in a
NESS from 
a unifying perspective  for a general Markovian dynamics on a discrete set of
states. Since overdamped Langevin systems can always be discretized,
 this case is a very general one.
We follow the concepts introduced in \cite{baie09,seif09} which were
briefly
reviewed in their continuum version in \cite{spec10}. Earlier related work for
a Markovian dynamics on a discrete set of states making somewhat more
explicit assumptions on observable and rates for spin models include 
\cite{ricc03,cris03,chat04,lipp05,diez05,deol07,corb07,lipp08} and 
for ageing in supercooled liquids \cite{bert07}.

\subsubsection{Equivalent correlation functions in a NESS}
\label{sec-equi1}
We consider a class of NESSs with rates $\wmn(h)$ that depend on a perturbation
$h$. The stationary 
distribution of the master equation dynamics (\ref{eq:me})
obeys
\beq
\sum_n L_{mn}(h)p^s_n(h)=0
\label{eq:master-ness}
\ee
with the generator
\beq
L_{mn}(h)\equiv \wnm(h)-\delta_{mn}\sum_kw_{mk}(h) .
\label{eq:gen}
\ee
For fixed $h$, any dynamic information is fully contained in the propagator
\beq
G_{kl}(\tau)\equiv {\rm prob}[n(\tau)=k|n(0)=l]
\ee
for which the master equation (\ref{eq:me}) implies the evolution
\beq
\partial_\tau G_{kl}=\sum_m L_{km}G_{ml}(\tau)=\sum_m G_{km}(\tau)L_{ml} .
\label{eq:prop}
\ee
In a NESS, denoted in the following with $\langle ... \rangle^s$, 
 two point correlation functions for 
state variables of the form $A(\tau)=\sum_mA_m\delta_{n(\tau)m}$ are given by
\beq
\langle A(\tau_2)B(\tau_1)\rangle^s = \sum_{mn} A_mG_{mn}(\tau_2-\tau_1)B_np^s_n
\ee if $\tau_2>\tau_1$. Using (\ref{eq:gen}) and (\ref{eq:prop}), 
a time derivative with respect to the earlier time can thus be written as an
ordinary two-point 
correlation function
\beq
\partial_{\tau_1} \langle A(\tau_2)B(\tau_1)\rangle^s =  \langle
A(\tau_2)C(\tau_1)\rangle^s
\ee
with 
\begin{eqnarray}
C_n &=& B_n\sum_m \wnm-\sum_m B_m\wmn p^s_m/p^s_n\nonumber\\
&=&\sum_m(B_n-B_m)p^s_m\wmn/p^s_n .
\end{eqnarray}

Besides state variables as observables we will also need current
 variables of the type
\beq
\label{eq:Dfdt}
D(\tau)\equiv \sum_j\delta(\tau-\tau_j)d_{\njm \njp}
\ee
that yield $ d_{\njm \njp}$ whenever a corresponding transition takes place. Their NESS average is
given by
$
\langle D(\tau)\rangle^s = \sum_{mn}p^s_m\wmn d_{mn} .
$ 
If $D(\tau)$  shows up in a correlation function at the earlier time $\tau_1$,
 we get 
\begin{eqnarray}
\langle A(\tau_2)D(\tau_1)\rangle^s &=&
\sum_{mkl}A_mG_{ml}(\tau_2-\tau_1)p^s_kw_{kl}d_{kl}\\
&=&\langle
A(\tau_2)E(\tau_1)\rangle^s
\end{eqnarray}
with 
\beq
E_n=\sum_kp^s_kw_{kn}d_{kn}/p^s_n .
\label{eq:Efdt}\ee
These relations imply, in particular, that the formal current variable
\beq
\dot B(\tau)\equiv  \sum_j\delta(\tau-t_j)(B_{\njp}-B_{\njm})
\ee
obeys
\beq
\langle A(\tau_2)\dot B(\tau_1)\rangle^s = \partial_{\tau_1}\langle A(\tau_2)
B(\tau_1)\rangle^s
\ee
which demonstrates, quite expectedly, that time derivatives  can be pulled in
and out  a NESS correlation function straigthforwardly.

The fact that NESS correlation functions can have the
same value  if the variable at the earlier time is written differently gives
rise to an equivalence 
relation denoted by
\beq
O^{(1)}(\tau)\cong O^{(2)}(\tau)\ee
 if
\beq
\langle A(\tau_2)O^{(1)}(\tau_1)\rangle^s= \langle A(\tau_2)O^{(2)}(\tau_1)\rangle^s
\ee
holds for all $ A$ and times $\tau_1<\tau_2$ \cite{seif09}. 
For the variables defined above we obviously have
\beq
D(\tau)\cong E(\tau)~~~{\rm and} ~~ \dot B(\tau)\cong C(\tau) 
\ee
 which summarizes how  current variables and time derivatives can be replaced by state
 variables in NESS correlation functions. 
This freedom will explain why apparently so differently looking FDTs can be
derived for a NESS.

\subsubsection{Equivalent forms of the FDT}
\label{sec-equi}
The apparent plethora of FDTs can be rationalized by starting with an expression for the
response function using the path weight (\ref{eq:dis-action}). In the presence 
of a time-dependent
perturbation $h(\tau)$, the mean value of the observable $A(\tau)$
is given by \cite{baie09,seif09}
\begin{equation}
  \begin{split}
    \mean{A(\tau)} &= \sum_{n(\tau)} A(\tau)p[n(\tau);h(\tau)|n_0]p^s_{n_0} \\
    &= \sum_{n(\tau)}
    A(\tau)\frac{p[n(\tau);h(\tau)|n_0]}{p[n(\tau)|n_0]}p[n(\tau)|n_0]p^s_{n_0} .
  \end{split}
\label{eq:response}
\end{equation}
 The response
function
\begin{equation}
  R_A(\tau_2-\tau_1) \equiv \left.\delta \langle A(\tau_2)\rangle/\delta h(\tau_1)\right|_{h=0}
\equiv \langle A(\tau_2) B^p(\tau_1)\rangle^s 
\end{equation}
can be expressed by a two-point correlation function
in the unperturbed NESS by evaluating (\ref{eq:response}) with  the action (\ref{eq:dis-action})
as
\begin{eqnarray}
  B^p(\tau_1) &=& - \left. {\delta{\cal A}[n(\tau);h(\tau)]}/\delta
  h(\tau_1)\right|_{h=0} \\
& =&  -\sum_kw_{n(\tau)k}\alpha_{n(\tau)k} +
  \sum_j\delta(t-\tau_j)\alpha_{n_j^-n_j^+}.
\label{eq:Bp}\end{eqnarray}
where
\beq
\alpha_{mn}\equiv \partial_h\ln\wmn(h)_{|h=0} .
\label{eq:alpha}
\ee
This form of the conjugate variable (with the superscript$^p$ alluding to the
derivation through the path weight) is convenient since it allows to determine
the response function by measuring a correlation function that requires only
knowledge about how the rates depend on the control parameter which is easily 
availabe in simulations. The more formal aspect that the first term in (\ref{eq:Bp})
arises from the time-symmetric part of the action and the second one from its
time-asymmetric one is emphasized and further exploited in \cite{baie09,baie09a,baie09b}.

A second equivalent representation of the conjugate variable is obtained by
replacing (by following the scheme (\ref{eq:Dfdt}-\ref{eq:Efdt})) 
the second  (current) part in $B^p$ by its equivalent  state variable form
which leads  to $B^p\cong B^a$ with
\begin{eqnarray}
B^a_n&=&    -\sum_kw_{nk}\alpha_{nk} + \sum_k w_{kn}\alpha_{kn}p^s_k/p^s_n 
\label{eq:Ba1}\\
&=& \sum_k \left. \partial_h L_{nk}(h)\right|_{h=0}
p^s_k/p^s_n .
\label{eq:Ba2}
\end{eqnarray}
The last equality follows from expanding (\ref{eq:gen}) in $h$ and the definition ({\ref{eq:alpha}).
This expression for the conjugate variable involving only state variables can
also 
be derived by straightforward time-dependent perturbation theory of the
Fokker-Planck equation as originally derived by Agarwal (hence, the superscipt$^a$) 
\cite{agar72}. 
Using this expression, however, requires knowledge of the stationary
distribution
which for interacting systems with many degrees of freedom is not  easily
available
in either simulations or experiments. 

Finally, as a third, arguably physically most transparent form of the conjugate
variable, it is easy to check explicitly that
$
-\partial_h\dot s\cong B^a 
$ by
expanding (\ref{eq:master-ness}) in $h$  and following the recipe of how to pull a
time-derivative
 into a correlation function given in the previous subsection.
Consequently, one has  \cite{seif09}
\begin{eqnarray}
  R_A(\tau_2-\tau_1)&=&-\langle\partial_h\dot s(\tau_1)\rangle^s
\label{eq:fdt-s}
\\
&=& \langle\partial_h\dot s\m(\tau_1)\rangle^s -
\langle\partial_h\dot s\t(\tau_1)\rangle^s  .
\label{eq:fdt-sm}
\end{eqnarray}
The first form expresses the response function as a time-derivative of a
correlation function where the observable $A(\tau_2)$ is correlated with the
$h$-derivative of the 
stochastic entropy at $\tau_1$.  This form of the conjugate variable is 
actually  unique if one
wants to write the 
response function as a time derivative of a correlation function. Moreover, it
allows 
a physically transparent interpretation by splitting it into the sum of 
medium and total entropy production as shown  in the second line.

\subsubsection{Comparison with equilibrium FDT}

For a comparison with the equilibrium FDT assume that 
 the steady state is a
genuine equilibrium state for $h=0$. In fact, two classes of such systems should
be distinguished. 

First, if the system is not only in equilibrium at $h=0$ but
also at small $h$,  the stationary distribution is given by the
Boltzmann distribution
\begin{equation}
  \label{eq:5}
  p\eq_n(h)=\exp\{-[E_n(h)-\F(h)]/T\},  
\end{equation}
where $E_n(h)$ is the internal energy and $\F(h)$ the $h$-dependent free energy of the system. The
stochastic entropy obeys $s_n(h)=-\ln p\eq_n(h)=[E_n(h)-\F(h)]/T$. Along an
individual trajectory, $\F(h)$ is constant and hence we have
\begin{equation}
 T \partial_h{\dot s_n(h)}_{|h=0} = 
 \partial_h{\dot E_n(h)}_{|h=0}
\end{equation}
Inserting this equivalence into  (\ref{eq:fdt-s}), the FDT 
acquires its well-known equilibrium
form (\ref{eq:fdt-eq},\ref{eq:fdt-B})
involving the observable conjugated to $h$ with respect to
energy.

Second, a system may be in equilibrium at $h=0$ but driven into a NESS even at
constant small $h$. The paradigmatic example is a perturbation through shear
flow for which there is no corresponding $E(h)$ for any $h\neq0$. For such
systems, the equilibrium  FDT can still be written  in the form (\ref{eq:fdt-s}) but also in the
pure state form with $B^a$ from (\ref{eq:Ba1},\ref{eq:Ba2}).

\subsubsection{Systems with local detailed balance}

A further comparison between the equilibrium and the NESS-FDT is 
instructive
for systems for which the perturbation enters the ratio of the rates in the
form of a local detailed balance condition
\beq
\frac{\wmn(h)}{\wnm(h)} = \frac{\wmn(0)}{\wnm(0)}
\exp [h d_{mn}/T] ,
\ee 
where $d_{mn}=-d_{nm}$ is the distance conjugate to the
field covered by the transition $m\to n$.
For $h=0$, the system is supposed to be in genuine equilibrium with averages denoted by
$\langle ... \rangle\eq$; for $h=h_0\not = 0$ a genuine
NESS denoted by  $\langle ... \rangle^s$ is reached. In equilibrium, using 
the global detailed balance condition (\ref{eq:db}),
one easily verifies $\partial_h\dot s\m \cong B^a$ and hence 
one has the equilibrium FDT  
\beq
 T R\eq_A(\tau_2-\tau_1) = \langle A(\tau_2) B^a(\tau_1)\rangle\eq =  \langle
 A(\tau_2) \partial_h\dot s\m (\tau_1) \rangle\eq  .
 \ee
On the other hand, the NESS-FDT  in the form (\ref{eq:fdt-sm})
always holds.
Since for 
such systems 
\beq
\partial_h\dot s\m = \sum_j\delta(\tau-\tau_j)d_{\njm,\njp}/T
\ee
is independent of $h$, the recipe for getting the FDT in a NESS from the
equilibrium FDT is to keep as a first term the observables showing up
in the correlation
function but to evaluate the latter under NESS conditions and to 
subtract an expression involving the total entropy production \cite{seif09}. 

\subsubsection{Generalized Green-Kubo relations}

In equilibrium, the Green-Kubo relations express transport coefficients like 
conductivity or viscosity by time-integrals over equilibrium correlation functions
of the corresponding currents. Based on the FDT derived above, it is possible to
derive similar relations between transport coefficients in a NESS and 
appropriate current-current correlation functions \cite{seif10}
as illustrated for a simple models of molecular motors in \cite{verl11a}.
This approach of studying the linear response of a NESS
should be distinguished from extensions of the Onsager symmetry
relations to the non-linear response coefficients of an equilibrium system
\cite{andr07b}.

\subsection{Colloidal particle on a ring as paradigm}
The overdamped particle driven along a periodic potential, see Fig. \ref{fig-ring}, 
as discussed in Sect. \ref{sec-ness} can serve as
paradigm for illustrating the different versions of the FDT \cite{seif09}. 

\subsubsection{Equivalent correlation functions}

The equivalence relation introduced in Sect. \ref{sec-equi1}  for variables occuring in
correlation functions in a NESS exists for continuous variables as well.
For a discretized position variable, jump rates can easily be derived from
discretizing the path integral. Going then through the steps as in
Sect. \ref{sec-equi} shows that the equivalence
\beq
\dot x \cong 2 v^s(x) - \mu F(x) 
\label{eq:dotx}
\ee 
can be used in a NESS correlation function at the earlier 
time.\footnote{In equilibrium, this
equivalence becomes $\dot x \cong \mu \partial_xV(x)$
where the crucial sign difference compared to naively ignoring the noise
in the Langevin equation (\ref{Lv}) should be noted.}
The mean local
velocity $v^s(x)$ has been introduced in (\ref{eq:sc}). 
 Sometimes, the
generalization
\beq
g(x)\dot x \cong g(x) [ 2 v^s(x) - \mu F(x)] - \mu T \partial_x g(x)
\label{eq:gxness}
\ee
is useful which can be derived 
similarly.\footnote{Applied to a NESS correlation function with $A(\tau_2)=1$,
this relation leads to
$\langle g(x) \dot x\rangle^s=\langle
g(x)[2v^s(x)-\mu F(x)]\rangle^s -\mu T \langle \partial_xg(x)\rangle^s= \langle g(x)v^s(x)\rangle^s$
which corresponds to (\ref{eq:gdotx}) applied to a NESS.}

\subsubsection{Three equivalent forms}

First, consider a NESS generated by a force $f_0$ which is further
perturbed
by an additional delta-like force impuls acting at time $\tau_1$. The response
function  can be written as correlation function in
the three equivalent forms
\begin{eqnarray}
   T R_A(\tau_2-\tau_1)_{|h_0\not = 0}&=&\langle A(\tau_2) [\dot x-\mu
   F(x)]_{|\tau_1}/2\rangle^s
\label{eq:fdt-ring1}\\
&=& \langle A(\tau_2) [v^s(x)-\mu
   F(x)]_{|\tau_1}\rangle^s
\label{eq:fdt-ring2}\\
&=& \langle A(\tau_2) [\dot x-v^s(x)]_{|\tau_1}\rangle^s .
\label{eq:fdt-ring3}
\end{eqnarray}
The first form follows from
applying perturbation theory to the path integral
expression. The advantage of this expression is that it does not require
explicit knowledge of the stationary distribution. By replacing the velocity
 with the corresponding
state variable as shown  in (\ref{eq:dotx})
one obtains the second form. 
This expression can also easily directly obtained
from perturbation theory of the Fokker-Plank equation as in the original
derivation \cite{agar72}.
Finally, a simple linear combination of the first two lines leads to the
third form originally first derived in \cite{spec06}. In this form, both the
additive correction to the equilibrium form 
$ T R_A(\tau_2-\tau_1)= \langle A(\tau_2) \dot x(\tau_1)\rangle\eq$
and the significance of a
locally co-moving frame become apparent.

\subsubsection{Generalized Einstein relation}
The Einstein relation (\ref{eqE}) connecting the bare mobility $\mu$ 
embedded in a viscous fluid with its diffusion constant $D$ has arguably
been 
the first form of an FDT which was based on a microscopic understanding of thermal 
motion. This relation has many manifestations in more complex soft matter
systems as reviewed in \cite{frey05}. If such a particle is in a periodic
potential $V(x)$, the diffusion coefficient
\beq
D[V(x)]= \lim_{t\to\infty}[\langle x^2(t)\rangle - \langle x(t)\rangle^2]/2t
\label{eq:Dv}
\ee
and the effective mobility
\beq
\mu[V(x)]=\partial_f \langle\dot x\rangle_{|f=0}
\label{eq:muV}
\ee
still obey $D[V(x)]=T\mu[V(x)]$ even though both terms are exponentially suppressed
if the barriers exceed the thermal energy.

In the presence of a non-zero base force, an effective diffusion constant
$D[V(x),f]$
and a mobility $\mu[V(x),f]$ as in (\ref{eq:Dv}) and (\ref{eq:muV}) evaluated at 
a finite force, respectively,
are still defined. The effective mobility is the time-integrated response
function. Hence, the generalized Einstein relation between $D[V(x),f]$ and
$\mu[V(x),f]$  follows
from integrating (\ref{eq:fdt-ring3}) from $\tau_2=-\infty$ to 
$\tau_2=\tau_1$ as \cite{spec06}
\beq
T\mu[V(x),f]=D[V(x),f] + \int_0^\infty d\tau [\langle \dot x(\tau)-\langle \dot
x\rangle][v^s(x(0))-\langle \dot x\rangle\rangle^s] 
\label{eq:ERG}
\ee
which shows how the ``violation'' of the usual Einstein relation
can be expressed as an integral over velocity correlation functions. This relation is a simple
example of a Green-Kubo relation generalized to a NESS \cite{seif10}.
Its form for other simple geometries like two-dimensional motion in the
presence of a magnetic field has been studied in \cite{baie11} and for a discrete model
showing anomalous diffusion in \cite{vill11}.
\subsubsection{Experiments}
The  generalized Einstein relation (\ref{eq:ERG}) has been measured experimentally in
\cite{blic07}. Significantly, in this experiment, the extra integral term
in (\ref{eq:ERG}) can be about four times as big as $T\mu[V(x),f]$ which 
shows clearly that this experiment probes a genuine NESS far from any linear
response regime of an equilibrium system. Still, the description of the
colloidal motion by a Markovian Brownian motion 
with unaltered thermal noise and a drift obviously remains a faithful
representation. 
The very fact that around  a critical force $f\simeq {\rm max}|\partial_xV(x)|$ the
diffusion coefficient becomes quite large is known as giant diffusion 
\cite{reim01,reim02}.

The time-resolved version of this FDT has been studied experimentally in
\cite{mehl10} where it was shown that even though the different correlation
functions (\ref{eq:fdt-ring1}-\ref{eq:fdt-ring3}) are theoretically equivalent their statistics can be vastly
different. Not surprisingly, the variant (\ref{eq:fdt-ring2}) 
involving only state functions
shows better convergence properties than the ones requiring  $\dot x$.
The response not to a force but to a change in the amplitude of the periodic
potential was studied experimentally in \cite{gome09,gome11}.

\subsection{Sheared suspensions}
\label{sec-shearsusp}
For studying the relation between
fluctuations and response in interacting non-equilibrium
systems, a colloidal suspension in shear flow provides a paradigmatic
case. Such a system follows a dynamics as introduced in
Sect. \ref{sec-264} with $\bu=\dot \gamma y{\bf e}_x$ (or the
corresponding underdamped version) and some pair interaction $V$.

One obvious question is to investigate the generalized Einstein relation
between the self-diffusion coefficient $D_{ij}(\dot \gamma)$ and the mobility 
$\mu_{ij}(\dot \gamma)$
of a tagged particle which both become tensorial quantities in such an
anisotropic system. Szamel \cite{szam04} studied these quantitities analytically using
the memory-function formalism. Kr\"uger and Fuchs \cite{krue09a} have studied this
relation analytically and numerically near the glass transition. Our
numerical study in the fluid phase \cite{land10} revealed that for moderate densitities
the results can surpringly well be expressed as an effective temperature since
the ratios $D_{ii}(\dot \gamma)/\mu_{ii}(\dot \gamma)$ of the diagonal
elements become isotropic with a roughly quadratic 
increase with shear rate. This effective temperature which turns out to be
the kinetic one can be rationalized by comparing this interacting system to
a harmonically bound single particle in shear flow \cite{land11}.
The response to a
perturbation in the shear rate has been investigated in \cite{spec09} and the one
to a static external long wave-length perturbation in \cite{zhan11,szam11}.

Further studies of the general FDT for sheared suspensions include the integration
through transient formalism \cite{krue09,krue09b}. One advantage of this approach
is that all quantities can be expressed in terms of (albeit complicated)
equilibrium correlation functions. The response to
a time-dependent additional shear strain has been studied numerically in \cite{spec09}
using essentially the form (\ref{eq:fdt-s}) that makes the excess compared to the
equilibrium case explicit. The relation between the violation of the equilibrium FDT and
energy dissipation using field variables has been addressed in \cite{hara09}.

\section{Biomolecular systems}
\label{sec-9}

\subsection{Overview}

Single molecules and (small) biomolecular networks constitute a paradigmatic
class of systems to which the concepts of stochastic thermodynamics can be applied.
Conformational changes of single molecules have become observable through a variety of
methods often summarized as single molecule techniques \cite{rito06,selv07,deni08}. 
There are essentially  two ways of exposing such a molecule
that is embedded in an aqueous solution of well-defined temperature
containing different
solutes at specified concentrations 
to non-equilibrium conditions.
 First,  one can apply a (possibly time-dependent) mechanical force 
if one end is  connected via polymeric spacers to the tip of an AFM or to
beads in an optical tweezer. This set-up allows to study, e.g., force-induced
unfolding of proteins. Another source of non-equilibrium are
unbalanced  chemical reactions catalyzed by the
enzyme under study. In combination with a mechanical force, this set-up allows
 in particular to resolve individual steps of a molecular motor and to
measure force-velocity curves of such a molecular machine 
 \cite{howard,schl03}.

The fluctuating
conformations of biomolecules in non-equilibrium
can be described in two ways \cite{rito08,bust04,kuma10}. 
First, one can model the observable degrees of freedom like the end to
end-distance of a protein by a continuous degree of freedom subject to
a Langevin equation. Such an approach is particularly
appropriate for studying
force-induced
un- and refolding of biopolymers, 
in particular,
with the perspective of
 recovering  free energy differences and even 
landscapes from  non-equilibrium experiments as reviewed in 
Sect. \ref{sec9-3} below.

Second, one
can identify discrete, distinguishable states between which (sudden) transitions
take place as it has often been done to model molecular motors
\cite{fish99,lipo00,qian00a,bust01,bake04,andr06c,wang07,gasp07,liep07,liep07a,liep08,liep09,lipo09,lipo09a,lau07a,kolo07,astu10}. 
In most of these works the focus has been on elucidating the
cycles involved in the action of the motor and on deriving force-velocity curves and their
dependence on ATP and ADP concentrations.
A combination of both types of models
has been used for 
describing molecular motors by a Langevin dynamics in a ratchet 
potential that depends explicitly on the current chemical state of the motor
as reviewed in \cite{juel97,astu02,reim02a,parr02}.

From the perspective of stochastic thermodynamics, one would like to
formulate a first law, discuss entropy production and derive the
corresponding fluctuation theorems on the single molecule level.
Within a discrete state description, a first law along
an individual trajectory has been
discussed for single enzymes in \cite{schm06a,seif05,min05},
for molecular motors in \cite{bake04,liep07,liep07a,liep08,liep09,lipo09,lipo09a},
and for small biochemical reaction networks in \cite{gasp04,shib00,schm06}.
Fluctuation theorems without explicit reference to a first law
were discussed for such systems in \cite{andr06c,lau07a,seif04,jarz05,qian06a,laco08,laco09}.

From a theoretical point of view, there are essentially two new aspects
that enter the stochastic thermodynamics of biomolecular systems
 beyond a  naive combination of the stochastic thermodynamics of
colloids as developed
in Sect. \ref{sec2} and the discrete dynamics as introduced in Sect. \ref{sec6}. 
First, the rates are not arbitrary as in Sect. \ref{sec6} 
but are rather constrained by
thermodynamic consistency \cite{schn76,hill,qian08a}
as discussed in detail below. Second,
 each of the states visited along a stochastic trajectory 
contains many microstates. Transitions between these (unobserved) microstates
are fast so that thermal equilibrium is reached within each state. Transitions
between the states, however, are slower, observable and can be driven by
external forces, flows or chemical gradients. As a consequence,
 each of the states 
described by stochastic thermodynamics carries an intrinsic entropy 
arising from the coupling to the fast polymeric degrees of freedom and to
those of the heat bath. This effect must
be taken into account in any consistent identification of heat on the
single trajectory level \cite{schm06a,lipo09,seki07,seif11}. Some of
the 
earlier studies quoted above missed this contribution and, hence, 
failed to identify the dissipated heat correctly.

How these systems can be described from the perspective of  a thermodynamic
engine will be pursued in Sect. \ref{sec-10}.

\subsection{Biopolymers with continuous degrees of freedom}
\label{sec9-2}
\def\wm{w^{{\rm mech}}_{\rho}}

\def\F{{\cal F}}
\def\E{{\cal E}}
\def\S{{\cal S}}
\def\bx{{\bf x}}
\def\cx{{\cal C}_\bx}

 In this section, we  show how
a continuum description fits into the framework developed for systems without
relevant internal degrees of freedom like a colloidal particle. This
systematic
presentation is largely original, even though some of the essential concepts
are implicit in most of the works dealing with free energy recovery
discussed in Sect. \ref{sec9-3} below.

\subsubsection{Thermodynamic states from a microscopic model}

A biopolymer contains a large number of coupled microscopic degrees
of freedom most of  which will not be accessible in experiments. 
Still, these microscopic degrees of freedom affect 
 processes  on a larger scale that are 
described by stochastic thermodynamics.
The microscopic configurational degrees of freedom collectively denoted by
$\xi$ are subject to a microscopic potential energy $\Phi(\xi,\l)$ containing
the interactions within the molecule (and possibly with some of the
 surrounding solvent and solute 
 molecules). The dependence on $\l$ allows for an external potential
like an AFM or an optical tweezer whose position can be controlled
through $\l$.

Under non-equilibrium conditions,  an external force (or field or flow) is applied
to the molecule leading to conformational changes  apparent through, e.g., 
 a changing end-to-end
distance. Such a quantity is an example of 
  a meso-scale description that involves a
certain number of variables denoted by ${\bx}$. Each such  state 
effectively
comprises many micro states. Formally, one can  split all microstates
$\{\xi\}$ in  classes
$\cx$ such that each $\xi$ belongs to exactly one $\cx$. The dynamics of $\bx$
is supposed to be slow and observable whereas equilibration among the micro
states
making up one state $\bx$ is fast. Under this crucial assumption, the
conditioned probability $p(\xi|\bx,\l)$ that a microstate is occupied is given by
\beq
p(\xi|\bx,\l) = \exp[-(\Phi(\xi,\l)-F(\bx,\l))/T]
\label{eq:pxl}
\ee
with the constrained free energy
\beq
F(\bx,\l)\equiv E(\bx,\l)-TS(\bx,\l) \equiv -T \ln \sum_{\xi\in \cx}
\exp[-\Phi(\xi,\l)/T], 
\label{eq:cfe}
\ee
  the constrained  intrinsic entropy 
\beq
S(\bx,\l)\equiv -\partial_TF(\bx,\l) = - \sum_{\xi\in \cx} p(\xi|\bx,\l)\ln p(\xi|\bx,\l) 
\label{eq:Sint}
\ee
 and  constrained internal energy 
\beq
E(\bx,\l)= \sum_{\xi\in \cx}\Phi(\xi,\l)p(\xi|\bx,\l).
\label{eq:cie}
\ee

This model includes but is more general than a more conventional description
of the configurational potential in the
additive form
\beq
\Phi(\bx,\xi,\l)=\Phi^0(\bx,\l)+\Phi^{\rm int}(\bx,\xi,\l) + \Phi^{\rm med}(\xi,\l)
\label{eq:Phi}
\ee made up by a system, a coupling (of arbitrary stength)  and a potential for the degrees of freedom
of the medium,
which all may depend on the control parameter. Here, all degrees of freedom are split into 
those of the system $\bx$ and those of the heat bath $\xi$. By replacing $\Phi(\xi,\l)$
in (\ref{eq:pxl},\ref{eq:cfe}) with
$\Phi(\bx,\xi,\l)$ and summing without constraint
over all $\xi$, the relations (\ref{eq:pxl}-\ref{eq:cie}) remain true
for the potential (\ref{eq:Phi}).

\subsubsection{First law}

For a time-dependent $\lambda(\tau)$, representing, e.g., the center of 
 a moving laser trap, the
increment
in applied work 
reads
\begin{eqnarray}
\dbar w&=& \sum_{\xi\in \cx}
\partial_\l\Phi(\xi,\l) p(\xi|\bx,\l) ~ d\l\\
&=&\partial_\l F(\bx,\l)d\l=dF(\bx,\l) -\bg F(\bx,\l)d\bx  .
\label{eq:work-c}
\end{eqnarray} 
In the first equality, for a changing external parameter the work arising from the microscopic interaction
$\Phi(\bx,\l)$ is expressed 
as an average over all
microstates contributing to the state with (fixed) $\bx$. The second 
equality follows with (\ref{eq:pxl}). 
Compared to the expression in the colloidal case (\ref{eq:work}), the essential difference here
is that $F(\bx,\l)$ is a free energy rather than a bare potential, i.e., internal
energy.
Consequently, the first law that, of course, should involve internal energies rather than
free energies becomes
\beq
\dbar w = dE(\bx,\lambda) + \dbar q= dF(\bx,\lambda) + T dS(\bx, \lambda) + \dbar q .
\label{eq:w-cc}
\ee
This relation together with  (\ref{eq:work-c}) implies for the increment in heat
\beq
\dbar q=-\bg F(\bx,\lambda) d\bx - T dS(\bx,\lambda),
\ee
which makes the contribution to heat that arises from the intrinsic entropy
$S(\bx,\l)$ clear.

\begin{figure}[t]
\includegraphics[width=6cm]{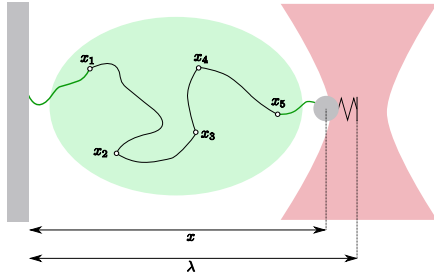}
\caption{Schematic view of a protein stretched by a bead in a laser trap.
In a meso-scale description, the configuration is characterized by the
positions ${\bf x}=\{x_1,x_2,x_3,x_4,x_5,x)\}$ where $x$ is the position of the
bead. The control parameter  $\l$ denotes the center of the trap.}
\label{fig-granada}
\end{figure}

For a practical evaluation of the work, one would have to know $F(\bx,\l)$,
which, in general, has a complicated $\l$-dependence if the microscopic
potential $\Phi(\xi,\l)$ is genuinely $\l$-dependent. However, if the external
potential
couples only to the slow degrees of freedom $\bx$
as typically assumed,  see Fig. \ref{fig-granada}, a significant
simplification occurs. In this case, one can write
\beq
F(\bx,\l) = F^0(\bx)+ V(\bx,\l)=E^0(\bx)-TS^0(\bx)+ V(\bx,\l),
\label{eq:F-simple}
\ee
where the quantities with superscript$^0$ are the thermodynamic
potentials (\ref{eq:cfe}-\ref{eq:cie})
of the molecule for constrained slow variables $\bx$
in the absence of the external potential. As a consequence
\beq
\dbar w=\partial_\l V(\bx,\l) d\l
\ee which becomes trivial for the typical case of a harmonic potential
$V(\bx,\l)=k(x_i-\l)^2/2$, with $x_i$ the relevant coordinate
for the coupling and $k$ the effective stiffness of the AFM tip or optical
trap
centered at $\l(\tau)$.

\subsubsection{Dynamics}
For the dynamics of the slow degrees of freedom one has the Langevin equation
\beq
\dot\bx = \bm[-\bg F(\bx,\l)] + \bz  
\label{eq:Lv-c}
\ee
with the noise correlations as in (\ref{eq:znoise}). Likewise, the
Fokker-Planck equation reads
\beq
\partial_\tau p(\bx,\tau) =  \bg(\bm\bg F(\bx,\l) p(\bx,\tau)+T\bm\bg
p(\bx,\tau)),
\label{eq:fp2}
\ee

 Compared to the
discussion in Sect. \ref{sect:many} the key point here is that whenever states
carry intrinsic entropy, the gradient of the free energy (rather than of internal
energy) has to show up in the
Langevin and Fokker-Planck  equations since for any fixed $\l$, the system
has to reach equilibrium with the Boltzmann factor
\beq
p\eq(\bx,\l)=\exp[-(F(\bx,\l)-\F(\l)/T]
\ee
with the $\l$-dependent free energy
\beq
\F(\l)\equiv -T \ln \int d\bx \exp[-F(\bx,\l)/T] .
\label{eq:F-c}
\ee

\subsubsection{Entropy production}

The stochastic entropy along the trajectory $\bx(\tau)$ becomes
\beq
s(\tau)\equiv -\ln p(\bx(\tau),\tau)
\ee
where $p(\bx,\tau)$ follows from solving the Fokker-Planck equation
(\ref{eq:fp2})
 with
an appropriate initial condition. 
For such a system with intrinsic entropy,
the total entropy production along a trajectory during time $t$
\beq
\Delta s\t\equiv \int_0^t d\tau \dot s\t = \int_0^t d\tau 
[\dot s(\tau)+\dot S(\bx,\l)  + \dot q/T] \ee
contains three contributions rather than two as in the case
without relevant intrinsic degrees of freedom.

\subsubsection{FTs}

In principle, the  FTs holds true in the presence of
intrinsic entropy as well provided the latter is taken into account properly.
The crucial point is that the master functional $R_1$ defined in (\ref{eq:R})
when using as conjugate
process the time-reversed one becomes
\beq
R_1=\Delta s\intr + q/T
\label{eq:R1c}
\ee
where
\beq
\Delta s\intr \equiv S(\bx_t,\l_t)-S(\bx_0,\l_0) 
\ee  is the change in intrinsic entropy along the forward trajectory.
 This result follows from evaluating the action in the path weight
corresponding
to the Langevin equation (\ref{eq:Lv-c}) and by using the first law
(\ref{eq:w-cc}) integrated
along the trajectory.

As a consequence, the FTs involving entropy production 
discussed in Sects. \ref{sect:ftent} and
\ref{sect:ftuni} essentially hold 
true with the replacement
\beq
\Delta s\m \to \Delta s\m + \Delta s\intr .
\label{eq:replace}
\ee 
All FTs involving total entropy production hold true
 unmodified.  

The FTs involving work as defined in
(\ref{eq:work-c}) hold true as well. In particular, the
JR stands with $\Delta \F=\F(\l_t)-\F(\l_0)$ where the free energies
have been defined in
(\ref{eq:F-c}). The reason why the intrinsic entropy does not spoil these
relations is the fact that both the work and the force in the
Langevin equation are determined by the free energy
 $F(\bx,\l)$ very much in the same way as the bare protential $V(x,\l)$
determines the corresponding quantities in the colloidal case.
Loosely speaking, the results of the simpler case hold true provided one
replaces the potential $V(x,\l)$ by the 
free energy (i.e, potential of mean force)
$F(\bx,\l)$.

In the
presence of intrinsic entropy, the FDT (\ref{eq:fdt-sm})
must be modified accordingly by replacing $\dot s\m$ by $\dot s\m + \dot
s\intr$.
\subsection{Free energy recovery from non-equilibrium data}
\label{sec9-3}
\subsubsection{Hummer-Szabo relation (HSR) and variants}
From a practical perspective, arguably the most relevant FT for biomolecules
 is the Hummer-Szabo relation \cite{humm01,humm05}. As a kind of JR resolved along a
reaction coordinate, it allows to determine the free energy landscape
$F^0(\bx)$ from 
non-equilibrium work measurements through an external potential $V(\bx,\l)$ 
 conditioned on a fixed value of $\bx$. It
reads
\begin{eqnarray}
\exp[-F^0(\bx)/T]&=&\exp[(V(\bx,\l_t)-\F(\l_0))/T]\nonumber\\
&~&\times \langle\exp[-w/T]\delta(\bx_t-\bx) \rangle .
\label{eq:hsr}
\end{eqnarray}
The right hand side is evaluated by measuring the accumulated work $w$
as a function of position $\bx_t$ irrespective of the particular $t$.

A concise derivation \cite{seif07} of the HSR
starts with the IFT (\ref{eq:gift}). With the necessary 
replacement (\ref{eq:replace}), 
the initial equilibrium distribution
$p_0(\bx,\l_0)=\exp[-(F(\bx_0,\l_0)-\F(\l_0)/T],$
the free choice $p_1(\bx_t)=\delta(\bx_t-\bx)$, the first law (\ref{eq:w-cc})
and the assumption (\ref{eq:F-simple}), it follows within a couple of lines.

A variant of the HSR not requiring the position histograms in (\ref{eq:hsr})
can be derived for a harmonic coupling $V(\bx,\l_t)$ \cite{humm10}. Moreover,
similarly as the CFT generalizes the JR by including information from the
time-reversed process,  bidirectional variants of the HSR have been derived
and tested in model calculations \cite{minh08,nico10}.

\subsubsection{Experiments}

The first experimental application of the JR to biomolecules was the
determination of the free energy involved in partially unfolding RNA
hairpins \cite{liph02}. The CFT was first applied in another experiment
measuring the free energies in RNA hairpins and  RNA three-helix junctions
\cite{coll05}. In a series of experiments, Ritort and
co-workers have used the CFT to determine the free energy involved
in unfolding DNA hairpins \cite{juni09,moss09,mano09,moss09a}. The free
energy involved in unfolding the multidomain protein, titin, has been
measured in \cite{harr07a} using a simplified variant of the
JR leading to some criticism \cite{frid08,harr08}. The CFT has
been used to determine free energy changes induced by
mechanically unfolding coiled-coil structures \cite{born08,gebh10}
and different topological variants of a 
protein \cite{shan10}. Axis-dependent anisotropy in protein
unfolding was investigated using the HSR in \cite{nome07}.
The free energy landscape derived from the HSR has been compared to equilibrium measurements in
\cite{gupt11}

\subsubsection{Numerical work}

As relatively scarce  real experimental studies using the FTs still are, as large
is the number of ``numerical'' experiments illustrating the use of the
HSR and its variants for recovering free energy landscapes.
The following brief list is necessarily incomplete.
Model calculations for a single coordinate deal with the
advantage of applying a periodic
force protocol \cite{brau04}, with random forcing \cite{minh06},
comparison with ``inherent structures'' \cite{impa07b},
with motion in a periodic potential \cite{berk08}, and with recovering  an
unknown spatially dependent mobility \cite{kosz06}. Multidimensional landscapes were
reconstructed in \cite{minh07}.
Monte-Carlo or molecular dynamic simulations were used for an off-lattice
model protein \cite{impa07}, for a protein domain
in \cite{mitt09} and for a membrane protein in \cite{prei07}.

An important line of research in this context is to 
find  methods
 for dealing with the error caused
by having only finite (and even noisy) data for evaluating  the 
non-linear averages involved in the JR and the HSR.
Some of the  papers dealing with this issue are
\cite{mara08,sun03,humm01a,zuck02,park03,gore03,zuck04,park04,ytre04,ober05,jarz06,mara06,pres06,west06,lech07,vaik08,hahn09,nico09,ober09,goet09,lind09,zima09,poho10,minh11,minh11a,vaik11,davy11,pala11,kund11}.

\subsection{Enzymes and molecular motors with discrete states}
In this section, we show how the general framework for a Markovian dynamics on
a discrete set of states can
be adapted to describe the
stochastic thermodynamics of enzymes and molecular motors in a thermal
environment starting again from a microscopic model.
Apart from keeping track of the intrinsic entropy of the states
the essential point is
to
incorporate the
enzymatic reactions involving solute molecules consistently.

\subsubsection{Thermodynamic states}
\def\enz{^{\rm enz}}
The enzyme is in an aqueous solution which consists of 
molecules of type $i$ with concentrations  $\{c_i\}$ and chemical potentials
$\{\mu_i\}$ enclosed in a volume $V$ at a  temperature
$T$. It  exhibits a set of states  such that equilibration
among microstates corresponding to the same  state
is fast whereas transitions between these states are assumed to be
slower and observable. Under these conditions, one can assign to
each state $n$ of the enzyme a free energy $\Fn$, an internal energy $\En$, and
an intrinsic entropy $\Sn$.
 As explicitly discussed in \cite{seif11}, 
these quantities follow from any microscopic model that specifies the
energy of the microstates of enzyme and solution. They
obey the usual thermodynamic
relation 
\beq
\Fn=\En-T\Sn 
\label{eq:fes}
\ee
despite the fact that the enzyme is small. Moreover, there is no need 
to assume that the interaction between enzyme and solution is somehow
weak. In general, these thermodynamic variables of the enzyme
depend on the concentrations of the various solutes.

\subsubsection{First law}
\label{sec-742}

In this section,  we  discuss the first law for
the three classes of (i) pure conformational changes, (ii) enzymatic
reactions including binding and release of solutes,
 and (iii) motor proteins.

{\sl (i) Conformational changes:}
If the enzyme   jumps from state $m$ to state $n$, the change in 
internal energy
\beq
\Delta E\enz \equiv \En-\Em= -q
\label{eq:qe}
\ee
must be identified with an amount $q$ of heat being released into
(or, if
negative, being taken up from) the surrounding heat bath since there is
no external work involved.

{\sl (ii) Enzymatic reactions:}
The more interesting case are enzymatic transitions that involve
binding of solute molecules $A_i$, their transformation while bound, and  finally 
their release from the enzyme.
Quite generally, one
 considers  transitions 
written as
\beq
\mr + \sum_i \ri A_i  \rightleftharpoons 
\nr + \sum_i \si A_i
\label{reactm}
\ee
where $1\leq \rho \leq N_\rho$ labels the possible transitions.
Here, $\mr$ and $\nr$ denote the states of the enzyme
before and after the reaction, respectively. For $\si=0$,
this scheme describes pure binding of solutes, and
for $\ri=0$  release of bound solutes. A transformation (like bound ATP to 
bound ADP + P$_i$) can also be described by the above scheme
with  $\ri=\si=0$ and the understanding that the enzyme states
contain the bound solutes, see \cite{seif11} for a more detailed
discussion.

The free energy difference 
involved in such a  transition 
\beq \dfr=\der-T\dsr \equiv\dfer +\dfsr
\label{eq:dsr}
\ee has two contributions
where 
\beq
\dfer=\deer-T\dser\equiv \Fe_\nr-\Fe_\mr
\ee
denotes the free energy change of the enzyme and
\beq
\dfsr= \desr-T\dssr\equiv \sum_i(\si-\ri)\mu_i\equiv \Delta \mu_\rho 
\label{eq:dmr}
\ee
denotes the free energy change attributed to the solution in this reaction.
Both free energy contributions can as usually be split
into internal energy and intrinsic entropy.

As in the case of pure conformational changes, one assigns a
first law type energy balance to each reaction of type $\rho$  
({\ref{reactm}). Once an initial
state
is prepared, in the closed system (enzyme plus solution) there is obviously
no source of external work. Neither does the system perform  any work.
Hence, the heat released in this transition is given by minus the change of internal
energy of the combined system \cite{seif11}
\beq
q\rr =  -\der =-\deer
-\dmr - T\dssr .
\label{eq:heatrelax}
\ee
This relation shows  that the enzyme and the solution are treated on the
same footing since only their combined  change in internal energy
enters. Since
the heat is released into the solution acting as a thermal bath, the 
configurational change of the enzyme
as well as  binding and releasing solute molecules contribute to the same bath.

{\sl (iii)  Molecular motors:}
Essentially the same formalism
applies to an enzyme acting as a molecular motor often described by such
discrete states. 
Most generally, if the motor undergoes a forward 
transition of type $\rho$ 
as in
(\ref{reactm}) it may advance a distance $d_\rho$ in the direction of
the applied force $f$ (or, if $f<0$, opposite to it). The special cases $d_\rho =0$ (pure
chemical step) or  $\si=\ri=0$ (pure mechanical step) are allowed.
 For $d_\rho \not = 0$, the 
mechanical work 
\beq
\wm\equiv fd\rr
\ee
is  applied to (or, if negative, delivered by)  the
motor.
 
The motor operates in  an
environment where the concentration of molecules
like ATP, ADP or P$_{\rm i}$ are essentially fixed. The first
law for a single transition of type $\rho$ becomes \cite{seif11}
\beq
q_\rho = \wm - \Delta E_\rho 
= fd_\rho -\deer -\dmr -T\dssr  .
\label{eq:qmot}
\ee

\subsubsection{Role of chemiostats: Genuine NESS conditions}
\label{sec-chem}

The more recent form (\ref{eq:qmot}) of the first law
differs from an expression discussed previously for molecular motors
\cite{bake04,liep07,liep07a,liep08,liep09,lipo09}. 
There, in the present notation and sign convention,
the first law for a step like in (\ref{reactm})
reads 
\beq
\bar q_\rho = \wm -\deer -\dmr .
\ee
The difference between the two expressions for the heat 
\beq
\bar q_\rho - q_\rho =  T \dssr
\ee involves the entropy change in the
solution resulting from the reaction. 

The physical origin of the two different forms
arises from the fact that in the older work
the enzyme is thought to be coupled to ``chemiostats'' providing and
accepting molecules at an energetic cost (or benefit) given by
their chemical potential. Introducing
the notion of a ``chemical work'' 
\beq
w_\rho^{\rm chem} \equiv  - \Delta \mu_\rho
\ee
the first law is then written in the form
\beq
\wm + w_\rho^{\rm chem} = \deer + \bar q_\rho  .
\label{eq:first-chem}
\ee

The concept of chemiostats is supposed to guarantee that the concentration
of solute molecules remains strictly constant. Physically, it could be
implemented by an ATP buffer of ATP-regeneration scheme that involves
additional
enzymes. From the perspective of stochastic thermodynamics
as long as one focusses on single tranistions, however,
it would be more appropriate to treat these additional enzymes and the
chemical reactions they catalyze in the same way as the reaction involving
the motor protein. It turns out that if these additional reactions 
operate quasistatically, then $\bar q_\rho$ is the dissipated heat that 
under steady state conditions would enter
an ensemble average \cite{seif11}. Therefore, choosing the heat $\bar q\rho$
is appropriate whenever one deals with strict NESS conditions while not
wanting to consider
the heat involved in enforcing these conditions explicitly as an extra
contribution.
On the trajectory level for a single motor protein, there seems to be no
sensible way for assigning $\bar q_\rho$ instead of $q_\rho$ to an individual
transition.

\def\wrp{w_\rho^+}
\def\wrm{w_\rho^-}
\def\wrpm{w_\rho^\pm}
\def\dsrt{\Delta s^{\rm tot}_\rho}
\def\dst{\Delta s^{\rm tot}}
\def\dsmr{\Delta S^{\rm med}_\rho}
\def\prp{p_\nr}
\def\prm{p_\mr}

\subsubsection {Stochastic trajectory and ensemble}

A trajectory  of  the enzyme can be characterized by the sequence of
jump times
$\{\tau_j\}$ and the sequence of reactions $\{\rho_j^{\sigma_j}\}$ where $\rho_j$
denotes the corresponding reactions (\ref{reactm}) and $\sigma_j= \pm$ characterizes the
direction in which the reaction takes place. 

An ensemble
is defined by specifying (i) the initial probability $p_n(0)$ for finding
the enzyme in state $n$ and (ii) the set of rates $\wrpm$ 
with which the reactions (\ref{reactm})
take place in either direction. Both inputs will then determine the
probability
$p_n(\tau)$ to find the enzyme in state $n$ at time $\tau$.

\subsubsection{Rates and local detailed balance}

An identification of entropy production along the trajectory requires
some input from the rates determining the transitions. For the simple
case of pure conformational changes, $m\rightleftharpoons n$,
choosing rates that obey 
\beq
\frac{\wmn}{\wnm}  =\exp[-(F_n\enz-F_m\enz)/T] 
\label{eq:wrho-conf}
\ee
is required by thermodynamic consistency. Indeed, only this choice guarantees
that irrespectively of the initial conditions the  ensemble 
will eventually reach thermal equilibrium, 
\beq
p_n(\tau)\to p\eq_n \equiv \exp[-(\Fn-\F\enz)/T]  ,
\ee
with the free energy of the enzyme
\beq
\F\enz\equiv -T \ln\sum_n\exp[-\Fn/T] .
\ee Fixing the ratio of the rates still leaves  one
free parameter per pair of states which can only be determined
from knowing the dynamics of the underlying more microscopic model.

The corresponding relation for transitions that involve enzymatic
reactions (\ref{reactm}),
\beq
\frac{w_\rho^+}{w_\rho^-}  =\exp[- \Delta F_\rho/T]= \exp[- (\dfer + \Delta
\mu_\rho)/T], 
\label{eq:wrho}
\ee 
 and for transitions of molecular motors, 
\beq
\frac{w_\rho^+}{w_\rho^-}  = \exp[-(\dfer + \Delta
\mu_\rho -  \wm)/T] , 
\label{eq:wrhom}
\ee are somewhat less obvious.
Essentially, three types  of justifications for choosing such ratios
can be given.

First, even though microreversibility is often invoked it seem
unclear  how to obtain these ratios  rigorously using this concept
if chemical reactions are involved. 

Second, one can derive (\ref{eq:wrho})
using the following argument. For any enzymatic reaction there will be concentrations
$\{c_i\eq\}$ of the solutes
such that the enzyme will reach equilibrium. For these particular 
equilibrium concentrations,
a choice of rates respecting (\ref{eq:wrho}) is 
mandatory as in the case of pure conformational changes.
If one now assumes that (i) the reaction rates obey the mass action law
and that  (ii)  the concentrations and the chemical potentials are related
by the ideal solution expression, $\mu_i(c_i) = \mu_i\eq+T\ln(c_i/c_i\eq)$,
 then the form ({\ref{eq:wrho}) follows.

Third, more recently it has been shown that by requiring a consistent
stochastic thermodynamic description on the trajectory level, one can 
indeed derive these conditions on the rates  using rather mild assumptions \cite{seif11}.

In all cases by invoking the respective first laws
(\ref{eq:qe},\ref{eq:heatrelax},\ref{eq:qmot}), 
the ratio of the rates can also be written in the form
\beq
\frac{w_\rho^+}{w_\rho^-}  =\exp[\dsr + q_\rho/T ]
\label{eq:wrho-heat}
\ee 
showing  that this ratio is determined by the
change of intrinsic and medium entropy involved in this transition.
This important relation should be compared with (\ref{eq:q-rev}) in the
colloidal case where the continuum version of such a ratio involves only
the dissipated heat since there is no relevant intrinsic entropy 
change given here by $\dsr$. Similarly, for a biopolymer within a continuum
description,
the  relation (\ref{eq:R1c}) shows the contribution of intrinsic entropy.

\subsubsection{Entropy production and FTs}

The total entropy production involved in one forward transition $\rho$ 
at time $\tau$ can be derived from the general expression (\ref{eq:dstot-tau})
and by using the ratio of the rates (\ref{eq:wrho-heat})  as
\beq
\dsrt(\tau) =  \ln\frac{p_\mr(\tau) \wrp}{p_\nr(\tau)\wrm} =\Delta s_\rho + \dsr +q_\rho/T  .
\ee
It consists of three contributions. The first is the change in
stochastic entropy,
\beq
\Delta s_\rho(\tau) = - \ln[p_\nr(\tau)/p_\mr(\tau)] .
\label{eq:stoch}
\ee
 The second denotes the change in the intrinsic
entropy (\ref{eq:dsr}) of the system which consists here of enzyme and
surrounding solution. The third term arises from the dissipated
heat (\ref{eq:heatrelax},\ref{eq:qmot}). 

Summing over all reactions taking place  up to
time $t$ and adding the concomitant change in
stochastic entropy, $\Delta s =-\ln p_{n_t}(t)+\ln p_{n_0}(0)$,
 one obtains the total entropy production along a trajectory
\beq
\Delta s\t= \Delta s + \sum_j \sigma_j[\Delta S_{\rho_j}(\tau_j)+q_{\rho_j}(\tau_j)/T]  ,
\label{eq:dst-c}\ee
where $\sigma_j=\pm1$ denotes the direction in which the transition $\rho_j$ takes
place
at time $\tau_j$.

The arguably most relevant situation for an enzyme modelled by discrete
  states is a NESS generated by non-equilibrated solute concentrations
and/or an applied external force in the case of a motor protein. In such a
NESS, one has the SSFT (\ref{eq:dft2}) for the total entropy production as defined in 
(\ref{eq:dst-c}).

\subsubsection{Time-dependent rates and work}
So far, it was implicitly assumed
that the rates are time-independent. 
Time-dependent rates could arise  either since the concentrations of the solutes
are externally modulated (or, in a finite system, depleted due to the action of the
enzymes) or since the forces applied to motor proteins 
are time-dependent. The ratio of the rates
is then still constrained by (\ref{eq:wrho}, \ref{eq:wrhom}).
 However, under such time-dependent external conditions
characterized by a parameter $\l(\tau)$, 
the thermodynamic state variables $E_n,S_n$ and $F_n$ can become
time-dependent as well.
In consequence, there are contributions to the first law and to entropy
production even if the enzyme remains in the same state. Specifically, if
the enzyme remains in state $n$, in analogy to (\ref{eq:work-c}) the first law  becomes
\beq
\dbar w_n=\partial_\l F_n d\l = (\partial_\l E_n - T \partial_\l S_n)d\l
=\partial_\l E_nd\l + \dbar q_n  .
\ee
Hence,  there is exchanged heat,
$\dbar q_n=-TdS_n$, even if the system remains in the same state whenever the
intrinsic
entropy depends on a changing external parameter (like, e.g., the concentration
of the solutes).

These expressions of heat and work resemble those of quasistatic processes as
they should since it is implicitly assumed that the distribution of
 microstates that contribute
to the state $n$ adapts (almost) instantaneously to thermal equilibrium. 
Consequently, these contributions to work and heat do not enter the FTs
non-trivially.

\subsubsection{Experiments: F1-ATPase}
Apart from free energy reconstructions described above, experimental
work using the concepts of stochastic thermodynamics is still scarce.
For the F1-ATPase, two groups have published work pointing in this direction.
In an intriguing example  of exploiting the SSFT, the
torque exterted by the F1-ATPase on a bead in an optical trap
 could be measured without
knowing the friction constant of the bead \cite{haya10}.
The implicite assumption, however, with this type of analysis is 
that no further dissipative mechanisms exist. In another study of this
molecule \cite{toya10,toya11}, it was inferred that this motor transfers 
almost the full free energy from
ATP hydrolysis into loading the elastic element connecting the motor with
the bead.

\subsubsection{Biochemical reaction networks}
The formalism described above for a single enzyme can easily be extended to
networks involving several types of (different) enzymes \cite{schm06a} 
or ordinary chemical
reactions networks using chemical master equations \cite{gasp04,schm06}.
Specific examples for which the distribution of entropy
production has been calculated are \cite{xiao08,xiao09,rao11}.

\section{Autonomous  isothermal   machines}
\label{sec-10}
\subsection{General aspects}
Enzymes and molecular motors as described in Sect.
\ref{sec-9}  from the stochastic thermodynamics 
perspective 
  provide a
 paradigm for isothermal machines. In contrast to
heat engines, which in their classical form are the archetypical thermodynamic 
machines and which in their stochastic version will be described in
Sect. \ref{sec-11}, 
isothermal machines do not transform heat but rather chemical energy 
into mechanical work (or vice versa) while the temperature of the surrounding 
medium remains constant.

A general classification relevant to machines is whether or not they operate
autonomously. In the stochastic setting, an autonomous machine will typically
correspond to a NESS driven by externally imposed time-independent boundary
conditions. Any molecular motor is a typical example of such an autonomous 
isothermal machine, since 
single molecule assays typically provide  conditions of constant
non-equilibrium concentrations of ``fuel'' molecules like ATP.
For a  non-autonomous machine, some time-dependent external
control is required that ``leads'' the machine through its cycle. 
Building reliable artificial molecular motors in the lab still
constitutes a major challenge as reviewed in \cite{brow06,kay07,bath07,heuv07,balz09,cosk12}}.

In this section, we present  a systematic theory for isothermal
autonomous 
machines in the discrete state version based on the representation in terms of
cycles of the underlying network. In principle, this approach includes earlier
models based on continuous coordinates diffusing in a ratchet potential that 
depends on the chemical state of the motor since any continuum model can be 
discretized. The emphasis here, however, is on the basic principles, and, in
particular, 
how the thermodynamic constraints are implemented consistently.

An important quantity for any type of machine is its efficiency defined as
the ratio between the power delivered by the machine and the rate of
``fuel'' consumption. Thermodynamics constrains this efficiency by 1 for isothermal
machines and by the Carnot efficiency for thermal heat engines operating between heat baths of different temperature. In both
cases working at the highest possible efficiency comes at the cost of
zero delivered power since reaching the thermodynamic bound requires a
quasi-static, i.e., infinitely slow operation. A practically more relevant
question then is about efficiency at maximum power (EMP). We will see that in
this thermodynamic framework rather general expressions for power, efficiency and
efficiency at maximum power emerge.  

It would be interesting to pursue these issues also for
periodically driven machines, which are one step more complex than the autonomous ones.
While
there is a vast literature
on how to generate transport by periodic modulation of system parameters as reviewed in
\cite{reim02a,hang09,sini09,astu11} the problem of efficiency and efficiency at maximum power,
however,  seems not to have been 
addressed systematically for such stochastic machines. One reason is that even making
explicit statements about the periodic steady state is much harder than  
for the NESS engine at constant external parameters.

\subsection{General framework for autonomous machines: Cycle representation
and entropy production}
\label{sec10-2}
An autonomously operating device or 
 machine can be modelled as a Markov process
on a network in a steady state. Transitions between different states in
this NESS depend on  rates that reflect both the coupling of
the machine to reservoirs with different chemical
potentials for solutes or
particles and external forces or loads. These non-equilibrium conditions
can be expressed by generalized thermodynamic forces or ``affinities''
$\F_k$ as listed  in Tab. \ref{table-aff1}.

\begin{table}[t]
\caption{Affinities and generalized distance for isothermal machines} 
~~\\~~
\begin{indented}
\item[]
\begin{tabular}{cccccc}
process & affinity $\F_k$  & gen. distance $d^k$   \\
\hline\hline
linear motion& force~~$f/T $& linear distance $d$ \\
\hline
rotation &torque~~$N/T$ & angle $\Delta \phi$   \\
\hline
particle transport &$ -\Delta \mu/T$ & (typically) 1 \\
\hline
chemical  reaction&  $ - \Delta \mu/T$ & (typically) 1 \\
\end{tabular}
\end{indented}
\label{table-aff1}
\end{table}

For a systematic presentation it is useful first to recall the representation
of a NESS in terms of cycle currents \cite{schn76,hill}, see
Fig. \ref{fig-cycle}.  
Rather than summing over the
individual transitions $(mn)$ or reactions $\rho$
 as we have done so far, in a NESS probability  currents
can be expressed by a sum  over directed cycle currents
\beq
j\a \equiv\jpa -\jma=\jpa(1-\jma/\jpa)=\jpa(1-  \prod_{\rho\in a}\wrm/\wrp)
\ee
where $a$ labels the cycles and $\jpa$ and $\jma$ denote the 
inverse mean times required
for completing the cycle in forward and backward direction, 
respectively.\footnote[1]{The directed cycle current $j_a\equiv\langle \jj_a\rangle$ is the 
mean of the fluctuating current $\jj_a$ introduced in
Sect. \ref{sec6-4}.}
These forward and backward (probability) currents
 can be expressed in a diagrammatic way by the transition
rates of the whole network (not just those of the respective cycle). 
However, the ratio $\jma/\jpa$, 
is given by the ratio
between the product of all backward rates and the product of all forward
rates contributing to the cycle $a$.

Thermodynamic consistency as formulated in (\ref{eq:cycl-aff}) or
 (\ref{eq:wrho-heat}) allows to 
express this ratio
\beq
\prod_{\rho\in a}\wrm/\wrp = \exp(-\sa) \label{eq:ratio} 
\ee
 by the
sum 
\beq
\sa \equiv \sum_{\rho\in a}\left(q_\rho/T +\dsr \right) \equiv  q\a/T + \sum_{\rho\in
  a}\dsr = \bar q_a/T
\label{eq:s-cycle}
\ee
of the entropy changes in the reservoirs and heat
baths associated with this cycle. The last equality recalls the definition
of heat under strict steady state conditions which includes the quasi-static 
refilling of the reservoirs as discussed in Sect. \ref{sec-chem}.
With the first law (\ref{eq:first-chem}) summed along a cycle, 
this entropy change can also be written as
\beq
\sa= (\wma + \wca)/T .
\ee This representation alluding to the definition of a ``chemical work''
introduced in  Sect. \ref{sec-chem}
becomes convenient when discussing the efficiency.

Alternatively, expressed in terms of affinities, the entropy change associated with
a cycle becomes
\beq
\sa=\sum_kd^k_a\F_k,
\label{eq:sa}
\ee
where $d^k_a$ is a generalized distance conjugate to the force $\F_k$
as listed in Tab. \ref{table-aff1}. To each affinity $\F_k$,
there corresponds a conjugate flux or current 
\beq
J_k\equiv \sum_a(\jpa-\jma) d^k_a = \sum\a \jpa [1-\exp(-\sa)]d^k_a
\label{eq:Jk}
\ee
describing the rate with which the respective quantity is ``processed'' by the
machine.

The mean entropy production rate can be written as
\begin{eqnarray}
\sigma &=& \sum_a(\jpa-\jma) \sa\\ &=&\sum_a\jpa[1-\exp(-\sa)]\sa
= \sum_kJ_k\F_k .
\label{eq:sigma}
\end{eqnarray}
These expressions are exact and do not imply any linear response assumption
as the final bilinear form may  suggest.

\subsection{Power and efficiency}
\subsubsection{Input and output power}
A device or machine is supposed to deliver some output from consuming
some input. Characteristically for nano-machines, the role of output
and
input can easily be reversed as it depends on the signs of the
corresponding affinities. Input has to be offered to the machine with a positive
affinity $\F_i>0$ whereas output is associated with a current or flux
that is opposite to an applied affinity $\F_o<0$. 

Quite
generally, for an isothermal machine 
 in a NESS, the total rate of production of output and input,
$\Po$ and $\Pii$, respectively, is given by the product between a pair
of corresponding flux and affinity according to
\begin{eqnarray}
\Poi&=&\eoi  \Joi (T\Foi)\\
           &=& \eoi \sum\a \jpa [1-\exp(-\sa)]d^{o,i}_a  (T\Foi)  
\label{eq:pow-gen}
\end{eqnarray} where $\eps_o\equiv -1$ and $\eps_i\equiv 1$ reflect 
the fact that
the output is delivered against an external load $\F_o<0$.
Expressed in terms of a cycle-specific work input 
\beq
\wia\equiv \F_id^i_a
\ee and work output
\beq
\woa\equiv - \F_od^o_a ,
\ee the power can also be written as \cite{seif11a}
\beq
\Poi = \sum\a \jpa [1-\exp(\wia-\woa)] T \woia  .
\label{eq:pow-gen-two}
\ee

In the contribution of each cycle, the expressions
(\ref{eq:pow-gen}, \ref{eq:pow-gen-two}) separate a system specific kinetic prefactor,
$\jpa=\jpa(\{\F_k\})$, from the remaining thermodynamic quantities.

\subsubsection{Efficiency and efficency at maximum power (EMP)}
The 
efficiency of a machine is defined as the ratio
\beq
\eta \equiv \Po/\Pii.
\label{eq:eff}
\ee
It has occasionally been argued that the traditional definition of efficiency
(\ref{eq:eff}) should be modified for molecular motors pulling cargo in order
to include the ``work'' required for overcoming Stokes friction even in the
absence of an external force \cite{dere99,wang02a}. More recently, a
``sustainable''
efficiency has been suggested as an alternative concept \cite{gave10,more11}.
 In this review, we keep
the traditional expression  (\ref{eq:eff}).

For the paradigmatic case of just two non-zero affinities, the entropy production
rate (\ref{eq:sigma})  becomes
\beq
\sigma=(\Pii-\Po)/T \geq 0
\ee
implying that efficiency of isothermal machines is bounded by $0\leq\eta\leq1$.
Working at the highest possible efficiency comes at the cost of
zero delivered power since reaching the thermodynamic bound requires a
quasi-static, i.e., infinitely slow operation. A practically more relevant
question then is about efficiency at maximum power.

 The notion of efficiency at maximum power (EMP) 
requires  one or several  parameters $\{\l_i\}$ 
with respect to  the variation of which 
$\Po$ can become maximal,  i.e., $\Po^*\equiv \max_{\{\l_i\}} \Po\equiv
\Po(\{\l^*_i\})$.  
EMP is then given by
\beq
\eta^*\equiv \Po^*/\Pii(\{\l^*_i\}) .
\ee
In general, the result for EMP will depend strongly both on the
choice and the allowed range of the variational parameters  $\{\l_i\}$ 
\cite{seif11a,gave10a} which is a fact occasionally
ignored when statements about {\it the} EMP are made. In particular, one
should distinguish variation with respect to the externally imposed
affinities from those with respect to structural or intrinsic parameters of
the
machine. Examples for the latter are  the topology of the network 
and  common prefactors
for forward and backward rates that leave their ratio and thus the
thermodynamics invariant.

\subsection{Linear response: Relation to phenomenological
  irreversible
thermodynamics}
\label{sec-lr}
At this point, it is instructive to consider 
 a machine operating close to
equilibrium and to cast the results into the framework of
linear  irreversible thermodynamics \cite{groot,pott09}. This theory truncates an
expansion of the fluxes in the first order of the affinities, i.e., assumes
that
\beq J_k=\sum_l L_{kl}\F_l
\ee
with the Onsager coefficients $ L_{kl}$. By expanding (\ref{eq:Jk}) for small
affinities
and using (\ref{eq:sa}),
we obtain for the Onsager coefficients the cycle representation
\beq
 L_{kl}= \sum_a \jaeq d^k_ad^l_a.
\ee
Here, $\jaeq\equiv\jpa(\{\F_k\}=0)$ is the equilibrium forward current
of a cycle $a$.
In this approach, the Onsager
symmetry $L_{kl}=L_{lk}$ is satisfied
automatically. 
Similarly, the rate of total entropy production (\ref{eq:sigma})
becomes
\beq
\sigma\approx \sum_a\jpa (\sa)^2=\sum_{kl}L_{kl}\F_k\F_l.
\ee
In this lowest order, power input and output (\ref{eq:pow-gen}) become
\beq
\Poi=\eoi \Joi(T\Foi)= \eoi T \sum_kL_{\{o,i\}k}\F_k\Foi.
\label{eq:power}
\ee

In the paradigmatic case of  two affinities, for fixed input
affinity $\F_i>0$  and choosing the output affinity
as
variational parameter $\F_o$ , maximum power is reached for
\beq
\F_o^* = -L_{oi}\F_i/2L_{oo}\ee
leading to an EMP of \cite{kede65}
\beq
\eta^*=L_{oi}^2/[4L_{oo}L_{ii}-2L_{oi}L_{io}]\leq1/2 .
\label{eq:etastar}
\ee
The upper bound imposed by the positivity of entropy
production is realized for a degenerate matrix of Onsager coefficients,
\beq
L_{oo}L_{ii}=L_{oi}L_{io} ,
\ee
which implies that $J_o\sim J_i$. Possible realizations of this structural
condition
 are (i) all unicyclic machines and (ii) tightly coupled
multicyclic machines. These two classes and the third one
of weakly coupled multicyclic machines  will be defined and
discussed in the next sections.

\subsection{Unicyclic machines}
Unicyclic machines consist of only one cycle which allows to drop the cycle
index
$a$
in this section, see Fig. \ref{fig-cycle} for an example.
In general, the power delivered and used by a
 unicyclic
motor becomes with (\ref{eq:pow-gen-two}) 
\beq
\Poi =Tj^+[1-e^{\wo-\wi}]\woi .
\label{eq:pow}
\ee Its efficiency $\eta\equiv w_o/w_i$ depends thus  trivially only on the
externally imposed affinities $\F_k$  and the intrinsic properties $d^{o,i}$
but is independent of the detailed kinetics. 
In the regime $0<\wo<\wi$,
the motor  will work as intended. 
For $\wo=\wi$, the motor has optimal efficiency $\eta = 1$ 
but does not deliver
any power  since it then cycles as often in forward as in backward
direction.

The concept of EMP requires to  identify
 the admissible variational parameters. A simple and physically transparent
 choice is
to fix the input $\wi$ and vary the
output $\wo$, e.g.,  by changing the applied force or
torque in the case of a molecular motor.
 The condition $d\Po/d\wo=0$ leads to the 
implicit relation \cite{seif11a}
\beq
\wi = \wo^* + \ln [1+ \wo^*/(1+\xo\wo^*)] 
\label{eq:win}
\ee
for the optimal 
output $\wo^*$
at fixed input $\wi$ with
\beq
\xo\equiv  d\ln j^+/d\wo \approx x_o\eq + O(\wi,\wo).
\ee These expressions can easily be evaluated for any unicyclic machine
 with
specified rates which determine the non-universal $j^+$.

The linear response regime
is defined by the condition $\wo<\wi\ll1$. 
By expanding  (\ref{eq:win}), one obtains
\beq
\eta^*=\wo^*/\wi\approx 1/2 + (\xo\eq+1/2)\wi/8 + O(\wi^2)
\ee which shows how system specific features like the coefficient
$\xo\eq$ enter EMP beyond the universal value  1/2. This expression proves that,
depending on the value of the non-universal parameter $\xo\eq$, EMP may well
rise beyond the linear response regime as found first in a case study 
of molecular motors \cite{schm08a}. 
These results seem to be at variance with another study along
similar lines where a universal bound of 1/2 was found for EMP \cite{gave10}. The
difference, however, is that the latter authors constrain the optimization
to a parameter space that leaves the stationary distribution invariant
which seems to be a somewhat artificial condition.
Further analytical and numerical results
for EMP of unicyclic machines  using $\wo$ or  both, $\wi$ and $\wo$, as
variation parameters can be found in \cite{seif11a}.
Bounds on EMP on simple unicyclic machines have been
derived in \cite{vdb12}.

\subsection{Multicyclic machines: Strong vs weak coupling}

A discussion of multicylic machines along similar lines
does not require much additional conceptual effort \cite{seif11a}. The crucial distinction
becomes the one between ``strong'' (or tight) and ``weak'' (or loose) coupling
first introduced within the
phenomenological linear response treatment in
\cite{kede65} and later stressed by van den Broeck and co-workers mostly in the
context of heat engines as reviewed in the next section.
In a strongly coupled multicylic machine, any cycle containing the
input transition also contains the output transition (assuming for simplicity that
input and output affect only one transition each).  For such strongly coupled machines exactly the
same formalism as for unicyclic machines applies with the only caveat
that $j^+$ appearing there is now given by 
$
j^+\equiv  \sum_a \jpa  
$ where the sum runs over all cycles that include input and output transition
and the $\jpa$ are the corresponding forward cycle currents.
Thus, such strongly coupled machines obey the same relations for efficiency
and
EMP as unicyclic machines.

In the weak coupling case, there are cycles containing the input but not the output
transition. Running through such a cycle the machine "burns" the input without
delivering any output which clearly decreases the efficiency.  In
particular, it turns out that in the linear response regime, EMP is less that 1/2,
but may still rise when moving deeper into the non-equilibrium regime \cite{seif11a}.

\subsection{Efficiency and EMP of molecular motors}

One important class of potential applications of the theory just described are 
molecular motors that transform chemical energy into mechanical energy (or vice versa).
As mentioned in the introductory section above, two general types of models must be
distinguished for these systems.

For dynamics on a discrete set of states, efficiency (rather than EMP) has been investigated
recently for various models \cite{kawa11,efre11}. Genuine EMP has been studied for both the
simplest unicyclic and a simple multicycle network
in \cite{schm08a}. It would be interesting to do so for more intricate models like, e.g., 
the one   introduced in  \cite{liep07a},
but also for artificial swimmers like the one discussed 
in \cite{gole10}. 

Traditionally,  efficiency of molecular motors has been studied within
 ratchet models where the motor undergoes
a continuous motion in a periodic potential that depends on the current
chemical state of the motor \cite{magn94,parm99,juel97,astu02,reim02a,parr02}. 
Dissipation then involves both, the continuous degree of freedom which should
be treated along the lines discussed in Sect. \ref{sec2} and the discrete
switching of the potential due to an enzymatic event. Model systems of
this type have been investigated in \cite{gasp07,wang05a,qian08,boks09,gerr10}. 
For a recent study of EMP in such a continuum description, see \cite{golu12}.

There is a second motivation for studying such models combining discrete with continuous
dynamics. In the typical  experimental set-up for measuring the efficiency
of a molecular motor under load, an external force or  torque is applied to a micron-sized bead
that is connected to the molecular motor like in the recent example of the rotary motor F1-ATPase
\cite{toya10,toya11,toya11a}. The discrete nano-sized steps of the motor become visible only
through monitoring the biased Brownian motion of the bead which clearly is continuous. For a
comprehensive description, both dissipation in the discrete steps of the motor and the one associated
with the continuous motion of the bead should be combined \cite{zimm12}.

\section{Efficiency of stochastic heat engines}
\label{sec-11}

\subsection{Carnot, Curzon-Ahlborn and beyond}

In classical thermodynamics, a heat engine, delivering work $-W$ by
extracting heat $-Q_1$ from a hot bath at temperature
$T_1$ and releasing heat $Q_2$ into a cold bath
at temperature $T_2$, has an efficiency
\beq
\eta\equiv |W|/|Q_1| \leq \eta_C\equiv 1-T_2/T_1
\label{eq:carnot}
\ee
limited by the Carnot efficiency $\eta_C$ which provides a universal bound
that follows from combining the first with the second law. Reaching
the upper bound comes at the price of zero power since this
condition requires a quasistatic, i.e., infinitely slow operation.
A practially more relevant efficiency is the one at maximum power, $\eta^*$,
which becomes well-defined only if the parameter space available for the
maximization is specified.
For macroscopic thermodynamics, introduction of this problem is often
attributed to Curzon and Ahlborn \cite{curz75} even though their result
has been described earlier, see \cite {novi58} and the comment
made in Ref.1 of \cite{gave10a}. Subsequent work for macroscopic engines
pursued under the label of finite-time thermodynamics is reviewed in
\cite{chen99,sala01,hoff03,andr11}.

Curzon and Ahlborn (CA) assume  ordinary heat conduction between the baths and 
the engine 
that is supposed to operate without further internal losses.
By optimizing the power with respect to (i) the temperature difference 
 responsible for the heat exchange between
the
baths and the machine, and (ii)
the duration of the two isothermal steps (while fixing a constant ratio
between the time allocated to isothermal and adiabatic steps) they obtained
for EMP
 the expression 
\beq
\eta_{CA}\equiv 1-(T_2/T_1)^{1/2} \approx  \eta_C/2 + \eta_C^2/8 + O(\eta_C^3).
\label{eq:ca}
\ee which is independent of the thermal conductivities between 
baths and engine.

 Whether or not the CA result  can claim more
universality
that under the original ``endoreversible''
assumptions, or is even  a bound on EMP, is a subtle, if not even ill-defined,
issue
since maximum power depends crucially on the admissible parameter space.
Beyond the original assumption there are conditions like for a cascade
of intermediate engines \cite{vdb05,cisn07} and 
for ``weak symmetric dissipation'' \cite {espo10c}
where CA can be shown to hold for a rather reasonable choice of variational
conditions. Numerical simulations of finite-time Carnot cycles for a 
weakly interacting gas have been analyzed for efficiency and EMP in
\cite{izum08,izum09,izum09a}.

A related question is the range of universal validity of the
expansion in (\ref{eq:ca}) for EMP. For tightly coupled machines defined 
by an output work flux that
is proportional to the heat flux taken from the hot reservoir,
the leading term, $\eta_C/2$,  
 follows
from simple linear
irreversible thermodynamics for fixed input and variable output
 \cite{kede65}. Such a result will hold both for macroscopic as for
small engines.

The question of efficiency and EMP  are indeed as relevant and applicable to
small engines or devices as they are for macroscopic ones. The new aspect
concerns the role of fluctuations not in the sense that a fluctuating
efficiency is defined which might lead to ill-defined results given the fact that
sometimes the heat taken from the hot bath would be zero or even negativ.
One rather keeps the definition (\ref{eq:carnot}) but now $W$ and $Q$ are
mean values that are determined by averaging over the 
fluctuations.\footnote{The inequality (\ref{eq:carnot}) is a trivial
consequence of the IFT
for total entropy production, if the latter quantities are expressed by fluctuating work
and heat contributions \cite{sini11}.}
A main advantage of a stochastic approach compared to the macroscopic
phenomenological one
is the fact that a thermodynamically consistent kinetics 
valid 
beyond the linear response regime can easily be imposed. 

Whether the CA result  has any relevance
to these small thermal engines has been one of the main issues in the field
especially since 
the coefficient 1/8 in the second term of an expansion of efficiency at
maximum power in $\eta_C$ was found in quite different systems
 \cite{schm08,tu08}. For autonomous machine,  i.e., in the  steady state regime, 
the 1/8 is indeed  universal if the system possesses an additional (left-right) symmetry 
\cite{espo09}. 
It should be
stressed, however, that getting this coefficient requires a second
(intrinsic or structural) variational parameter beyond the output control
required for getting the 1/2. For such steady state machines, beyond the
second term in the expansion (\ref{eq:ca}), the full CA result is irrelevant.

On the other hand, for small cyclic machines, which can be
treated formally in a
spirit
closer to CA's original approach, obtaining
 the factor  1/8 requires a
symmetry in the exchange with
the
hot and cold bath \cite{espo09}. Moreover, for such a machine it is possible
to
obtain the CA result over the full temperature
range for certain  conditions \cite{schm08,espo09}.

In the following, we decribe paradigmatically how small heat engines or devices fit into the
stochastic 
framework from which  these and further results for both autonomous (steady
state) and periodically driven  machines can easily be derived.
We focus on both
the formal similarities and differences  with the isothermal machines and 
the 
issue of efficiency of maximum 
power.\footnote{Universality of the efficiency if other quantities
are optimized has been studied in \cite{sanc10a}.} Even though we restrict the
following discussion to heat engines, similar concepts can be applied to
refrigerators, see, e.g., \cite{toma12}.

\subsection{Autonomous heat engines}

For understanding both the general issues and the necessary
 ramifications of the comparably simpler framework introduced in 
Sect. \ref{sec-10}
for the isothermal case, it is helpful to have a few specific examples in mind.

\subsubsection{B\"uttiker-Landauer (BL) and Feynman ratchet}

Transport of a colloidal particle in a periodic potential can be induced by
an external force at constant temperature  as discussed in Sect. \ref{sec2}.
As an alternative, in the absence of an external force, a spatially periodic
temperature profile (out of phase with the potential) will also lead to
net motion  as discussed by B\"uttiker \cite{buet87}, van Kampen \cite{kamp88}
and Landauer \cite{land88}. This set-up
is one example of noise-induced transport which is comprehensively reviewed
by Reimann \cite{reim02a}. From a more thermodynamic perspective, and in the presence of
an additional opposing external force, such a B\"uttiker-Landauer (BL) ratchet
is a simple example for a stochastic heat engine that transforms heat into
mechanical work. One can then ask for the efficiency of such a device which
is a subtle question especially when using  the overdamped limit for a discontinuous 
temperature profile \cite{seki97,dere99a,mats00,hond00,seki00,benj08}. The optimization of such a
device 
for maximal power
has been studied both for variation of the external force \cite{asfa04,gome06}
and for variation of
the intrinsic potential \cite{asfa08,berg09}. These issues become technically simpler in
discretized versions \cite{tu08,jarz99a,vela01,zhan10,chen11} as in the  simple model
sketched and described in Fig. \ref{fig-ratchet}. 
This system  can also be seen as a simplified (one degree of freedom)
version of
the famous Feyman ratchet \cite{feyn63} that as a paradigm for rectification of thermal
noise
has its own conceptual subtleties \cite{parr96,seki97}.
The Feynman ratchet  has inspired various model systems which have been
 analyzed both
analytically and in numerical simulations \cite{vdb04,zhen10}.

From an experimental perspective, realizing such 
a ratchet in aqueous solution
is not straightforward since one needs significant temperature differences 
on rather small length scales as realized in single particle studies of
thermophoresis, see, e.g.,  \cite{duhr06}.

\begin{figure}[t]
\includegraphics[width=6cm]{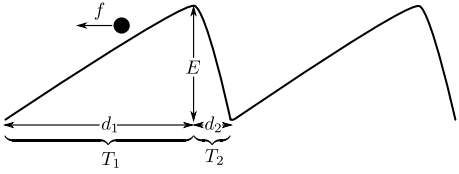}
\caption{ In a  BL ratchet,  a particle
preferentially
climbs a potential barrier with height $E$ over a distance $d_1$ while in
contact with a heat bath at $T_1$. It slides down a distance $d_2$ on the cold
side with $T_2=T_1-\Delta T$.
This temperature-difference  driven motion to the right persists for
a small enough force $f<0$ pulling to the left.
}
\label{fig-ratchet}
\end{figure}

\subsubsection{Electronic devices}

In electronic devices, 
 temperature differences can more easily be imposed as it is done, e.g., 
in thermoelectrics where temperature differences are exploited to transport
electrons against an electro-chemical potential. 
For such systems, thermodynamic considerations 
have been emphasized
by Linke and co-workers who pointed out that such machines can indeed operate
at the Carnot
limit \cite{hump02,hump05}. More recent studies based on simple models
for quantum dots have addressed in particular the issue of efficiency at maximum power
\cite{espo09b,sanc11,espo12a}. The simple paradigm discussed in \cite{espo09b} is 
sketched in Fig. \ref{fig-thermo}. 

A particularly intriguing aspect of such devices is
 the observation that in the
presence of a magnetic field the Onsager-Casimir symmetry relations, in principle, 
seem to allow
Carnot efficiency at finite power \cite{bene11}. This issue deserves further study
through the analysis of microscopic models as the
one suggested in \cite{sait11}.

\begin{figure}[t]
\includegraphics[width=6cm]{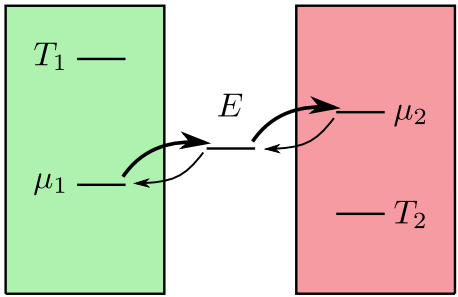}
\caption{ In a thermoelectric device,  simplified here as a quantum dot with a
single relevant energy level $E$, electrons are transported on average
from a hot reservoir with
$\mu_1$ and $T_1$ to a cold reservoir with $\mu_2= \mu_1 + \Delta \mu>\mu_1$ and
$T_2=T_1-\Delta T$ using
heat from the hot reservoir.
}
\label{fig-thermo}
\end{figure}

\begin{figure}[h]
\includegraphics[width=6cm]{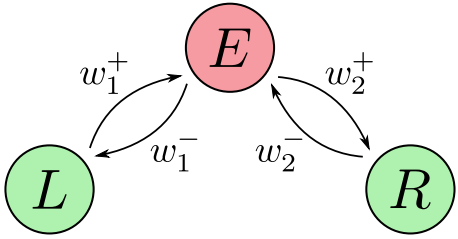}
\caption{Common three-state diagram for the simplified BL ratchet (Fig. \ref{fig-ratchet}) and the
thermo-electric device (Fig. \ref{fig-thermo}). For the BL-ratchet L and R refer to the particle
sitting in the mininum  and E corresponds to the particle 
being
 on the barrier top. For the electronic device, L and R refer to the
electron being in the left or 
right reservoir. E corresponds to the electron
sitting
 on  the quantum dot.
 In both cases,
 the state R should be identified with L after the electron or particle
has been transported from left to right thus completing the cycle. 
The log-ratio of the transition rates is given in Tab. \ref{table-rates}.
}
\label{fig-net}
\end{figure}

\begin{table}[b]
\caption{Ratio of rates for the devices shown in Figs. \ref{fig-ratchet}
and \ref{fig-thermo} with their network representation Fig. \ref{fig-net}.}
~~\\~~
\begin{indented}\item[]
\begin{tabular}{ccc}
~~&$\ln w_1^+/w_1^-$&$\ln w_2^+/w_2^-$\\
\hline
\hline
BL ratchet &$-(E+|f|d_1)/T_1$& $(E-|f|d_2)/T_2$\\
\hline
thermo-electric device&$(\mu_1-E)/T_1$& $(E-\mu_2)/T_2$\\ 
\end{tabular}
\end{indented}
\label{table-rates}
\end{table}

\begin{table}[b]
\caption{Characteristic quantities  for the two example of unicyclic
heat engines shown in Figs. \ref{fig-ratchet} and \ref{fig-thermo}.} 
~~\\~~
\begin{tabular}{ccccccc}
 & relevant affinities & $\F_k$ in &conjugate
 &Input & Output\\
&$\F_k$ &linear response&distance $d^k$&$-\bar q^{(1)}=T_2\wi$& $T_2\wo$\\
\hline
BL  & $1/T_2-1/T_1$&  $\Delta T/T^2_2$& $E$  &$ E+|f|d_1$ &\\
 ratchet &  $f/T_1,f/T_2$& $f/T_2,f/T_2$&$d_1,d_2$&  & $ |f|(d_1+d_2)$ \\
\hline
thermoelectric &  $1/T_2-1/T_1 $&  $\Delta T/T^2_2$&$E$ &$   E-\mu_1$ &   \\
 device & $\mu_1/T_1-\mu_2/T_2$ &$-(\Delta \mu/T_2 + \mu_1 \Delta T/T_2^2)$ &
 1 & &  
$  \mu_2-\mu_1$\\
\end{tabular}
\label{table-aff2}
\end{table}

Likewise, any photo-electric 
device is also coupled to a  reservoir of rather high temperature since the 
photons being absorbed from the sun come with the black-body distribution of the sun's
temperature. Therefore, photo-electric devices are amenable to a thermodynamic 
description focussing of efficiency and EMP, see, e.g.,  \cite{rutt09}.

\subsubsection{General theory}

For any discrete  autonomous 
heat engines in contact with heat baths of at least two different
temperatures, it must be specified  for
each transition  at which temperature it takes place,
i.e., with which heat bath the machine is in contact at this particular
transition. 
As in the isothermal case, 
the assumption of local detailed
balance implies thermodynamic constraints on the
ratios of forward and backward rates as given 
in Fig. \ref{fig-net} and Tab.
\ref{table-rates} for the two examples introduced
above.

The thermodynamic constraints imply that 
 the total entropy production along a cycle still
fulfills (\ref{eq:ratio}).  For the representation (\ref{eq:sa}),
one needs the affinities with the corresponding
conjugate distances entering the conjugate fluxes given 
for the two examples in Tab. \ref{table-aff2}.
The general  differences compared to the isothermal case arise from
the presence of (at least) two different temperatures.
First, there is a new affinity associated with the two
heat baths with energy flow as the
corresponding flux. Second, if matter is transported between baths of different
chemical potential and different temperature, the 
corresponding affinity  involves the two temperatures. As a consequence,
in linear response, the latter affinity carries both a $\Delta \mu$ and a $\Delta T$
term. 
Finally, a force applied to a particle in a thermal ratchet subject to two different
temperature requires to introduce even  two affinities with this force. 
As in the isothermal case, for an autonomous heat engine
 the total 
entropy production rate  can be expressed by affinities and conjugate fluxes
according to 
(\ref{eq:sigma}).

On the cycle level, the total entropy change becomes 
\beq
\sa = \bar q^{(1)}_a/T_1 + \bar q^{(2)}_a/T_2 ,
\label{eq:saq}
\ee 
where we use the heat as appropriate under NESS conditions which fulfills
first laws of the type
\beq
{\wma}^{(1,2)}+{\wca}^{(1,2)} =\bar q^{(1,2)}_a + \Delta E^{(1,2)}_a
\ee where $\Delta E^{(1)}_a=-\Delta E^{(2)}_a$ is the change in internal energy 
of the system arising from the transitions
associated with the respective heat baths labelled by superscripts. 
With this  relation, the total
entropy change along a cycle  (\ref{eq:saq})
can
also be expressed by the heat extracted from the hot reservoir  as 
\beq
\sa= - \bar q^{(1)}_a\eta_C/T_2 + (\wma+\wca)/T_2 \equiv \wia\eta_C -\woa ,
\ee
where the work terms refer to the sum of the contributions from the
respective bath contacts.
The  definition of dimensionless input $w_{i,a}$  is motivated by the fact that for a heat engine the
input is the heat extracted from the hotter bath. Dimensionless output 
denoted by  $\woa$ is either mechanical
or
chemical work delivered by the machine.

The power of the machine can now be expressed analogously to the isothermal
case as
\beq
\Poi = \eoi \sum\a \jpa [1-\exp(-\wia\eta_C+\woa)] T_2 \woia  .
\label{eq:pow-gen-three}
\ee
where the occurence of the
Carnot efficiency  $\eta_C$ in the exponent compared to the
isothermal case (\ref{eq:pow-gen-two}) is crucial. 
In the affinity representation, the difference to the
isothermal case is even  more drastic 
 since output and input power
can no longer be written as simple products of a pair of conjugate flux $\Joi$ 
and affinity $T\Foi$ as in 
(\ref{eq:pow-gen}). One could have
anticipated this complication since  with two baths it is not obvious
which temperature should be chosen in (\ref{eq:pow-gen}) when trying to
generalize to the non-isothermal case.

\subsubsection{EMP for unicyclic machines}

For unicyclic machines (and  hence dropping the cycle index $a$), maximizing the power $\Po$
with
respect to the output $\wo$, which would physically correspond to varying the
external force or chemical potentials, leads to the analogue of (\ref{eq:win})
in the form  
\beq
\wi =(\wo^* + \ln [1+ \wo^*/(1+\xo\wo^*)])/\eta_C . 
\label{eq:winc}
\ee
This relation implies immediately the universal
$
\eta^*\equiv \wo^*/\wi\approx \eta_C/2 + O( \eta_C^2)  
$ in the linear response regime which can also easily be obtained from a 
phenomenological treatment analogously to the one
presented in Sect. \ref{sec-lr} \cite{vdb05}.
 
Maximizing the power with respect to the input variable $\wi$ leads to
\beq
\wi^*= [\wo^* + \ln (1-\eta_C/x_i)]/\eta_C
\label{eq:wonc}
\ee
with 
\beq
x_i\equiv  d\ln j^+/d\wi \approx x_i\eq + O(\eta_C,\wo).
\ee
As the respective column in Tab. \ref{table-aff2} show, $\wi$ 
involves an intrinsic parameter of the machine like the 
relevant energy level. 
Combining the relations (\ref{eq:winc},\ref{eq:wonc})
and varying both $\wo$ and $\wi$ leads to the EMP 
\begin{eqnarray}
\eta^{**} &=& \eta_C/[1-(\xo+x_i/\eta_C) \ln(1-\eta_C/x_i)]
\label{eq:emp}
\\
&\approx& \eta_C/2-[(2\xo\eq+1)/x_i\eq] \eta_C^2/8 + O(\eta_C^3) 
\end{eqnarray}
which shows that the second order coefficient is system-specific. 

It can be checked that
for a unicyclic device with spatial symmetry, for which the current $j$
reverses sign when the affinities $\F_k$ change sign, the square-bracket
prefactor of
the second term is indeed -1 thus recovering the overall 1/8 as previously
derived by extending the phenomonological irreversible thermodynamics
approach to second order \cite{espo09}.

For an explicit evaluation of the EMP (\ref{eq:emp}) one needs the 
specific form of $\xoi=\xoi(\wo,\wi)$ which requires assumptions on the specific
rates beyond the constraints imposed by thermodynamics
exploited so far. 
 For the mechanical 
BL ratchet, it is interesting to note that even for $d_1\not= d_2$,
an explicit calculation for $w_2^+=w_1^-=1$ (and the other rates as given in
Tab. \ref{table-rates})
 recovers the coefficient 1/8
despite the obvious breaking of the left-right symmetry.
The case $d_2=0$ is discussed for the full temperature range in  \cite{tu08},
where, not surprisingly, deviations from the CA result are found.
The
 thermoelectric device is treated in \cite{espo09,espo09b}.
For a 
photo-electric device, explicit results can be found in \cite{rutt09}, 
where also the role of non-radiative transition is discussed.
Further examples of EMP in a three and five state network have been discussed
in
\cite{gave10a}.

\subsection{Periodically driven heat engines}

The autonomous machines just discussed  reach a NESS since they
are permanently connected to both heat baths. For a 
periodically driven  heat engine,
contact with either one bath or, in an adiabatic step, with none, is periodically
enforced externally. 

\subsubsection{Brownian heat engine}

\begin{figure}[t]
\includegraphics[width=6cm]{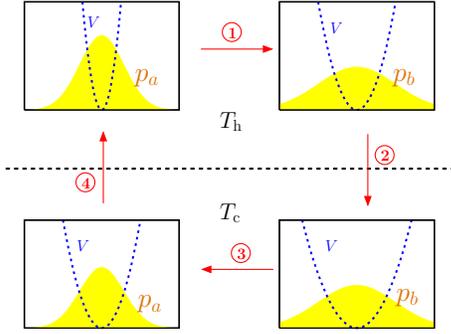}
\caption{Paradigm for a stochastic heat engine based on a colloidal particle
in a time-dependent harmonic laser trap in consecutive contact with a hot $(T_h)$ and cold  ($T_c)$ bath
\cite{schm08}. The steps
1 and 3 are isothermal; the steps 2 and 4 are instantanous and adiabatic during which the
distributions are $p_b$ and $p_a$, respectively.}
\label{fig-ss08}
\end{figure}

Within stochastic thermodynamics such a model
was introduced in \cite{schm08} for a Brownian particle in a time-dependent
potential, see Fig. \ref{fig-ss08}. Optimizing for both, the potential and
the time-interval spent in the isothermal transitions, EMP for fixed $T_{1,2}$
was shown to be
\beq
\eta^*=\eta_C/(2-\alpha\eta_C)\approx \eta_C+ (\alpha/4)\eta_C^2+ O(\eta_C^3)  
\label{eq:ca-ss}
\ee
where 
\beq
\alpha\equiv 1/[1+(\mu(T_1)/\mu(T_2))^{1/2}]
\ee
is a system-specific 
coefficient given by the temperature-dependence of the mobility
$\mu(T)$. If the
latter
is independent of temperature, one recovers the coefficient 1/8.
Since $0\leq\alpha\leq1$,  the expression (\ref{eq:ca-ss})
 implies the bounds
\beq
\eta_C/2\leq \eta^*\leq \eta_C/(2-\eta_C)
\ee
on EMP later also derived under the assumption of ``weak'' dissipation which
leads to a quite similar formalism  \cite{espo10c,izum11}. Further 
ramifications and classifications of such bounds have been discussed 
by Wang and Tu \cite{wang11,wang11a} and in \cite{aper12}.
The Onsager coefficients for a linear response treatment of this heat engine 
have been defined and calculated in  \cite{izum10}.

A micron-sized heat engine based on the Stirling version of the scheme  shown in Fig.
\ref{fig-ss08}
has just been realized experimentally \cite{blic12}. The colloidal particle as
``working fluid'' in a laser trap acting as the analogue of a piston
can be heated locally thereby realizing the contact with a hot bath.
By varying the cycle time both a maximum in the power at finite time and
the approach to the maximal efficiency in the quasi-static limit could
be demonstrated. For further interesting comments on this experiment,
see \cite{horo12}.

\subsubsection{Quantum dot}

A similar analysis can be applied to a finite-time Carnot cycle of a quantum
dot that can be connected to two different reservoirs similar as the set-up
shown in Fig. \ref{fig-thermo} \cite{espo10}. For a cyclic engine,
 the energy level $E(\tau)$ 
is controlled
in both the isothermal steps when the dot is connected to either one bath and
in the adiabatic steps when it is disconnected. Optimizing for the protocol $E(\tau)$ as well as for
the duration  of isothermal and adiabatic steps, one finds for EMP an expression
similar
to (\ref{eq:ca-ss}). In the limit of weak dissipation,
i.e., for small
deviations from the respective thermal population of the energy level, and
symmetric conditions, the coefficient
$\alpha$ turns out to be $\alpha=\eta_{CA}/\eta_C$ and hence one can here
recover
the CA result over the full temperature range.  
The distributions of work and heat for such a simple two-state engine have been
calculated in \cite{chvo09}.

\section{Concluding perspective}
\label{sec-12}

After this long exposition it may be appropriate to recall the basic assumption
of this approach,
to summarize the main achievements and to raise a few open general issues. 

\subsection {Summary}
Stochastic thermodynamics applies to systems where 
 a few  observable degrees of freedom like the positions of colloidal 
particles or the gross conformations of biomolecules are in non-equilibrium 
due to the action of possibly time-dependent external
forces, fields, flow or unbalanced chemical reactants. The unobserved degrees
of freedom like those making up the aqueous solution, however,  are 
assumed to be fast and thus
always in  the constrained equilibrium imposed by the instantaneous
values of the observed slow degrees of freedom. Then internal energy,
 intrinsic entropy and free energy are  well-defined and, if a microscopic
Hamiltonian was given, in principle, 
computable for fixed values of the slow variables. This assumption
is sufficient to identify a first-law like energy balance 
along any fluctuating trajectory recording
the 
changing  state of the slow variables.

Entropy change along such a trajectory consists of three parts: heat exchanged with 
the bath, intrinsic entropy of the states and stochastic entropy. The
latter requires in addition an ensemble from which this trajectory is taken.
If the same trajectory is taken from a different ensemble it leads to a different
stochastic entropy.
Thermodynamic consistency of the Markovian dynamics generating the
trajectory imposes a local-detailed 
balance condition constraining either the noise in a Langevin-type continuum
dynamics or the ratio of transition rates in a discrete dynamics.

At their core, the fluctuation theorems are mathematical identities
derived from properties of the weight of stochastic paths under time
reversal or other transformations. They acquire physical meaning by
associating the mathematical ingredients with the thermodynamic
quantities identified within stochastic thermodynamics. The detailed
fluctuation theorems then express a symmetry of the distribution function
 for thermodynamic
quantities.
An open question is whether the probability distributions of 
work, heat and entropy production can be grouped into ''universality
classes'' characterized, e.g., by the asymptotics of such distributions,
and, if yes, which specific features of a system determine this
class. The  more generally applicable integral theorems
often can be expressed as refinements of the second law for transitions
between certain states. Still, these integral fluctuation theorems should
not be considered a ``proof'' of the second law since irreversibility
has been implemented consistently from the beginning by choosing a
stochastic dynamics including the local detailed balance condition.

Conceptually, a major step has been to include feedback schemes 
based on perfect or imperfect measurements into
this framework which requires surprisingly little additional effort
due to the strong formal similarity of stochastic entropy with information.
Still missing is a full integration of  measurement apparatus
and the erasure process into the thermodynamic balance of the
efficiency for specific information machines.

The crucial ingredient for the developments summarized so far is the
notion of an individual trajectory and the concomitant concept of
distributions for thermodynamic quantities which represents
 the main difference compared to classical thermodynamics.

New insights, however, have emerged from this approach even when 
focussing on averages
and correlation functions as we have done in the second part of the
review. The general fluctuation-dissipation theorem for non-equilibrium steady
states shows how the response of any observable to a perturbation 
can be expressed as a sum of two correlation functions involving 
entropy production. In which cases this additive relation between
response and correlation can be reformulated as a multiplicative one
using the concept of an effective temperature is still not understood
despite some insights gained from specific case studies.

Our discussion of molecular motors, machines and
devices centered on the notion of efficiency and efficiency at
maximum power. Despite the fundamental difference of isothermal
engines operating at one temperature as do all cellular ones from
genuine heat engines like thermoelectric devices involving two 
baths of different temperature, a common framework exists based on
the representation of entropy production in terms of cycles of the 
underlying network of states. Clearly, more realistic networks
need to be studied in the future, in particular, for applications and
modelling of specific biophysical systems but the basic concepts
seem to be identified.  One particulary intriguing perspective
comes from the recent analysis
of the energetic cost of sensory adaptation 
using the concepts of stochastic thermodynamics \cite{lan12}.

\subsection {Beyond a Markovian dynamics: Memory effects and coarse-graining}

The identification of states, of work and 
of internal energy, i.e., of the ingredients entering the first law
on the level of trajectories is logically independent of the 
assumption of a Markovian dynamics connecting these states. The
crucial step is the splitting of all degrees of freedom in slow
and fast ones, the latter always being in a constraint equilibrium
imposed by the instantaneous values of the slow ones.
Likewise, the identification of entropy production only
requires the notion of an ensemble which determines stochastic entropy
along an individual trajectory.
Any dynamics guaranteeing that for fixed external parameters
compatible with genuine equilibrium, this equilibrium will
be reached for an arbitrary initial distribution of the slow
variables could qualify as a thermodynamically consistent one. 

\subsubsection{Continuous states}

A popular choice for a non-Markovian dynamics obeying these constraints
is Langevin dynamics with a memory kernel that, for thermodynamics
consistency, determines the correlations of the coloured 
noise.\footnote{Stochastic thermodynamics for a 
non-Markovian dynamics not obeying such a constraint
has been explored for generalized Langevin equations
in \cite{fara04,zamp05}, for
delayed Langevin systems in \cite{jian11},
and for Poissonian shot noise in \cite{baul09}.}
Under this assumption, the notions of stochastic thermodynamics are
well-defined and the various fluctuation theorems hold true as shown quite 
generally in 
\cite{zamp05,spec07a,ohku07,ohku09,aron10}.
Some of these papers contains illustrations
for model systems as do the references \cite{mai07,pugl09,hase11a}.
One  specific  motivation to explore such a dynamics
 arises from the recent fascinating
experimental data that show how hydrodynamic effects
lead to a frequency dependent mobility for colloidal motion \cite{fran11}.

A somewhat different and more subtle situation occurs if not all
slow variables are accessible in the experiment or the simulation.
The effective dynamics for the observable ones will then no longer be
Markovian and  its specific form in the case on non-harmonic
interactions between the slow one is typically not accessible. The
proper identification of, e.g., entropy production is then difficult
if not impossible. Still, one might be inclined to infer an apparent
entropy production by applying the rules for Markovian dynamics and
to check whether this quantity obeys the FT. In a recent study using two
coupled driven colloidal particles it turned out that the 
apparent entropy production based on the observation of just one
particle shows an FT-like symmetry but with a different prefactor for
a surprisinglingly large range of parameter values. However, 
there are also clear
cases for which not even an effective FT can be identified \cite{mehl12}.
This type of coarse-graining in the context of the FDT for a NESS has been explored 
in \cite{cris12}. 

\subsubsection{Discrete states}

For an underlying dynamics on a discrete set of states following
a Markovian master
equation, one option for coarse-graining is to group several states into
new ``meso-states'' or  aggregated states. Typically, 
the dynamics between these meso-states is then no longer Markovian. 
One question is whether one can then distinguish genuine equilibrium 
from a NESS if only the coarse-grained trajectory is accessible. 
For a three state system coarse-grained
into a two-state system, this
issue has been addressed in \cite{qian00,aman10} and, for more general
cases,
in \cite{rold10,rold12}.

Coarse-graining of a discrete network becomes systematically possible
if states among which the transitions are much faster are grouped together.
From the perspective of stochastic thermodynamics,
entropy production and fluctuation theorems
this approach has been followed in \cite{raha07,li08,pugl10,szab10,nico11c,hinr11,espo12,alta12}.

\subsection {Coupling of non-equilibrium steady states: A zeroth law?}

Besides the first and the second law, classical thermodynamics is founded
on a zeroth law stating that the notion of temperature and chemical potential
for equilibrium systems is transitive, i.e., if a system A is in separate
equilibrium with system B and system C, then upon contact of B and C 
neither heat nor particle flow will occur between these two systems.
A natural question is whether a similar equilibration
also occurs for non--equilibrium systems brought in contact such that they
can exchange energy or particles. Do then quantities exist resembling
temperature or chemical potential that govern ``equilibration'' between
such steady states? On a phenomenological level this question has been
introduced within the context  of steady state thermodynamics by Oono
and Paniconi \cite{oono98} and further refined by Sasa and  Tasaki \cite{sasa06}. 
For simple
one-dimensional model systems like zero-range processes in contact a non-equilibrium
chemical potential is indeed well-defined \cite{bert07a}. For 
two-dimensional driven lattice gases in
contact, numerical work has revealed that in a large parameter range
such a putative zeroth law and a corresponding thermodynamic structure
is approximately valid \cite{prad10,prad11}.

\subsection{Final remark}

From its very beginnings, thermodynamics fascinated scientists
by posing deep conceptual issues that needed to be resolved for understanding 
and optimizing quite practical matters like the design of heat engines.
With the experimentally realized micron-sized heat engine \cite{blic12}
discussed in one of the last sections of
this review, these latest developments have brought us back to
the very origins of classical thermodynamics albeit on quite different 
time and length scales and, quite importantly,
 with a much refined view on individual
trajectories. Indeed, without the spectacular advances in experimental 
techniques concerning tracking and manipulation of single particles and 
molecules, stochastic thermodynamics could still have been conceived
as a theoretical framework but would have not reached the broader appeal
that it has gained over the last fifteen years. Whether the next decade
of research in the field will be dominated by  specific applications, 
most likely for biomolecular networks and devices facilitating transport
of all sorts, or by further conceptual work exploring the ultimate limits 
of a thermodynamic approach to non-equilibrium beyond the Markovian 
paradigm into feedback-driven, information-processing,
strongly interacting systems  remains to 
be seen.

\ack

I thank T. Speck for a long-standing enjoyable collaboration on several
 topics
treated in this review. Many of my graduate students, in particular,
 D. Abreu, E. Dieterich, B. Lander, T. Schmiedl and E. Zimmermann have contributed through their thesis works to
my understanding of special topics discussed here. I have enjoyed a productive
interaction with two experimental groups in Stuttgart headed by C. Bechinger
and J. Wrachtrup and their students and post-docs, especially, V. Blickle, J. Mehl and
C. Tietz. Discussion on fundamental aspects with C. van den Broeck, M. Esposito, P. H\"anggi, 
C. Jarzynski, J. Parrondo, F. Ritort, K. Sekimoto,
H. Spohn, H. Wagner and R. Zia have always been most inspiring. Funding through DFG and
ESF is gratefully acknowledged.

\section*{References}


\begin{thebibliography}{999}

\bibitem{evan93}
D.~J. {E}vans, E.~G.~D. {C}ohen, and G.~P. {M}orriss.
\newblock Probability of second law violations in shearing steady states.
\newblock {\em Phys.\ Rev.\ Lett.}, 71:2401, 1993.

\bibitem{gall95}
G.~{G}allavotti and E.~G.~D. {C}ohen.
\newblock Dynamical ensembles in nonequilibrium statistical mechanics.
\newblock {\em Phys.\ Rev.\ Lett.}, 74:2694, 1995.

\bibitem{kurc98}
J.~Kurchan.
\newblock Fluctuation theorem for stochastic dynamics.
\newblock {\em J.\ Phys.\ A:\ Math.\ Gen.}, 31:3719, 1998.

\bibitem{lebo99}
J.~L. Lebowitz and H.~Spohn.
\newblock {A {G}allavotti-{C}ohen-type symmetry in the large deviation
  functional for stochastic dynamics}.
\newblock {\em J.\ Stat.\ Phys.}, 95:333, 1999.

\bibitem{evan94}
D.~J. {E}vans and D.~J. Searles.
\newblock Equilibrium microstates which generate second law violating steady
  states.
\newblock {\em Phys.\ Rev.\ E}, 50:1645, 1994.

\bibitem{jarz97}
C.~{J}arzynski.
\newblock Nonequilibrium equality for free energy differences.
\newblock {\em Phys.\ Rev.\ Lett.}, 78:2690, 1997.

\bibitem{jarz97a}
C.~{J}arzynski.
\newblock Equilibrium free-energy differences from nonequilibrium measurements:
  A master-equation approach.
\newblock {\em Phys.\ Rev.\ E}, 56:5018, 1997.

\bibitem{croo99}
G.~E. {C}rooks.
\newblock Entropy production fluctuation theorem and the nonequilibrium work
  relation for free energy differences.
\newblock {\em Phys.\ Rev.\ E}, 60:2721, 1999.

\bibitem{croo00}
G.~E. {C}rooks.
\newblock Path-ensemble averages in systems driven far from equilibrium.
\newblock {\em Phys.\ Rev.\ E}, 61:2361, 2000.

\bibitem{humm01}
G.~Hummer and A.~{S}zabo.
\newblock Free energy reconstruction from nonequilibrium single-molecule
  pulling experiments.
\newblock {\em Proc.\ Natl.\ Acad.\ Sci.\ U.S.A.}, 98:3658, 2001.

\bibitem{boch77}
G.~N. Bochkov and Y.~E. Kuzovlev.
\newblock General theory of thermal fluctuations in nonlinear systems.
\newblock {\em Sov. Phys. JETP}, 45:125--130, 1977.

\bibitem{boch79}
G.~N. Bochkov and Y.~E. Kuzlovlev.
\newblock Fluctuation-disspation relations for nonequilibrium processes in open
  systems.
\newblock {\em Sov. Phys. JETP}, 49:543--551, 1979.

\bibitem{hata01}
T.~{H}atano and S.~{S}asa.
\newblock Steady-state thermodynamics of {L}angevin systems.
\newblock {\em Phys.\ Rev.\ Lett.}, 86:3463, 2001.

\bibitem{seki97}
K.~Sekimoto.
\newblock Kinetic characterisation of heat bath and the energetics of thermal
  ratchet models.
\newblock {\em J.\ Phys.\ Soc.\ Jpn.}, 66:1234--1237, 1997.

\bibitem{seki98}
K.~Sekimoto.
\newblock {L}angevin equation and thermodynamics.
\newblock {\em Prog.\ Theor.\ Phys.\ Supp.}, 130:17, 1998.

\bibitem{seki10}
K.~Sekimoto.
\newblock {\em Stochastic Energetics}.
\newblock Springer-Verlag, Berlin, Heidelberg, 2010.

\bibitem{maes03}
C.~Maes.
\newblock On the origin and use of fluctuation relations for entropy.
\newblock {\em S\'em. Poincar\'e}, 2:29, 2003.

\bibitem{maes03b}
C.~Maes and K.~Netocn\'y.
\newblock Time-reversal and entropy.
\newblock {\em J.\ Stat.\ Phys.}, 110:269, 2003.

\bibitem{qian02b}
H.~Qian.
\newblock Mesoscopic nonequilibrium thermodynamics of single macromolecules and
  dynamic entropy-energy compensation.
\newblock {\em Phys.\ Rev.\ E}, 65:016102, 2002.

\bibitem{seif05a}
U.~Seifert.
\newblock Entropy production along a stochastic trajectory and an integral
  fluctuation theorem.
\newblock {\em Phys.\ Rev.\ Lett.}, 95:040602, 2005.

\bibitem{seif07}
U.~Seifert.
\newblock {Stochastic thermodynamics: Principles and perspectives}.
\newblock {\em Eur. Phys. J. B}, 64:423--431, 2008.

\bibitem{vdb85}
C.~van~den Broeck.
\newblock Stochastic thermodynamics.
\newblock In W.~Ebeling and H.~Ulbricht, editors, {\em Selforganization by
  Nonlinear Irreversible Processes. Proceedings of the Third International
  Conference, K\"uhlungsborn, GDR, March 18-22, 1985}, pages 57--61, Berlin,
  1985. Springer.

\bibitem{mou86}
{C}.~Y. Mou, J.-L. Luo, and G.~Nicolis.
\newblock Stochastic thermodynamics of nonequilibrium steady states in chemical
  reaction systems.
\newblock {\em J.\ Chem.\ Phys.}, 84:7011, 1986.

\bibitem{evan08}
D.~J. {E}vans and G.~P. {M}orriss.
\newblock {\em Statistical mechanics of nonequilibrium liquids}.
\newblock Cambridge Univ. Press, Cambridge, 2nd edition, 2008.

\bibitem{espo09c}
M.~Esposito, U.~Harbola, and S.~Mukamel.
\newblock Nonequilibrium fluctuations, fluctuation theorems, and counting
  statistics in quantum systems.
\newblock {\em Rev. Mod. Phys.}, 81:1665--1702, 2009.

\bibitem{camp11}
M.~Campisi, P.~H\"anggi, and P.~Talkner.
\newblock Colloquium. quantum fluctuation relations: Foundations and
  applications.
\newblock {\em Rev. Mod. Phys}, 83:771--791, 2011.

\bibitem{bust05}
C.~Bustamante, J.~Liphardt, and F.~Ritort.
\newblock The nonequilibrium thermodynamics of small systems.
\newblock {\em Physics Today}, 58(7):43, 2005.

\bibitem{jarz11}
C.~{J}arzynski.
\newblock Equalities and inequalities: Irreversibility and the second law of
  thermodynamics at the nanoscale.
\newblock {\em Ann. Rev. Cond. Mat. Phys.}, 2:329--351, 2011.

\bibitem{vdb10}
C.~van~den Broeck.
\newblock The many faces of the second law.
\newblock {\em J. Stat. Mech.}, page P10009, 2010.

\bibitem{qian06}
H.~Qian.
\newblock Open-system nonequilibrium steady state: Statistical thermodynamics,
  fluctuations, and chemical oscillations.
\newblock {\em J.\ Phys.\ Chem.\ B}, 110:15063--15074, 2006.

\bibitem{kurc07}
J.~Kurchan.
\newblock Non-equilibrium work relations.
\newblock {\em J.\ Stat.\ Mech.:\ Theor.\ Exp.}, page P07005, 2007.

\bibitem{impa07a}
A.~Imparato and L.~Peliti.
\newblock Work and heat probability distributions in out-of-equilibrium
  systems.
\newblock {\em C. R. Physique}, 8:556--566, 2007.

\bibitem{klag12}
R.~Klages, W.~Just, and C.~{J}arzynski, editors.
\newblock {\em Nonequilibrium Statistical Physics of Small Systems: Fluctuation
  relations and beyond}.
\newblock Reviews of Nonlinear Dynamics and Complexity. Wiley-VCH, Weinheim,
  2012.

\bibitem{rito08}
F.~Ritort.
\newblock Nonequilibrium fluctuations in small systems: From physics to
  biology.
\newblock {\em Adv.\ Chem.\ Phys.}, 137:31--123, 2008.

\bibitem{cili10}
S.~Ciliberto, S.~Joubaud, and A.~Petrosyan.
\newblock Fluctuations in out-of-equilibrium systems: from theory to
  experiment.
\newblock {\em J.\ Stat.\ Mech.:\ Theor.\ Exp.}, page P12003, 2010.

\bibitem{harr07}
R.~J. Harris and G.~M. Sch\"utz.
\newblock Fluctuation theorems for stochastic dynamics.
\newblock {\em J.\ Stat.\ Mech.:\ Theor.\ Exp.}, page P07020, 2007.

\bibitem{evan02}
D.~J. {E}vans and D.~J. Searles.
\newblock The fluctuation theorem.
\newblock {\em Adv.\ Phys.}, 51:1529 -- 1585, 2002.

\bibitem{gall99a}
G.~{G}allavotti.
\newblock {\em Statistical mechanics. A short treatise.}
\newblock Springer, Berlin Heidelberg, 1999.

\bibitem{rond07}
L.~Rondoni and C.~Mej\'{i}a-Monasterio.
\newblock Fluctuations in nonequilibrium statistical mechanics: models,
  mathematical theory, physical mechanisms.
\newblock {\em Nonlinearity}, 20:R1--R37, 2007.

\bibitem{zamp07}
F.~Zamponi.
\newblock Is it possible to experimentally verify the fluctuation relation? {A}
  review of theoretical motivations and numerical evidence.
\newblock {\em J.\ Stat.\ Mech.:\ Theor.\ Exp.}, page P02008, 2007.

\bibitem{sevi08}
E.M. Sevick, R.~Prabhakar, S.~R.Williams, and D.~J. Searles.
\newblock Fluctuation theorems.
\newblock {\em Annu. Rev. Phys. Chem.}, 59:603--633, 2008.

\bibitem{atta09}
P.~Attard.
\newblock The second entropy: a general theory for non-equilibrium
  thermodynamics and statistical mechanics.
\newblock {\em Annu. Rep. Prog. Chem. Sect. C}, 105:63--173, 2009.

\bibitem{jou01}
D.~Jou, J.~Casas-V\'azquez, and G.~Lebon.
\newblock {\em Extended irreversible thermodynamics}.
\newblock Springer, New York, Dordrecht, Heidelberg, London, 3rd enlarged
  edition, 2001.

\bibitem{otti04}
H.~C. \"Ottinger.
\newblock {\em Beyond Equilibrium Thermodynamics}.
\newblock Wiley, Hoboken, New Jersey and Canada, 2004.

\bibitem{regu05}
D.~Reguera, J.~M. Rub\'{i}, and J.~M.~G. Vilar.
\newblock The mesoscopic dynamics of thermodynamic systems.
\newblock {\em J.\ Phys.\ Chem.\ B}, 109:21502, 2005.

\bibitem{oono98}
Y.~Oono and M.~Paniconi.
\newblock Steady state thermodynamics.
\newblock {\em Prog.\ Theor.\ Phys.\ Suppl.}, 130:29, 1998.

\bibitem{sasa06}
S.~I. {S}asa and H.~Tasaki.
\newblock {Steady State Thermodynamics}.
\newblock {\em J.\ Stat.\ Phys.}, 125:125--224, 2006.

\bibitem{frey05}
E.~Frey and K.~Kroy.
\newblock {B}rownian motion: a paradigm of soft matter and biological physics.
\newblock {\em Ann. d. Physik}, 14:20--50, 2005.

\bibitem{chou11}
T.~Chou, K.~Mallick, and R.~K.~P. Zia.
\newblock Non-equilibrium statistical mechanics: from a paradigmatic model to
  biological transport.
\newblock {\em Rep. Prog. Phys.}, 74:116601, 2011.

\bibitem{gardiner}
C.~W. Gardiner.
\newblock {\em Handbook of Stochastic Methods}.
\newblock Springer-Verlag, Berlin, 3rd edition, 2004.

\bibitem{risken}
H.~Risken.
\newblock {\em The {F}okker-{P}lanck Equation}.
\newblock Springer-Verlag, Berlin, {2nd} edition, 1989.

\bibitem{chai01a}
M.~Chaichian and A.~Demichev.
\newblock {\em Path integrals in physics. Volume I: Stochastic processes and
  quantum mechanics}.
\newblock Institute of Physics publishing, Bristol and Philadelphia, 2001.

\bibitem{chai01b}
M.~Chaichian and A.~Demichev.
\newblock {\em Path integrals in physics. Volume II: Quantum field theory,
  statistical physics and other modern applications}.
\newblock Institute of Physics, Bristol and Philadelphia, 2001.

\bibitem{tail08}
J.~Tailleur, J.~Kurchan, and V.~Lecomte.
\newblock Mapping out-of-equilibrium into equilibrium in one-dimensional
  transport models.
\newblock {\em J. Phys. A Math. Theor.}, 41:505001, 2008.

\bibitem{vila08}
J.~M.~G. Vilar and J.~M. Rubi.
\newblock Failure of the work-{H}amiltonian connection for free-energy
  calculations.
\newblock {\em Phys.\ Rev.\ Lett.}, 100:020601, 2008.

\bibitem{peli08}
L.~Peliti.
\newblock Comment on '{F}ailure of the work-{H}amiltonian connection for
  free-energy calculations'.
\newblock {\em Phys.\ Rev.\ Lett.}, 101:098903, 2008.

\bibitem{peli08a}
L.~Peliti.
\newblock On the work{H}amiltonian connection in manipulated systems.
\newblock {\em J. Stat. Mech.}, page P05002, 2008.

\bibitem{horo08}
J.~Horowitz and C.~{J}arzynski.
\newblock Comment on '{F}ailure of the work-{H}amiltonian connection for
  free-energy calculations'.
\newblock {\em Phys.\ Rev.\ Lett.}, 101:098901, 2008.

\bibitem{spec08}
T.~Speck, J.~Mehl, and U.~Seifert.
\newblock Role of external flow and frame invariance in stochastic
  thermodynamics.
\newblock {\em Phys.\ Rev.\ Lett.}, 100:178302, 2008.

\bibitem{jarz08}
C.~{J}arzynski.
\newblock Nonequilibrium work relations: foundations and applications.
\newblock {\em Eur. Phys. J. B}, 64:331--340, 2008.

\bibitem{merh10}
N.~Merhav and Y.~Kafri.
\newblock Statistical properties of entropy production derived from fluctuation
  theorems.
\newblock {\em J.\ Stat.\ Mech.:\ Theor.\ Exp.}, page P12022, 2010.

\bibitem{cohe04}
E.~G.~D. {C}ohen and D.~Mauzerall.
\newblock A note on the {J}arzynski equality.
\newblock {\em J.\ Stat.\ Mech.:\ Theor.\ Exp.}, page P07006, 2004.

\bibitem{jarz04}
C.~{J}arzynski.
\newblock Nonequilibrium work theorem for a system strongly coupled to a
  thermal environment.
\newblock {\em J.\ Stat.\ Mech.:\ Theor.\ Exp.}, page P09005, 2004.

\bibitem{jarz07}
C.~{J}arzynski.
\newblock Comparison of far-from-equilibrium work relations.
\newblock {\em C. R. Physique}, 8:495--506, 2007.

\bibitem{horo07}
J.~Horowitz and C.~{J}arzynski.
\newblock Comparison of work fluctuation relations.
\newblock {\em J.\ Stat.\ Mech.:\ Theor.\ Exp.}, page P11002, 2007.

\bibitem{mazo99}
O.~Mazonka and C.~{J}arzynski.
\newblock Exactly solvable model illustrating far-from-equilibrium predictions.
\newblock cond-mat/9912121, 1999.

\bibitem{spec05}
T.~Speck and U.~Seifert.
\newblock Dissipated work in driven harmonic diffusive systems: General
  solution and application to stretching rouse polymers.
\newblock {\em Eur.\ Phys.\ J.\ B}, 43:543, 2005.

\bibitem{spec04}
T.~Speck and U.~Seifert.
\newblock Distribution of work in isothermal nonequilibrium processes.
\newblock {\em Phys.\ Rev.\ E}, 70:066112, 2004.

\bibitem{herm91}
J.~Hermans.
\newblock Simple analysis of noise and hysteresis in (slow-growth) free energy
  simulations.
\newblock {\em J.\ Phys.\ Chem.}, 95:9029, 1991.

\bibitem{wood91}
R.~H. Wood, W.~C.~F. M{\"u}hlbauer, and P.~T. Thompson.
\newblock Systematic errors in free energy perturbation calculations due to a
  finite sample of configuration space: Sample-size hysteresis.
\newblock {\em J.\ Phys.\ Chem.}, 95:6670, 1991.

\bibitem{hend01}
D.~A. Hendrix and C.~{J}arzynski.
\newblock A "fast growth" method of computing free energy differences.
\newblock {\em J.\ Chem.\ Phys.}, 114:5974, 2001.

\bibitem{spec11}
T.~Speck.
\newblock Work distribution for the driven harmonic oscillator with
  time-dependent strength: exact solution and slow driving.
\newblock {\em J. Phys. A: Math. Theor.}, 44:305001, 2011.

\bibitem{enge09}
A.~Engel.
\newblock Asymptotics of work distributions in nonequilibrium systems.
\newblock {\em Phys.\ Rev.\ E}, 80:021120, 2009.

\bibitem{nick11}
D.~Nickelsen and A.~Engel.
\newblock Asymptotics of work distributions: The pre-exponential factor.
\newblock {\em Eur.\ Phys.\ J.\ B}, 82:207--218, 2011.

\bibitem{saha11}
A.~Saha, J.~K. Bhattacharjee, and S.~Chakraborty.
\newblock Work probability distribution and tossing a biased coin.
\newblock {\em Phys.\ Rev.\ E}, 83:011104, 2011.

\bibitem{pere11}
C.~Perez-Espigares, A.~B. Kolton, and J.~Kurchan.
\newblock An infinite family of second law-like inequalities.
\newblock {\em arXiv: 1110.0967}, 2011.

\bibitem{trep04}
E.~H. Trepagnier, C.~{J}arzynski, F.~Ritort, G.~E. {C}rooks, C.~J. Bustamante,
  and J.~Liphardt.
\newblock Experimental test of {H}atano and {S}asa's nonequilibrium
  steady-state equality.
\newblock {\em Proc.\ Natl.\ Acad.\ Sci.\ U.S.A.}, 101:15038, 2004.

\bibitem{spec05a}
T.~Speck and U.~Seifert.
\newblock Integral fluctuation theorem for the housekeeping heat.
\newblock {\em J.\ Phys.\ A:\ Math.\ Gen.}, 38:L581--L588, 2005.

\bibitem{cher06}
V.~Y. Chernyak, M.~Chertkov, and C.~{J}arzynski.
\newblock Path-integral analysis of fluctuation theorems for general {L}angevin
  processes.
\newblock {\em J.\ Stat.\ Mech.:\ Theor.\ Exp.}, page P08001, 2006.

\bibitem{chet08a}
R.~Chetrite and K.~Gawedzki.
\newblock Fluctuation relations for diffusion processes.
\newblock {\em Comm. Math. Phys.}, 282:469--518, 2008.

\bibitem{ge08a}
H.~Ge and D.-Q. Jiang.
\newblock Generalized {J}arzynski's equality of inhomogeneous multidimensional
  diffusion processes.
\newblock {\em J.\ Stat.\ Phys.}, 131:675--689, 2008.

\bibitem{liu09a}
F.~Liu and Z.-C. Ou-Yang.
\newblock Generalized integral fluctuation theorem for diffusion processes.
\newblock {\em Phys.\ Rev.\ E}, 79:060107, 2009.

\bibitem{sugh11}
Y.~Sughiyama and M.~Ohzeki.
\newblock Extended {J}arzynski equality in general {L}angevin systems.
\newblock {\em Physica E}, 43:790--793, 2011.

\bibitem{schm06a}
T.~Schmiedl, T.~Speck, and U.~Seifert.
\newblock Entropy production for mechanically or chemically driven
  biomolecules.
\newblock {\em J.\ Stat.\ Phys.}, 128:77, 2007.

\bibitem{garc10}
R.~Garc\'{i}a-Garc\'{i}a, D.~Dominguez, V.~Lecomte, and A.~B. Kolton.
\newblock Unifying approach for fluctuation theorems from joint probability
  distributions.
\newblock {\em Phys.\ Rev.\ E}, 82:030104(R), 2010.

\bibitem{mara08}
P.~Maragakis, F.~Ritort, C.~Bustamante, M.~Karplus, and G.~E. {C}rooks.
\newblock Bayesian estimates of free energies from nonequilibrium work data in
  the presence of instrument noise.
\newblock {\em J. Chem. Phys.}, 129:024102, 2008.

\bibitem{juni09}
I.~Junier, A.~Mossa, M.~Manosas, and F.~Ritort.
\newblock Recovery of free energy branches in single molecule experiments.
\newblock {\em Phys.\ Rev.\ Lett.}, 102:070602, 2009.

\bibitem{schu05}
S.~Schuler, T.~Speck, C.~Tietz, J.~Wrachtrup, and U.~Seifert.
\newblock Experimental test of the fluctuation theorem for a driven two-level
  system with time-dependent rates.
\newblock {\em Phys.\ Rev.\ Lett.}, 94:180602, 2005.

\bibitem{baie06}
M.~Baiesi, T.~Jacobs, C.~Maes, and N.~S. Skantzos.
\newblock Fluctuation symmetries for work and heat.
\newblock {\em Phys.\ Rev.\ E}, 74:021111, 2006.

\bibitem{tiet06}
C.~Tietz, S.~Schuler, T.~Speck, U.~Seifert, and J.~Wrachtrup.
\newblock Measurement of stochastic entropy production.
\newblock {\em Phys.\ Rev.\ Lett.}, 97:050602, 2006.

\bibitem{shar09}
B.~H. Shargel and T.~Chou.
\newblock Fluctuation theorems for entropy production and heat dissipation in
  periodically driven markov chains.
\newblock {\em J.\ Stat.\ Phys.}, 137:165--188, 2009.

\bibitem{koma08}
T.~S. Komatsu and N.~Kakagawa.
\newblock Expression for the stationary distribution in nonequilibrium steady
  states.
\newblock {\em Phys.\ Rev.\ Lett.}, 100:030601, 2008.

\bibitem{koma08a}
T.~S. Komatsu, N.~Nakagawa, and S.~{S}asa.
\newblock Steady-state thermodynamics for heat conduction: Microscopic
  derivation.
\newblock {\em Phys.\ Rev.\ Lett.}, 100:230602, 2008.

\bibitem{koma09}
T.~S. Komatsu, N.~Nakagawa, S.~{S}asa, and H.~Tasa.
\newblock Representation of nonequilibrium steady states in large mechanical
  systems.
\newblock {\em J.\ Stat.\ Phys.}, 134:401--423, 2009.

\bibitem{koma11}
T.~S. Komatsu, N.~Nakagawa, S.~{S}asa, and H.~Tasaki.
\newblock Entropy and nonlinear thermodynamic relation for heat conducting
  steady states.
\newblock {\em J.\ Stat.\ Phys.}, 142:127--153, 2011.

\bibitem{cola11}
M.~Colangeli, C.~Maes, and B.~Wynants.
\newblock A meaningful expansion around detailed balance.
\newblock {\em J. Phys. A Math. Theor.}, 44:095001, 2011.

\bibitem{impa06}
A.~Imparato and L.~Peliti.
\newblock Fluctuation relations for a driven {B}rownian particle.
\newblock {\em Phys.\ Rev.\ E}, 74:026106, 2006.

\bibitem{turi07}
K.~Turitsyn, M.~Chertkov, V.Y. Chernyak, and A.~Puliafito.
\newblock Statistics of entropy production in linearized stochastic systems.
\newblock {\em Phys.\ Rev.\ Lett.}, 98(180603):180603, 2007.

\bibitem{prad10a}
P.~Pradhan and U.~Seifert.
\newblock Nonexistence of classical diamagnetism and nonequilibrium fluctuation
  theorems for charged particles on a curved surface.
\newblock {\em EPL}, 89:37001, 2010.

\bibitem{kuma09}
N.~Kumar and K.~V. Kumar.
\newblock Classical {L}angevin dynamics of a charged particle moving on a
  sphere and diamagnetism: A surprise.
\newblock {\em EPL}, 86:17001, 2009.

\bibitem{espo10b}
M.~Esposito and C.~van~den Broeck.
\newblock Three detailed fluctuation theorems.
\newblock {\em Phys.\ Rev.\ Lett.}, 104:090601, 2010.

\bibitem{espo10d}
M.~Esposito and C.~van~den Broeck.
\newblock Three faces of the second law. {I.} {M}aster equation formulation.
\newblock {\em Phys.\ Rev.\ E}, 82:011143, 2010.

\bibitem{vdb10a}
C.~van~den Broeck and M.~Esposito.
\newblock Three faces of the second law. {II.} {F}okker-{P}lanck formulation.
\newblock {\em Phys.\ Rev.\ E}, 82:011144, 2010.

\bibitem{garc12}
R.~Garc\'{i}a-Garc\'{i}a, V.~Lecomte, A.~B. Kolton, and D.~Dominguez.
\newblock Joint probability distributions and fluctuation theorems.
\newblock {\em J.\ Stat.\ Mech.:\ Theor.\ Exp.}, page P02009, 2012.

\bibitem{verl12}
G.~Verley, R.~Ch\'etrite, and D.~Lacoste.
\newblock Inequalities generalizing the second law of thermodynamics for
  transitions between non-stationary states.
\newblock {\em Phys.\ Rev.\ Lett.}, 108:120601, 2012.

\bibitem{kard07}
M.~Kardar.
\newblock {\em Statistical Physics of Fields}.
\newblock Cambridge Univ. Press, Cambridge, 2007.

\bibitem{baie05}
M.~Baiesi and C.~Maes.
\newblock Enstrophy dissipation in two-dimensional turbulence.
\newblock {\em Phys.\ Rev.\ E}, 72:056314, 2005.

\bibitem{mall11}
K.~Mallick, M.~Moshe, and H.~Orland.
\newblock A field-theoretic approach to non-equilibrium work identities.
\newblock {\em J. Phys. A: Math. Theor.}, 44:095002, 2011.

\bibitem{rama10}
S.~Ramaswamy.
\newblock The mechanics and statistics of active matter.
\newblock {\em Ann. Rev. Cond. Mat. Phys.}, 1:323--345, 2010.

\bibitem{must10}
V.~Mustonen and M.~L\"assig.
\newblock Fitness flux and ubiquity of adaptive evolution.
\newblock {\em Proc.\ Natl.\ Acad.\ Sci.\ U.S.A.}, 107:4248--4251, 2010.

\bibitem{foge09}
H.~C. Fogedby and A.~Imparato.
\newblock Heat distribution function for motion in a general potential at low
  temperature.
\newblock {\em J. Phys. A: Math. Theor.}, 42:475004, 2009.

\bibitem{chat10}
D.~Chatterjee and B.~J. Cherayil.
\newblock Exact path-integral evaluation of the heat distribution function of a
  trapped {B}rownian oscillator.
\newblock {\em Phys.\ Rev.\ E}, 82:051104, 2010.

\bibitem{chat11}
D.~Chatterjee and B.~J. Cherayil.
\newblock Single-molecule thermodynamics: the heat distribution function of a
  charged particle in a static magnetic field.
\newblock {\em J.\ Stat.\ Mech.:\ Theor.\ Exp.}, page P03010, 2011.

\bibitem{wang02}
G.~M. Wang, E.~M. Sevick, E.~Mittag, D.~J. Searles, and D.~J. {E}vans.
\newblock Experimental demonstration of violations of the second law of
  thermodynamics for small systems and short time scales.
\newblock {\em Phys.\ Rev.\ Lett.}, 89:050601, 2002.

\bibitem{wang05}
G.~M. Wang, J.~C. Reid, D.~M. Carberry, D.~R.~M. Williams, E.~M. Sevick, and
  D.~J. {E}vans.
\newblock Experimental study of the fluctuation theorem in a nonequilibrium
  steady state.
\newblock {\em Phys.\ Rev.\ E}, 71:046142, 2005.

\bibitem{zon03}
R.~van Zon and E.~G.~D. {C}ohen.
\newblock Stationary and transient work-fluctuation theorems for a dragged
  {B}rownian particle.
\newblock {\em Phys.\ Rev.\ E}, 67:046102, 2003.

\bibitem{zon03a}
R.~van Zon and E.~G.~D. {C}ohen.
\newblock Extension of the fluctuation theorem.
\newblock {\em Phys.\ Rev.\ Lett.}, 91:110601, 2003.

\bibitem{zon04a}
R.~van Zon and E.~G.~D. {C}ohen.
\newblock Extended heat-fluctuation theorems for a system with deterministic
  and stochastic forces.
\newblock {\em Phys.\ Rev.\ E}, 69:056121, 2004.

\bibitem{impa07c}
A.~Imparato, L.~Peliti, G.~Pesce, G.~Rusciano, and A.~Sasso.
\newblock Work and heat probability distribution of an optically driven
  {B}rownian particle: Theory and experiments.
\newblock {\em Phys. Rev. E}, 76(5):050101, Nov 2007.

\bibitem{saha09}
A.~Saha, S.~Lahiri, and A.~M. Jayannavar.
\newblock Entropy production theorems and some consequences.
\newblock {\em Phys.\ Rev.\ E}, 80:011117, 2009.

\bibitem{jaya07}
A.~M. Jayannavar and M.~Sahoo.
\newblock Charged particle in a mangetic field: {J}arzynski equality.
\newblock {\em Phys.\ Rev.\ E}, 75:032102, 2007.

\bibitem{jime09}
J.~I. Jimenez-Aquino, R.~M. Velasco, and F.~J. Uribe.
\newblock Fluctuation relations for a classical harmonic oscillator in an
  electromagnetic field.
\newblock {\em Phys.\ Rev.\ E}, 79:061109, 2009.

\bibitem{jime10}
J.~I. Jimenez-Aquino, F.~J. Uribe, and R.~M. Velasco.
\newblock Work-fluctuation theorems for a particle in an electromagnetic field.
\newblock {\em J. Phys. A Math. Theor.}, 43:255001, 2010.

\bibitem{jime10a}
J.~I. Jimenez-Aquino.
\newblock Entropy production for a charged particle in an eletromagnetic field.
\newblock {\em Phys.\ Rev.\ E}, 82:051118, 2010.

\bibitem{jime11}
J.~I. Jimenez-Aquino.
\newblock Work-fluctuation theorem for a charged harmonic oscillator.
\newblock {\em J. Phys. A Math. Theor.}, 44:295002, 2011.

\bibitem{saha08}
A.~Saha and A.~M. Jayannavar.
\newblock Nonequilibrium work distributions for a trapped {B}rownian particle
  in a time-dependent magnetic field.
\newblock {\em Phys.\ Rev.\ E}, 77:022105, 2008.

\bibitem{zon04}
R.~van Zon, S.~Ciliberto, and E.~G.~D. {C}ohen.
\newblock Power and heat fluctuation theorems for electric circuits.
\newblock {\em Phys.\ Rev.\ Lett.}, 92:130601, 2004.

\bibitem{garn05}
N.~Garnier and S.~Ciliberto.
\newblock Nonequilibrium fluctuations in a resistor.
\newblock {\em Phys.\ Rev.\ E}, 71:060101(R), 2005.

\bibitem{joub08}
S.~Joubaud, N.~B. Garnier, and S.~Ciliberto.
\newblock Fluctuations of the total entropy production in stochastic systems.
\newblock {\em EPL}, 82:30007, 2008.

\bibitem{falc09}
C.~Falcon and E.~Falcon.
\newblock Fluctuations of energy flux in a simple dissipative
  out-of-equilibrium system.
\newblock {\em Phys.\ Rev.\ E}, 79:041110, 2009.

\bibitem{bona09}
M.~Bonaldi, L.~Conti, P.~De Gregorio, L.~Rondoni, G.~Vedovato, A.~Vinante,
  M.~Bignotto, M.~Cerdonio, P.~Falferi, N.~Liguori, S.~Longo, R.~Mezzena,
  A.~Ortolan, G.~A. Prodi, F.~Salemi, L.~Taffarello, S.~Vitale, and J.-P.
  Zendri.
\newblock Nonequilibrium steady-state fluctuations in actively cooled
  resonators.
\newblock {\em Phys.\ Rev.\ Lett.}, 103:010601, 2009.

\bibitem{berg08}
J.~Berg.
\newblock Out-of-equilibrium dynamics of gene expression and the {J}arzynski
  equation.
\newblock {\em Phys.\ Rev.\ Lett.}, 100:188101, 2008.

\bibitem{carb04}
D.~M. Carberry, J.~C. Reid, G.~M. Wang, E.~M. Sevick, D.~J. Searles, and D.~J.
  {E}vans.
\newblock Fluctuations and irreversibility: An experimental demonstration of a
  second-law-like theorem using a colloidal particle held in an optical trap.
\newblock {\em Phys.\ Rev.\ Lett.}, 92:140601, 2004.

\bibitem{khan11}
M.~Khan and A.~K. Sood.
\newblock Irreversibility-to-reversibility crossover in transient response of
  an optically trapped particle.
\newblock {\em EPL}, 94:60003, 2011.

\bibitem{gome11a}
J.~R. Gomez-Solano, A.~Petrosyan, and S.~Ciliberto.
\newblock Heat fluctuations in a nonequilibrium bath.
\newblock {\em Phys.\ Rev.\ Lett.}, 106:200602, 2011.

\bibitem{blic06}
V.~Blickle, T.~Speck, L.~Helden, U.~Seifert, and C.~Bechinger.
\newblock Thermodynamics of a colloidal particle in a time-dependent
  nonharmonic potential.
\newblock {\em Phys.\ Rev.\ Lett.}, 96:070603, 2006.

\bibitem{sun03}
S.~X. Sun.
\newblock Generating generalized distributions from dynamical simulation.
\newblock {\em J.\ Chem.\ Phys.}, 118:5769, 2003.

\bibitem{gamm98}
L.~Gammaitoni, P.~H\"anggi, P.~Jung, and F.~Marchesoni.
\newblock Stochastic resonance.
\newblock {\em Rev. Mod. Phys.}, 70:223--287, 1998.

\bibitem{jop08}
P.~Jop, A.~Petrosyan, and S.~Ciliberto.
\newblock Work and dissipation fluctuations near the stochastic resonance of a
  colloidal particle.
\newblock {\em EPL}, 81(5):50005, 2008.

\bibitem{impa08}
A.~Imparato, P.~Jop, A.~Petrosyan, and S.~Ciliberto.
\newblock Probability density functions of work and heat near the stochastic
  resonance of a colloidal particle.
\newblock {\em J.\ Stat.\ Mech.:\ Theor.\ Exp.}, page P10017, 2008.

\bibitem{iwai01}
T.~Iwai.
\newblock Study of stochastic resonance by method of stochastic energetics.
\newblock {\em Physica A}, 300:350--358, 2001.

\bibitem{dan05}
D.~Dan and A.~M. Jayannavar.
\newblock Bona fide stochastic resonance: a view point from stochastic
  energetics.
\newblock {\em Physica A}, 345:404--410, 2005.

\bibitem{lahi09}
S.~Lahiri and A.~M. Jayannavar.
\newblock Total entropy production fluctuation theorems in a nonequilibrium
  time-periodic steady state.
\newblock {\em Eur. Phys. J. B}, 69:87--92, 2009.

\bibitem{spec07}
T.~Speck, V.~Blickle, C.~Bechinger, and U.~Seifert.
\newblock Distribution of entropy production for a colloidal particle in a
  nonequilibrium steady state.
\newblock {\em EPL}, 79:30002, 2007.

\bibitem{mari06}
A.~Gomez-Marin and I.~Pagonabarraga.
\newblock Test of the fluctuation theorem for stochastic entropy production in
  a nonequilibrium steady state.
\newblock {\em Phys.\ Rev.\ E}, 74:061113, 2006.

\bibitem{mehl08}
J.~Mehl, T.~Speck, and U.~Seifert.
\newblock Large deviation function for entropy production in driven
  one-dimensional systems.
\newblock {\em Phys.\ Rev.\ E}, 78:011123, 2008.

\bibitem{nemo11}
T.~Nemoto and S.~{S}asa.
\newblock Variational formula for experimental determination of high-order
  correlations of current fluctuations in driven systems.
\newblock {\em Phys.\ Rev.\ E}, 83:030105, 2011.

\bibitem{doua05}
F.~Douarche, S.~Ciliberto, A.~Petrosyan, and I.~Rabbiosi.
\newblock An experimental test of the {J}arzynski equality in a mechanical
  experiment.
\newblock {\em Europhys.\ Lett.}, 70:593, 2005.

\bibitem{doua05a}
F.~Douarche, S.~Ciliberto, and A.~Petrosyan.
\newblock Estimate of the free energy difference in mechanical systems from
  work fluctuations: experiments and models.
\newblock {\em J.\ Stat.\ Mech.:\ Theor.\ Exp.}, page P09011, 2005.

\bibitem{doua06}
F.~Douarche, S.~Joubaud, N.~B. Garnier, A.~Petrosyan, and S.~Ciliberto.
\newblock Work fluctuation theorems for harmonic oscillators.
\newblock {\em Phys.\ Rev.\ Lett.}, 97:140603, 2006.

\bibitem{joub07a}
S.~Joubaud, N.~B. Garnier, and S.~Ciliberto.
\newblock Fluctuation theorems for harmonic oscillators.
\newblock {\em J.\ Stat.\ Mech.:\ Theor.\ Exp.}, page P09018, 2007.

\bibitem{fara02}
J.~Farago.
\newblock Injected power fluctuations in {L}angevin equation.
\newblock {\em J.\ Stat.\ Phys.}, 107:781, 2002.

\bibitem{tani07a}
T.~Taniguchi and E.~G.~D. {C}ohen.
\newblock {O}nsager-{M}achlup theory for nonequilibrium steady states and
  fluctuation theorems.
\newblock {\em J.\ Stat.\ Phys.}, 126:1--41, 2007.

\bibitem{tani08}
T.~Taniguchi and E.~G.~D. {C}ohen.
\newblock Inertial effects in nonequilibrium work fluctuations by a path
  integral approach.
\newblock {\em J.\ Stat.\ Phys.}, 130:1--26, 2008.

\bibitem{tani07}
T.~Taniguchi and E.~G.~D. {C}ohen.
\newblock Nonequilibrium steady state thermodynamics and fluctuations for
  stochastic systems.
\newblock {\em J.\ Stat.\ Phys.}, 130:633, 2008.

\bibitem{cohe08}
E.~G.~D. {C}ohen.
\newblock Properties of nonequilibrium steady states: a path integral approach.
\newblock {\em J.\ Stat.\ Mech.:\ Theor.\ Exp.}, page P07014, 2008.

\bibitem{minh09a}
D.~D.~L. Minh and A.~B. Adib.
\newblock Path integral analysis of {J}arzynski's equality: Analytical results.
\newblock {\em Phys.\ Rev.\ E}, 79:021122, 2009.

\bibitem{lev10}
B.~Lev and A.~D. Kiselev.
\newblock Energy representation for nonequilibrium {B}rownian-like systems:
  Steady states and fluctuation relations.
\newblock {\em Phys.\ Rev.\ E}, 82:031101, 2010.

\bibitem{fing07}
A.~Fingerle.
\newblock Relativistic fluctuation theorems.
\newblock {\em C. R. Physique}, 8:696--713, 2007.

\bibitem{iso11}
S.~Iso and S.~Okazawa.
\newblock Stochastic equations in black hole backgrounds and non-equilibrium
  fluctuation theorems.
\newblock {\em Nucl. Phys. B}, 851:380--419, 2011.

\bibitem{sabh11}
S.~Sabhapandit.
\newblock Work fluctuations for a harmonic oscillator driven by an external
  random force.
\newblock {\em EPL}, 96:20005, 2011.

\bibitem{gome10}
J.~R. Gomez-Solano, L.~Bellon, A.~Petrosyan, and S.~Ciliberto.
\newblock Steady-state fluctuation relations for systems driven by an external
  random force.
\newblock {\em EPL}, 89(6):60003, 2010.

\bibitem{visc06}
P.~Visco.
\newblock Work fluctuations for a {B}rownian particle between two thermostats.
\newblock {\em J.\ Stat.\ Mech.:\ Theor.\ Exp.}, page P06006, 2006.

\bibitem{tome10}
T.~Tom\'e and M.~J. de~Oliveira.
\newblock Entropy production in irreversible systems described by a
  {F}okker-{P}lanck equation.
\newblock {\em Phys.\ Rev.\ E}, 82:021120, 2010.

\bibitem{foge11}
H.~C. Fogedby and A.~Imparato.
\newblock A bound particle coupled to two thermostats.
\newblock {\em J.\ Stat.\ Mech.:\ Theor.\ Exp.}, page P05015, 2011.

\bibitem{auma01}
S.~Aumaitre, S.~Fauve, S.~McNamara, and P.~Poggi.
\newblock Power injected in dissipative systems and the fluctuation theorem.
\newblock {\em Eur. Phys. J. B}, 19:449--460, 2001.

\bibitem{feit04}
K.~Feitosa and N.~Menon.
\newblock Fluidized granular medium as an instance of the fluctuation theorem.
\newblock {\em Phys.\ Rev.\ Lett.}, 92:164301, 2004.

\bibitem{kuma11}
N.~Kumar, S.~Ramaswamy, and A.~K. Sood.
\newblock Symmetry properties of the large-deviation function of the velocity
  of a self-propelled polar particle.
\newblock {\em Phys.\ Rev.\ Lett.}, 106:118001, 2011.

\bibitem{naer11}
A.~Naert.
\newblock Experimental study of work exchange with a granular gas: the
  viewpoint of the fluctuation theorem.
\newblock {\em EPL}, 97:20010, 2012.

\bibitem{wils11}
L.~G. Wilson, A.~W. Harrison, W.~C.~K. Poon, and A.~M. Puertas.
\newblock Microrheology and the fluctuation theorem in dense colloids.
\newblock {\em EPL}, 93:58007, 2011.

\bibitem{visc05}
P.~Visco, A.~Puglisi, A.~Barrat, E.~Trizac, and F.~Van Wijland.
\newblock Injected power and entropy flow in a heated granular gas.
\newblock {\em Europhys. Lett.}, 72:55--61, 2005.

\bibitem{pugl05a}
A.~Puglisi, P.~Visco, A.~Barrat, E.~Trizac, and F.~van Wijland.
\newblock Fluctuations of internal energy flow in a vibrated granular gas.
\newblock {\em Phys.\ Rev.\ Lett.}, 95:110202, 2005.

\bibitem{pugl06}
A.~Puglisi, P.~Visco, E.~Trizac, and F.~Van Wijland.
\newblock Dynamics of a tracer granular perticle as a nonequilibrium markov
  process.
\newblock {\em Phys.\ Rev.\ E}, 73:021301, 2006.

\bibitem{pugl06a}
A.~Puglisi, L.~Rondoni, and A.~Vulpiani.
\newblock Relevance of initial and final conditions for the fluctuation
  relation in markov processes.
\newblock {\em J.\ Stat.\ Mech.:\ Theor.\ Exp.}, page P08010, 2006.

\bibitem{sarr10}
A.~Sarracino, D.~Villamaina, G.~Gradenigo, and A.~Puglisi.
\newblock Irreversible dynamics of a massive intruder in dense granular fluids.
\newblock {\em EPL}, page 34001, 2010.

\bibitem{chon10}
S.-H. Chong, M.~Otsuki, and H.~Hayakawa.
\newblock Generalized {G}reen-{K}ubo relation and integral fluctuation theorem
  for driven dissipative systems without microscopic time reversibility.
\newblock {\em Phys.\ Rev.\ E}, 81:041130, 2010.

\bibitem{droc11}
J.~A. Drocco, C.~J.~Olson Reichhardt, and C.~Reichhardt.
\newblock Characterizing plastic depinning dynamics with the fluctuation
  theorem.
\newblock {\em Eur.\ Phys.\ J.\ E}, 34:117, 2011.

\bibitem{grad11}
G.~Gradenigo, U.~Marini~Bettolo Marconi, A.~Puglisi, and A.~Sarracino.
\newblock Non-equilibrium fluctuations in a driven stochastic lorentz gas.
\newblock {\em arXiv: 1111.5798}, 2011.

\bibitem{cili98}
S.~Ciliberto and C.~Laroche.
\newblock An experimental test of the {G}allavotti-{C}ohen fluctuation theorem.
\newblock {\em J.\ Phys.\ IV\ France}, 08:215--219, 1998.

\bibitem{cili04}
S.~Ciliberto, N.~Garnier, S.~Hernandez, C.~Lacpatia, J.-F. Pinton, and G.~Ruiz
  Chavarria.
\newblock Experimental test of the {G}allavotti-{C}ohen fluctuation theorem in
  turbulent flows.
\newblock {\em Physica\ A}, 340:240, 2004.

\bibitem{shan05}
X.-D. Shang, P.~Tong, and K.-Q. Xia.
\newblock Test of steady state fluctuation theorem in turbulent
  {R}ayleigh-{B}\'enard convection.
\newblock {\em Phys.\ Rev.\ E}, 72:015301R, 2005.

\bibitem{gilb04}
T.~Gilbert.
\newblock Entropy fluctuations in shell models of turbulence.
\newblock {\em Europhys. Lett.}, 67:172--178, 2004.

\bibitem{belu11}
M.~Belushkin, R.~Livi, and G.~Foffi.
\newblock Hydrodynamics and the fluctuation theorem.
\newblock {\em Phys.\ Rev.\ Lett.}, 106:210601, 2011.

\bibitem{gold01}
W.~I. Goldburg, Y.~Y. Goldschmidt, and H.~Kellay.
\newblock Fluctuation and dissipation in liquid-crystal electroconvection.
\newblock {\em Phys.\ Rev.\ Lett.}, 87:245502, 2001.

\bibitem{joub09}
S.~Joubaud, G.~Huillard, A.~Petrosyan, and S.~Ciliberto.
\newblock Work fluctuations in a nematic liquid crystal.
\newblock {\em J.\ Stat.\ Mech.:\ Theor.\ Exp.}, page P01033, 2009.

\bibitem{cado08}
O.~Cadot, A.~Boudaoud, and C.~Touzé.
\newblock Statistics of power injection in a plate set into chaotic vibration.
\newblock {\em Eur. Phys. J. B}, 66:399--407, 2008.

\bibitem{suzu11}
R.~Suzuki, H.~R. Jiang, and M.~Sano.
\newblock Validity of fluctuation theorem on self-propelling particles.
\newblock {\em arXiv: 1104.5607}, 2011.

\bibitem{vankampen}
N.~G. van Kampen.
\newblock {\em Stochastic Processes in Physics and Chemistry}.
\newblock North-Holland, Amsterdam, 1981.

\bibitem{schn76}
J.~{S}chnakenberg.
\newblock Network theory of microscopic and macroscopic behavior of master
  equation systems.
\newblock {\em Rev.\ Mod.\ Phys.}, 48:571, 1976.

\bibitem{hill}
T.~L. Hill.
\newblock {\em Free Energy Transduction and Biochemical Cycle Kinetics}.
\newblock Dover, Mineola, New York, 2nd edition, 1989.

\bibitem{zia06}
R.K.P. Zia and B.~Schmittmann.
\newblock A possible classification of nonequilibrium steady states.
\newblock {\em J.\ Phys.\ A:\ Math.\ Gen.}, 39:L407, 2006.

\bibitem{zia07}
R.~K.~P. Zia and B.~Schmittmann.
\newblock Probability currents as principal characteristics in the statistical
  mechanics of non-equilibrium steady states.
\newblock {\em J.\ Stat.\ Mech.:\ Theor.\ Exp.}, page P07012, 2007.

\bibitem{evan04}
R.~M.~L. {E}vans.
\newblock Rules for transition rates in nonequilibrium steady states.
\newblock {\em Phys.\ Rev.\ Lett.}, 92:150601, 2004.

\bibitem{baul08}
A.~Baule and R.~M.~L. {E}vans.
\newblock Invariant quantities in shear flow.
\newblock {\em Phys.\ Rev.\ Lett.}, 101:240601, 2008.

\bibitem{baul10}
A.~Baule and R.~M.~L. {E}vans.
\newblock Nonequilibrium statistical mechanics of shear flow: invariant
  quantities and current relations.
\newblock {\em J.\ Stat.\ Mech.:\ Theor.\ Exp.}, page P03030, 2010.

\bibitem{mont11}
C.~Monthus.
\newblock Non-equilibrium steady states: maximization of the {S}hannon entropy
  associated with the distribution of dynamical trajectories in the presence of
  constraints.
\newblock {\em J.\ Stat.\ Mech.:\ Theor.\ Exp.}, 2011:P03008, 2011.

\bibitem{poll11}
M.~Polettini.
\newblock Macroscopic constraints for the minimum entropy production principle.
\newblock {\em Phys.\ Rev.\ E}, 84:051117, 2011.

\bibitem{plat11}
T.~Platini.
\newblock Measure of the violation of the detailed balance criterion: A
  possible definition of a 'distance' from equilibrium.
\newblock {\em Phys.\ Rev.\ E}, 83:011119, 2011.

\bibitem{maes11}
C.~Maes, K.~Netocn\'y, and B.~Wynants.
\newblock Monotonic return to steady nonequilibrium.
\newblock {\em Phys.\ Rev.\ Lett.}, 107:010601, 2011.

\bibitem{poll12}
M.~Polettini.
\newblock Nonequilibrium thermodynamics as a gauge theory.
\newblock {\em EPL}, 97:30003, 2012.

\bibitem{luo84}
J.~L. Luo, C.~van~den Broeck, and G.~Nicolis.
\newblock Stability-criteria and fluctuations around nonequilibrium states.
\newblock {\em Z. Phys. B Cond. Mat.}, 56:165--170, 1984.

\bibitem{saga11b}
T.~Sagawa and H.~Hayaka.
\newblock Geometrical expression of excess entropy production.
\newblock {\em Phys.\ Rev.\ E}, 84:051110, 2011.

\bibitem{ge10}
H.~Ge and H.~Qian.
\newblock Physical origins of entropy production, free energy dissipation, and
  their mathematical representations.
\newblock {\em Phys.\ Rev.\ E}, 81:051133, 2010.

\bibitem{altl10}
A.~Altland, A.~De Martino, R.~Egger, and B.~Narohzny.
\newblock Fluctuation relations and rare realizations of transport.
\newblock {\em Phys.\ Rev.\ Lett.}, 105:170601, 2010.

\bibitem{altl10a}
A.~Altland, A.~De Martino, R.~Egger, and B.~Narohzny.
\newblock Transient fluctuation relations for time-dependent particle
  transport.
\newblock {\em Phys. Rev. B}, 82:115323, 2010.

\bibitem{liu09}
F.~Liu, Y.-P. Luo, M.-C. Huang, and Z.-C. Ou-Yang.
\newblock A generalized integral fluctuation theorem for general jump
  processes.
\newblock {\em J. Phys. A Math. Theor.}, 42:332003, 2009.

\bibitem{monn04}
T.~Monnai.
\newblock Fluctuation theorem in ratchet system.
\newblock {\em J. Phys. A: Math. Gen.}, 37:L75--79, 2004.

\bibitem{seif05}
U.~Seifert.
\newblock Fluctuation theorem for a single enzym or molecular motor.
\newblock {\em Europhys.\ Lett.}, 70:36, 2005.

\bibitem{saka06}
H.~Sakaguchi.
\newblock Efficiency and fluctuation in tight-coupling model of molecuar motor.
\newblock {\em J. Phys. Soc. Japan}, 75:063001, 2006.

\bibitem{bere08}
A.~M. Berezhkovskii and S.~M. Bezrukov.
\newblock Counting translocations of strongly repelling particles through
  single channels: fluctuation theorem for membrane transport.
\newblock {\em Phys.\ Rev.\ Lett.}, 100:038104, 2008.

\bibitem{sun10}
B.~Sun, D.~G. Grier, and A.~Y. Grosberg.
\newblock Minimal model for {B}rownian vortexes.
\newblock {\em Phys.\ Rev.\ E}, 82:021123, 2010.

\bibitem{kuma11a}
N.~Kumar, C.~van~den Broeck, M.~Esposito, and K.~Lindenberg.
\newblock Thermodynamics of a stochastic twin elevator.
\newblock {\em Phys.\ Rev.\ E}, 84:051134, 2011.

\bibitem{aver11}
D.~V. Averin and J.~P. Pekola.
\newblock Statistics of the dissipated energy in driven single-electron
  transitions.
\newblock {\em EPL}, 96:67004, 2011.

\bibitem{doro11}
S.~Dorosz and M.~Pleimling.
\newblock Entropy production in the nonequilibrium steady states of interacting
  many-body systems.
\newblock {\em Phys.\ Rev.\ E}, 83:031107, 2011.

\bibitem{doro09}
S.~Dorosz and M.~Pleimling.
\newblock Fluctuation ratios in the absence of microscopic time reversibility.
\newblock {\em Phys.\ Rev.\ E}, 79:030102(R), 2009.

\bibitem{bena11}
D.~Ben-Avraham, S.~Dorosz, and M.~Pleimling.
\newblock Realm of validity of the {C}rooks relation.
\newblock {\em Phys.\ Rev.\ E}, 83:041129, 2011.

\bibitem{andr10}
B.~Andrae, J.~Cremer, T.~Reichenbach, and E.~Frey.
\newblock Entropy production of cyclic population dynamics.
\newblock {\em Phys.\ Rev.\ Lett.}, 104:218102, 2010.

\bibitem{cleu06}
B.~Cleuren, C.~van~den Broeck, and R.~Kawai.
\newblock Fluctuation theorem for the effusion of an ideal gas.
\newblock {\em Phys.\ Rev.\ E}, 74:021117, 2006.

\bibitem{mara05}
R.~Marathe and A.~Dhar.
\newblock Work distribution functions for hysteresis loops in a single-spin
  system.
\newblock {\em Phys.\ Rev.\ E}, 72:066112, 2005.

\bibitem{eina09}
M.~Einax and P.~Maass.
\newblock Work distributions for ising chains in a time-dependent magnetic
  field.
\newblock {\em Phys.\ Rev.\ E}, 80:020101(R), 2009.

\bibitem{ohze10}
M.~Ohzeki and H.~Nishimori.
\newblock Nonequilibrium relations for spin glasses with gauge symmetry.
\newblock {\em J. Phys. So}, 79:084003, 2010.

\bibitem{chat07}
C.~Chatelain.
\newblock A temperature-extended {J}arzynski relation: application to the
  numerical calculation of surface tension.
\newblock {\em J.\ Stat.\ Mech.:\ Theor.\ Exp.}, page P04011, 2007.

\bibitem{croc05}
L.~Crochik and T.~Tom\'e.
\newblock Entropy production in the majority-vote model.
\newblock {\em Phys.\ Rev.\ E}, 72:057103, 2005.

\bibitem{deol11}
M.~J. de~Oliveira.
\newblock Irreversible models with {B}oltzmann-{G}ibbs probability distribution
  and entropy production.
\newblock {\em J.\ Stat.\ Mech.:\ Theor.\ Exp.}, page P12012, 2011.

\bibitem{tome12}
T.~Tom\'e and M.~J. de~Oliveira.
\newblock Entropy production in nonequilibrium systems at stationary states.
\newblock {\em Phys.\ Rev.\ Lett.}, 108:020601, 2012.

\bibitem{bara12}
A.~C. Barato and H.~Hinrichsen.
\newblock Entropy production of a bound nonequilibrium interface.
\newblock {\em J. Phys. A Math. Theor.}, 45:115005, 2012.

\bibitem{spec11a}
T.~Speck and J.~P. Garrahan.
\newblock Space-time phase transitions in driven kinetically constrained
  lattice models.
\newblock {\em Eur.\ Phys.\ J.\ B}, 79:1--6, 2011.

\bibitem{gasp04}
P.~Gaspard.
\newblock Fluctuation theorem for nonequilibrium reactions.
\newblock {\em J.\ Chem.\ Phys.}, 120:8898, 2004.

\bibitem{andr04}
D.~Andrieux and P.~Gaspard.
\newblock Fluctuation theorem and {O}nsager reciprocity relations.
\newblock {\em J.\ Chem.\ Phys.}, 121:6167, 2004.

\bibitem{andr06}
D.~Andrieux and P.~Gaspard.
\newblock Fluctuation theorem for transport in mesoscopic systems.
\newblock {\em J.\ Stat.\ Mech.:\ Theor.\ Exp.}, page P01001, 2006.

\bibitem{andr07c}
D.~Andrieux and P.~Gaspard.
\newblock Fluctuation theorem for currents and {S}chnakenberg network theory.
\newblock {\em J.\ Stat.\ Phys.}, 127:107--131, 2007.

\bibitem{touc09}
H.~Touchette.
\newblock The large deviation approach to statistical mechanics.
\newblock {\em Phys. Rep.}, 478:1--69, 2009.

\bibitem{andr07a}
D.~Andrieux and P.~Gaspard.
\newblock Network and thermodynamic conditions for a single macroscopic current
  fluctuation theorem.
\newblock {\em C. R. Physique}, 8:579--590, 2007.

\bibitem{fagg11}
A.~Faggionato and D.~Di Pietro.
\newblock {G}allavotti-{C}ohen-type symmetry related to cycle decompositions
  for markov chains and biochemical applications.
\newblock {\em J.\ Stat.\ Phys.}, 143:11--32, 2011.

\bibitem{bara10}
A.~C. Barato, R.~Chetrite, H.~Hinrichsen, and D.~Mukamel.
\newblock Entropy production and fluctuation relations for a {KPZ} interface.
\newblock {\em J.\ Stat.\ Mech.:\ Theor.\ Exp.}, page P10008, 2010.

\bibitem{bara11}
A.~C. Barato, R.~Chetrite, H.~Hinrichsen, and D.~Mukamel.
\newblock A gallavotii-{C}ohen-{E}vans-{M}orriss like symmetry for a class of
  {M}arkov jump processes.
\newblock {\em J.\ Stat.\ Phys.}, 146:294--313, 2012.

\bibitem{sini07}
N.~A. Sinitsyn and I.~Nemenman.
\newblock The berry phase and the pump flux in stochastic chemical kinetics.
\newblock {\em EPL}, 77:58001, 2007.

\bibitem{cher09}
V.~Y. Chernyak, M.~Chertkov, S.~V. Malinin, and R.~Teodorescu.
\newblock Non-equilibrium thermodynamics and topology of currents.
\newblock {\em J.\ Stat.\ Phys.}, 137:109147, 2009.

\bibitem{ohku08}
J.~Ohkubo.
\newblock The stochastic pump current and the non-adiabatic geometrical phase.
\newblock {\em J.\ Stat.\ Mech.:\ Theor.\ Exp.}, 2008:P02011, 2008.

\bibitem{ohku10}
J.~Ohkubo and T.~Eggel.
\newblock Noncyclic and nonadiabatic geometric phase for counting statistics.
\newblock {\em J.\ Phys.\ A:\ Math.\ Gen.}, 43:425001, 2010.

\bibitem{sing10}
N.~Singh and B.~Wynants.
\newblock Dynamical fluctuations for periodically driven diffusions.
\newblock {\em J.\ Stat.\ Mech.:\ Theor.\ Exp.}, page P03007, 2010.

\bibitem{sini11a}
N.~A. Sinitsyn, A.~Akimov, and V.~Y. Chernyak.
\newblock Supersymmetry and fluctuation relations for currents in closed
  networks.
\newblock {\em Phys.\ Rev.\ E}, 83:021107, 2011.

\bibitem{hurt11}
P.~I. Hurtado, C.~Perez-Espigares, J.~J. del Pozo, and P.~L. Garrido.
\newblock Symmetries in fluctuations far from equilibrium.
\newblock {\em Proc.\ Natl.\ Acad.\ Sci.\ U.S.A.}, 108:7704--7709, 2011.

\bibitem{sait08}
K.~Saito and U.~Yasuhiro.
\newblock Symmetry in full counting statistics, fluctuation theorem, and
  relations among nonlinear transport coefficients in the presence of a
  magnetic field.
\newblock {\em Phys. Rev. B}, 78:115429, 2008.

\bibitem{sanc10}
R.~S\'anchez, R.~L\'opez, D.~S\'anchez, and M.~{B}\"uttiker.
\newblock Mesoscopic coulomb drag, broken detailed balance, and fluctuation
  relations.
\newblock {\em Phys.\ Rev.\ Lett.}, 104:076801, 2010.

\bibitem{monn10}
T.~Monnai.
\newblock Derivation of quantum master equation with counting fields by
  monitoring a probe.
\newblock {\em Phys.\ Rev.\ E}, 82:051113, 2010.

\bibitem{utsu10}
Y.~Utsumi, D.~S. Golubev, M.~Marthaler, K.~Saito, T.~Fujisawa, and G.~Sch\"on.
\newblock Bidirectional single-electron counting and the fluctuation theorem.
\newblock {\em Phys. Rev. B}, 81:125331, 2010.

\bibitem{golu11}
D.~S. Golubev, Y.~Utsumi, M.~Marthaler, and G.~Sch\"on.
\newblock Fluctuation theorem for a double quantum dot coupled to a
  point-contact electrometer.
\newblock {\em Phys. Rev. B}, 84:075323, 2011.

\bibitem{krau11}
T.~Krause, G.~Schaller, and T.~Brandes.
\newblock Incomplete current fluctuation theorems for a four-terminal model.
\newblock {\em Phys. Rev. B}, 84:195113, 2011.

\bibitem{nico11}
L.~Nicolin and D.~Segal.
\newblock Non-equilibrium spin-boson model: Counting statistics and the heat
  exchange fluctuation theorem.
\newblock {\em J.\ Chem.\ Phys.}, 135:164106, 2011.

\bibitem{nico11a}
L.~Nicolin and D.~Segal.
\newblock Quantum fluctuation theorem for heat exchange in the strong coupling
  regime.
\newblock {\em Phys. Rev. B}, 84:161414, 2011.

\bibitem{cuet11}
G.~B. Cuetara, M.~Esposito, and P.~Gaspard.
\newblock Fluctuation theorems for capacitively coupled electronic currents.
\newblock {\em Phys. Rev. B}, 84:165114, 2011.

\bibitem{gane11}
S.~Ganeshan and N.~A. Sinitsyn.
\newblock Fluctuation relations for current components in mesoscopic electric
  circuits.
\newblock {\em Phys. Rev. B}, 84:245405, 2011.

\bibitem{derr07}
B.~Derrida.
\newblock Non-equilibrium steady states: fluctuations and large deviations of
  the density and of the current.
\newblock {\em J.\ Stat.\ Mech.:\ Theor.\ Exp.}, page P07023, 2007.

\bibitem{schm07}
T.~Schmiedl and U.~Seifert.
\newblock Optimal finite-time processes in stochastic thermodynamics.
\newblock {\em Phys.\ Rev.\ Lett.}, 98:108301, 2007.

\bibitem{koni05}
M.~de~Koning.
\newblock Optimizing the driving function for nonequilibrium free-energy
  calculations in the linear regime: A variational approach.
\newblock {\em J.\ Chem.\ Phys.}, 122:104106, 2005.

\bibitem{geig10}
P.~Geiger and C.~Dellago.
\newblock Optimum protocol for fast-switching free-energy calculations.
\newblock {\em Phys. Rev. E}, 81(2):021127, 2010.

\bibitem{then08}
H.~Then and A.~Engel.
\newblock Computing the optimal protocol for finite-time processes in
  stochastic thermodynamics.
\newblock {\em Phys.\ Rev.\ E}, 77:041105, 2008.

\bibitem{aure11}
E.~Aurell, C.~Mejia-Monasterio, and P.~Muratore-Ginanneschi.
\newblock Optimal protocols and optimal transport in stochastic thermodynamics.
\newblock {\em Phys.\ Rev.\ Lett.}, 106:250601, 2011.

\bibitem{aure12}
E.~Aurell, C.~Mej\'{i}a-Monasterio, and P.~Muratore-Ginanneschi.
\newblock Boundary layers in stochastic thermodynamics.
\newblock {\em Phys.\ Rev.\ E}, 85:020103, 2012.

\bibitem{schm08}
T.~Schmiedl and U.~Seifert.
\newblock Efficiency at maximum power: An analytically solvable model for
  stochastic heat engines.
\newblock {\em EPL}, 81:20003, 2008.

\bibitem{aure12a}
E.~Aurell, K.~Gawedzki, C.~Mej\'{i}a-Monasterio, R.~Mohayaee, and
  P.~Muratore-Ginanneschi.
\newblock Refined second law of thermodynamics for fast random processes.
\newblock {\em arXiv: 1201.3207}, 2012.

\bibitem{gome08}
A.~Gomez-Marin, T.~Schmiedl, and U.~Seifert.
\newblock Optimal protocols for minimal work processes in underdamped
  stochastic thermodynamics.
\newblock {\em J.\ Chem.\ Phys.}, 129:024114, 2008.

\bibitem{espo10a}
M.~Esposito, R.~Kawai, K.~Lindenberg, and C.~van~den Broeck.
\newblock Finite time thermodynamics for a single level quantum dot.
\newblock {\em EPL}, 89:20003, 2010.

\bibitem{mura12}
P.~Muratore-Ginanneschi, C.~Mej{\'i}a-Monasterio, and L.~Peliti.
\newblock Heat release by controlled continuous-time markov jump processes.
\newblock {\em arXiv: 1203.4062}, 2012.

\bibitem{cove06}
T.~M. Cover and J.~A. Thomas.
\newblock {\em Elements of information theory}.
\newblock Telecommunications and signal processing. Wiley, Hoboken, NJ, and
  Canada, 2006.

\bibitem{kawa07}
R.~Kawai, J.~M.~R. Parrondo, and C.~van~den Broeck.
\newblock Dissipation: The phase-space perspective.
\newblock {\em Phys.\ Rev.\ Lett.}, 98:080602, 2007.

\bibitem{parr09}
J.~M.~R. Parrondo, C.~van~den Broeck, and R.~Kawai.
\newblock Entropy production and the arrow of time.
\newblock {\em New J. Phys.}, 11:073008, 2009.

\bibitem{gome08a}
A.~Gomez-Marin, J.~M.~R. Parrondo, and C.~Van den Broeck.
\newblock Lower bounds on dissipation upon coarse-graining.
\newblock {\em Phys.\ Rev.\ E}, 78:011107, 2008.

\bibitem{vaik09}
S.~Vaikuntanathan and C.~{J}arzynski.
\newblock Dissipation and lag in irreversible processes.
\newblock {\em EPL}, 87:60005, 2009.

\bibitem{blyt08}
R.~A. Blythe.
\newblock Reversibility, heat dissipation, and the importance of the thermal
  environment in stochastic models of nonequilibrium steady states.
\newblock {\em Phys.\ Rev.\ Lett.}, 100:010601, 2008.

\bibitem{espo11}
M.~Esposito and C.~van~den Broeck.
\newblock Second law and {L}andauer principle far from equilibrium.
\newblock {\em EPL}, 95:40004, 2011.

\bibitem{feng08}
E.~H. Feng and G.~E. {C}rooks.
\newblock Length of time's arrow.
\newblock {\em Phys.\ Rev.\ Lett.}, 101:090602, 2008.

\bibitem{feng09}
E.~H. Feng and G.~E. {C}rooks.
\newblock Far-from-equilibrium measurements of thermodynamic length.
\newblock {\em Phys.\ Rev.\ E}, 79:012104, 2009.

\bibitem{croo11}
G.~E. {C}rooks and D.~A. Sivak.
\newblock Measures of trajectory ensemble disparity in nonequilibrium
  statistical dynamics.
\newblock {\em J.\ Stat.\ Mech.:\ Theor.\ Exp.}, page P06003, 2011.

\bibitem{andr08}
D.~Andrieux and P.~Gaspard.
\newblock Nonequilibrium generation of information in copolymerization
  processes.
\newblock {\em Proc.\ Natl.\ Acad.\ Sci.\ U.S.A.}, 105:9516--9521, 2008.

\bibitem{jarz08a}
C.~{J}arzynski.
\newblock The thermodynamics of writing a random polymer.
\newblock {\em Proc.\ Natl.\ Acad.\ Sci.\ U.S.A.}, 105:9451--9452, 2008.

\bibitem{leff03}
H.~S. Leff and A.~F. Rex.
\newblock {\em {M}axwell's Demon : Entropy, Classical and Quantum Information,
  Computing}.
\newblock IOP, Bristol and Philadelphia, 2003.

\bibitem{maru09}
K.~Maruyama, F.~Nori, and V.~Vedral.
\newblock Colloquium: The physics of {M}axwell's demon and information.
\newblock {\em Rev. Mod. Phys.}, 81(1):1, 2009.

\bibitem{hond07}
T.~Hondou.
\newblock Equation of state in a small system: Violation of an assumption of
  {M}axwell's demon.
\newblock {\em EPL}, 80:50001, 2007.

\bibitem{saga10}
T.~Sagawa and M.~Ueda.
\newblock Generalized {J}arzynski equality under nonequilibrium feedback
  control.
\newblock {\em Phys. Rev. Lett.}, 104:090602, 2010.

\bibitem{saga12}
T.~Sagawa and M.~Ueda.
\newblock Nonequilibrium thermodynamics of feedback control.
\newblock {\em Phys. Rev. E}, 85:021104, 2012.

\bibitem{abre11a}
D.~Abreu and U.~Seifert.
\newblock Thermodynamics of genuine non-equilibrium states under feedback
  control.
\newblock {\em Phys.\ Rev.\ Lett.}, 108:030601, 2012.

\bibitem{horo10}
J.~M. Horowitz and S.~Vaikuntanathan.
\newblock Nonequilibrium detailed fluctuation theorem for repeated discrete
  feedback.
\newblock {\em Phys. Rev. E}, 82:061120, 2010.

\bibitem{ponm10}
M.~Ponmurugan.
\newblock Generalized detailed fluctuation theorem under nonequilibrium
  feedback control.
\newblock {\em Phys. Rev. E}, 82(3):031129, Sep 2010.

\bibitem{lahi12}
S.~Lahiri, S.~Rana, and A.~M. Jayannavar.
\newblock Fluctuation theorems in the presence of information gain and
  feedback.
\newblock {\em J. Phys. A: Math. Theor.}, 45:065002, 2012.

\bibitem{baue12}
M.~Bauer, D.~Abreu, and U.~Seifert.
\newblock Efficiency of a {B}rownian information machine.
\newblock {\em J. Phys. A Math. Theor.}, 45:162001, 2012.

\bibitem{cao09}
F.~J. Cao and M.~Feito.
\newblock Thermodynamics of feedback controlled systems.
\newblock {\em Phys. Rev. E}, 79:041118, 2009.

\bibitem{kim04}
K.-H. Kim and H.~Qian.
\newblock Entropy production of {B}rownian macromolecules with inertia.
\newblock {\em Phys.\ Rev.\ Lett.}, 93:120602, 2004.

\bibitem{kim07}
K.-H. Kim and H.~Qian.
\newblock Fluctuation theorems for a molecular refrigerator.
\newblock {\em Phys.\ Rev.\ E}, 75:022102, 2007.

\bibitem{muna12}
T.~Munakata and M.~L. Rosinberg.
\newblock Entropy production and fluctuation theorems under feedback control:
  the molecular refrigerator model revisited.
\newblock {\em arXiv: 1202.0974}, 2012.

\bibitem{suzu09}
H.~Suzuki and Y.~Fujitani.
\newblock One-dimensional shift of a {B}rownian particle under the feedback
  control.
\newblock {\em Journal of the Physical Society of Japan}, 78:074007, 2009.

\bibitem{fuji10}
Y.~Fujitani and H.~Suzuki.
\newblock {J}arzynski equality modified in the linear feedback system.
\newblock {\em Journal of the Physical Society of Japan}, 79(10):104003, 2010.

\bibitem{cao04}
F.~J. Cao, L.~Dinis, and J.~M.~R. Parrondo.
\newblock Feedback control in a collective flashing ratchet.
\newblock {\em Phys.\ Rev.\ Lett.}, 93:040603, 2004.

\bibitem{dini05}
L.~Dinis, J.~M.~R. Parrondo, and F.~J. Cao.
\newblock Closed-loop control strategy with improved current for a flashing
  ratchet.
\newblock {\em Europhys.\ Lett.}, 71:536--541, 2005.

\bibitem{feit09}
M.~Feito and F.~J. Cao.
\newblock Optimal operation of feedback flashing ratchets.
\newblock {\em J.\ Stat.\ Mech.:\ Theor.\ Exp.}, page P010331, 2009.

\bibitem{feit07}
M.~Feito and F.~J. Cao.
\newblock Information and maximum power in a feedback controlled {B}rownian
  ratchet.
\newblock {\em Eur. Phys. J. B}, 59:63--68, 2007.

\bibitem{touc00}
H.~Touchette and S.~Lloyd.
\newblock Information-theoretic limits of control.
\newblock {\em Phys.\ Rev.\ Lett.}, 84:1156, 2000.

\bibitem{cao09a}
F.J. Cao, M.~Feito, and H.~Touchette.
\newblock Information and flux in a feedback controlled {B}rownian ratchet.
\newblock {\em Physica A}, 388(2-3):113 -- 119, 2009.

\bibitem{gran11}
L.~Granger and H.~Kantz.
\newblock Thermodynamic cost of measurements.
\newblock {\em Phys. Rev. E}, 84:061110, 2011.

\bibitem{abre11}
D.~Abreu and U.~Seifert.
\newblock Extracting work from a single heat bath through feedback.
\newblock {\em Europhys.\ Lett.}, 94:10001, 2011.

\bibitem{horo11}
J.~M. Horowitz and J.~M.~R. Parrondo.
\newblock Thermodynamic reversibility in feedback processes.
\newblock {\em EPL}, 95(1):10005, 2011.

\bibitem{horo11a}
J.~M. Horowitz and J.~M.~R. Parrondo.
\newblock Designing optimal discrete-feedback thermodynamic engines.
\newblock {\em New J. Phys.}, 13:123019, 2011.

\bibitem{dill09}
R.~Dillenschneider and E.~Lutz.
\newblock Memory erasure in small systems.
\newblock {\em Phys.\ Rev.\ Lett.}, 102:210601, 2009.

\bibitem{toya10a}
S.~Toyabe, T.~Sagawa, M.~Ueda, E.~Muneyuki, and M.~Sano.
\newblock Experimental demonstration of information-to-energy conversion and
  validation of the generalized {J}arzynski equality.
\newblock {\em Nature Phys.}, 6:988, 2010.

\bibitem{beru12}
A.~B{\'e}rut, A.~Arakelyan, A.~Petrosyan, S.~Ciliberto, R.~Dillenschneider, and
  E.~Lutz.
\newblock Experimental verification of {L}andauer's principle linking
  information and thermodynamics.
\newblock {\em Nature}, 483:187--189, 2012.

\bibitem{alla02}
A.E. Allahverdyan and T.M. Nieuwenhuizen.
\newblock A mathematical theorem as the basis for the second law: {T}homsons
  formulation applied to equilibrium.
\newblock {\em Physica A}, 305:542--552, 2002.

\bibitem{sato02a}
K.~Sato.
\newblock An example of a mechanical system whose ensemble average energy,
  starting with a microcanonical ensemble, decreases after an operation.
\newblock {\em J. Phys. Soc. Japan}, 71:1065--1066, 2002.

\bibitem{mara10}
R.~Marathe and J.~M.~R. Parrondo.
\newblock Cooling classical particles with a microcanonical {S}zilard engine.
\newblock {\em Phys.\ Rev.\ Lett.}, 104(24):245704, 2010.

\bibitem{vaik11a}
S.~Vaikuntanathan and C.~{J}arzynski.
\newblock Modeling {M}axwell's demon with a microcanonical szilard engine.
\newblock {\em Phys. Rev. E}, 83:061120, 2011.

\bibitem{kubo}
R.~{K}ubo, M.~Toda, and N.~Hashitsume.
\newblock {\em Statistical Physics II}.
\newblock Springer-Verlag, Berlin, 2nd edition, 1991.

\bibitem{agar72}
G.~S. Agarwal.
\newblock Fluctuation-dissipation theorems for systems in non-thermal
  equilibrium and applications.
\newblock {\em Z.\ Physik}, 252:25, 1972.

\bibitem{boch81}
G.~N. Bochkov and Y.~E. Kuzovlev.
\newblock Nonlinear fluctuation-dissipation relations and stochastic models in
  nonequilibrium thermodynamics {I.} {G}eneralized fluctuation-dissipation
  theorem.
\newblock {\em Physica\ A}, 106:443--479, 1981.

\bibitem{boch81a}
G.~N. Bochkov and Y.~E. Kuzovlev.
\newblock Nonlinear fluctuation-dissipation relations and stochastic models in
  nonequilibrium thermodynamics {II.} {K}inetic potential and variational
  principles for nonlinear irreversible processes.
\newblock {\em Physica\ A}, 106:480--520, 1981.

\bibitem{hang82}
P.~H\"anggi and H.~Thomas.
\newblock Stochastic processes: Time evolution, symmetries and linear response.
\newblock {\em Phys.\ Rep.}, 88:207, 1982.

\bibitem{marc08}
U.~Marconi, A.~Puglisi, L.~Rondoni, and A.~Vulpiani.
\newblock Fluctuation-dissipation: Response theory in statistical physics.
\newblock {\em Phys. Rep.}, 461:111--195, 2008.

\bibitem{cugl97}
L.~F. Cugliandolo, D.~S. Dean, and J.~Kurchan.
\newblock Fluctuation-dissipation theorems and entropy production in
  relaxational systems.
\newblock {\em Phys.\ Rev.\ Lett.}, 79:2168, 1997.

\bibitem{hara05}
T.~Harada and S.~{S}asa.
\newblock Equality connecting energy dissipation with a violation of the
  fluctuation-response relation.
\newblock {\em Phys.\ Rev.\ Lett.}, 95:130602, 2005.

\bibitem{hara06}
T.~Harada and S.~{S}asa.
\newblock Energy dissipation and violation of the fluctuation-response relation
  in nonequilibrium {L}angevin systems.
\newblock {\em Phys.\ Rev.\ E}, 73:026131, 2006.

\bibitem{hara09}
T.~Harada.
\newblock Macroscopic expression connecting the rate of energy dissipation with
  the violation of the fluctuation response relation.
\newblock {\em Phys.\ Rev.\ E}, 79:030106, 2009.

\bibitem{spec06}
T.~Speck and U.~Seifert.
\newblock Restoring a fluctuation-dissipation theorem in a nonequilibrium
  steady state.
\newblock {\em Europhys.\ Lett.}, 74:391, 2006.

\bibitem{spec09}
T.~Speck and U.~Seifert.
\newblock Extended fluctuation-dissipation theorem for soft matter in
  stationary flow.
\newblock {\em Phys. Rev. E}, 79:040102(R), 2009.

\bibitem{chet08}
R.~Chetrite, G.~Falkovich, and K.~Gawedzki.
\newblock Fluctuation relations in simple examples of non-equilibrium steady
  states.
\newblock {\em J. Stat. Mech.: Theor. Exp.}, page P08005, 2008.

\bibitem{chet09}
R.~Chetrite and K.~Gawedzki.
\newblock Eulerian and lagrangian pictures of non-equilibrium diffusions.
\newblock {\em J. Stat. Phys.}, 137:890--916, 2009.

\bibitem{baie09}
M.~Baiesi, C.~Maes, and B.~Wynants.
\newblock Fluctuations and response of nonequilibrium states.
\newblock {\em Phys.\ Rev.\ Lett.}, 103:010602, 2009.

\bibitem{baie09a}
M.~Baiesi, C.~Maes, and B.~Wynants.
\newblock Nonequilibrium linear response for {M}arkov dynamics, {I}: Jump
  processes and overdamped diffusion.
\newblock {\em J.\ Stat.\ Phys.}, 137:1094--1116, 2009.

\bibitem{baie09b}
M.~Baiesi, E.~Boksenbojm, C.~Maes, and B.~Wynants.
\newblock Nonequilibrium linear response for markov dynamics, {II}: Inertial
  dynamics.
\newblock {\em J.\ Stat.\ Phys.}, 139:492--505, 2010.

\bibitem{pros09}
J.~Prost, J.-F. Joanny, and J.~M.~R. Parrondo.
\newblock Generalized fluctuation-dissipation theorem for steady-state systems.
\newblock {\em Phys.\ Rev.\ Lett.}, 103:090601, 2009.

\bibitem{seif09}
U.~Seifert and T.~Speck.
\newblock Fluctuation-dissipation theorem in nonequilibrium steady states.
\newblock {\em EPL}, 89:10007, 2010.

\bibitem{chet11}
R.~Chetrite and S.~Gupta.
\newblock Two refreshing views of fluctuation theorems through kinematics
  elements and exponential martingale.
\newblock {\em J.\ Stat.\ Phys.}, 143:543--584, 2011.

\bibitem{verl11}
G.~Verley, R.~Ch\'etrite, and D.~Lacoste.
\newblock Modified fluctuation-dissipation theorem near non-equilibrium states
  and applications to the glauber-ising chain.
\newblock {\em J.\ Stat.\ Mech.:\ Theor.\ Exp.}, page P10025, 2011.

\bibitem{feng11}
H.~Feng and J.~Wang.
\newblock Potential and flux decomposition for dynamical systems and
  non-equilibrium thermodynamics: Curvature, gauge field, and generalized
  fluctuation-dissipation theorem.
\newblock {\em J.\ Chem.\ Phys.}, 135:234511, 2011.

\bibitem{cugl11}
L.~F. Cugliandolo.
\newblock The effective temperature.
\newblock {\em J. Phys. A Math. Theor.}, 44:483001, 2011.

\bibitem{bert02}
L.~Berthier and J.-L. Barrat.
\newblock Nonequilibrium dynamics and fluctuation-dissipation relation in a
  sheared fluid.
\newblock {\em J.\ Chem.\ Phys.}, 116:6228, 2002.

\bibitem{oher04}
C.~S. O'Hern, A.~J. Liu, and S.~R. Nagel.
\newblock Effective temperatures in driven systems: static versus
  time-dependent relations.
\newblock {\em Phys.\ Rev.\ Lett.}, 93:165702, 2004.

\bibitem{fiel02}
S.~Fielding and P.~Sollich.
\newblock Observable dependence of fluctuation-dissipation relations and
  effective temperatures.
\newblock {\em Phys.\ Rev.\ Lett.}, 88:050603, 2002.

\bibitem{sreb04}
Y.~Srebro and D.~Levine.
\newblock Exactly solvable model for driven dissipative systems.
\newblock {\em Phys.\ Rev.\ Lett.}, 93:240601, 2004.

\bibitem{shok06}
Y.~Shokef, G.~Bunin, and D.~Levine.
\newblock Fluctuation-dissipation relations in driven dissipative systems.
\newblock {\em Phys.\ Rev.\ E}, 73:046132, 2006.

\bibitem{mart09}
K.~Martens, E.~Bertin, and M.~Droz.
\newblock Dependence of the fluctuation-dissipation temperature on the choice
  of observable.
\newblock {\em Phys. Rev. Lett.}, 103(26):260602, 2009.

\bibitem{mart10}
K.~Martens, E.~Bertin, and M.~Droz.
\newblock Entropy-based characterizations of the observable dependence of the
  fluctuation-dissipation temperature.
\newblock {\em Phys.\ Rev.\ E}, 81:061107, 2010.

\bibitem{cugl94}
L.~F. Cugliandolo, J.~Kurchan, and G.~Parisi.
\newblock Off equilibrium dynamics and aging in unfrustrated systems.
\newblock {\em J.\ Phys.\ I}, 4:1641, 1994.

\bibitem{cugl97a}
L.~F. Cugliandolo, J.~Kurchan, and L.~Peliti.
\newblock Energy flow, partial equilibration, and effective temperatures in
  systems with slow dynamics.
\newblock {\em Phys.\ Rev.\ E}, 55:3898, 1997.

\bibitem{zamp05}
F.~Zamponi, F.~Bonetto, L.F. Cugliandolo, and J.~Kurchan.
\newblock A fluctuation theorem for non-equilibrium relaxational systems driven
  by external forces.
\newblock {\em J.\ Stat.\ Mech.:\ Theor.\ Exp.}, page P09013, 2005.

\bibitem{cala04}
P.~Calabrese and A.~Gambassi.
\newblock On the definition of a unique effective temperature for
  non-equilibrium critical systems.
\newblock {\em J.\ Stat.\ Mech.:\ Theor.\ Exp.}, page P07013, 2004.

\bibitem{mart01}
P.~Martin, A.~J. Hudspeth, and F.~J{\"u}licher.
\newblock Comparison of a hair bundle's spontaneous oscillations with its
  response to mechanical stimulation reveals the underlying active process.
\newblock {\em Proc.\ Natl.\ Acad.\ Sci.\ U.S.A.}, 98:14380, 2001.

\bibitem{mizu07}
D.~Mizuno, C.~Tardin, C.~F. Schmidt, and F.~C. MacKintosh.
\newblock Nonequilibrium mechanics of active cytoskeletal networks.
\newblock {\em Science}, 315:370--373, 2007.

\bibitem{levi09}
A.~J. Levine and F.~C. MacKintosh.
\newblock The mechanics and fluctuation spectrum of active gels.
\newblock {\em J. Phys. Chem. B}, 113:3820--3830, 2009.

\bibitem{lego02}
L.~Le Goff, F.~Amblard, and E.~M. Furst.
\newblock Motor-driven dynamics in actin-myosin networks.
\newblock {\em Phys.\ Rev.\ Lett.}, 88:018101, 2002.

\bibitem{kiku09}
N.~Kikuchi, A.~Ehrlicher, D.~Koch, J.~K\"as, S.~Ramaswamy, and M.~Rao.
\newblock Buckling, stiffening, and negative dissipation in the dynamics of a
  biopolymer in an active medium.
\newblock {\em Proc.\ Natl.\ Acad.\ Sci.\ U.S.A.}, 106:19776, 2009.

\bibitem{loi08}
D.~Loi, S.~Mossa, and L.~F. Cuglia.
\newblock Effective temperature of active matter.
\newblock {\em Phys.\ Rev.\ E}, 77:051111, 2008.

\bibitem{mann01}
J.-B. Manneville, P.~Bassereau, S.~Ramaswamy, and J.~Prost.
\newblock Active membrane fluctuations studied by micropipet aspiration.
\newblock {\em Phys.\ Rev.\ E}, 64:021908, 2001.

\bibitem{betz09}
T.~Betz, M.~Lenz, J.-F. Joanny, and C.~Sykes.
\newblock {ATP}-dependent mechanics of red blood cells.
\newblock {\em Proc.\ Natl.\ Acad.\ Sci.\ U.S.A.}, 106:15320--15325, 2009.

\bibitem{beni11}
E.~Ben-Isaac, Y.~Park, G.~Popescu, F.~L.~H. Brown, N.~S. Gov, and Y.~Shokef.
\newblock Effective temperature of red-blood-cell membrane fluctuations.
\newblock {\em Phys.\ Rev.\ Lett.}, 106:238103, 2011.

\bibitem{radu09}
R.~Raduenz, D.~Rings, K.~Kroy, and F.~Cichos.
\newblock Hot {B}rownian particles and photothermal correlation spectroscopy.
\newblock {\em J. Phys. Chem. A}, 113:1674--1677, 2009.

\bibitem{ring10}
D.~Rings, R.~Schachoff, M.~Selmke, F.~Cichos, and K.~Kroy.
\newblock Hot {B}rownian motion.
\newblock {\em Phys.\ Rev.\ Lett.}, 105:090604, 2010.

\bibitem{ring11}
D.~Rings, M.~Selmke, F.~Cichos, and K.~Kroy.
\newblock Theory of hot {B}rownian motion.
\newblock {\em Soft Matter}, 7:3441--3452, 2011.

\bibitem{chak11}
D.~Chakraborty, M.~V. Gnann, D.~Rings, J.~Glaser, F.~Otto, F.~Cichos, and
  K.~Kroy.
\newblock Generalised {E}instein relation for hot {B}rownian motion.
\newblock {\em EPL}, 96:60009, 2011.

\bibitem{joly11}
L.~Joly, S.~Merabia, and J.-L. Barrat.
\newblock Effective temperatures of a heated {B}rownian particle.
\newblock {\em EPL}, 94:50007, 2011.

\bibitem{prad08}
P.~Pradhan, Y.~Kafri, and D.~Levine.
\newblock Nonequilibrium fluctuation theorems in the presence of local heating.
\newblock {\em Phys.\ Rev.\ E}, 77:041129, 2008.

\bibitem{spec10}
T.~Speck.
\newblock Driven soft matter: Entropy production and the
  fluctuation-dissipation theorem.
\newblock {\em Progr. Theor. Phys. Suppl.}, 184:248--261, 2010.

\bibitem{ricc03}
F.~Ricci-Tersenghi.
\newblock Measuring the fluctuation-dissipation ratio in glassy systems with no
  perturbing field.
\newblock {\em Phys.\ Rev.\ E}, 68:065104, 2003.

\bibitem{cris03}
A.~Crisanti and F.~Ritort.
\newblock Violation of the fluctuation-dissipation theorem in glassy systems:
  basic notions and the numerical evidence.
\newblock {\em J.\ Phys.\ A:\ Math.\ Gen.}, 36:R181, 2003.

\bibitem{chat04}
C.~Chatelain.
\newblock On universality in ageing ferromagnets.
\newblock {\em J.\ Stat.\ Mech.:\ Theor.\ Exp.}, page P06006, 2004.

\bibitem{lipp05}
E.~Lippiello, F.~Corberi, and M.~Zannetti.
\newblock Off-equilibrium generalization of the fluctuation dissipation theorem
  for ising spins and measurement of the linear response function.
\newblock {\em Phys.\ Rev.\ E}, 71:036104, 2005.

\bibitem{diez05}
G.~Diezemann.
\newblock Fluctuation-dissipation relations for markov processes.
\newblock {\em Phys.\ Rev.\ E}, 72:011104, 2005.

\bibitem{deol07}
M.~J. de~Oliveira.
\newblock Fluctuation-dissipation relation for stochastic dynamics without
  detailed balance.
\newblock {\em Phys.\ Rev.\ E}, 76:011114, 2007.

\bibitem{corb07}
F.~Corberi, E.~Lippiello, and M.~Zannetti.
\newblock Fluctuation dissipation relations far from equilibrium.
\newblock {\em J.\ Stat.\ Mech.:\ Theor.\ Exp.}, page P07002, 2007.

\bibitem{lipp08}
E.~Lippiello, F.~Corberi, A.~Sarracino, and M.~Zannetti.
\newblock Nonlinear response and fluctuation-dissipation relations.
\newblock {\em Phys.\ Rev.\ E}, 78:041120, 2008.

\bibitem{bert07}
L.~Berthier.
\newblock Efficient measurement of linear susceptibilities in molecular
  simulations: Application to aging supercooled liquids.
\newblock {\em Phys.\ Rev.\ Lett.}, 98:220601, 2007.

\bibitem{seif10}
U.~Seifert.
\newblock Generalized {E}instein or {G}reen-{K}ubo relations for active
  biomolecular transport.
\newblock {\em Phys.\ Rev.\ Lett.}, 104:138101, 2010.

\bibitem{verl11a}
G.~Verley, K.~Mallick, and D.~Lacoste.
\newblock Modified fluctuation-dissipation theorem for non-equilibrium steady
  states and applications to molecular motors.
\newblock {\em EPL}, 93:10002, 2011.

\bibitem{andr07b}
D.~Andrieux and P.~Gaspard.
\newblock A fluctuation theorem for currents and non-linear response
  coefficients.
\newblock {\em J. Stat. Mech.}, page P02006, 2007.

\bibitem{baie11}
M.~Baiesi, C.~Maes, and B.~Wynants.
\newblock The modified {S}utherland{E}instein relation for diffusive
  non-equilibria.
\newblock {\em Proc. Roy. Soc. A}, 467:2792--2809, 2011.

\bibitem{vill11}
D.~Villamaina, A.~Sarracino, G.~Gradenigo, A.~Puglisi, and A.~Vulpiani.
\newblock On anomalous diffusion and the out-of-equilibrium response function
  in one-dimensional models.
\newblock {\em J.\ Stat.\ Mech.:\ Theor.\ Exp.}, page L01002, 2011.

\bibitem{blic07}
V.~Blickle, T.~Speck, C.~Lutz, U.~Seifert, and C.~Bechinger.
\newblock {E}instein relation generalized to nonequilibrium.
\newblock {\em Phys.\ Rev.\ Lett.}, 98:210601, 2007.

\bibitem{reim01}
P.~Reimann, C.~van~den Broeck, H.~Linke, P.~H{\"a}nggi, M.~Rubi, and
  A.~P{\'e}rez-Madrid.
\newblock Giant acceleration of free diffusion by use of tilted periodic
  potentials.
\newblock {\em Phys.\ Rev.\ Lett.}, 87:010602, 2001.

\bibitem{reim02}
P.~Reimann, C.~Van den Broeck, H.~Linke, P.~Hanggi, J.~M. Rubi, and
  A.~Perez-Madrid.
\newblock Diffusion in tilted periodic potentials: Enhancement, universality,
  and scaling.
\newblock {\em Phys.\ Rev.\ E}, 65(3):031104, 2002.

\bibitem{mehl10}
J.~Mehl, V.~Blickle, U.~Seifert, and C.~Bechinger.
\newblock Experimental accessibility of generalized fluctuation-dissipation
  relations for nonequilibrium steady states.
\newblock {\em Phys.\ Rev.\ E}, 82:032401, 2010.

\bibitem{gome09}
J.~R. Gomez-Solano, A.~Petrosyan, S.~Ciliberto, R.~Chetrite, and K.~Gawedzki.
\newblock Experimental verification of a modified fluctuation-dissipation
  relation for a micron-sized particle in a nonequilibrium steady state.
\newblock {\em Phys.\ Rev.\ Lett.}, 103:040601, 2009.

\bibitem{gome11}
J.~R. Gomez-Solano, A.~Petrosyan, S.~Ciliberto, and C.~Maes.
\newblock Fluctuations and response in a non-equilibrium micron-sized system.
\newblock {\em J. Stat. Mech.}, page P01008, 2011.

\bibitem{szam04}
G.~Szamel.
\newblock Self-diffusion in sheared colloidal suspensions: Violation of
  fluctuation-dissipation relation.
\newblock {\em Phys.\ Rev.\ Lett.}, 93:178301, 2004.

\bibitem{krue09a}
M.~Kr\"uger and M.~Fuchs.
\newblock Non-equilibrium relation between mobility and diffusivity of
  interacting {B}rownian particles under shear.
\newblock {\em Progr. Theor. Phys. Suppl.}, 184:172--186, 2010.

\bibitem{land10}
B.~Lander, U.~Seifert, and T.~Speck.
\newblock Mobility and diffusion of a tagged particle in a driven colloidal
  suspension.
\newblock {\em Europhys.\ Lett.}, 92(58001), 2010.

\bibitem{land11}
B.~Lander, U.~Seifert, and T.~Speck.
\newblock Effective confinement as origin of the equivalence of kinetic
  temperature and fluctuation-dissipation ratio in a dense shear driven
  suspension.
\newblock {\em Phys.\ Rev.\ E}, 85:021103, 2012.

\bibitem{zhan11}
M.~Zhang and G.~Szamel.
\newblock Effective temperatures of a driven, strongly anisotropic {B}rownian
  system.
\newblock {\em Phys.\ Rev.\ E}, 83:061407, 2011.

\bibitem{szam11}
G.~Szamel and M.~Zhang.
\newblock Tagged particle in a sheared suspension: Effective temperature
  determines density distribution in a slowly varying external potential beyond
  linear response.
\newblock {\em EPL}, 96:50007, 2011.

\bibitem{krue09}
M.~Kr\"uger and M.~Fuchs.
\newblock Fluctuation dissipation relations in stationary states of interacting
  {B}rownian particles under shear.
\newblock {\em Phys.\ Rev.\ Lett.}, 102(13):135701, 2009.

\bibitem{krue09b}
M.~Kr\"uger and M.~Fuchs.
\newblock Nonequilibrium fluctuation-dissipation relations of interacting
  {B}rownian particles driven by shear.
\newblock {\em Phys. Rev. E}, 81:011408, 2010.

\bibitem{rito06}
F.~Ritort.
\newblock Single-molecule experiments in biological physics: methods and
  applications.
\newblock {\em J.\ Phys.:\ Condens.\ Matter}, 18:R531, 2006.

\bibitem{selv07}
P.~R. Selvin and T.~Ha.
\newblock {\em Single Molecule Techniques: A Laboratory Manual}.
\newblock Cold Spring Harbor Laboratory Press, New York, December 2007.

\bibitem{deni08}
A.~A. Deniz, S.~Mukhopadhyay, and E.~A. Lemke.
\newblock Single-molecule biophysics: at the interface of biology, physics and
  chemistry.
\newblock {\em J. Royal Soc. Interface}, 5:15--45, 2008.

\bibitem{howard}
J.~Howard.
\newblock {\em Mechanics of Motor Proteins and the Cytoskeleton}.
\newblock Sinauer, New York, 1 edition, 2001.

\bibitem{schl03}
M.~Schliwa.
\newblock {\em Molecular Motors}.
\newblock Wiley-VCH, Weinheim, 2003.

\bibitem{bust04}
C.~Bustamante, Y.~R. Chemla, N.~R. Forde, and D.~Izhaky.
\newblock Mechanical processes in biochemistry.
\newblock {\em Ann. Rev. of Biochemistry}, 73:705--748, 2004.

\bibitem{kuma10}
S.~Kumar and M.~S. Li.
\newblock Biomolecules under mechanical force.
\newblock {\em Phys. Rep.}, 486:1--74, 2010.

\bibitem{fish99}
M.~E. Fisher and A.~B. Kolomeisky.
\newblock The force exerted by a molecular motor.
\newblock {\em Proc.\ Natl.\ Acad.\ Sci.\ U.S.A.}, 96:6597, 1999.

\bibitem{lipo00}
R.~Lipowsky.
\newblock Universal aspects of the chemomechanical coupling for molecular
  motors.
\newblock {\em Phys.\ Rev.\ Lett.}, 85:4401, 2000.

\bibitem{qian00a}
H.~Qian.
\newblock A simple theory of motor protein kinetics and energetics. ii.
\newblock {\em Biophys. Chem.}, 83:35--43, 2000.

\bibitem{bust01}
C.~Bustamante, D.~Keller, and G.~Oster.
\newblock The physics of molecular motors.
\newblock {\em Accounts Chem. Res.}, 34:412--420, 2001.

\bibitem{bake04}
J.~E. Baker.
\newblock Free energy transduction in a chemical motor model.
\newblock {\em J.\ Theor.\ Biol.}, 228:467, 2004.

\bibitem{andr06c}
D.~Andrieux and P.~Gaspard.
\newblock Fluctuation theorems and the nonequilibrium thermodynamics of
  molecular motors.
\newblock {\em Phys.\ Rev.\ E}, 74:011906, 2006.

\bibitem{wang07}
H.~Wang and T.~C. Elston.
\newblock Mathematical and computational methods for studying energy
  transduction in protein motors.
\newblock {\em J.\ Stat.\ Phys.}, 128:35--76, 2007.

\bibitem{gasp07}
P.~Gaspard and E.~Gerritsma.
\newblock The stochastic chemomechanics of the {F1}-{ATP}ase molecular motor.
\newblock {\em J.\ Theor.\ Biol.}, 247:672--686, 2007.

\bibitem{liep07}
S.~Liepelt and R.~Lipowsky.
\newblock Steady-state balance conditions for molecular motor cycles and
  stochastic nonequilibrium processes.
\newblock {\em EPL}, 77:50002, 2007.

\bibitem{liep07a}
S.~Liepelt and R.~Lipowsky.
\newblock Kinesin's network of chemomechanical motor cycles.
\newblock {\em Phys.\ Rev.\ Lett.}, 98:258102, 2007.

\bibitem{liep08}
R.~Lipowsky and S.~Liepelt.
\newblock Chemomechanical coupling of molecular motors: Thermodynamics, network
  representations, and balance conditions.
\newblock {\em J.\ Stat.\ Phys.}, 130:39--67, 2008.

\bibitem{liep09}
S.~Liepelt and R.~Lipowsky.
\newblock Operation modes of the molecular motor kinesin.
\newblock {\em Phys.\ Rev.\ E}, 79:011917, 2009.

\bibitem{lipo09}
R.~Lipowsky and S.~Liepelt.
\newblock Chemomechanical coupling of molecular motors: Thermodynamics, network
  representations, and balance conditions (vol 130, pg 39, 2008).
\newblock {\em J.\ Stat.\ Phys.}, 135:777--778, 2009.

\bibitem{lipo09a}
R.~Lipowsky, S.~Liepelt, and A.~Valleriani.
\newblock Energy conversion by molecular motors coupled to nucleotide
  hydrolysis.
\newblock {\em J.\ Stat.\ Phys.}, 135:951--975, 2009.

\bibitem{lau07a}
A.~W.~C. Lau, D.~Lacoste, and K.~Mallick.
\newblock Non-equilibrium fluctuations and mechanochemical couplings of a
  molecular motor.
\newblock {\em Phys.\ Rev.\ Lett.}, 99:158102, 2007.

\bibitem{kolo07}
A.~B. Kolomeisky and M.~E. Fisher.
\newblock {Molecular Motors: A Theorist's Perspective}.
\newblock {\em Ann. Rev. Phys. Chem.}, 58:675--695, 2007.

\bibitem{astu10}
R.~D. Astumian.
\newblock Thermodynamics and kinetics of molecular motors.
\newblock {\em Biophys. J.}, 98:2401--2409, 2010.

\bibitem{juel97}
F.~J\"ulicher, A.~Ajdari, and J.~Prost.
\newblock Modeling molecular motors.
\newblock {\em Rev. Mod. Phys.}, 69(4):1269--1282, 1997.

\bibitem{astu02}
R.~D. Astumian and P.~H\"anggi.
\newblock {B}rownian motors.
\newblock {\em Physics Today}, 55(11):33, 2002.

\bibitem{reim02a}
P.~Reimann.
\newblock {B}rownian motors: noisy transport far from equilibrium.
\newblock {\em Phys.\ Rep.}, 361:57, 2002.

\bibitem{parr02}
J.~M.~R. Parrondo and B.~J.~De Cisneros.
\newblock Energetics of {B}rownian motors: a review.
\newblock {\em Applied Physics A}, 75:179, 2002.

\bibitem{min05}
Wei Min, Liang Jiang, Ji~Yu, S.~C. Kou, Hong Qian, and X.~Sunney Xie.
\newblock Nonequilibrium steady state of a nanometric biochemical system:
  Determinig the thermodynamic driving force from single enzyme turnover time
  traces.
\newblock {\em Nano Lett.}, 5:2373--2378, 2005.

\bibitem{shib00}
T.~Shibata.
\newblock A generalization of {C}lausius inequality for processes between
  nonequilibrium steady states in chemical reaction systems.
\newblock {\em cond-mat/0012404}, 2000.

\bibitem{schm06}
T.~Schmiedl and U.~Seifert.
\newblock Stochastic thermodynamics of chemical reaction networks.
\newblock {\em J.\ Chem.\ Phys.}, 126:044101, 2007.

\bibitem{seif04}
U.~Seifert.
\newblock Fluctuation theorem for birth-death or chemical master equations with
  time-dependent rates.
\newblock {\em J.\ Phys.\ A:\ Math.\ Gen.}, 37:L517, 2004.

\bibitem{jarz05}
C.~{J}arzynski.
\newblock Lag inequality for birth-death processes with time-dependent rates.
\newblock {\em J.\ Phys.\ A:\ Math.\ Gen.}, 38:L227, 2005.

\bibitem{qian06a}
H.~Qian and X.~S. Xie.
\newblock Generalized {H}aldane equation and fluctuation theorem in the
  steady-state cycle kinetics of single enzymes.
\newblock {\em Phys.\ Rev.\ E}, 74:010902, 2006.

\bibitem{laco08}
D.~Lacoste, A.~W.~C. Lau, and K.~Mallick.
\newblock Fluctuation theorem and large deviation function for a solvable model
  of a molecular motor.
\newblock {\em Phys.\ Rev.\ E}, 78:011915, 2008.

\bibitem{laco09}
D.~Lacoste and K.~Mallick.
\newblock Fluctuation theorem for the flashing ratchet model of molecular
  motors.
\newblock {\em Phys.\ Rev.\ E}, 80:021923, 2009.

\bibitem{qian08a}
D.~A. Beard and H.~Qian.
\newblock {\em Chemical Biophysics: Quantitative Analysis of Cellular Systems}.
\newblock Cambridge Univ. Press, Cambridge, 2008.

\bibitem{seki07}
K.~Sekimoto.
\newblock Microscopic heat from the energetics of stochastic phenomena.
\newblock {\em Phys.\ Rev.\ E}, 76:060103(R), 2007.

\bibitem{seif11}
U.~Seifert.
\newblock Stochastic thermodynamics of single enzymes and molecular motors.
\newblock {\em Eur. Phys. J. E}, 34:26, 2011.

\bibitem{humm05}
G.~Hummer and A.~{S}zabo.
\newblock Free energy surfaces from single-molecule force spectroscopy.
\newblock {\em Acc.\ Chem.\ Res.}, 38:504--513, 2005.

\bibitem{humm10}
G.~Hummer and A.~{S}zabo.
\newblock Free energy profiles from single-molecule pulling experiments.
\newblock {\em Proc.\ Natl.\ Acad.\ Sci.\ U.S.A.}, 107:21441--21446, 2010.

\bibitem{minh08}
D.~D.~L. Minh and A.~B. Adib.
\newblock Optimized free energies from bidirectional single-molecule force
  spectroscopy.
\newblock {\em Phys.\ Rev.\ Lett.}, 100:180602, 2008.

\bibitem{nico10}
P.~Nicolini, P.~Procacci, and R.~Chelli.
\newblock Hummer and {S}zabo-like potential of mean force estimator for
  bidirectional nonequilibrium pulling experiments/simulations.
\newblock {\em J. Phys. Chem. B}, 114:9546--9554, 2010.

\bibitem{liph02}
J.~Liphardt, S.~Dumont, S.~B. Smith, I.~Tinoco~Jr, and C.~Bustamante.
\newblock Equilibrium information from nonequilibrium measurements in an
  experimental test of {J}arzynski's equality.
\newblock {\em Science}, 296:1832, 2002.

\bibitem{coll05}
D.~Collin, F.~Ritort, C.~{J}arzynski, S.B. Smith, I.~Tinoco, and C.~Bustamante.
\newblock Verification of the {C}rooks fluctuation theorem and recovery of
  {RNA} folding free energies.
\newblock {\em Nature}, 437:231, 2005.

\bibitem{moss09}
A.~Mossa, M.~Manosas, N.~Forns, J.~M. Huguet, and F.~Ritort.
\newblock Dynamic force spectroscopy of {DNA} hairpins: {I.} {F}orce kinetics
  and free energy landscapes.
\newblock {\em J.\ Stat.\ Mech.:\ Theor.\ Exp.}, page P02060, 2009.

\bibitem{mano09}
M.~Manosas, A.~Mossa, N.~Forns, J.~M. Huguet, and F.~Ritort.
\newblock Dynamic force spectroscopy of {DNA} hairpins: {II.} {I}rreversibility
  and dissipation.
\newblock {\em J.\ Stat.\ Mech.:\ Theor.\ Exp.}, P02061, 2009.

\bibitem{moss09a}
A.~Mossa, S.~de~Lorenzo, J.~M. Huguet, and F.~Ritort.
\newblock Measurement of work in single-molecule pulling experiments.
\newblock {\em J. Chem. Phys.}, 130:234116, 2009.

\bibitem{harr07a}
N.~C. Harris, Y.~Song, and C.~H. Kiang.
\newblock Experimental free energy surface reconstruction from single-molecule
  force spectroscopy using {J}arzynski's equality.
\newblock {\em Phys.\ Rev.\ Lett.}, 99:068101, 2007.

\bibitem{frid08}
R.~W. Friddle.
\newblock Experimental free energy surface reconstruction from single-molecule
  force spectroscopy using {J}arzynski's equality - comment.
\newblock {\em Phys.\ Rev.\ Lett.}, 100:019801, 2008.

\bibitem{harr08}
N.~C. Harris, Y.~Song, and C.~H. Kiang.
\newblock Experimental free energy surface reconstruction from single-molecule
  force spectroscopy using {J}arzynski's equality - reply.
\newblock {\em Phys.\ Rev.\ Lett.}, 100:019802, 2008.

\bibitem{born08}
T.~Bornschl\"ogl and M.~Rief.
\newblock Single-molecule dynamics of mechanical coiled-coil unzipping.
\newblock {\em Langmuir}, 24:1338--1342, 2008.

\bibitem{gebh10}
J.~C.~M. Gebhardt, T.~Bornschl\"ogl, and M.~Rief.
\newblock Full distance-resolved folding energy landscape of one single protein
  molecule.
\newblock {\em Proc.\ Natl.\ Acad.\ Sci.\ U.S.A.}, 107:2013--2018, 2010.

\bibitem{shan10}
E.~A. Shank, C.~Cecconi, J.~W. Dil, S.~Marqusee, and C.~Bustamante.
\newblock The folding cooperativity of a protein is controlled by its chain
  topology.
\newblock {\em Nature}, 465:637, 2010.

\bibitem{nome07}
R.~A. Nome, J.~M. Zhao, W.~D. Hoff, and N.~F. Scherer.
\newblock Axis-dependent anisotropy in protein unfolding from integrated
  nonequilibrium single-molecule experiments, analysis, and simulation.
\newblock {\em Proc.\ Natl.\ Acad.\ Sci.\ U.S.A.}, 104:20799--20804, 2007.

\bibitem{gupt11}
N.~A. Gupta, V.~Abhilash, K.~Neupane, H.~Yu, F.~Wang, and M.~T. Woodside.
\newblock Experimental validation of free-energy-landscape reconstruction from
  non-equilibrium single-molecule force spectroscopy measurements.
\newblock {\em Nature Physics}, 7:631--634, 2011.

\bibitem{brau04}
O.~Braun, A.~Hanke, and U.~Seifert.
\newblock Probing molecular free energy landscapes by periodic loading.
\newblock {\em Phys.\ Rev.\ Lett.}, 93:158105, 2004.

\bibitem{minh06}
D.~D.~L. Minh.
\newblock Free-energy reconstruction from experiments performed under different
  biasing programs.
\newblock {\em Phys.\ Rev.\ E}, 74:061120, 2006.

\bibitem{impa07b}
A.~Imparato, S.~Luccioli, and A.~Torcini.
\newblock Reconstructing the free-energy landscape of a mechanically unfolded
  model protein.
\newblock {\em Phys.\ Rev.\ Lett.}, 99:168101, 2007.

\bibitem{berk08}
R.~Berkovich, J.~Klafter, and M.~Urbakh.
\newblock Analyzing friction forces with the {J}arzynski equality.
\newblock {\em J. Phys.: Condens. Matter}, 20:354008, 2008.

\bibitem{kosz06}
I.~Kosztin, B.~Barz, and L.~Janosi.
\newblock Calculating potentials of mean force and diffusion coefficients from
  nonequilibrium processes without {J}arzynskis equality.
\newblock {\em J.\ Chem.\ Phys.}, 124:064106, 2006.

\bibitem{minh07}
D.~D.~L. Minh.
\newblock Multidimensional potentials of mean force from biased experiments
  along a single coordinate.
\newblock {\em J.\ Phys.\ Chem.\ B}, 111:4137--4140, 2007.

\bibitem{impa07}
A.~Imparato and L.~Peliti.
\newblock The distribution function of entropy flow in stochastic systems.
\newblock {\em J.\ Stat.\ Mech.:\ Theor.\ Exp.}, page L02001, 2007.

\bibitem{mitt09}
S.~Mitternacht, S.~Luccioli, A.~Torcini, A.~Imparato, and A.~Irb\"ack.
\newblock Changing the mechanical unfolding pathway of ${FnIII}_{10}$ by tuning
  the pulling strength.
\newblock {\em Biophys.\ J.}, 96:429--441, 2009.

\bibitem{prei07}
J.~Preiner, H.~Janovjak, C.~Rankl, H.~Knaus, D.~A. Cisneros, A.~Kedrov,
  F.~Kienberger, D.~J. Muller, and P.~Hinterdorfer.
\newblock Free energy of membrane protein unfolding derived from
  single-molecule force measurements.
\newblock {\em Biophys.\ J.}, 93:930--937, 2007.

\bibitem{humm01a}
G.~Hummer.
\newblock Fast-growth thermodynamic integration: Error and efficiency analysis.
\newblock {\em J.\ Chem.\ Phys.}, 114:7330, 2001.

\bibitem{zuck02}
D.~M. Zuckerman and T.~B. Woolf.
\newblock Theory of a systematic computational error in free energy
  differences.
\newblock {\em Phys.\ Rev.\ Lett.}, 89:180602, 2002.

\bibitem{park03}
S.~Park, F.~Khalili-Araghi, E.~Tajkhorshid, and K.~Schulten.
\newblock Free energy calculation from steered molecular dynamics simulations
  using {J}arzynski's equality.
\newblock {\em J.\ Chem.\ Phys.}, 119:3559, 2003.

\bibitem{gore03}
J.~Gore, F.~Ritort, and C.~Bustamante.
\newblock Bias and error in estimates of equilibrium free-energy differences
  from nonequilibrium measurements.
\newblock {\em Proc.\ Natl.\ Acad.\ Sci.\ U.S.A.}, 100:12564, 2003.

\bibitem{zuck04}
D.~M. Zuckerman and T.~B. Woolf.
\newblock Systematic finite-sampling inaccuracy in free energy differences and
  other nonlinear quantities.
\newblock {\em J.\ Stat.\ Phys.}, 114:1303--1323, 2004.

\bibitem{park04}
S.~Park and K.~Schulten.
\newblock Calculating potentials of mean force from steered molecular dynamics
  simulations.
\newblock {\em J.\ Chem.\ Phys.}, 120:5946, 2004.

\bibitem{ytre04}
F.~M. Ytreberg and D.~M. Zuckerman.
\newblock Single-ensemble nonequilibrium path-sampling estimates of free energy
  differences.
\newblock {\em J.\ Chem.\ Phys.}, 120:10876, 2004.

\bibitem{ober05}
H.~Oberhofer, C.~Dellago, and P.~L. Geissler.
\newblock Biased sampling of nonequilibrium trajectories: Can fast switching
  simulations outperform conventional free energy calculation methods?
\newblock {\em J.\ Phys.\ Chem.\ B}, 109:6902, 2005.

\bibitem{jarz06}
C.~{J}arzynski.
\newblock Rare events and the convergence of exponentially averaged work
  values.
\newblock {\em Phys.\ Rev.\ E}, 73:046105, 2006.

\bibitem{mara06}
P.~Maragakis, M.~Spichty, and M.~Karplus.
\newblock Optimal estimates of free energies from multistate nonequilibrium
  work data.
\newblock {\em Phys.\ Rev.\ Lett.}, 96:100602, 2006.

\bibitem{pres06}
S.~Presse and R.~Silbey.
\newblock Ordering of limits in the {J}arzynski equality.
\newblock {\em J.\ Chem.\ Phys.}, 124:054117, 2006.

\bibitem{west06}
D.~K. West, P.~D. Olmsted, and E.~Paci.
\newblock Free energy for protein folding from nonequilibrium simulations using
  the {J}arzynski equality.
\newblock {\em J.\ Chem.\ Phys.}, 125:204910, 2006.

\bibitem{lech07}
W.~Lechner and C.~Dellago.
\newblock On the efficiency of path sampling methods for the calculation of
  free energies from non-equilibrium simulations.
\newblock {\em J.\ Stat.\ Mech.:\ Theor.\ Exp.}, page P04001, 2007.

\bibitem{vaik08}
S.~Vaikuntanathan and C.~{J}arzynski.
\newblock Escorted free energy simulations: Improving convergence by reducing
  dissipation.
\newblock {\em Phys.\ Rev.\ Lett.}, 100:190601, 2008.

\bibitem{hahn09}
A.~M. Hahn and H.~Then.
\newblock Using bijective maps to improve free energy estimates.
\newblock {\em Phys.\ Rev.\ E}, 79:011113, 2009.

\bibitem{nico09}
P.~Nicolini and R.~Chelli.
\newblock Improving fast-switching free energy estimates by dynamicsl freezing.
\newblock {\em Phys.\ Rev.\ E}, 80:041124, 2009.

\bibitem{ober09}
H.~Oberhofer and C.~Dellago.
\newblock Efficient extraction of free energy profiles from nonequilibrium
  experiments.
\newblock {\em J. Comp. Chem.}, 30:1726, 2009.

\bibitem{goet09}
M.~Goette and H.~Grubm\"uller.
\newblock Accuracy and convergence of free energy differences calculated from
  nonequilibrium switching processes.
\newblock {\em J. Comp. Chem.}, 30:447--456, 2009.

\bibitem{lind09}
G.~E. Lindberg, T.~C. Berkelbach, and W.~Feng.
\newblock Optimizing the switching function for nonequilibrium free-energy
  calculations: An one-the-fly approach.
\newblock {\em J.\ Chem.\ Phys.}, 130:174705, 2009.

\bibitem{zima09}
E.~N. Zimanyi and R.~J. Silbey.
\newblock The work-{H}amiltonian connection and the usefulness of the
  {J}arzynski equality for free energy calculations.
\newblock {\em J. Chem. Phys.}, 130:171102, 2009.

\bibitem{poho10}
A.~Pohorille, C.~{J}arzynski, and C.~Chipot.
\newblock Good practices in free-energy calculations.
\newblock {\em J. Phys. Chem. B}, 114:10235--10253, 2010.

\bibitem{minh11}
D.~D.~L. Minh and J.~D. Chodera.
\newblock Estimating equilibrium ensemble averages using multiple time slices
  from driven nonequilibrium processes: Theory and application to free
  energies, moments, and thermodynamic length in single-molecule pulling
  experiments.
\newblock {\em J. Chem. Phys.}, 134:024111, 2011.

\bibitem{minh11a}
D.~D.~L. Minh and S.~Vaikuntanathan.
\newblock Density-dependent analysis of nonequilibrium paths improves free
  energy estimates {II.} {A} {F}eynman{K}ac formalism.
\newblock {\em J.\ Chem.\ Phys.}, 134:034117, 2011.

\bibitem{vaik11}
S.~Vaikuntanathan and C.~{J}arzynski.
\newblock Escorted free energy simulations.
\newblock {\em J.\ Chem.\ Phys.}, 134:054107, 2011.

\bibitem{davy11}
A.~Davydov.
\newblock Inequalities for non-equilibrium fluctuations of work.
\newblock {\em J.\ Stat.\ Phys.}, 142:394--402, 2011.

\bibitem{pala11}
M.~Palassini and F.~Ritort.
\newblock Improving free-energy estimates from unidirectional work.
\newblock {\em Phys.\ Rev.\ Lett.}, 107:060601, 2011.

\bibitem{kund11}
A.~Kundu, S.~Sabhapandit, and A.~Dhar.
\newblock Application of importance sampling to the computation of large
  deviations in nonequilibrium processes.
\newblock {\em Phys.\ Rev.\ E}, 83:031119, 2011.

\bibitem{haya10}
K.~Hayashi, H.~Ueno, R.~Iino, and H.~Noji.
\newblock Fluctuation theorem applied to {F1}-{ATP}ase.
\newblock {\em Phys.\ Rev.\ Lett.}, 104:218103, 2010.

\bibitem{toya10}
S.~Toyabe, T.~Okamoto, T.~Watanabe-Nakayama, H.~Taketani, S.~Kudo, and
  E.~Muneyuki.
\newblock Nonequilibrium energetics of a single {F1}-{ATP}ase molecule.
\newblock {\em Phys.\ Rev.\ Lett.}, 104:198103, 2010.

\bibitem{toya11}
S.~Toyabe, T.~Watanabe-Nakayama, T.~Okamoto, S.~Kudo, and E.~Muneyuki.
\newblock Thermodynamic efficiency and mechanochemical coupling of
  {F1}-{ATP}ase.
\newblock {\em Proc.\ Natl.\ Acad.\ Sci.\ U.S.A.}, 108:17951--17956, 2011.

\bibitem{xiao08}
T.~J. Xiao, Z.~H. Hou, and H.~W. Xin.
\newblock Entropy production and fluctuation theorem along a stochastic limit
  cycle.
\newblock {\em J. Chem. Phys}, 129:114506, 2008.

\bibitem{xiao09}
T.~J. Xiao, Z.~H. Hou, and H.~W. Xin.
\newblock Stochastic thermodynamics in mesoscopic chemical oscillation systems.
\newblock {\em J. Phys. Chem. B}, 113:9316, 2009.

\bibitem{rao11}
T.~Rao, T.~Xiao, and Z.~Hou.
\newblock Entropy production in a mesoscopic chemical reaction system with
  oscillatory and excitable dynamics.
\newblock {\em J.\ Chem.\ Phys.}, 134:214112, 2011.

\bibitem{brow06}
W.~R. Browne and B.~L. Feringa.
\newblock Making molecular machines work.
\newblock {\em Nature Nanotechnology}, 1:25--35, 2006.

\bibitem{kay07}
E.~R. Kay, D.~A. Leigh, and F.~Zerbetto.
\newblock Synthetic molecular motors and mechanical machines.
\newblock {\em Angew. Chemie - Int. Edition}, 46:72--191, 2007.

\bibitem{bath07}
J.~Bath and A.~J. Turberfield.
\newblock {DNA} nanomachines.
\newblock {\em Nature Nanotechnology}, 2:275--284, 2007.

\bibitem{heuv07}
M.~G.~L. van~den Heuvel and C.~Dekker.
\newblock Motor proteins at work for nanotechnology.
\newblock {\em Science}, 317:333--336, 2007.

\bibitem{balz09}
V.~Balzani, A.~Credi, and M.~Venturi.
\newblock Light powered molecular machines.
\newblock {\em Chem. Soc. Rev.}, 38:1542--1550, 2009.

\bibitem{cosk12}
A.~Coskun, M.~Banaszak, R.~D. Astumian, J.~F. Stoddart, and B.~A. Grzybowski.
\newblock Great expectations: can artificial molecular machines deliver on
  their promise?
\newblock {\em Chem. Soc. Rev.}, 41:19--30, 2012.

\bibitem{hang09}
P.~H\"anggi and F.~Marchesoni.
\newblock Artificial {B}rownian motors: Controlling transport on the nanoscale.
\newblock {\em Rev. Mod. Phys}, 81:387--442, 2009.

\bibitem{sini09}
N.~A. Sinitsyn.
\newblock The stochastic pump effect and geometric phases in dissipative and
  stochastic systems.
\newblock {\em J. Phys. A Math. Theor.}, 42:193001, 2009.

\bibitem{astu11}
R.~D. Astumian.
\newblock Stochastic conformational pumping: A mechanism for free-energy
  transduction by molecules.
\newblock {\em Ann. Rev. Biophys.}, 40:289--313, 2011.

\bibitem{seif11a}
U.~Seifert.
\newblock Efficiency of autonomous soft nano-machines at maximum power.
\newblock {\em Phys.\ Rev.\ Lett.}, 106:020601, 2011.

\bibitem{dere99}
I.~Derenyi, M.~Bier, and R.~D. Astumian.
\newblock Generalized efficiency and its application to microscopic engines.
\newblock {\em Phys.\ Rev.\ Lett.}, 83:903, 1999.

\bibitem{wang02a}
H.~Wang and G.~F. Oster.
\newblock The {S}tokes efficiency for molecular motors and its applications.
\newblock {\em Europhys.\ Lett.}, 57:134, 2002.

\bibitem{gave10}
B.~Gaveau, M.~Moreau, and L.~S. Schulman.
\newblock Stochastic thermodynamics and sustainable efficiency in work
  production.
\newblock {\em Phys.\ Rev.\ Lett.}, 105:060601, 2010.

\bibitem{more11}
M.~Moreau, B.~Gaveau, and L.~S. Schulman.
\newblock Stochastic dynamics, efficiency and sustainable power production.
\newblock {\em Eur. Phys. J. B}, 62:67--71, 2011.

\bibitem{gave10a}
B.~Gaveau, M.~Moreau, and L.~S. Schulman.
\newblock Constrained maximal power in small engines.
\newblock {\em Phys.\ Rev.\ E}, 82:051109, 2010.

\bibitem{groot}
S.~R. de~Groot and P.~Mazur.
\newblock {\em Non-equilibrium thermodynamics}.
\newblock North-Holland, Amsterdam, 1962.

\bibitem{pott09}
N.~Pottier.
\newblock {\em Nonequilibrium Statistical Physics: Linear Irreversible
  Processes}.
\newblock Oxford University Press, New York, 2009.

\bibitem{kede65}
O.~Kedem and S.~R. Caplan.
\newblock Degree of coupling and its relation to efficiency of energy
  conversion.
\newblock {\em Trans. Faraday Soc.}, 61:1897, 1965.

\bibitem{schm08a}
T.~Schmiedl and U.~Seifert.
\newblock Efficiency of molecular motors at maximum power.
\newblock {\em EPL}, 83:30005, 2008.

\bibitem{vdb12}
C.~van~den Broeck.
\newblock Efficiency of isothermal molecular machines at maximum power.
\newblock {\em arXiv: 1201.6396}, 2012.

\bibitem{kawa11}
K.~Kawaguchi and M.~Sano.
\newblock Efficiency of free energy transduction in autonomous systems.
\newblock {\em J. Phys. Soc. Japan}, 80:083003, 2011.

\bibitem{efre11}
A.~Efremov and Z.~Wang.
\newblock Universal optimal working cycles of molecular motors.
\newblock {\em Phys. Chem. Chem. Phys.}, 13:6223--6233, 2011.

\bibitem{gole10}
R.~Golestanian.
\newblock Synthetic mechanochemical molecular swimmer.
\newblock {\em Phys.\ Rev.\ Lett.}, 105:018103, 2010.

\bibitem{magn94}
M.~O. Magnasco.
\newblock Molecular combustion motors.
\newblock {\em Phys.\ Rev.\ Lett.}, 72:2656--2659, 1994.

\bibitem{parm99}
A.~Parmeggiani, F.~J\"ulicher, A.~Ajdari, and J.~Prost.
\newblock Energy transduction of isothermal ratchets: Generic aspects and
  specific examples close to and far from equilibrium.
\newblock {\em Phys.\ Rev.\ E}, 60:2127, 1999.

\bibitem{wang05a}
H.~Wang.
\newblock Chemical and mechanical efficiencies of molecular motors and
  implications for motor mechanisms.
\newblock {\em J. Phys. Cond. Mat.}, 17:S3997--S4014, 2005.

\bibitem{qian08}
M.~Qian, X.~Zhang, R.~J. Wilson, and J.~Feng.
\newblock Efficiency of {B}rownian motors in terms of entropy production rate.
\newblock {\em EPL}, 84, 2008.

\bibitem{boks09}
E.~Boksenbojm and B.~Wynants.
\newblock The entropy and efficiency of a molecular motor model.
\newblock {\em J. Phys. A: Math. Theor.}, 42, 2009.

\bibitem{gerr10}
E.~Gerritsma and P.~Gaspard.
\newblock Chemomechanical coupling and stochastic thermodynamics of the
  {F1}-{ATP}ase molecuar motor with an applied external torque.
\newblock {\em Biophys. Rev. Lett.}, 5:163--208, 2010.

\bibitem{golu12}
N.~Golubeva, A.~Imparato, and L.~Peliti.
\newblock Efficiency of molecular machines with continuous phase space.
\newblock {\em EPL}, 97:60005, 2012.

\bibitem{toya11a}
S.~Toyabe, H.~Ueno, and E.~Muneyuki.
\newblock Recovery of state-specific potential of molecular motor from
  single-molecule trajectory.
\newblock {\em EPL}, 97:40004, 2012.

\bibitem{zimm12}
E.~Zimmermann and U.~Seifert.
\newblock preprint.
\newblock 2012.

\bibitem{curz75}
F.~L. Curzon and B.~Ahlborn.
\newblock Efficiency of a {C}arnot engine at maximum power output.
\newblock {\em Am.\ J.\ Phys.}, 43:22, 1975.

\bibitem{novi58}
I.~I. Novikov.
\newblock The efficiency of atomic power stations.
\newblock {\em J. Nucl. Energy II}, 7:125--128, 1958.

\bibitem{chen99}
L.~Chen, C.~Wu, and F.~Sun.
\newblock Finite time thermodynamic optimization of entropy generation
  minimization of energy systems.
\newblock {\em J. Non-Equilib. Thermodyn.}, 24:327--359, 1999.

\bibitem{sala01}
P.~Salamon, J.~D. Nulton, G.~Siragusa, T.~R. Andersen, and A.~Limon.
\newblock Principles of control thermodynamics.
\newblock {\em Energy}, 26:307--319, 2001.

\bibitem{hoff03}
K.~H. Hoffmann, J.~Burzler, A.~Fischer, M.~Schaller, and S.~Schubert.
\newblock Optimal process paths for endoreversible systems.
\newblock {\em J.\ Noneq.\ Thermodyn.}, 28:233--268, 2003.

\bibitem{andr11}
B.~Andresen.
\newblock Current trends in finite-time thermodynamics.
\newblock {\em Angew. Chem. Int. Ed.}, 50:2690--2704, 2011.

\bibitem{vdb05}
C.~van~den Broeck.
\newblock Thermodynamic efficiency at maximum power.
\newblock {\em Phys.\ Rev.\ Lett.}, 95:190602, 2005.

\bibitem{cisn07}
B.~Jim\'{e}nez de~Cisneros and A.~C. Hern\'{a}ndez.
\newblock Collective working regimes for coupled heat engines.
\newblock {\em Phys.\ Rev.\ Lett.}, 98:130602, 2007.

\bibitem{espo10c}
M.~Esposito, R.~Kawai, K.~Lindenberg, and C.~van~den Broeck.
\newblock Efficiency at maximum power of low-dissipation carnot engines.
\newblock {\em Phys.\ Rev.\ Lett.}, 105:150603, 2010.

\bibitem{izum08}
Y.~Izumida and K.~Okuda.
\newblock Molecular kinetic analysis of a finite-time {C}arnot cycle.
\newblock {\em EPL}, 83:60003, 2008.

\bibitem{izum09}
Y.~Izumida and K.~Okuda.
\newblock {O}nsager coefficients of a finite-time {C}arnot cycle.
\newblock {\em Phys.\ Rev.\ E}, 80:021121, 2009.

\bibitem{izum09a}
Y.~Izumida and K.~Okuda.
\newblock Numerical experiments of a finite-time thermodynamic cycle.
\newblock {\em Progr. Theor. Phys. Suppl.}, 178:163, 2009.

\bibitem{sini11}
N.~A. Sinitsyn.
\newblock Fluctuation relation for heat engines.
\newblock {\em J. Phys. A: Math. Gen.}, 44:405001, 2011.

\bibitem{tu08}
Z.~C. Tu.
\newblock Efficiency at maximum power of {F}eynman's ratchet as a heat engine.
\newblock {\em J. Phys. A: Math. Theor.}, 41:312003, 2008.

\bibitem{espo09}
M.~Esposito, K.~Lindenberg, and C.~van~den Broeck.
\newblock Universality of efficiency at maximum power.
\newblock {\em Phys. Rev. Lett.}, 102:130602, 2009.

\bibitem{sanc10a}
N.~S\'anchez-Salas, L.~L\'opez-Palacios, S.~Velasco, and A.~Calvo Hern\'andez.
\newblock Optimization criteria, bounds, and efficiencies of heat engines.
\newblock {\em Phys.\ Rev.\ E}, 82:051101, 2010.

\bibitem{toma12}
C.~de~Tom{\'a}s, A.~C. Hern\'andez, and J.~M.~M. Roco.
\newblock Optimal low symmetric dissipation {C}arnot engines and refrigerators.
\newblock {\em Phys.\ Rev.\ E}, 85:010104, 2012.

\bibitem{buet87}
M.~{B}\"uttiker.
\newblock Transport as a consequence of state-dependent diffusion.
\newblock {\em Z.\ Phys.\ B}, 68:161, 1987.

\bibitem{kamp88}
N.~G. van Kampen.
\newblock Relative stability in nonuniform temperature.
\newblock {\em IBM J. of Research and Development}, 32:107--111, 1988.

\bibitem{land88}
R.~{L}andauer.
\newblock Motion out of noisy states.
\newblock {\em J.\ Stat.\ Phys.}, 53:233, 1988.

\bibitem{dere99a}
I.~Derenyi and R.~D. Astumian.
\newblock Efficiency of {B}rownian heat engines.
\newblock {\em Phys.\ Rev.\ E}, 59:R6219--R6222, 1999.

\bibitem{mats00}
M.~Matsuo and S.~{S}asa.
\newblock Stochastic energetics of non-uniform temperature systems.
\newblock {\em Physica A}, 276:188, 2000.

\bibitem{hond00}
T.~Hondou and K.~Sekimoto.
\newblock Unattainability of {C}arnot efficiency in the {B}rownian heat engine.
\newblock {\em Phys.\ Rev.\ E}, 62:6021, 2000.

\bibitem{seki00}
K.~Sekimoto, F.~Takagi, and T.~Hondou.
\newblock {C}arnot's cycle for small systems: Irreversibility and cost of
  operations.
\newblock {\em Phys.\ Rev.\ E}, 62:7759, 2000.

\bibitem{benj08}
R.~Benjamin and R.~Kawai.
\newblock Inertial effects in buttiker-{L}andauer motor and refrigerator at the
  overdamped limit.
\newblock {\em Phys.\ Rev.\ E}, 77:051132, 2008.

\bibitem{asfa04}
M.~Asfaw and M.~Bekele.
\newblock Current, maximum power and optimized efficiency of a {B}rownian heat
  engine.
\newblock {\em Eur.\ Phys.\ J.\ B}, 38:457, 2004.

\bibitem{gome06}
A.~Gomez-Marin and J.~M. Sancho.
\newblock Tight coupling in thermal {B}rownian motors.
\newblock {\em Phys.\ Rev.\ E}, 74:062102, 2006.

\bibitem{asfa08}
M.~Asfaw.
\newblock Modeling an efficient {B}rownian heat engine.
\newblock {\em Eur.\ Phys.\ J.\ B}, 65:109--116, 2008.

\bibitem{berg09}
F.~Berger, T.~Schmiedl, and U.~Seifert.
\newblock Optimal potentials for temperature ratchets.
\newblock {\em Phys.\ Rev.\ E}, 79:031118, 2009.

\bibitem{jarz99a}
C.~{J}arzynski and O.~Mazonka.
\newblock {F}eynman's ratchet and pawl: An exactly solvable model.
\newblock {\em Phys.\ Rev.\ E}, 59:6448--6459, 1999.

\bibitem{vela01}
S.~Velasco, J.~M.~M. Roco, A.~Medina, and A.~C. Hernandez.
\newblock {F}eynman's ratchet optimization: maximum power and maximum
  efficiency regimes.
\newblock {\em J.\ Phys.\ D}, 34:1000--1006, 2001.

\bibitem{zhan10}
Y.~P. Zhang, J.~Z. He, X.~A. He, and J.~L. Xiao.
\newblock Thermodynamic performance characteristics of a {B}rownian microscopic
  heat engine driven by discrete and periodic temperature field.
\newblock {\em Comm. Theor. Phys.}, 54:857--862, 2010.

\bibitem{chen11}
L.~Chen, Z.~Ding, and F.~Sun.
\newblock Optimum performance analysis of {F}eynman's engine as cold and hot
  ratchets.
\newblock {\em J. Non-Equilib. Thermodyn.}, 36:155--177, 2011.

\bibitem{feyn63}
R.~P. {F}eynman, R.~B. Leighton, and M.~Sands.
\newblock {\em The {F}eynman Lectures on Physics. Vol. I}.
\newblock Addison-Wesley, Reading, MA, 1963.

\bibitem{parr96}
J.~M.~R. Parrondo and P.~Espanol.
\newblock Criticism of {F}eynman's analysis of the ratchet as an engine.
\newblock {\em Am. J. Phys.}, 64:1125--1130, 1996.

\bibitem{vdb04}
C.~van~den Broeck, R.~Kawai, and P.~Meurs.
\newblock Microscopic analysis of a thermal {B}rownian motor.
\newblock {\em Phys.\ Rev.\ Lett.}, 93:090601, 2004.

\bibitem{zhen10}
J.~Zheng, X.~Zheng, C.~Y. Yam, and G.~H. Chen.
\newblock Computer simulation of {F}eynman's ratchet and pawl system.
\newblock {\em Phys.\ Rev.\ E}, 81:061104, 2010.

\bibitem{duhr06}
S.~Duhr and D.~Braun.
\newblock Thermophoretic depletion follows {B}oltzmann distribution.
\newblock {\em Phys.\ Rev.\ Lett.}, 96:168301, 2006.

\bibitem{hump02}
T.~E. Humphrey, R.~Newbury, R.~P. Taylor, and H.~Linke.
\newblock Reversible quantum {B}rownian heat engines for electrons.
\newblock {\em Phys.\ Rev.\ Lett.}, 89:116801, 2002.

\bibitem{hump05}
T.~E. Humphrey and H.~Linke.
\newblock Reversible thermoelectric nanomaterials.
\newblock {\em Phys.\ Rev.\ Lett.}, 94:096601, 2005.

\bibitem{espo09b}
M.~Esposito, K.~Lindenberg, and C.~van~den Broeck.
\newblock Thermoelectric efficiency at maximum power in a quantum dot.
\newblock {\em Europhys.\ Lett.}, 85:60010, 2009.

\bibitem{sanc11}
R.~S\'anchez and M.~{B}\"uttiker.
\newblock Optimal energy quanta to current conversion.
\newblock {\em Phys. Rev. B}, 83:085428, 2011.

\bibitem{espo12a}
M.~Esposito, N.~Kumar, K.~Lindenberg, and C.~van~den Broeck.
\newblock Stochastically driven single level quantum dot: a nano-scale
  finite-time thermodynamic machine and its various operational modes.
\newblock {\em arXiv: 1201.0669}, 2012.

\bibitem{bene11}
G.~Benenti, K.~Saito, and G.~Casati.
\newblock Thermodynamic bounds on efficiency for systems with broken
  time-reversal symmetry.
\newblock {\em Phys.\ Rev.\ Lett.}, 106:230602, 2011.

\bibitem{sait11}
K.~Saito, G.~Benenti, G.~Casati, and T.~Prosen.
\newblock Thermopower with broken time-reversal symmetry.
\newblock {\em Phys. Rev. B}, 84:201306, 2011.

\bibitem{rutt09}
B.~Rutten, M.~Esposito, and B.~Cleuren.
\newblock Reaching optimal efficiencies using nanosized photoelectric devices.
\newblock {\em Phys. Rev. B}, 80:235122, 2009.

\bibitem{izum11}
Y.~Izumida and K.~Okuda.
\newblock Efficiency at maximal power of minimal nonlinear irreversible heat
  engines.
\newblock {\em EPL}, 97:10004, 2012.

\bibitem{wang11}
Y.~Wang and Z.~C. Tu.
\newblock Efficiency at maximum power output of linear irreversible
  {C}arnot-like heat engines.
\newblock {\em Phys.\ Rev.\ E}, 85:011127, 2012.

\bibitem{wang11a}
Y.~Wang and Z.~C. Tu.
\newblock Bounds of efficiency at maximum power for linear, superlinear and
  sublinear irreversible {C}arnot-like heat engines.
\newblock {\em arXiv: 1110.6493}, 2011.

\bibitem{aper12}
Y.~Apertet, H.~Ouerdane, C.~Goupil, and P.~Lecoeur.
\newblock Irreversibilities and efficiency at maximum power of heat engines:
  The illustrative case of a thermoelectric generator.
\newblock {\em Phys.\ Rev.\ E}, 85:031116, 2012.

\bibitem{izum10}
Y.~Izumida and K.~Okuda.
\newblock {O}nsager coefficients of a {B}rownian {C}arnot cycle.
\newblock {\em Eur. Phys. J. B}, 77:499--504, 2010.

\bibitem{blic12}
V.~Blickle and C.~Bechinger.
\newblock Realization of a micrometre-sized stochastic heat engine.
\newblock {\em Nature Phys.}, 8:143, 2012.

\bibitem{horo12}
J.~M. Horowitz and J.~M.~R. Parrondo.
\newblock Thermodynamics: A {S}tirling effort.
\newblock {\em Nature Phys.}, 8:108--109, 2012.

\bibitem{espo10}
M.~Esposito, R.~Kawai, K.~Lindenberg, and C.~van~den Broeck.
\newblock Quantum-dot {C}arnot engine at maximum power.
\newblock {\em Phys. Rev. E}, 81:041106, 2010.

\bibitem{chvo09}
P.~Chvosta, M.~Einax, V.~Holubec, A.~Ryabov, and P.~Maass.
\newblock Energetics and performance of a microscopic heat engine based on
  exact calculations of work and heat distributions.
\newblock {\em J.\ Stat.\ Mech.:\ Theor.\ Exp.}, page P03002, 2010.

\bibitem{lan12}
G.~Lan, P.~Sartori, S.~Neumann, V.~Sourjik, and Y.~Tu.
\newblock The {energyspeedaccuracy} trade-off in sensory adaptation.
\newblock {\em Nature Phys.}, 2012.

\bibitem{fara04}
J.~Farago.
\newblock Power fluctuations in stochastic models of dissipative systems.
\newblock {\em Physica A}, 331:69--89, 2004.

\bibitem{jian11}
H.~Jiang, T.~Xiao, and Z.~Hou.
\newblock Stochastic thermodynamics for delayed {L}angevin systems.
\newblock {\em Phys.\ Rev.\ E}, 83:061144, 2011.

\bibitem{baul09}
A.~Baule and E.~G.~D. {C}ohen.
\newblock Steady-state work fluctuations of a dragged particle under external
  and thermal noise.
\newblock {\em Phys.\ Rev.\ E}, 80:011110, 2009.

\bibitem{spec07a}
T.~Speck and U.~Seifert.
\newblock The {J}arzynski relation, fluctuation theorems, and stochastic
  thermodynamics for non-markovian processes.
\newblock {\em J.\ Stat.\ Mech.:\ Theor.\ Exp.}, page L09002, 2007.

\bibitem{ohku07}
T.~Ohkuma and T.~Ohta.
\newblock Fluctuation theorems for non-linear generalized {L}angevin systems.
\newblock {\em J.\ Stat.\ Mech.:\ Theor.\ Exp.}, page P10010, 2007.

\bibitem{ohku09}
J.~Ohkubo.
\newblock Posterior probability and fluctuation theorem in stochastic
  processes.
\newblock {\em J. Phys. Soc. Japan}, 78:123001, 2009.

\bibitem{aron10}
C.~Aron, G.~Biroli, and L.~F. Cugliandolo.
\newblock Symmetries of generating functionals of {L}angevin processes with
  colored multiplicative noise.
\newblock {\em J.\ Stat.\ Mech.:\ Theor.\ Exp.}, page P11018, 2010.

\bibitem{mai07}
T.~Mai and A.~Dhar.
\newblock Nonequilibrium work fluctuations for oscillators in non-markovian
  baths.
\newblock {\em Phys.\ Rev.\ E}, 75:061101, 2007.

\bibitem{pugl09}
A.~Puglisi and D.~Villamaina.
\newblock Irreversible effects of memory.
\newblock {\em EPL}, 88:30004, 2009.

\bibitem{hase11a}
H.~Hasegawa.
\newblock Classical open systems with nonlinear nonlocal dissipation and
  state-dependent diffusion: Dynamical responses and the {J}arzynski equality.
\newblock {\em Phys.\ Rev.\ E}, 84:051124, 2011.

\bibitem{fran11}
T.~Franosch, M.~Grimm, M.~Belushkin, F.~M. Mor, G.~Foffi, L.~Forro, and
  S.~Jeney.
\newblock Resonances arising from hydrodynamic memory in {B}rownian motion.
\newblock {\em Nature}, 478:85--88, 2011.

\bibitem{mehl12}
J.~Mehl, B.~Lander, C.~Bechinger, V.~Blickle, and U.~Seifert.
\newblock Role of hidden slow degrees of freedom in the fluctuation theorem.
\newblock {\em arXiv: 1205.0238}, 2012.

\bibitem{cris12}
A.~Crisanti, A.~Puglisi, and D.~Villamaina.
\newblock Non-equilibrium and information: the role of cross-correlations.
\newblock {\em arXiv:1202.0508}, 2012.

\bibitem{qian00}
H.~Qian and M.~Qian.
\newblock Pumped biochemical reactions, nonequilibrium circulation, and
  stochastic resonance.
\newblock {\em Phys.\ Rev.\ Lett.}, 84:2271, 2000.

\bibitem{aman10}
C.~P. Amann, T.~Schmiedl, and U.~Seifert.
\newblock Communications: Can one identify nonequilibrium in a three-state
  system by analyzing two-state trajectories?
\newblock {\em J.\ Chem.\ Phys.}, 132:041102, 2010.

\bibitem{rold10}
E.~Roldan and J.~M.~R. Parrondo.
\newblock Estimating dissipation from single stationary trajectories.
\newblock {\em Phys.\ Rev.\ Lett.}, 105:150607, 2010.

\bibitem{rold12}
E.~Roldan and J.~M.~R. Parrondo.
\newblock Entropy production and {K}ullback-{L}eibler divergence between
  stationary trajectories of discrete systems.
\newblock {\em Phys.\ Rev.\ E}, 85:031129, 2012.

\bibitem{raha07}
S.~Rahav and C.~{J}arzynski.
\newblock Fluctuation relations and coarse-graining.
\newblock {\em J.\ Stat.\ Mech.:\ Theor.\ Exp.}, page P09012, 2007.

\bibitem{li08}
Y.~Li, T.~Zhao, P.~Bhimalapuram, and A.~R. Dinner.
\newblock How the nature of an observation affects single-trajectory entropies.
\newblock {\em J.\ Chem.\ Phys.}, 128:074102, 2008.

\bibitem{pugl10}
A.~Puglisi, S.~Pigolotti, L.~Rondoni, and A.~Vulpiani.
\newblock Entropy production and coarse graining in markov processes.
\newblock {\em J. Stat. Mech.}, page P05015, 2010.

\bibitem{szab10}
G.~Szab{\'o}, T.~Tom{\'e}, and I.~Borsos.
\newblock Probability currents and entropy production in nonequilibrium lattice
  systems.
\newblock {\em Phys.\ Rev.\ E}, 82:011105, 2010.

\bibitem{nico11c}
G.~Nicolis.
\newblock Transformation properties of entropy production.
\newblock {\em Phys.\ Rev.\ E}, 83:011112, 2011.

\bibitem{hinr11}
H.~Hinrichsen, C.~Gogolin, and P.~Janotta.
\newblock Non-equilibrium dynamics, thermalization and entropy production.
\newblock {\em J. Phys. Conf. Ser.}, 297:012011, 2011.

\bibitem{espo12}
M.~Esposito.
\newblock Stochastic thermodynamics under coarse-graining.
\newblock {\em arXiv: 1112.5410}, 2012.

\bibitem{alta12}
B.~Altaner and J.~Vollmer.
\newblock Fluctuation preserving coarse graining for biochemical systems.
\newblock {\em arXiv: 1112.4745}, 2012.

\bibitem{bert07a}
E.~Bertin, K.~Martens, O.~Dauchot, and M.~Droz.
\newblock Intensive thermodynamic parameters in nonequilibrium systems.
\newblock {\em Phys.\ Rev.\ E}, 75:031120, 2007.

\bibitem{prad10}
P.~Pradhan, C.~P. Amann, and U.~Seifert.
\newblock Nonequilibrium steady states in contact: approximate thermodynamic
  structure and zeroth law for driven lattice gases.
\newblock {\em Phys.\ Rev.\ Lett.}, 105:150601, 2010.

\bibitem{prad11}
P.~Pradhan, R.~Ramsperger, and U.~Seifert.
\newblock Approximate thermodynamic structure for driven lattice gases in
  contact.
\newblock {\em Phys.\ Rev.\ E}, 84:041104, 2011.

\end{thebibliography}
\end{document}